\documentclass[
 aps,
 prd,
 twocolumn,
 superscriptaddress,
 amsmath,
 amssymb,
 showpacs,
 nofootinbib
]{revtex4-1}

\usepackage{graphicx}
\usepackage{dcolumn}

\begin{document}

\preprint{ \vbox{
  \hbox{Belle Preprint 2010-20}
  \hbox{KEK Preprint 2010-32}
% \hbox{hep-ex nnnn}
}}

%%%%% Title

\title{ 
  Study of the $K^{+} \pi^{+} \pi^{-}$ Final State in 
  $B^{+} \rightarrow J/\psi K^{+} \pi^{+} \pi^{-}$ and
  $B^{+} \rightarrow \psi^{\prime} K^{+} \pi^{+} \pi^{-}$ 
}

%%%%% Authors

\affiliation{Budker Institute of Nuclear Physics, Novosibirsk}
\affiliation{Faculty of Mathematics and Physics, Charles University, Prague}
\affiliation{University of Cincinnati, Cincinnati, Ohio 45221}
\affiliation{The Graduate University for Advanced Studies, Hayama}
\affiliation{Gyeongsang National University, Chinju}
\affiliation{Hanyang University, Seoul}
\affiliation{University of Hawaii, Honolulu, Hawaii 96822}
\affiliation{High Energy Accelerator Research Organization (KEK), Tsukuba}
\affiliation{Institute of High Energy Physics, Vienna}
\affiliation{Institute of High Energy Physics, Protvino}
\affiliation{Institute for Theoretical and Experimental Physics, Moscow}
\affiliation{J. Stefan Institute, Ljubljana}
\affiliation{Kanagawa University, Yokohama}
\affiliation{Institut f\"ur Experimentelle Kernphysik, Karlsruher Institut f\"ur Technologie, Karlsruhe}
\affiliation{Korea Institute of Science and Technology Information, Daejeon}
\affiliation{Korea University, Seoul}
\affiliation{Kyungpook National University, Taegu}
\affiliation{\'Ecole Polytechnique F\'ed\'erale de Lausanne (EPFL), Lausanne}
\affiliation{University of Maribor, Maribor}
\affiliation{Max-Planck-Institut f\"ur Physik, M\"unchen}
\affiliation{McGill University, Montr\'{e}al}
\affiliation{University of Melbourne, School of Physics, Victoria 3010}
\affiliation{Universit\'{e} de Montr\'{e}al, Montr\'{e}al}
\affiliation{Nagoya University, Nagoya}
\affiliation{Nara Women's University, Nara}
\affiliation{National Central University, Chung-li}
\affiliation{National United University, Miao Li}
\affiliation{Department of Physics, National Taiwan University, Taipei}
\affiliation{H. Niewodniczanski Institute of Nuclear Physics, Krakow}
\affiliation{Nippon Dental University, Niigata}
\affiliation{Niigata University, Niigata}
\affiliation{University of Nova Gorica, Nova Gorica}
\affiliation{Novosibirsk State University, Novosibirsk}
\affiliation{Osaka City University, Osaka}
\affiliation{Panjab University, Chandigarh}
\affiliation{Princeton University, Princeton, New Jersey 08544}
\affiliation{University of Science and Technology of China, Hefei}
\affiliation{Seoul National University, Seoul}
\affiliation{Sungkyunkwan University, Suwon}
\affiliation{School of Physics, University of Sydney, NSW 2006}
\affiliation{Tata Institute of Fundamental Research, Mumbai}
\affiliation{Excellence Cluster Universe, Technische Universit\"at M\"unchen, Garching}
\affiliation{Toho University, Funabashi}
\affiliation{Tohoku Gakuin University, Tagajo}
\affiliation{Tohoku University, Sendai}
\affiliation{Department of Physics, University of Tokyo, Tokyo}
\affiliation{Tokyo Metropolitan University, Tokyo}
\affiliation{Tokyo University of Agriculture and Technology, Tokyo}
\affiliation{IPNAS, Virginia Polytechnic Institute and State University, Blacksburg, Virginia 24061}
\affiliation{Wayne State University, Detroit, Michigan 48202}
\affiliation{Yonsei University, Seoul}
\author{H.~Guler}\affiliation{University of Hawaii, Honolulu, Hawaii 96822}\affiliation{McGill University, Montr\'{e}al}\affiliation{Universit\'{e} de Montr\'{e}al, Montr\'{e}al} 
\author{H.~Aihara}\affiliation{Department of Physics, University of Tokyo, Tokyo} 
\author{K.~Arinstein}\affiliation{Budker Institute of Nuclear Physics, Novosibirsk}\affiliation{Novosibirsk State University, Novosibirsk} 
\author{V.~Aulchenko}\affiliation{Budker Institute of Nuclear Physics, Novosibirsk}\affiliation{Novosibirsk State University, Novosibirsk} 
\author{T.~Aushev}\affiliation{\'Ecole Polytechnique F\'ed\'erale de Lausanne (EPFL), Lausanne}\affiliation{Institute for Theoretical and Experimental Physics, Moscow} 
\author{A.~M.~Bakich}\affiliation{School of Physics, University of Sydney, NSW 2006} 
\author{V.~Balagura}\affiliation{Institute for Theoretical and Experimental Physics, Moscow} 
\author{E.~Barberio}\affiliation{University of Melbourne, School of Physics, Victoria 3010} 
\author{K.~Belous}\affiliation{Institute of High Energy Physics, Protvino} 
\author{V.~Bhardwaj}\affiliation{Panjab University, Chandigarh} 
% ...
\collaboration{The Belle Collaboration}\noaffiliation

%%%%% Abstract

\begin{abstract}
  Using $535 \times 10^{6}$ $B$-meson pairs collected
  by the Belle detector at the KEKB $e^{+}e^{-}$ collider,
  we measure
  branching fractions of 
  $(7.16 \pm 0.10{\mathrm{(stat)}} 
         \pm 0.60{\mathrm{(syst)}}) \times 10^{-4}$
  for $B^{+} \to J/\psi K^{+} \pi^{+} \pi^{-}$ and
  $(4.31 \pm 0.20{\mathrm{(stat)}} 
         \pm 0.50{\mathrm{(syst)}}) \times 10^{-4}$
  for $B^{+} \to \psi^{\prime} K^{+} \pi^{+} \pi^{-}$.
  We perform
  amplitude analyses 
  to determine the resonant structure 
  of the $K^{+} \pi^{+} \pi^{-}$ final state 
  in $B^{+} \to J/\psi K^{+} \pi^{+} \pi^{-}$ and
  $B^{+} \to \psi^{\prime} K^{+} \pi^{+} \pi^{-}$
  and find that the $K_{1}(1270)$ 
  is a prominent component of both decay modes.
  There is significant interference 
  among the different intermediate states,
  which leads, in particular,
  to a striking 
  distortion of the $\rho$ line shape due to the $\omega$.
  Based on the results of the fit to the 
  $B^{+} \to J/\psi K^{+} \pi^{+} \pi^{-}$ data,
  the relative decay fractions of the $K_{1}(1270)$ 
  to $K \rho$, $K \omega$, and $K^{*}(892) \pi$
  are consistent with previous measurements,
  but the decay fraction to $K_{0}^{*}(1430)$ 
  is significantly smaller. 
  Finally, by floating the mass and width of the $K_{1}(1270)$
  in an additional fit of the
  $B^{+} \to J/\psi K^{+} \pi^{+} \pi^{-}$ data,
  we measure 
  a mass of 
  $(1248.1 \pm 3.3{\mathrm{(stat)}} 
           \pm 1.4{\mathrm{(syst)}})
  ~{\mathrm{MeV}}/c^{2}$
  and
  a width of
  $(119.5 \pm 5.2{\mathrm{(stat)}}
          \pm 6.7{\mathrm{(syst)}})
  ~{\mathrm{MeV}}/c^{2}$
  for the $K_{1}(1270)$.
\end{abstract}

%%%%% PACs


\maketitle

%%%%% Introduction

\section{Introduction}
\label{section_introduction}

The large number of $B$-meson decays observed at $B$ factories 
allows detailed studies of 
the intermediate-state resonances involved in these decays.
This paper analyzes the structure of
the $K^{+} \pi^{+} \pi^{-}$ final state in the decays
$B^{+} \rightarrow J/\psi K^{+} \pi^{+} \pi^{-}$ and
$B^{+} \rightarrow \psi^{\prime} K^{+} \pi^{+} 
\pi^{-}$.\footnote{Charge-conjugate modes are always implicit.}
Kaon excitations that decay to a $K \pi \pi$ final state 
are difficult to distinguish based on the mass of the 
$K \pi \pi$ system alone, 
owing to their
overlapping line shapes.{\footnote{In 
2001, the Belle Collaboration measured the
branching fraction for $B^{+} \rightarrow J/\psi K_{1}(1270)$
with $2\%$ of the data presented here.
The $K^{+} \pi^{+} \pi^{-}$ final state in 
$B^{+} \rightarrow J/\psi K^{+} \pi^{+} \pi^{-}$
was found to be dominated by the $K_{1}(1270)$, 
and no other structure was detected~\cite{abe:2001}.}}
In this analysis, data are therefore fitted in the three dimensions
$M^{2}(K\pi\pi)$, $M^{2}(K\pi)$, and $M^{2}(\pi\pi)$,
which are the squared invariant masses of the 
$K^{+} \pi^{+} \pi^{-}$, $K^{+} \pi^{-}$ and $\pi^{+} \pi^{-}$ systems, 
respectively.  
An unbinned maximum-likelihood fit is performed 
to extract maximal information from the data.
The fitting model accounts for
interferences among different intermediate states, 
as well as the spin-dependent angular distributions of the 
final state.
The large sample size,
combined with the clean environment afforded by the presence of
a $J/\psi$ or $\psi^{\prime}$ in the final state, makes 
it possible to distinguish 
the different kaon excitations 
that contribute to the $K^{+} \pi^{+} \pi^{-}$ final state.
The results provide information not only on intermediate-state
interactions but also on the structure of the kaon spectrum.
By performing an additional fit 
in which the mass and width of the $K_{1}(1270)$ are floated,
we measure the mass and width of the $K_{1}(1270)$.

Identifying the kaon excitations involved in 
$B^{+} \rightarrow J/\psi K^{+} \pi^{+} \pi^{-}$ and
$B^{+} \rightarrow \psi^{\prime} K^{+} \pi^{+} \pi^{-}$
can lead to a better understanding of the underlying theory.
For example, the breaking of $SU(3)$ flavor symmetry 
mixes the $1^{3}P_{1}$ and $1^{1}P_{1}$ states 
of the kaon system into the physical states 
$K_{1}(1270)$ and $K_{1}(1400)$ as
\begin{eqnarray}
K_{1}(1270) &=& K(1^{3}P_{1}) \sin \theta_{K} 
             +  K(1^{1}P_{1}) \cos \theta_{K},\\
K_{1}(1400) &=& K(1^{3}P_{1}) \cos \theta_{K} 
             -  K(1^{1}P_{1}) \sin \theta_{K}, 
\end{eqnarray}
where $\theta_{K}$ is the $^{3}P_{1}$-$^{1}P_{1}$ mixing angle.
The value of $\theta_{K}$ can be related to
the masses of the $K_{1}(1270)$ and $K_{1}(1400)$, 
to the strong decays of the $K_{1}(1270)$ and $K_{1}(1400)$, 
and to 
rates of weak decays to final states involving the
$K_{1}(1270)$ and 
$K_{1}(1400)$~\cite{suzuki:1993,suzuki:1994,blundell:1996}.
The measurements presented here 
can
lead to a better determination of $\theta_{K}$.

The data sample used in this study
was produced by the KEKB
asymmetric-energy
$e^{+}e^{-}$ collider~\cite{kekb}
and reconstructed by the Belle detector~\cite{belle_detector}.
It corresponds to an integrated luminosity of
$492~{\mathrm{fb}}^{-1}$
accumulated at the $\Upsilon(4S)$ resonance and contains
$535 \times 10^{6}$~$B\bar{B}$ meson pairs.

%%%%% Apparatus

\section{The Belle detector}
\label{section_apparatus}

The Belle detector~\cite{belle_detector}
is a large-solid-angle magnetic spectrometer.
A silicon vertex detector 
surrounds the interaction point 
and reconstructs decay vertices.
A $50$-layer central drift chamber (CDC) 
provides charged-particle tracking over
the laboratory polar-angle region
$17^{\circ} \le \theta \le 150^{\circ}$,
which corresponds to
$92\%$ of the solid angle in the $\Upsilon(4S)$ rest frame.
A system of $1188$ aerogel Cherenkov counters (ACC) 
and an array of 128 time-of-flight counters 
provide particle identification.
An electromagnetic calorimeter,
comprising $8736$ CsI(Tl) crystals,
records the energy deposited by photons, leptons, and hadrons.  
These subdetectors are surrounded by a superconducting solenoid 
$3.4~{\mathrm{m}}$ in diameter and $4.4~{\mathrm{m}}$ in length, 
which produces a
$1.5$-${\mathrm{T}}$ magnetic field parallel to the positron beam.
An iron flux return installed outside the coil 
is instrumented with large-area resistive-plate counters 
to identify muons and $K_{L}$ mesons.
Monte Carlo (MC) simulations{\footnote{Lists of four-vectors 
for a given decay chain are generated using EvtGen~\cite{evtgen}.
The detector response is then simulated using GEANT~\cite{geant},
combining randomly-triggered data with the simulated events.}} 
are used to 
determine the acceptance of the detector for the processes of interest.

%%%%% Events

\section{Event selection}
\label{section_events}

Electron candidates are identified by combining information 
from the CDC, electromagnetic calorimeter, and ACC.
Muon candidates are
identified by extrapolating charged-particle tracks 
from the silicon vertex detector and CDC into the 
$K_{L}/\mu$ detector.
To identify charged hadrons, 
momentum measurements from the CDC 
are combined with velocity information from the 
time-of-flight counters, ACC, and CDC 
($dE/dx$)~\cite{nakano:2002}.
The kaon identification efficiency is above $80\%$,
while the probability of misidentifying a pion as a kaon
is below $10\%$.

Low-momentum charged tracks that curl up in the CDC
can be reconstructed multiple times by the track finder.
To ensure that no track is included more than 
once, criteria similar to those of
Ref.~\onlinecite{thesis:hidekazu_kakuno:2003} are 
used.{\footnote{See Ref.~\onlinecite{thesis:hulya_guler:2008} 
for a detailed description of the event selection.}}

In studying a mode that has a $J/\psi$ or $\psi^{\prime}$ 
in the final state, a key strategy 
is to reconstruct the $J/\psi$ 
only in its decays to $e^{+}e^{-}$ or $\mu^{+}\mu^{-}$, 
and the $\psi^{\prime}$ only in its decays to 
$e^{+}e^{-}$, $\mu^{+}\mu^{-}$, or $J/\psi \pi^{+} \pi^{-}$.  
Although this choice abandons
all but $12\%$ of $J/\psi$'s and $5\%$ of $\psi^{\prime}$'s, 
it reduces continuum backgrounds to a negligible level. 
The decays $J/\psi \rightarrow \mu^{+} \mu^{-}$ and 
$\psi^{\prime} \rightarrow \mu^{+} \mu^{-}$ are reconstructed 
by combining oppositely-charged muon candidates.
The invariant-mass distribution is then fitted,
modeling the $J/\psi$ and $\psi^{\prime}$ as double Gaussians.
Muon pairs are discarded unless they have an invariant mass 
within $\pm 3\sigma$ of the fitted $J/\psi$ or $\psi^{\prime}$
mean, where $\sigma$ is the width of the narrower Gaussian.
Similarly, $J/\psi \rightarrow e^{+} e^{-}$ and 
$\psi^{\prime} \rightarrow e^{+} e^{-}$ decays are reconstructed 
by combining oppositely-charged electron candidates.
To account for energy losses due to 
final-state radiation or bremsstrahlung in the
detector, any photons detected within $50~{\mathrm{mrad}}$ of the 
initial direction of an electron candidate
are also included in the $e^{+}e^{-}$ 
invariant-mass calculation.
Electron pairs are discarded unless they have an invariant
mass within the range extending from $-4\sigma$ to $+3\sigma$ of the 
fitted $J/\psi$ or $\psi^{\prime}$ mean.  
This mass window is asymmetric about the mean so as
to include the radiative tails of the $J/\psi$ and $\psi^{\prime}$, 
which are not completely recovered by the photon addition.

Lepton track pairs that survive the mass requirements are fitted to a common
vertex, which is constrained within errors to the measured interaction point. 
This vertex is then fixed, and another fit is performed, constraining the 
dilepton invariant mass to the nominal $J/\psi$ or $\psi^{\prime}$ mass.
Since the observed widths of the $J/\psi$ and $\psi^{\prime}$ are 
dominated by measurement error, this procedure
improves the mass resolution of the $B$ candidate.

To reconstruct $\psi^{\prime} \rightarrow J/\psi \pi^{+} \pi^{-}$
decays, leptonic $J/\psi$ candidates are combined with 
a pair of 
oppositely-charged tracks that satisfy pion-identification criteria.
As the $\pi^{+} \pi^{-}$ invariant-mass distribution in
$\psi^{\prime} \rightarrow J/\psi \pi^{+} \pi^{-}$ decays is known to
peak at high values~\cite{coffman:1992}, the dipion invariant mass is
required to be greater than $0.4~{\mathrm{GeV}}/c^{2}$.
Unless they have an invariant mass within $\pm 3\sigma$ of the fitted 
$\psi^{\prime}$ mean,
$\psi^{\prime} \rightarrow J/\psi \pi^{+} \pi^{-}$ candidates
are discarded.

To reconstruct $B$-meson candidates,
each $J/\psi$ or $\psi^{\prime}$ candidate is combined with 
a kaon candidate and two oppositely-charged pion candidates.
Kaon and pion candidates are charged tracks that satisfy 
identification criteria for kaons and pions, respectively,
and have an impact parameter 
with respect to the fitted dilepton vertex of
$|dr| < 0.4~{\mathrm{cm}}$ and 
$|dz| < 1.5~{\mathrm{cm}}$.\footnote{The impact-parameter requirement 
is not applied to the pions in 
$\psi^{\prime} \rightarrow J/\psi \pi^{+} \pi^{-}$.}
Any pion candidate that is identified as the product of a
$K_{S}^{0} \rightarrow \pi^{+} \pi^{-}$ decay is 
discarded.{\footnote{To reconstruct 
$K_{S}^{0} \rightarrow \pi^{+} \pi^{-}$,
oppositely-charged pion candidates are combined, and
the $K_{S}^{0}$ selection criteria
of Ref.~\onlinecite{thesis:fang_fang:2003} are applied.
Both pions are vetoed if their combined invariant mass lies between 
$0.482~{\mathrm{GeV}}/c^{2}$ and
$0.510~{\mathrm{GeV}}/c^{2}$,
which corresponds roughly to a region extending from $-4\sigma$ to 
$+3\sigma$ around the nominal $K_{S}^{0}$ mass.}}

\subsection{$B$-Meson reconstruction}
\label{events:section_dembc}

Two kinematic variables can be used to identify $B$ mesons. 
First, the reconstructed mass of a true $B$ meson
is likely to fall near the nominal $B$ mass.
Second, as $B$ mesons are produced in the reaction 
\begin{equation}
e^{+} e^{-} \rightarrow \Upsilon(4S) \rightarrow B^{+} B^{-},
\end{equation}
the energy of each $B$ in the $\Upsilon(4S)$ frame
is half the total energy of the electron and positron beams in this 
frame.  Since beam-energy drifts can cause the mass of the
$\Upsilon(4S)$ 
to vary, it is customary to 
recast these kinematic variables in forms that are
readily corrected for drifts in the beam energy---namely, the energy
difference $\Delta E$ and beam-constrained mass $M_{\mathrm{bc}}$:
\begin{eqnarray}
\label{events:eq_deltae}
\Delta E = E^{*}(B) - E_{\mathrm{beam}}^{*}, \\
M_{\mathrm{bc}} = \sqrt{E_{\mathrm{beam}}^{*2} - P^{*2}(B)}.
\end{eqnarray}
Here, $P^{*}(B)$ is the momentum of the $B$ candidate in the
$\Upsilon(4S)$ frame, while
$E_{\mathrm{beam}}^{*}$ is 
half the energy of the $\Upsilon(4S)$
and is measured independently.
For a correctly-reconstructed $B$ meson, 
$\Delta E$ peaks at zero, and 
$M_{\mathrm{bc}}$ peaks at the nominal $B$ mass.

In the case of a multiparticle final state such as
$J/\psi K^{+} \pi^{+} \pi^{-}$ or $\psi^{\prime} K^{+} \pi^{+} \pi^{-}$,
multiple $B$ candidates can pose a challenge.
If a correctly-reconstructed $B$ candidate includes a low-momentum pion,
then an 
additional
$B$ candidate can be formed by replacing that
pion with a low-momentum pion from the other $B$.  As the exchange
does not significantly affect the energy or momentum of the $B$ 
candidate, both candidates can satisfy $\Delta E$ and $M_{\mathrm{bc}}$ 
criteria.
Multiple candidates can spoil branching-fraction measurements
and distort observed mass spectra.
To ensure that no $B$ decay is counted more than once,
a best-candidate selection is performed.
First, $B$ candidates are required to have
$|\Delta E| < 0.2~{\mathrm{GeV}}$ and 
$M_{\mathrm{bc}} > 5.27~{\mathrm{GeV}}/c^{2}$.
This leaves
$25\%$ of 
$B^{+} \rightarrow J/\psi K^{+} \pi^{+} \pi^{-}$ events
and $34\%$ of 
$B^{+} \rightarrow \psi^{\prime} K^{+} \pi^{+} \pi^{-}$ events
with multiple candidates; 
these events have 
a mean multiplicity of
$2.4$ and $2.7$, respectively.
If a given event has 
multiple $B$ candidates with the same final state,
the charged tracks that make up each $B$ candidate
are fitted to a common vertex.  
The candidate whose vertex fit has the smallest $\chi^{2}$ is selected.
According to MC studies, this procedure identifies the correct 
$B$ candidate in approximately $55\%$ of cases
where there are multiple candidates.

In the case of $B^{+} \rightarrow J/\psi K^{+} \pi^{+} \pi^{-}$,
the decay 
$B^{+} \rightarrow \psi^{\prime} K^{+}$ 
is vetoed by rejecting all $B$ candidates that have a 
$J/\psi \pi^{+} \pi^{-}$ invariant mass between
$3.675~{\mathrm{GeV}}/c^{2}$ and $3.695~{\mathrm{GeV}}/c^{2}$.{\footnote{The 
small contribution from $B^{+} \rightarrow X(3872) K^{+}$ 
is not vetoed.}}
According to MC studies, $1.1\%$
of $B^{+} \rightarrow \psi^{\prime} K^{+}$
events in which the $\psi^{\prime}$ decays to $J/\psi \pi^{+} \pi^{-}$
and the $J/\psi$ decays to $e^{+}e^{-}$ or $\mu^{+}\mu^{-}$
survive this veto.

\subsection{Signal and sideband regions}
\label{events:section_signal_sideband}

\begin{figure}[hbtp]
\centerline{
\scalebox{0.34}{
\rotatebox{270}{
\includegraphics*[268,35][570,723]
  {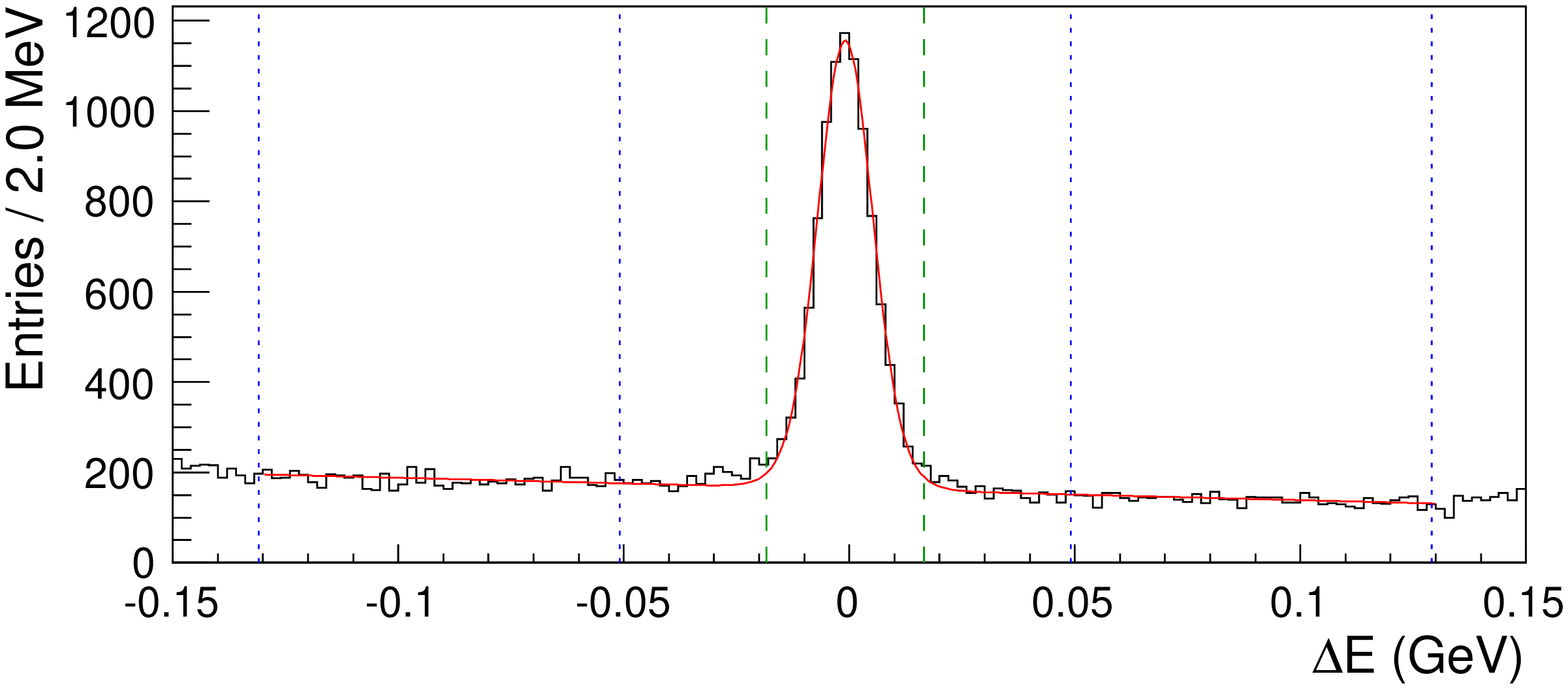}}}}
\vspace{1mm}
\centerline{
\scalebox{0.34}{
\rotatebox{270}{
\includegraphics*[268,35][570,723]
  {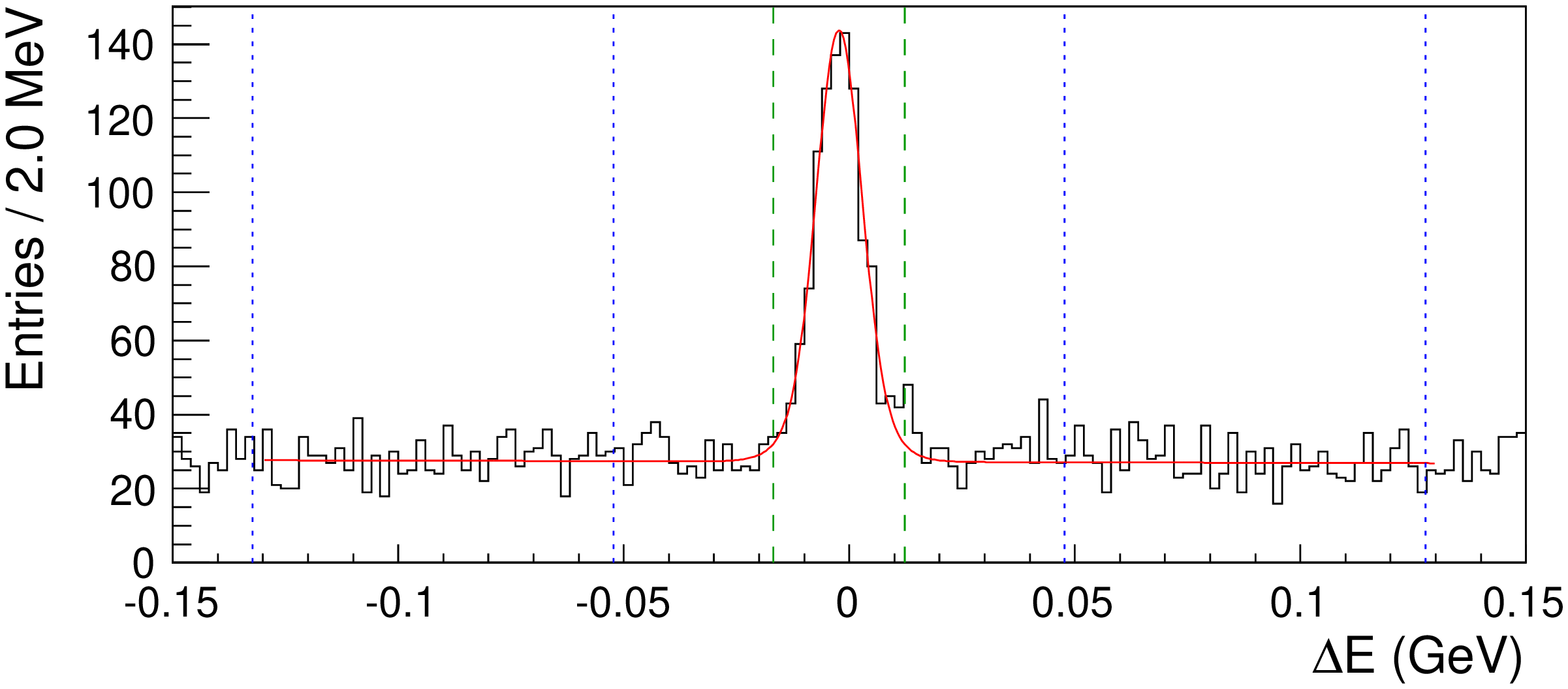}}}}
\caption{Data $\Delta E$ distributions for 
  $B^{+} \rightarrow J/\psi K^{+} \pi^{+} \pi^{-}$ (top) and 
  $B^{+} \rightarrow \psi^{\prime} K^{+} \pi^{+} \pi^{-}$ (bottom).
  The curves show the results of the fits described in the text.
  Dashed and dotted lines indicate the signal and sideband regions,
  respectively.}
\label{events:fig_de_data}
\end{figure}

Data distributions of $\Delta E$ for
$B^{+} \rightarrow J/\psi K^{+} \pi^{+} \pi^{-}$ and 
$B^{+} \rightarrow \psi^{\prime} K^{+} \pi^{+} \pi^{-}$ 
are shown in
Fig.~\ref{events:fig_de_data}.
The signal is modeled as a double Gaussian with a single mean, fixing
the width and relative height of the wider Gaussian to the
results of a MC fit.  
The background is modeled as a first-order polynomial.  
Based on these fits, the signal region is defined as
\begin{equation}
\label{events:eq_signal_region}
-3 \sigma_{\Delta E} 
< \Delta E - \mu_{\Delta E} <
+3\sigma_{\Delta E} 
,
\end{equation}
where $\mu_{\Delta E}$ is the mean of the signal peak,
and $\sigma_{\Delta E}$ is the width of the narrower Gaussian.
The sideband region, which is used to estimate the background under
the signal, is defined as
\begin{eqnarray}
\label{events:eq_sideband_region}
-0.13~{\mathrm{GeV}}
  & < \Delta E - \mu_{\Delta E} <
  &  -0.05~{\mathrm{GeV}}, \nonumber \\
 0.05~{\mathrm{GeV}}
  & < \Delta E - \mu_{\Delta E} <
  &   0.13~{\mathrm{GeV}}.
\end{eqnarray}
The sideband normalization factor is given by
\begin{equation}
\label{events:eq_f_b}
f_{B} ~=~ 
 \frac{\int\limits_{\mathrm{signal}} p_{\mathrm{bkg}} \,d\Delta E } 
      {\int\limits_{\mathrm{sideband}} p_{\mathrm{bkg}} \,d\Delta E} 
,
\end{equation}
where $p_{\mathrm{bkg}}$ is the polynomial representing the background.
The fraction of signal-region events that are background is estimated
as
\begin{equation}
\label{events:eq_n_b}
n_{B} ~=~ \frac{B}{S} ~ f_{B}
,
\end{equation}
where $S$ and $B$ are the numbers
of events in the signal and sideband regions, respectively.

%%%%% Transformations

\section{Coordinate transformations}
\label{section_transformations}

\begin{figure}[htbp]
\centerline{
\scalebox{0.35}{
\rotatebox{270}{
\includegraphics*[270,30][567,703]
 {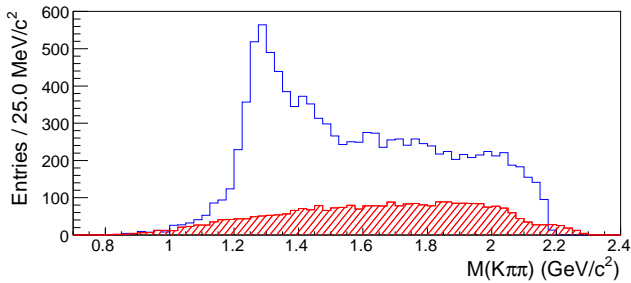}
}}}
\caption{Invariant mass of the 
  $K^{+} \pi^{+} \pi^{-}$ system
  in $B^{+} \rightarrow J/\psi   K^{+} \pi^{+} \pi^{-}$ data.
  Open and filled histograms show events in the signal
  and normalized sideband regions, respectively.}
\label{transformations:fig_mkpp_data_before}
\end{figure}

The data in the sideband region are used to model 
the background in the signal region.
Figure~\ref{transformations:fig_mkpp_data_before},
which shows the distribution of $M(K\pi\pi)$ 
for $B^{+} \rightarrow J/\psi K^{+} \pi^{+} \pi^{-}$ events 
in the signal and sideband regions,
reveals a problem:
signal and sideband data have different end points in $M(K\pi\pi)$.

\begin{figure}[htbp]
\centerline{
\scalebox{0.38}{
\rotatebox{270}{
\includegraphics*[58,32][560,570]
  {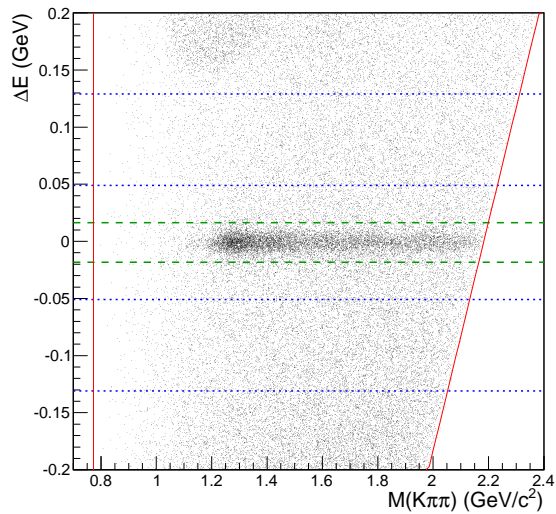}}}}
\caption{$\Delta E$ versus $M(K\pi\pi)$ 
  for $B^{+} \rightarrow J/\psi K^{+} \pi^{+} \pi^{-}$ data.
  Dashed lines outline the signal region,
  and dotted lines outline the sidebands.
  The solid lines indicate the minimum and maximum
  values of $M(K\pi\pi)$.}
\label{transformations:fig_demkpp_data_before}
\end{figure}

Plotting $\Delta E$ versus $M(K\pi\pi)$
reveals the cause of the discrepancy.  As 
Fig.~\ref{transformations:fig_demkpp_data_before} demonstrates,
the kinematically allowed range of $M(K\pi\pi)$ depends on $\Delta E$.
While the minimum value of $M(K\pi\pi)$ is 
$M(K\pi\pi)_{\mathrm{min}} = M_{K} + 2 M_{\pi}$,
the maximum value,
which is attained when
both the $K^{+} \pi^{+} \pi^{-}$ system and the $J/\psi$ are at rest in
the $B$-candidate's rest frame,
varies with $\Delta E$ as
$M(K\pi\pi)_{\mathrm{max}}  = 
\Delta E + M_{B} - M_{J/\psi}$.
Here, $M_{K}$, $M_{\pi}$, $M_{B}$ and $M_{J/\psi}$ 
stand for the nominal masses of the  subscripted particles.

\begin{figure}[htbp]
\centerline{
\scalebox{0.38}{
\rotatebox{270}{
\includegraphics*[58,32][560,570]
  {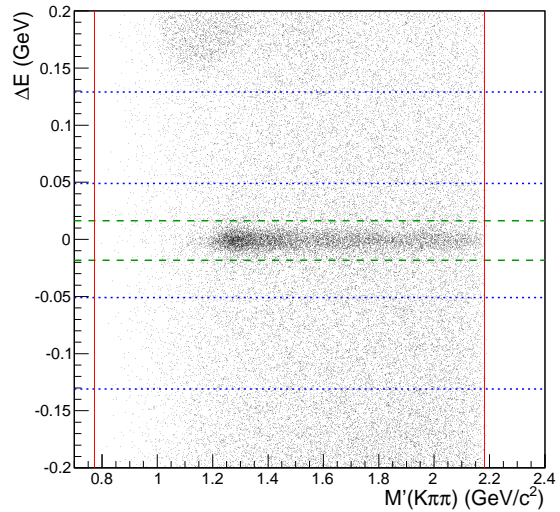}}}}
\caption{$\Delta E$ versus $M^{\prime}(K\pi\pi)$ 
  for $B^{+} \rightarrow J/\psi K^{+} \pi^{+} \pi^{-}$ data.
  The lines are defined as in 
  Fig.~\ref{transformations:fig_demkpp_data_before}.}
\label{transformations:fig_demkpp_data_after}
\end{figure}

Transforming $M(K\pi\pi)$ as follows 
removes its dependence on $\Delta E$:
\begin{eqnarray}
\label{transformations:eq_mkpp_transformation}
M^{\prime}(K\pi\pi) & = &  M(K\pi\pi)_{\mathrm{min}} 
\nonumber\\
{} & + & \left[ M(K\pi\pi) - M(K\pi\pi)_{\mathrm{min}} \right]
\nonumber\\
{} & \times &
    \frac{M(K\pi\pi)_{\mathrm{max}}^{0} - M(K\pi\pi)_{\mathrm{min}}}
         {M(K\pi\pi)_{\mathrm{max}}     - M(K\pi\pi)_{\mathrm{min}}}
.
\end{eqnarray}
Here, 
$M(K\pi\pi)_{\mathrm{max}}^{0} = M_{B} - M_{J/\psi}$ is the value of
$M(K\pi\pi)_{\mathrm{max}}$ at $\Delta E = 0$.
Figure~\ref{transformations:fig_demkpp_data_after} shows
$\Delta E$ versus the transformed coordinate $M^{\prime}(K\pi\pi)$.
While the minimum value of $M(K\pi\pi)$ is unaffected by the
transformation, the maximum value is changed such that 
the maximum of $M^{\prime}(K\pi\pi)$ at any $\Delta E$ 
is equal to the maximum of $M(K\pi\pi)$ at $\Delta E = 0$.
The range of $M(K\pi\pi)$ is compressed for positive values of
$\Delta E$ and stretched for negative values of 
$\Delta E$.{\footnote{Although correctly-reconstructed $B$ mesons should have 
$\Delta E = 0$ on average, systematic errors shift 
the observed mean of the signal $\Delta E$ peak
away from zero by $1$-$2~{\mathrm{MeV}}$.
For simplicity of presentation, this mean is assumed to be zero in
the equations of this section.
In fact, just as the signal and sideband regions are centered 
around the
measured $\Delta E$ mean, the transformations are also made about the
measured $\Delta E$ mean.}}   

\begin{figure}[htbp]
\centerline{
\scalebox{0.35}{
\rotatebox{270}{
\includegraphics*[270,30][567,703]
 {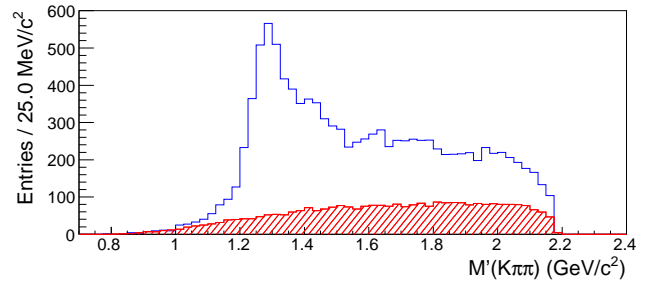}
}}}
\caption{Transformed invariant mass of the 
  $K^{+} \pi^{+} \pi^{-}$ system 
  in $B^{+} \rightarrow J/\psi   K^{+} \pi^{+} \pi^{-}$ data.
  Open and filled histograms show events in the signal
  and normalized sideband regions, respectively.}
\label{transformations:fig_mkpp_data_after}
\end{figure}

Figure~\ref{transformations:fig_mkpp_data_after} shows
$M^{\prime}(K\pi\pi)$ for events in the signal and sideband regions.
The problem of 
Fig.~\ref{transformations:fig_mkpp_data_before} has been solved:
the end points of the transformed signal and sideband distributions
match.

An important feature of the transformation is that it does not change 
$M(K\pi\pi)$ at $\Delta E = 0$.
Thus, although sideband and signal regions are both transformed, the
change is minimal in the signal region.

\begin{figure}[htbp]
\centerline{
\scalebox{0.38}{
\rotatebox{270}{
\includegraphics*[58,32][560,570]
  {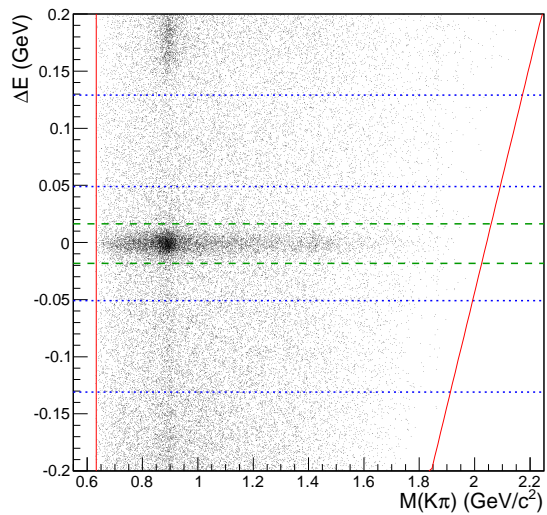}}}}
\caption{$\Delta E$ versus $M(K\pi)$ 
  for $B^{+} \rightarrow J/\psi K^{+} \pi^{+} \pi^{-}$ data.
  The lines are defined as in 
  Fig.~\ref{transformations:fig_demkpp_data_before}.
  The concentration of events near $0.9$~GeV$/c^{2}$ represents
  random particle combinations containing real $K^{*}(892)$s.}
\label{transformations:fig_demkp_data_before}
\end{figure}

\begin{figure}[htbp]
\centerline{
\scalebox{0.38}{
\rotatebox{270}{
\includegraphics*[58,32][560,570]
  {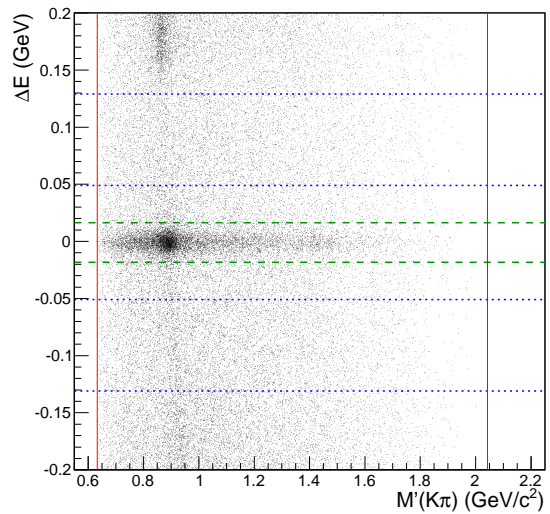}}}}
\caption {$\Delta E$ versus $M^{\prime}(K\pi)$
  for $B^{+} \rightarrow J/\psi K^{+} \pi^{+} \pi^{-}$ data.
  The lines are defined as in 
  Fig.~\ref{transformations:fig_demkpp_data_before}.}
\label{transformations:fig_demkp_data_after}
\end{figure}

Just as the range of $M(K\pi\pi)$ depends on $\Delta E$, 
the ranges of $M(K\pi)$ and $M(\pi\pi)$ also depend on $\Delta E$.
To correct for this dependence, 
transformations similar to 
Eq.~\ref{transformations:eq_mkpp_transformation}
are applied.
The variable $M(K\pi)$ is transformed as
\begin{eqnarray}
\label{transformations:eq_mkp_transformation}
M^{\prime}(K\pi) & = & M(K\pi)_{\mathrm{min}} 
\nonumber\\
{} & + & \left[ M(K\pi) - M(K\pi)_{\mathrm{min}} \right] 
\nonumber\\
{} & \times &
    \frac{M(K\pi\pi)_{\mathrm{max}}^{0} - M(K\pi\pi)_{\mathrm{min}}}
         {M(K\pi\pi)_{\mathrm{max}}     - M(K\pi\pi)_{\mathrm{min}}}
,
\end{eqnarray}
and the variable $M(\pi\pi)$ is transformed as
\begin{eqnarray}
\label{transformations:eq_mpp_transformation}
M^{\prime}(\pi\pi) & = & M(\pi\pi)_{\mathrm{min}} 
\nonumber\\
{} & + & \left[ M(\pi\pi) - M(\pi\pi)_{\mathrm{min}} \right] 
\nonumber\\
{} & \times &
    \frac{M(K\pi\pi)_{\mathrm{max}}^{0} - M(K\pi\pi)_{\mathrm{min}}}
         {M(K\pi\pi)_{\mathrm{max}}     - M(K\pi\pi)_{\mathrm{min}}}
.
\end{eqnarray}
Here, 
$M(K\pi)_{\mathrm{min}} = M_{K} + M_{\pi}$, and 
$M(\pi\pi)_{\mathrm{min}} = 2 M_{\pi}$.
Figures 
\ref{transformations:fig_demkp_data_before} and
\ref{transformations:fig_demkp_data_after} 
show $\Delta E$ versus $M(K\pi)$ and $M^{\prime}(K\pi)$. 
A similar effect is observed for 
$M(\pi\pi)$ and $M^{\prime}(\pi\pi)$.

\begin{figure}[htb]
\centerline{
\scalebox{0.35}{
\rotatebox{270}{
\includegraphics*[264,34][572,707]
 {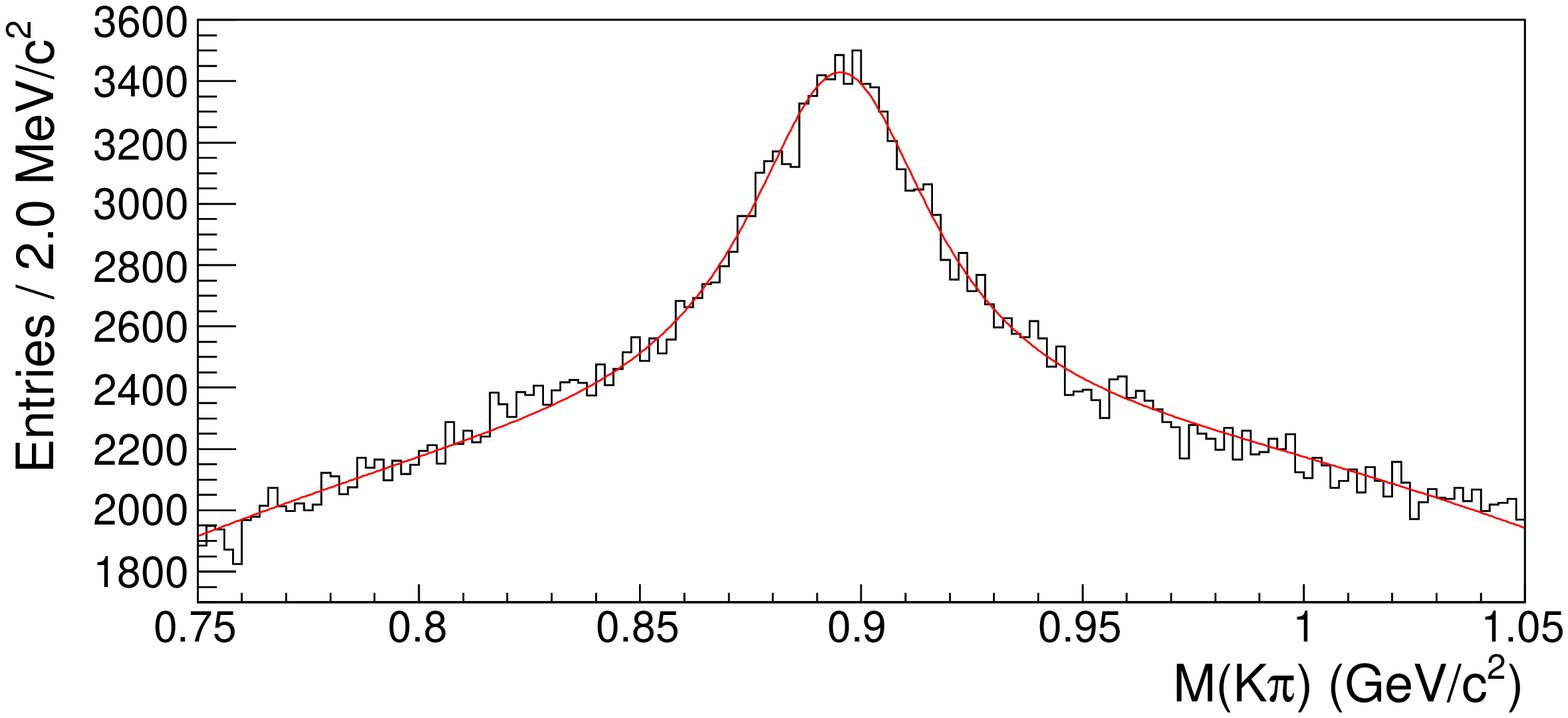}}}}
\vspace{1mm}
\centerline{
\scalebox{0.35}{
\rotatebox{270}{
\includegraphics*[264,34][572,707]
 {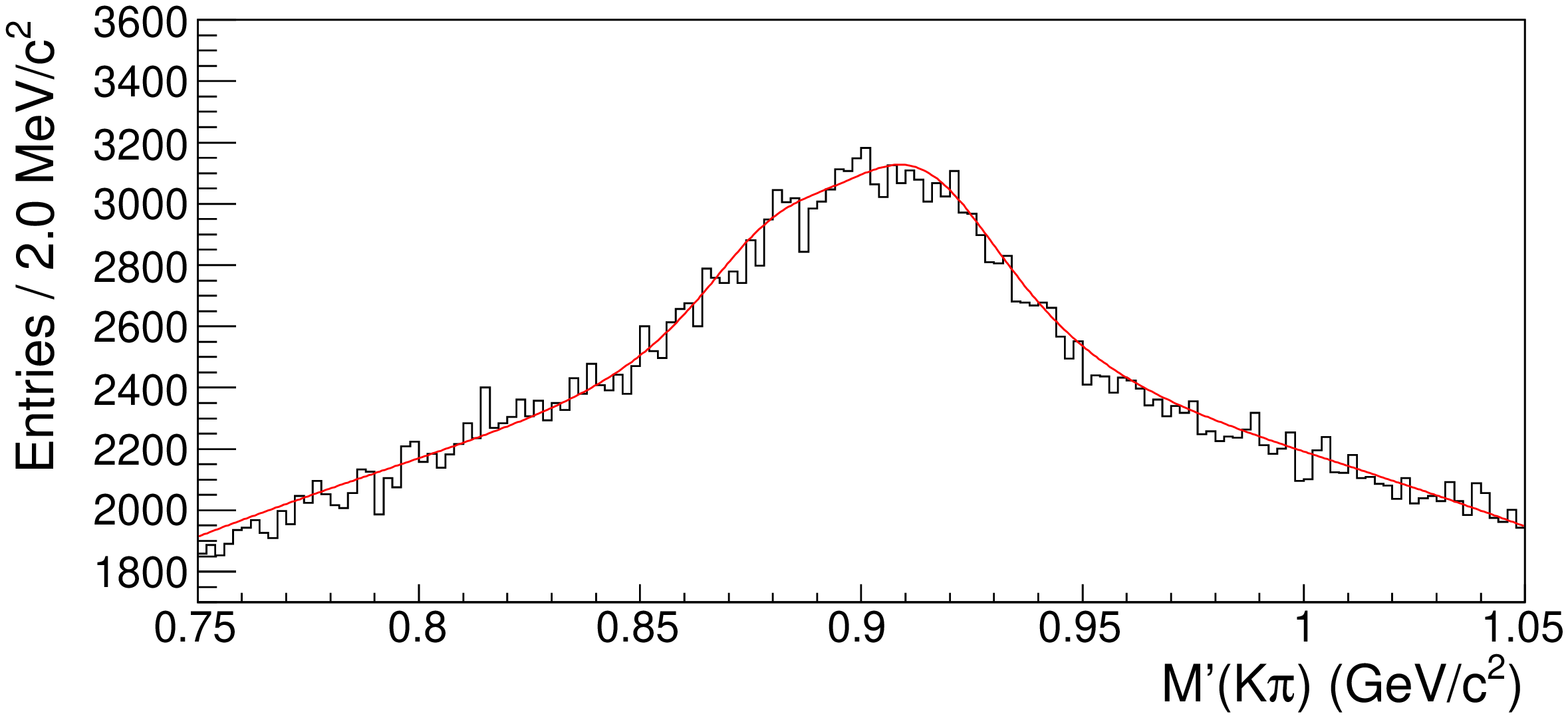}}}}
\caption {$M(K\pi)$ (top) and $M^{\prime}(K\pi)$ (bottom)
  distributions of the $K^{*}(892)$ peak in generic-MC sidebands.
  The curve in the top plot is the 
  result of a fit to a Breit-Wigner plus a polynomial background.
  This curve is transformed as described in the text and is 
  then superimposed on the $M^{\prime}(K\pi)$ distribution in the 
  bottom plot.}
\label{transformations:fig_peak_transformation}
\end{figure}

As Fig.~\ref{transformations:fig_demkp_data_after} illustrates,
transforming the $M(K\pi)$ coordinate distorts the shapes
of the $K^{*}(892)$ and $D^{0}$ backgrounds.
This distortion must be taken into account
in parametrizing the background
for the three-dimensional fits of Sec.~\ref{section_amplitude}
(i.e., in Eqs.~\ref{amplitude:eq_bkg_jkpp} 
and~\ref{amplitude:eq_bkg_pkpp}).
Modeling the distortion is straightforward.
First, the peak is described as a Breit-Wigner or Gaussian
in the untransformed coordinate, $M(K\pi)$.
Using Eq.~\ref{transformations:eq_mkp_transformation}, $M(K\pi)$ 
is then written as a function of $M^{\prime}(K\pi)$ and $\Delta E$.
The expression is numerically integrated over 
the relevant range of $\Delta E$ to obtain the shape of the 
peak as a function of $M^{\prime}(K\pi)$.  
Figure~\ref{transformations:fig_peak_transformation} 
demonstrates the transformation of the $K^{*}(892)$ background shape.

The data also contain $K^{0}_{S}$ and $\rho^{0}$ backgrounds, 
albeit less prominently.
The distortion of these peaks by the $M(\pi\pi)$ transformation
is modeled by 
describing the $K^{0}_{S}$ as a Gaussian and the $\rho^{0}$ as a
Breit-Wigner in $M(\pi\pi)$,
expressing $M(\pi\pi)$ as a function of
$M^{\prime}(\pi\pi)$ and $\Delta E$ 
with the help of
Eq.~\ref{transformations:eq_mpp_transformation}, 
and integrating this over the appropriate region of $\Delta E$.

As the main source of background in this analysis
is misreconstructed $B$-meson decays, 
the transformations were checked by analyzing
a generic-MC simulation of $\Upsilon(4S)$ decays to 
$B^{+} B^{-}$ and $B^{0} \bar{B}^{0}$,
with all known decay modes included.
The $M(K\pi\pi)$, $M(K\pi)$, and $M(\pi\pi)$ distributions were
found to display the same $\Delta E$-dependence 
in MC simulation as in data.
Excluding signal events from the MC sample,
the distributions of background events 
in the signal and sideband regions were 
compared with and without the transformations.
The transformed sidebands were found to 
reproduce the shape of the background in the signal region 
more accurately than the untransformed sidebands,
especially at high $M(K\pi\pi)$, $M(K\pi)$, and 
$M(\pi\pi)$.{\footnote{For details,
see Ref.~\onlinecite{thesis:hulya_guler:2008}.}}

Although some discrepancy was observed
between the background in the signal and sideband regions 
near the $K_{S}^{0}$ and $\rho$ masses, this is 
mostly independent of the transformation and 
is taken into consideration when calculating systematic errors.

The transformations of 
Eqs.~\ref{transformations:eq_mkpp_transformation}-\ref{transformations:eq_mpp_transformation}
were also applied to 
$B^{+} \rightarrow \psi^{\prime} K^{+} \pi^{+} \pi^{-}$,
with $M_{J/\psi}$ replaced with $M_{\psi^{\prime}}$.
The results of the checks were the same.

For simplicity, 
the variables
$M^{\prime}(K\pi\pi)$, $M^{\prime}(K\pi)$, and $M^{\prime}(\pi\pi)$
are henceforth referred to as
$M(K\pi\pi)$, $M(K\pi)$, and $M(\pi\pi)$, respectively.

%%%%% Inclusive

\section{Total branching fractions}
\label{section_inclusive}

Branching fractions for 
$B$-meson decays to
$J/\psi K^{+} \pi^{+} \pi^{-}$
and 
$\psi^{\prime} K^{+} \pi^{+} \pi^{-}$
final states are measured using a background-subtraction 
technique.{\footnote{Peaking backgrounds are not expected 
in these final states and were not seen in generic-MC simulation.}}
For each final state, data events in the signal and sideband regions 
are distributed into 
cubic 
bins in $M^{2}(K\pi\pi)$, $M^{2}(K\pi)$, and $M^{2}(\pi\pi)$.
The number of signal events observed in each bin is calculated as
\begin{equation}
\label{inclusive:eq_nsignal}
N_{i} = S_{i} - f_{B} \, B_{i}
,
\end{equation}
where $S_{i}$ and $B_{i}$ are the numbers of 
signal-region and sideband-region events,
respectively, that fall into the $i$th bin, 
and $f_{B}$ is the sideband normalization factor given by
Eq.~\ref{events:eq_f_b}.
The fraction of charged $B$ mesons that decay to the final state 
in question
can be expressed as
\begin{equation}
\label{inclusive:eq_branching_fraction}
{\mathcal{B}} = \frac{1}{N_{B}} \sum_{i} \frac{N_{i}}{\varepsilon_{i}}
,
\end{equation}
where $\varepsilon_{i}$ is the signal efficiency in bin $i$,
and $N_{B}$ is the total number of charged $B$ mesons in the data
sample.  Assuming equal rates for 
$\Upsilon(4S) \rightarrow  B^{+} B^{-}$ and 
$\Upsilon(4S) \rightarrow  B^{0} \bar{B}^{0}$,
$N_{B}$ is equal to the number of $B$ pairs produced,
which is measured independently.

\begin{figure*}[hbtp]
\centerline{
\scalebox{0.32}{
\rotatebox{270}{
\includegraphics*[270,37][569,691]
  {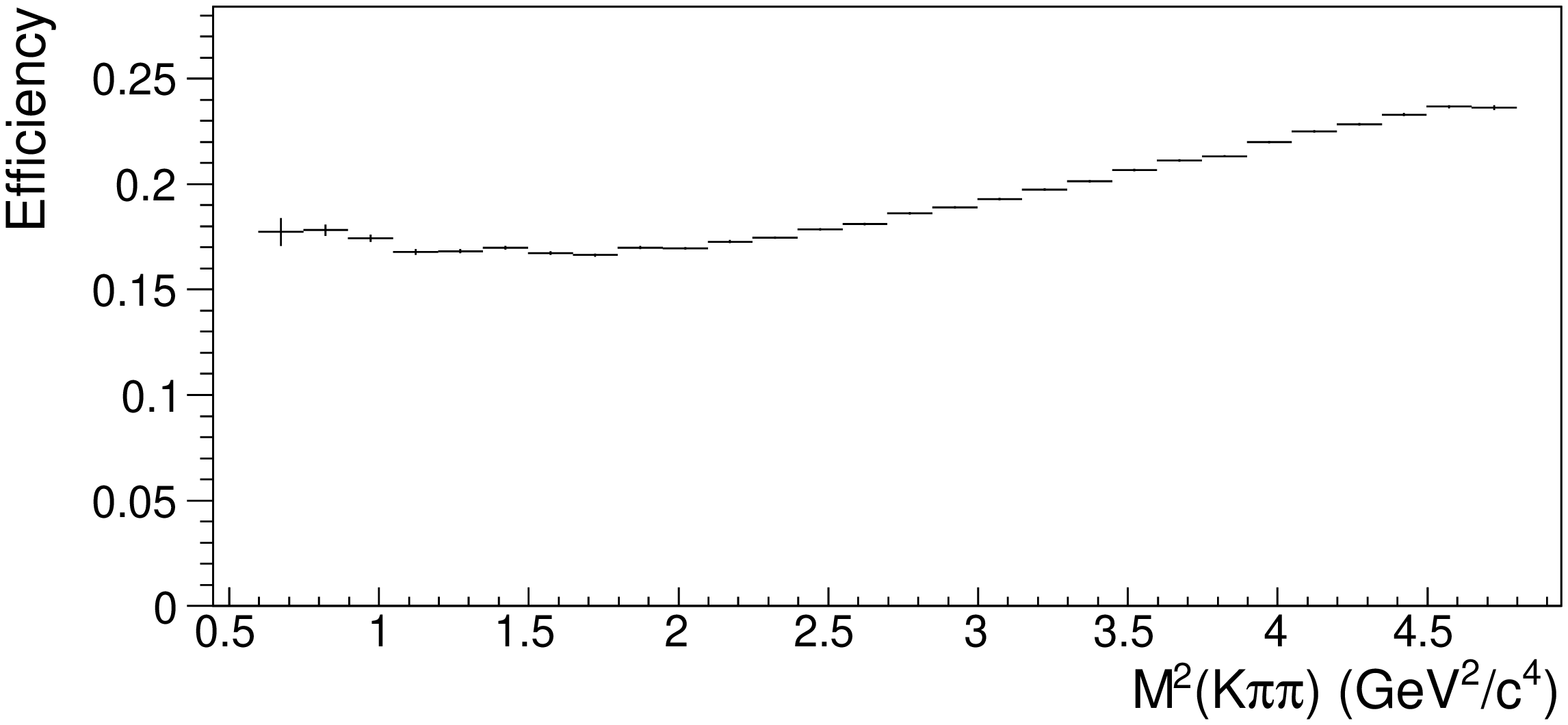}
}}
\hspace{4mm}
\scalebox{0.32}{
\rotatebox{270}{
\includegraphics*[270,37][569,691]
  {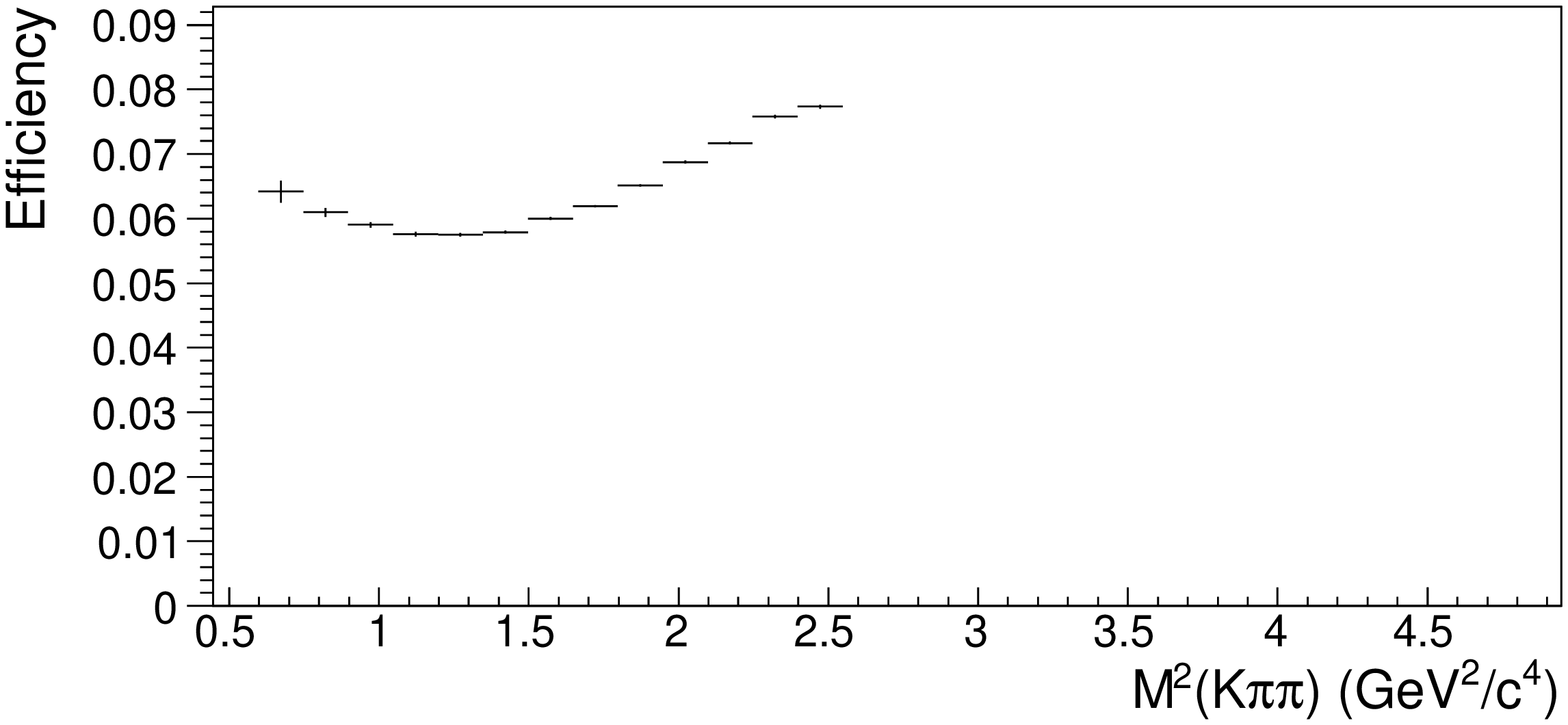}
}}}
\centerline{
\scalebox{0.32}{
\rotatebox{270}{
\includegraphics*[270,37][569,691]
  {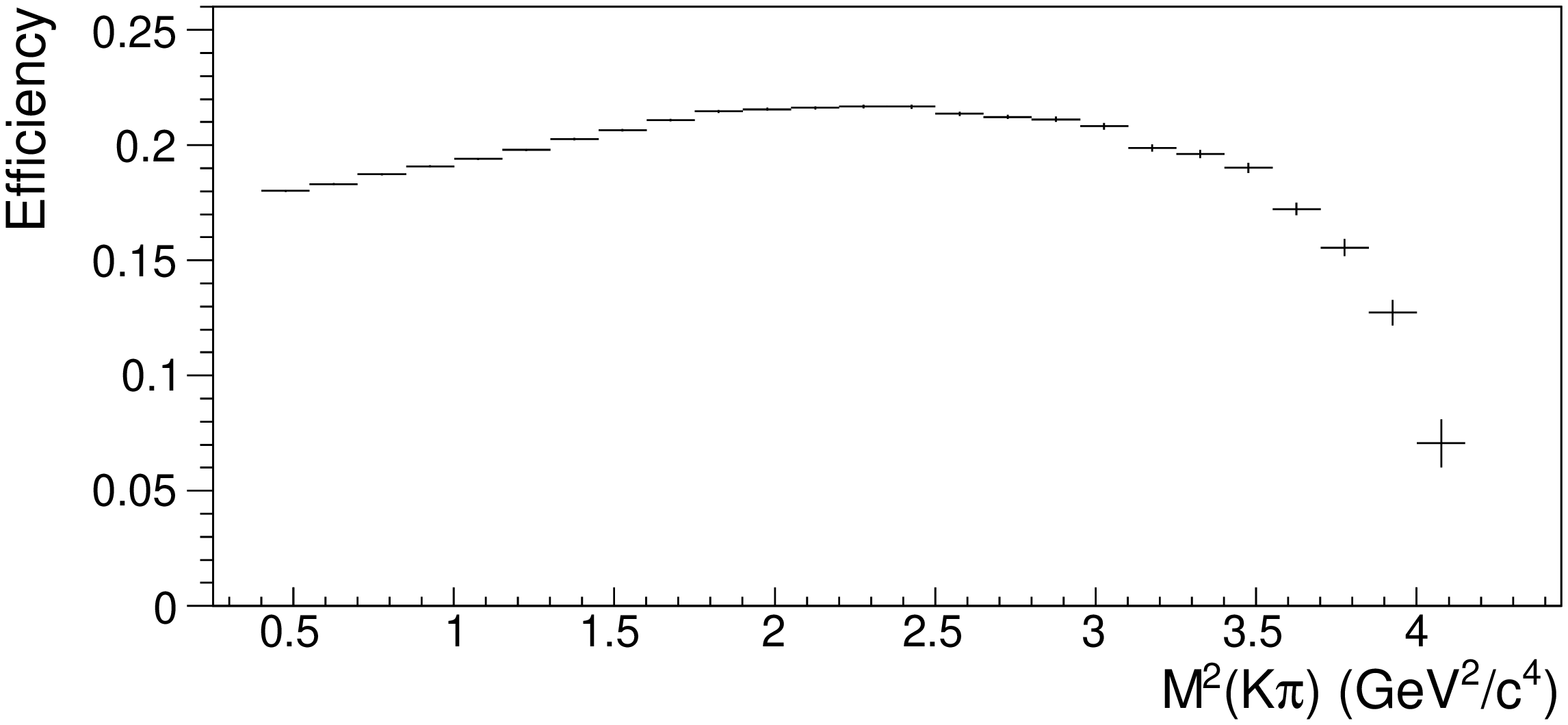}
}}
\hspace{4mm}
\scalebox{0.32}{
\rotatebox{270}{
\includegraphics*[270,37][569,691]
  {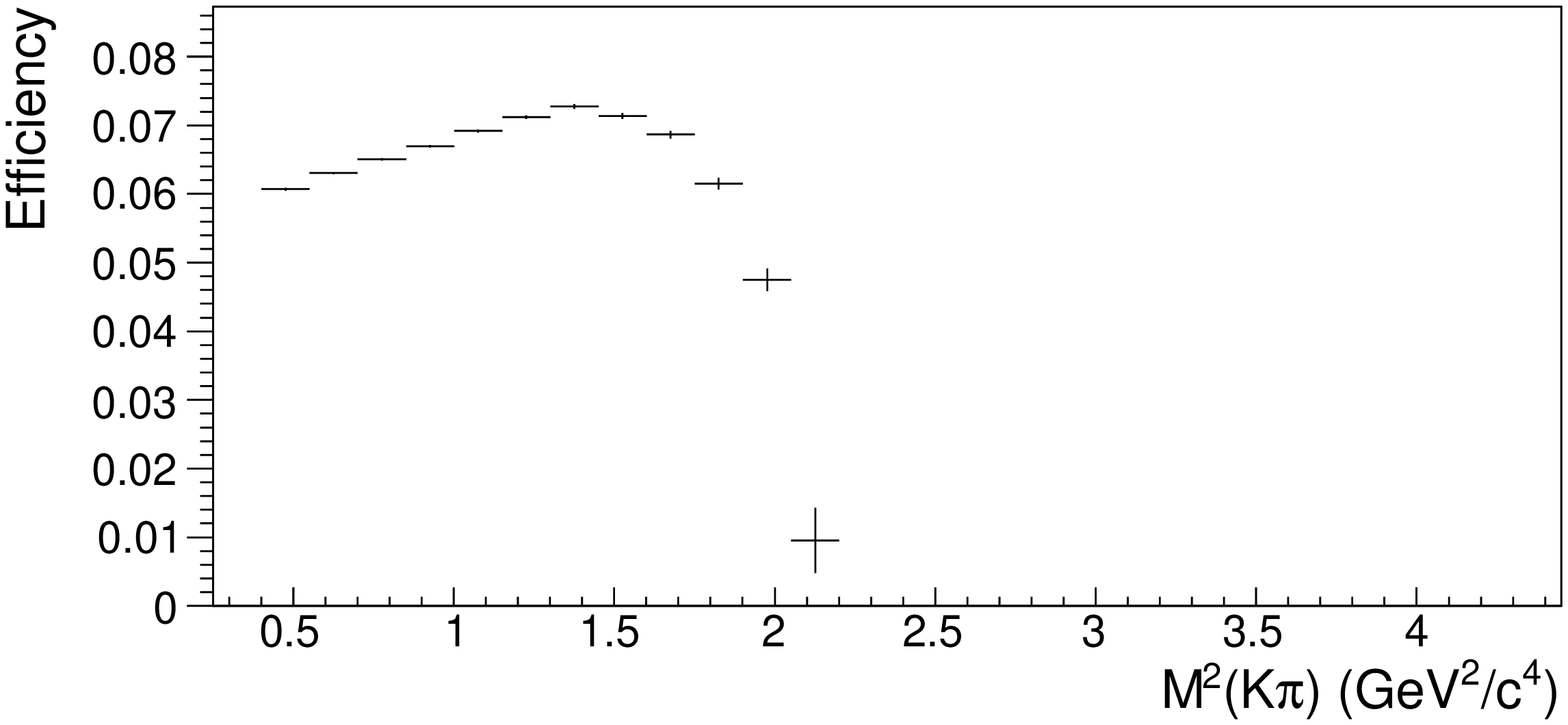}
}}}
\centerline{
\scalebox{0.32}{
\rotatebox{270}{
\includegraphics*[270,37][569,691]
  {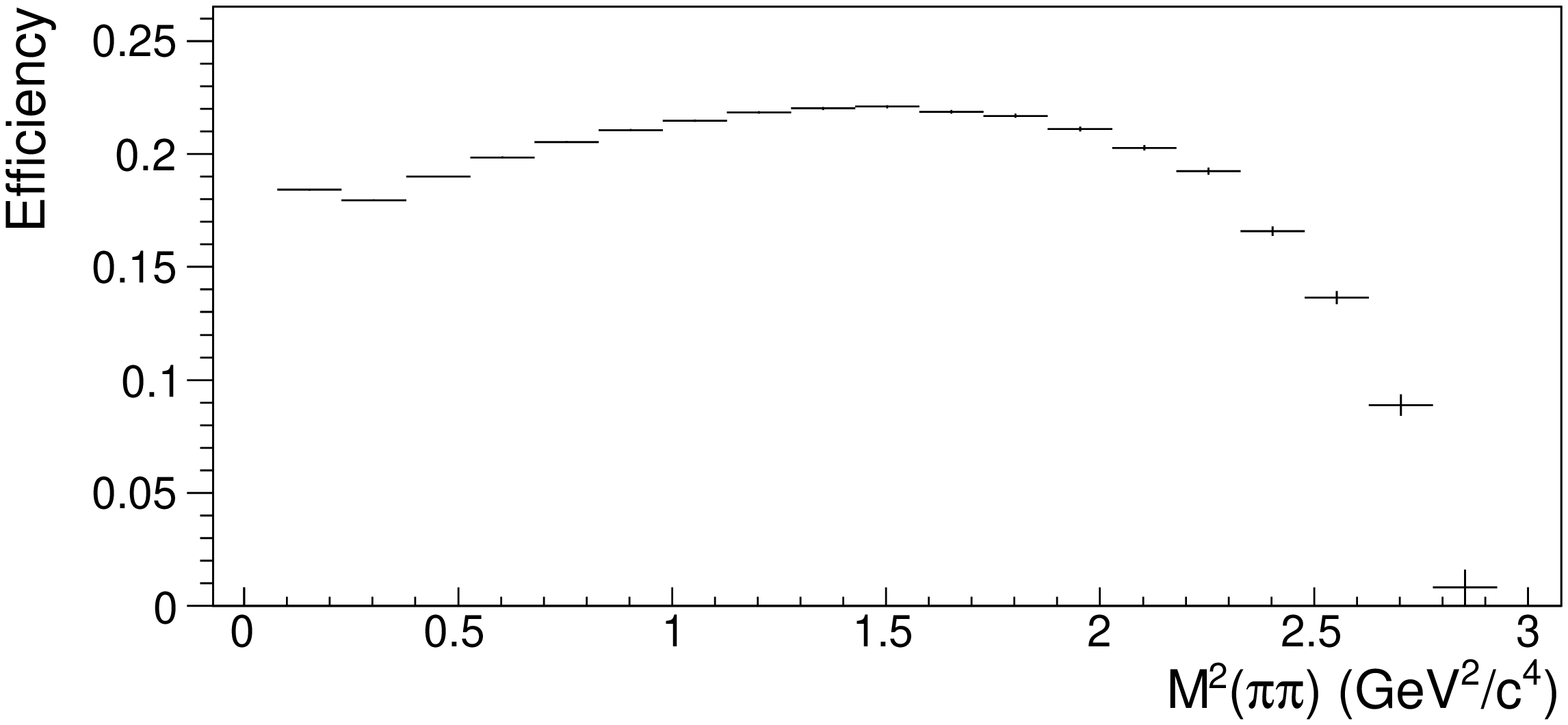}
}}
\hspace{4mm}
\scalebox{0.32}{
\rotatebox{270}{
\includegraphics*[270,37][569,691]
  {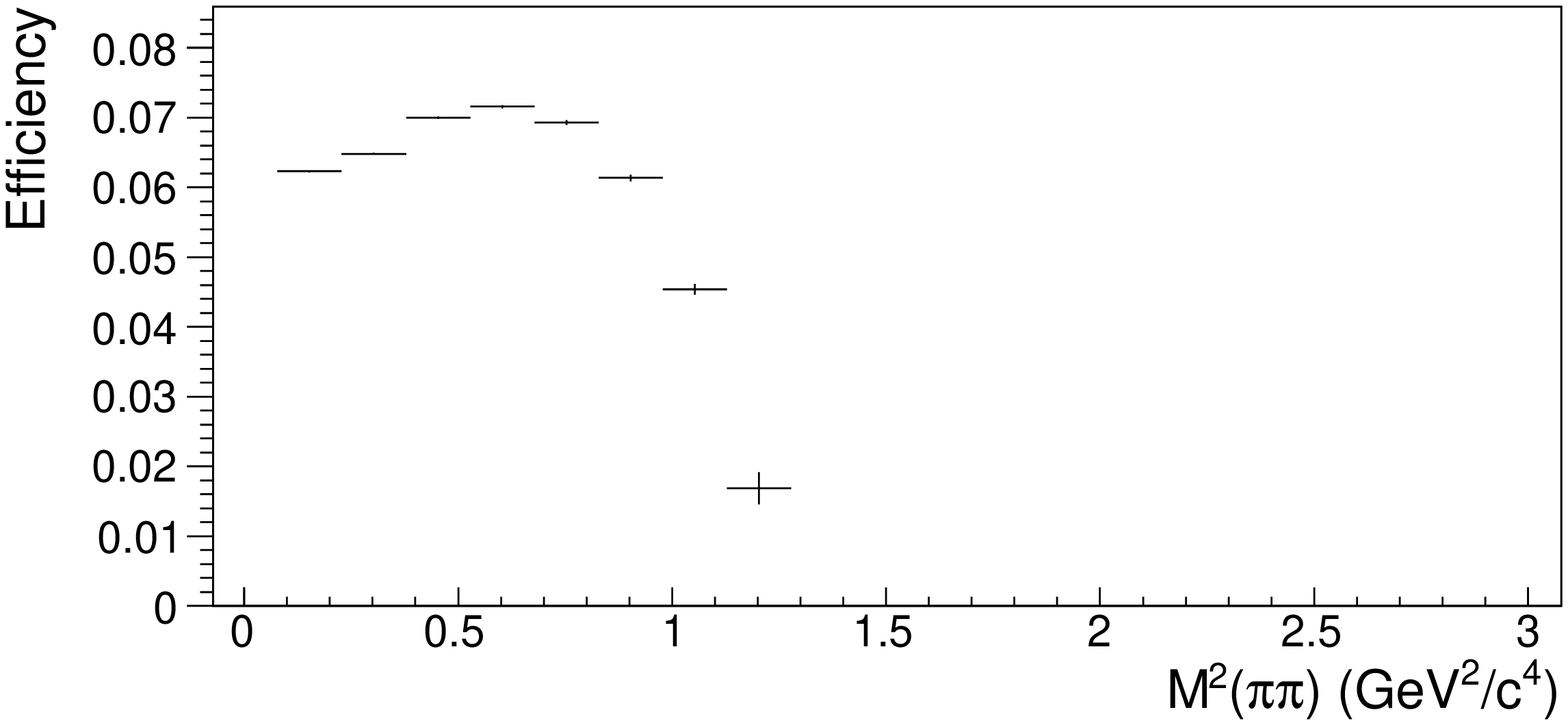}
}}}
\caption{Dependence of the signal efficiency 
  on the kinematic variables for 
  $B^{+} \rightarrow J/\psi K^{+} \pi^{+} \pi^{-}$ (left) and
  $B^{+} \rightarrow \psi^{\prime} K^{+} \pi^{+} \pi^{-}$ (right).}
\label{inclusive:fig_efficiency}
\end{figure*}

\begin{figure*}[hbtp]
\centerline{
\scalebox{0.32}{
\rotatebox{270}{
\includegraphics*[270,37][569,691]
  {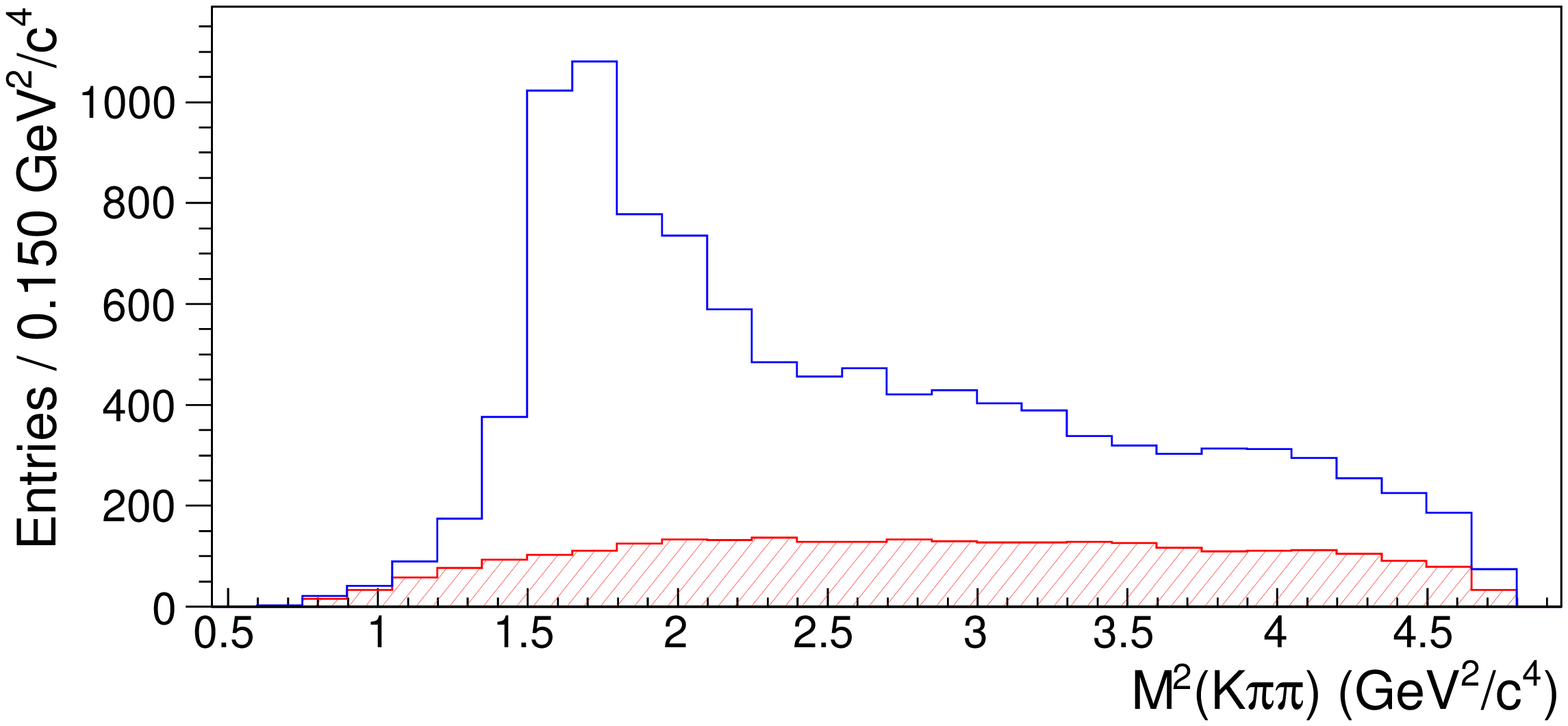}
}}
\hspace{4mm}
\scalebox{0.32}{
\rotatebox{270}{
\includegraphics*[270,37][569,691]
  {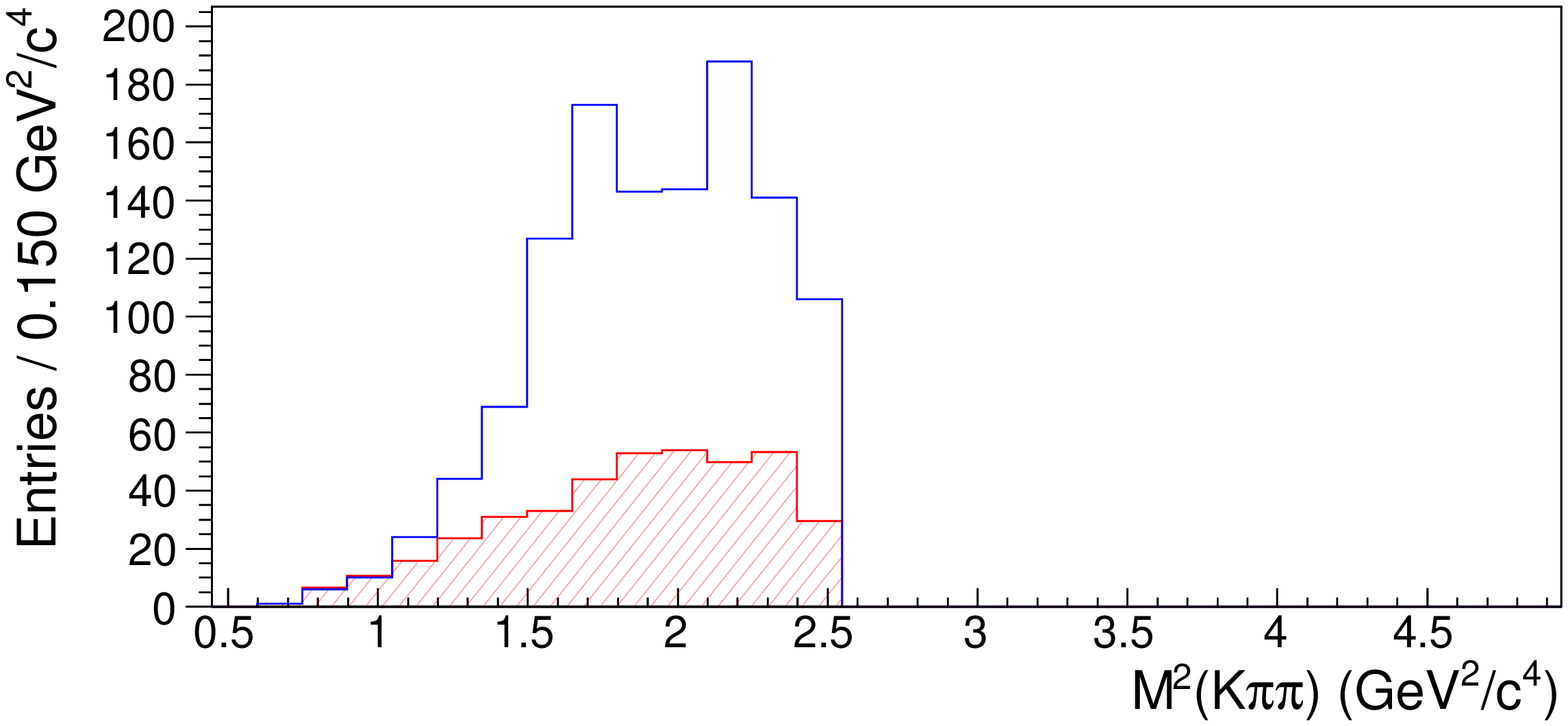}
}}}
\centerline{
\scalebox{0.32}{
\rotatebox{270}{
\includegraphics*[270,37][569,691]
  {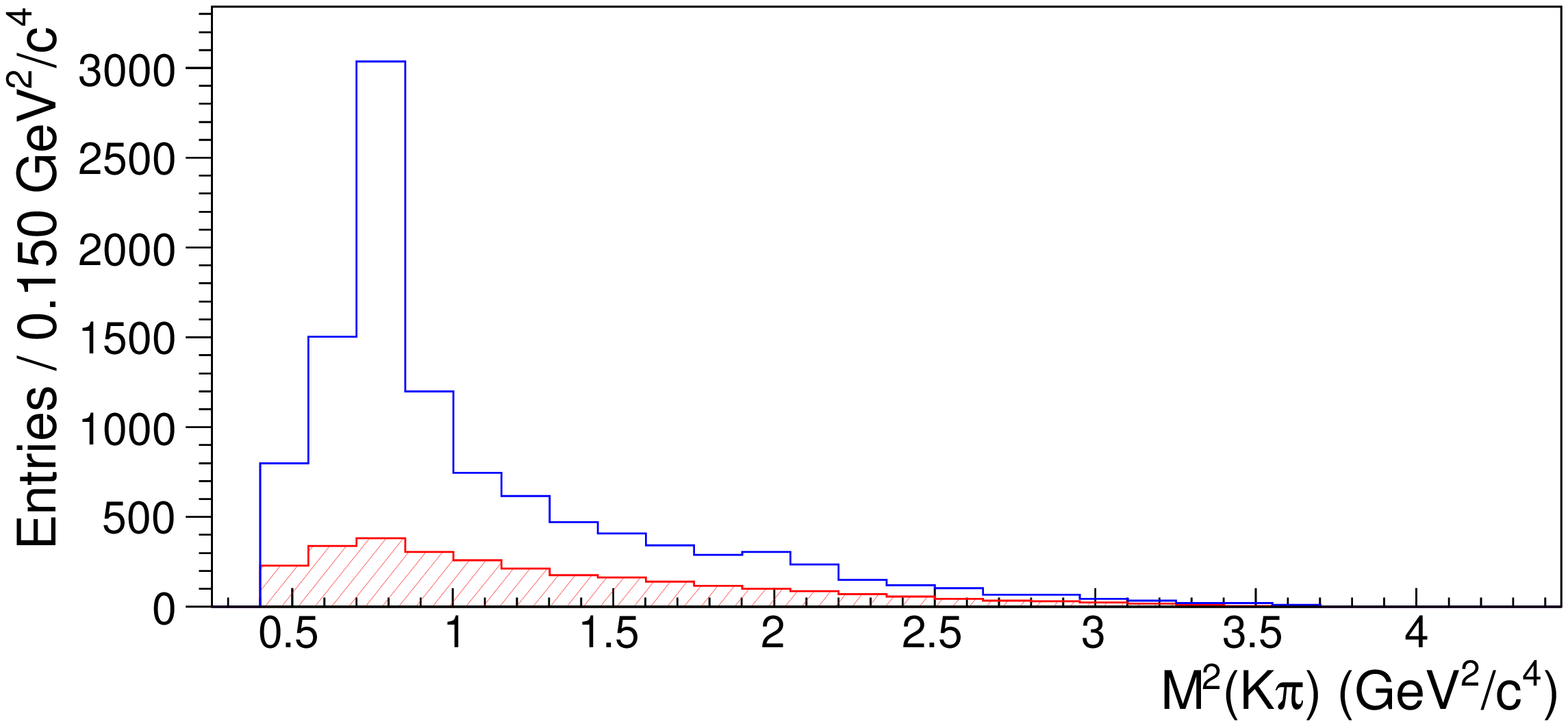}
}}
\hspace{4mm}
\scalebox{0.32}{
\rotatebox{270}{
\includegraphics*[270,37][569,691]
  {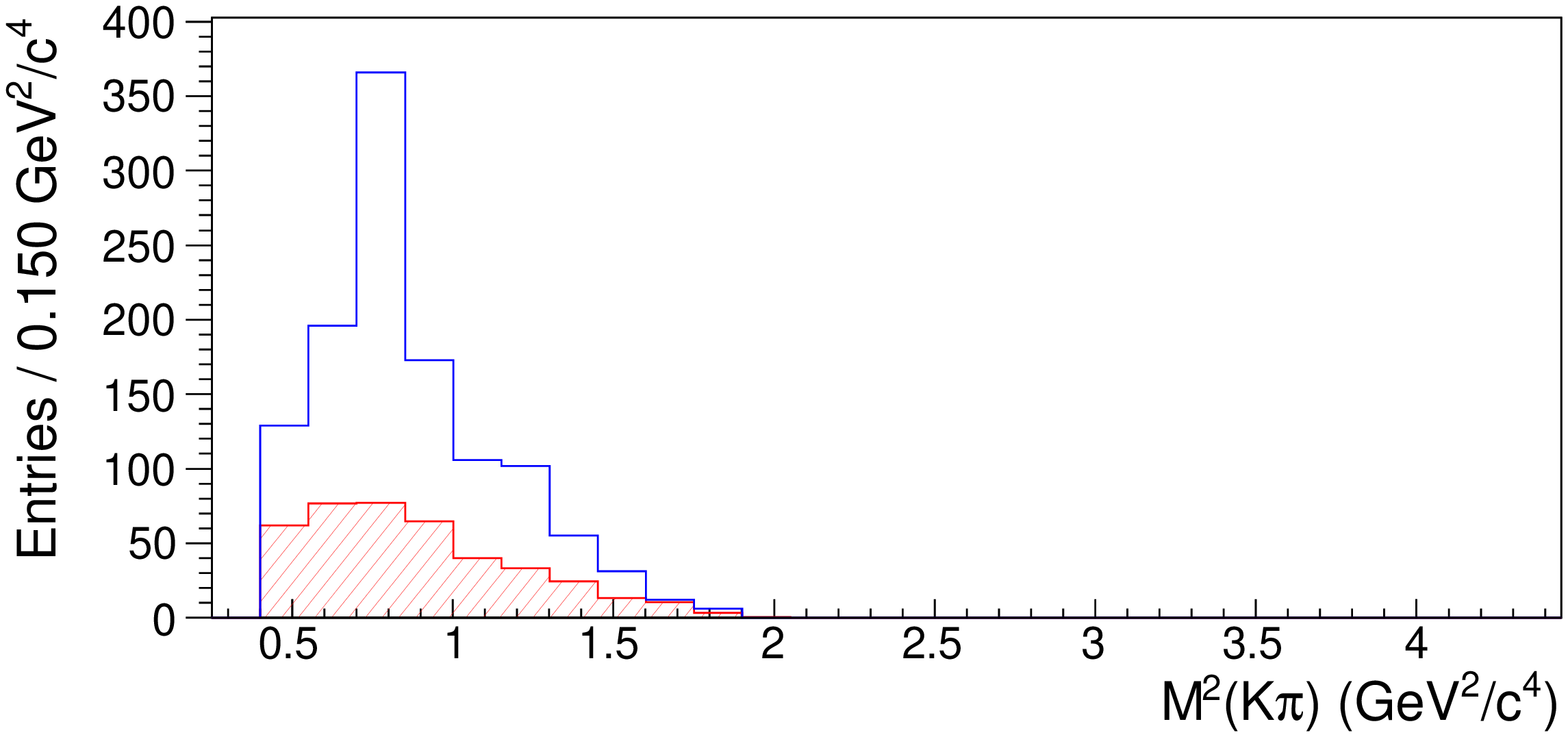}
}}}
\centerline{
\scalebox{0.32}{
\rotatebox{270}{
\includegraphics*[270,37][569,691]
  {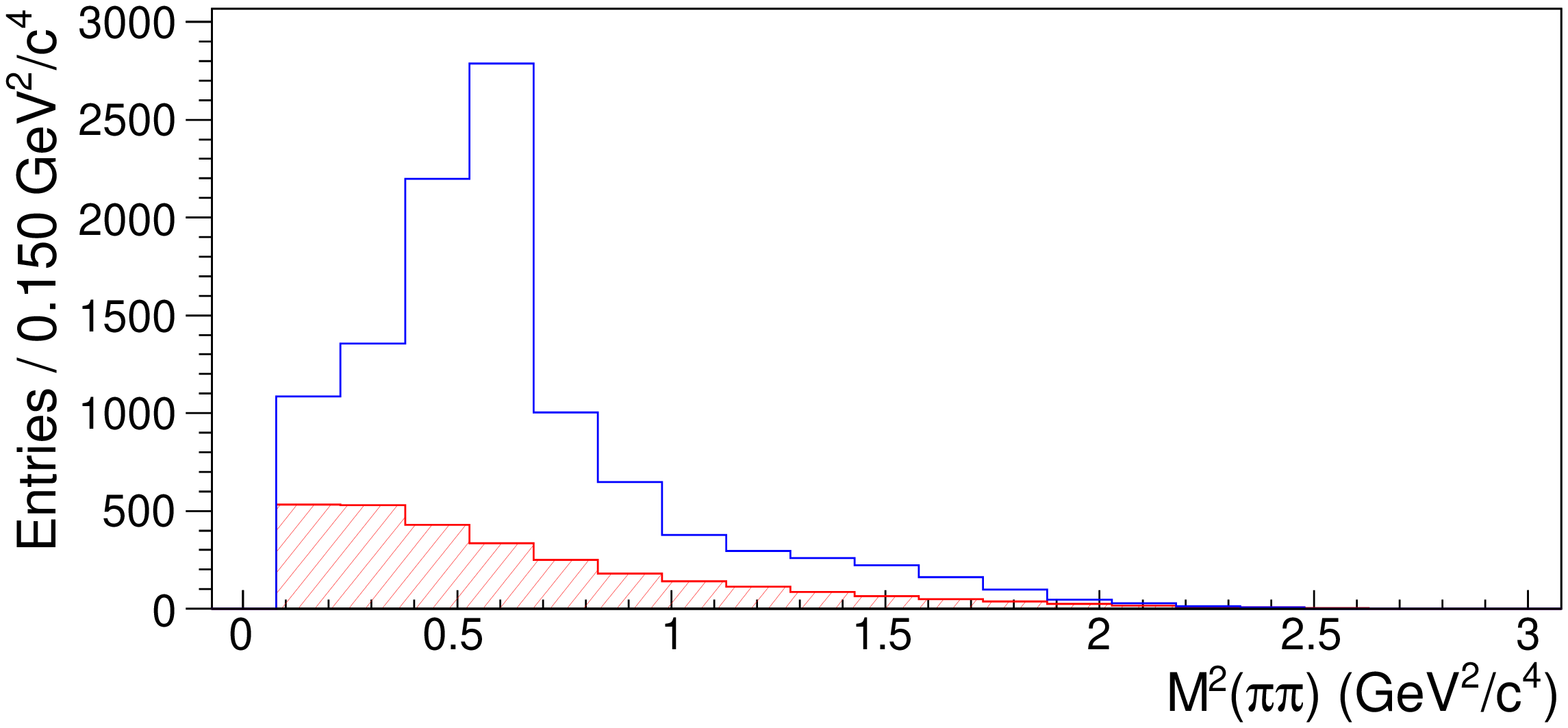}
}}
\hspace{4mm}
\scalebox{0.32}{
\rotatebox{270}{
\includegraphics*[270,37][569,691]
  {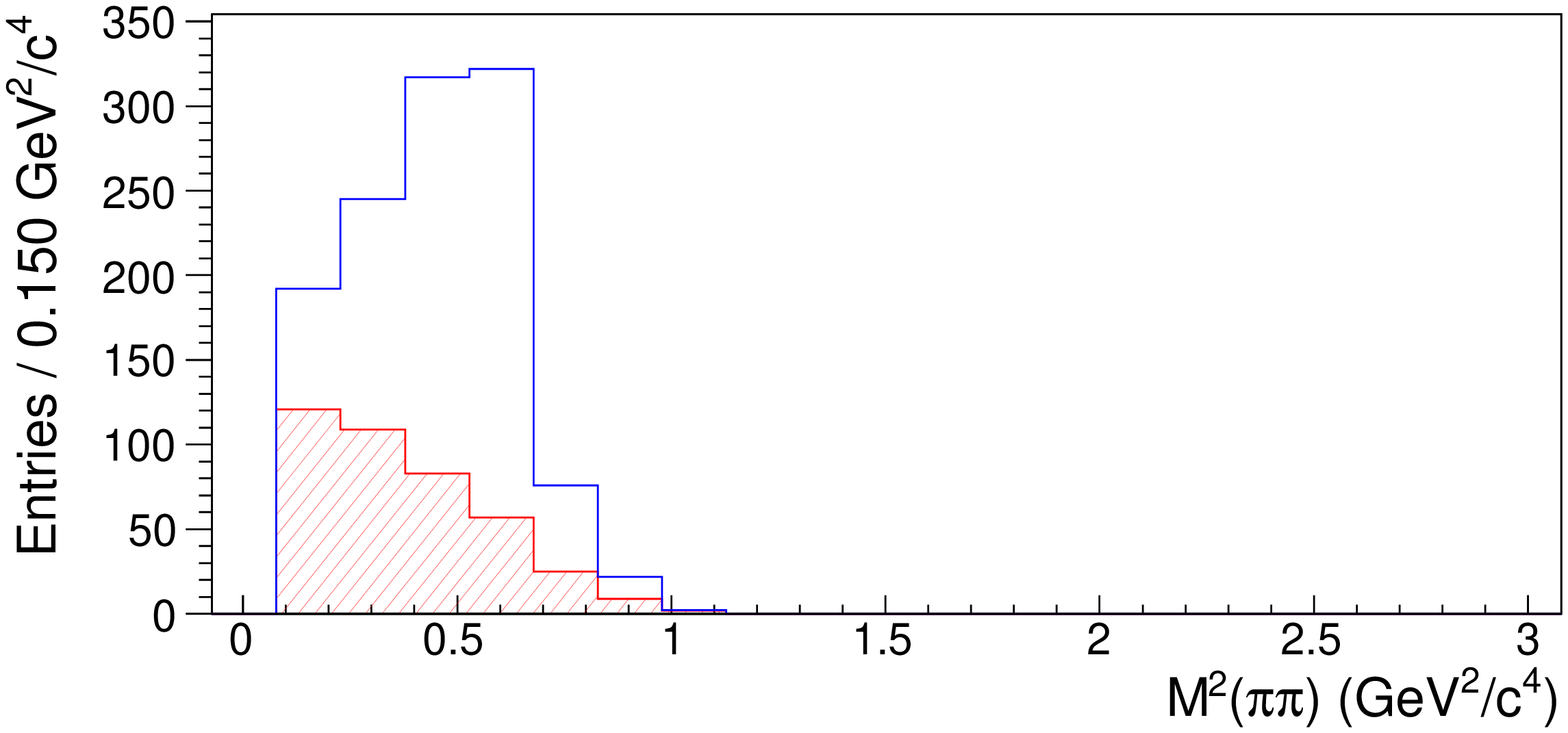}
}}}
\caption{Data distributions of
  $M^{2}(K\pi\pi)$, $M^{2}(K\pi)$, and $M^{2}(\pi\pi)$ 
  for 
  $B^{+} \rightarrow J/\psi K^{+} \pi^{+} \pi^{-}$ (left) and
  $B^{+} \rightarrow \psi^{\prime} K^{+} \pi^{+} \pi^{-}$ (right)
  in the signal region (open histograms) 
  and normalized sideband region (filled histograms).}
\label{inclusive:fig_data}
\end{figure*}

To determine the signal efficiency, 
we generate $10.7 \times 10^{6}$ nonresonant 
$B^{\pm}$ decays to each of the two final states of interest.
We then reconstruct these signal-MC events, 
applying the same event-selection requirements as with data.
We bin the generated events 
according to the generated values of
$M^{2}(K\pi\pi)$, $M^{2}(K\pi)$, and $M^{2}(\pi\pi)$,
and the reconstructed events 
according to the reconstructed
values of $M^{2}(K\pi\pi)$, $M^{2}(K\pi)$, and $M^{2}(\pi\pi)$.
The efficiency in each bin is 
the ratio of reconstructed to generated events in that bin.
Figure~\ref{inclusive:fig_efficiency} shows the dependence of the 
efficiency on the three variables.  
Figure~\ref{inclusive:fig_data} shows the corresponding data
distributions.
The overall efficiency is
$(19.85 \pm 0.01)\%${\footnote{Throughout this paper, 
when a single error is
presented, it is statistical; when two errors are presented,
the first is statistical, and the second is systematic.}}
for $B^{+} \rightarrow J/\psi K^{+} \pi^{+} \pi^{-}$ and
$(6.58 \pm 0.01)\%$ 
for $B^{+} \rightarrow \psi^{\prime} K^{+} \pi^{+} \pi^{-}$.
The number of efficiency-corrected signal events observed is
$(4.14 \pm 0.06) \times 10^{4}$   
for $B^{+} \rightarrow J/\psi K^{+} \pi^{+} \pi^{-}$ and
$(1.12 \pm 0.05) \times 10^{4}$   
for $B^{+} \rightarrow \psi^{\prime} K^{+} \pi^{+} \pi^{-}$.

This method of measuring branching fractions automatically
corrects for efficiency variations over the 
phase space.  It also makes no assumptions as to the 
shape of the signal in $\Delta E$.

\subsection{Systematic errors}
\label{inclusive:section_systematics}

The systematic error in the branching fractions
is estimated by adding in quadrature
various contributions,
which are assumed to be uncorrelated.
Where possible, a correction is applied.

Since we use MC simulation to determine the signal efficiency 
in Eq.~\ref{inclusive:eq_branching_fraction},
any discrepancy in 
signal-re\-con\-struc\-tion efficiency between data and simulation
will result in a systematic error.
Based on studies of the track-reconstruction 
efficiency,
we include
a systematic error of
$1.0\%$ for each lepton track, 
$1.4\%$ for each pion track, 
and $1.2\%$ for each kaon track,
adding linearly.
Based on studies of the lepton-identification 
efficiency,
which show that the simulation underestimates 
the lepton-identification efficiency,
we apply
a correction factor of 
$0.984 \pm 0.019$ for each electron track and
$0.962 \pm 0.031$ for each muon track.
Based on studies of the kaon identification 
efficiency,
we include
a systematic error of
$1\%$ for $B^{+} \rightarrow J/\psi K^{+} \pi^{+} \pi^{-}$ and
$2\%$ for $B^{+} \rightarrow \psi^{\prime} K^{+} \pi^{+} \pi^{-}$.

The bin size of $0.15~{\mathrm{GeV}}^{2}/c^{4}$ 
is chosen based on the dependence 
of the efficiency-corrected signal yield
on the bin size.
The error associated with this choice
is taken to be the rms of the signal yield
in the region between 
$0.1~{\mathrm{GeV}}^{2}/c^{4}$ and $0.2~{\mathrm{GeV}}^{2}/c^{4}$.

\begin{figure}[htbp]
\centerline{
\scalebox{0.34}{
\rotatebox{270}{
\includegraphics*[268,35][570,723]
  {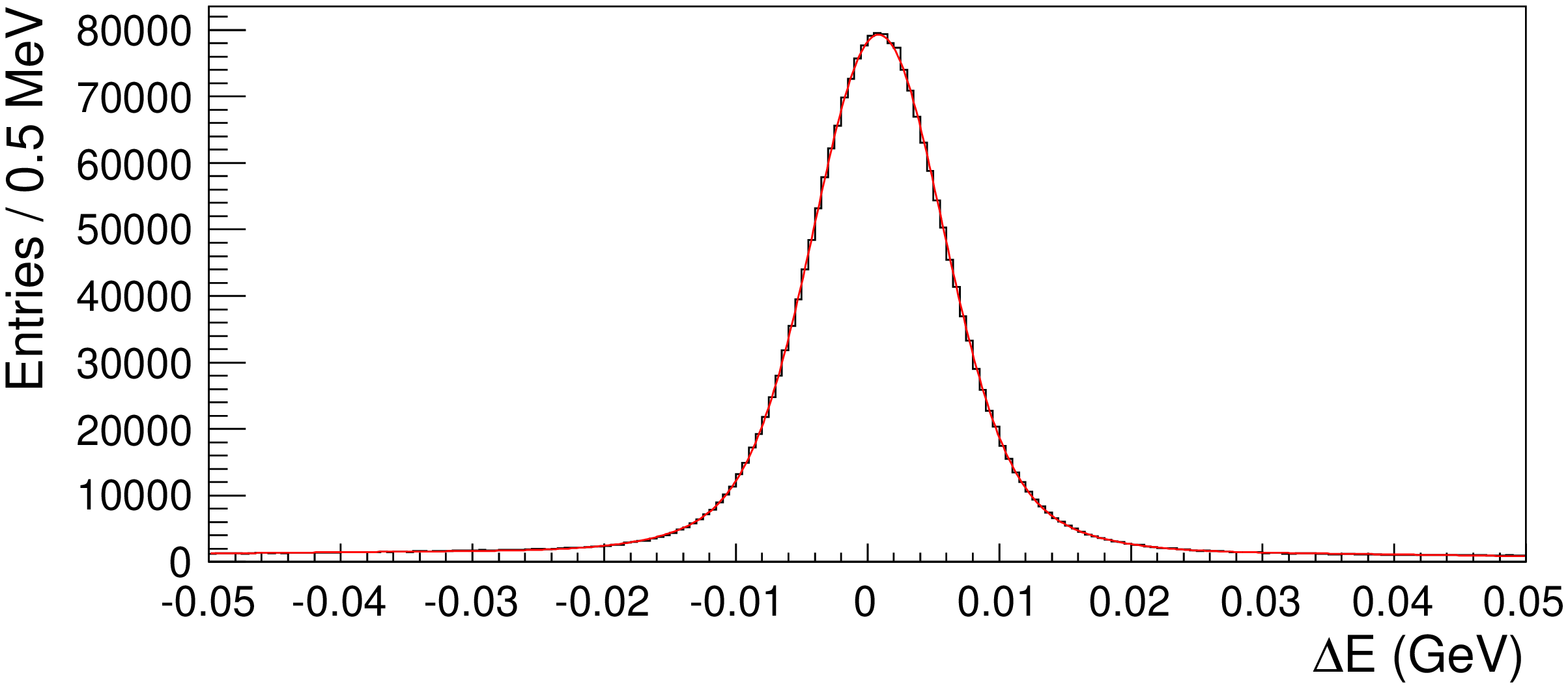}}}}
\vspace{1mm}
\centerline{
\scalebox{0.34}{
\rotatebox{270}{
\includegraphics*[268,35][570,723]
  {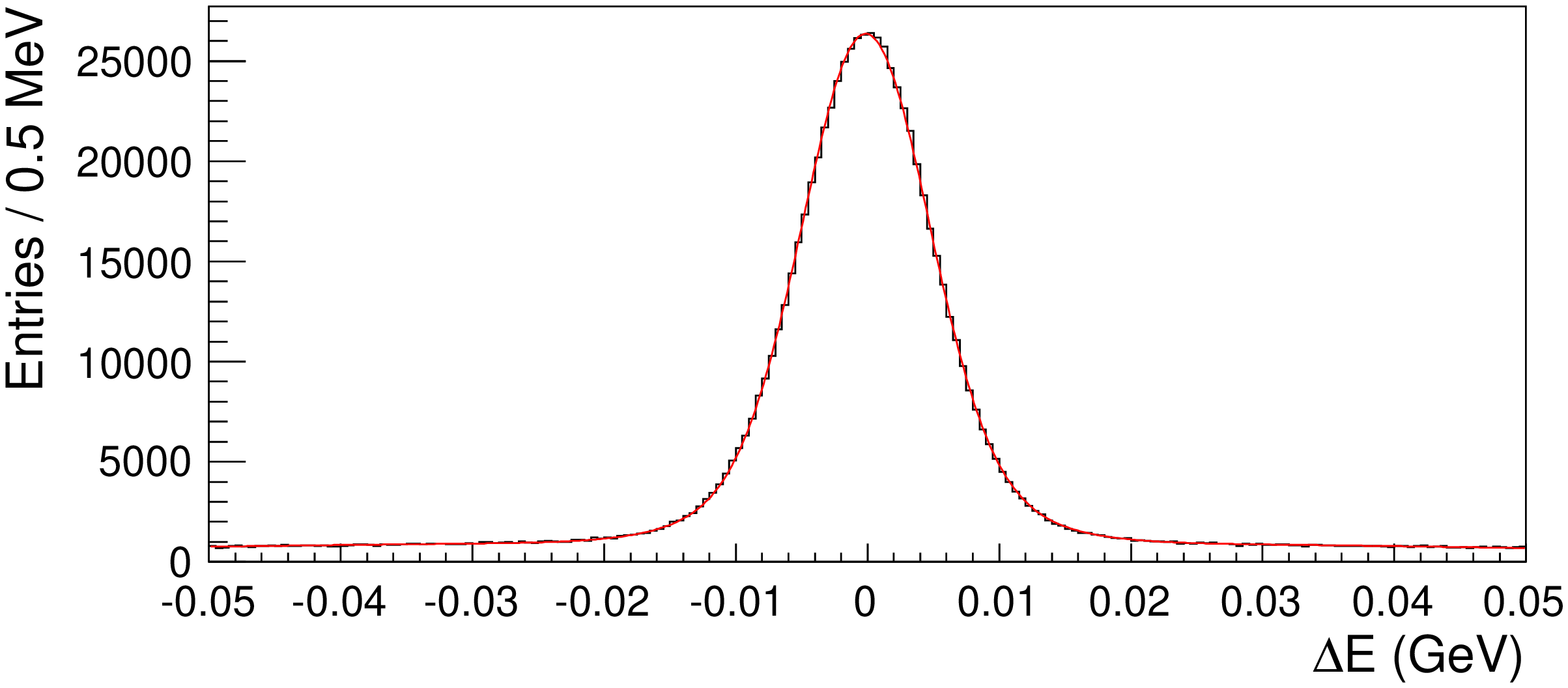}}}}
\caption{MC $\Delta E$ distributions for
  $B^{+} \rightarrow J/\psi K^{+} \pi^{+} \pi^{-}$ (top) and
  $B^{+} \rightarrow \psi^{\prime} K^{+} \pi^{+} \pi^{-}$ (bottom).}
\label{inclusive:fig_de_mc}
\end{figure}

In the $\Delta E$ distributions for
$B^{+} \rightarrow J/\psi K^{+} \pi^{+} \pi^{-}$ and
$B^{+} \rightarrow \psi^{\prime} K^{+} \pi^{+} \pi^{-}$ 
nonresonant MC simulation,
shown in Fig.~\ref{inclusive:fig_de_mc},
a small polynomial background can be seen under the peak.
Since all the events in the MC sample include a signal decay, this 
``background'' is made up of misreconstructed signal events.
Although these events are included as signal in the efficiency
calculation, they are removed by the background-subtraction procedure.
The fraction of signal that is subtracted
in this way is found to be
$(3.75 \pm 0.91)\%$ for 
$B^{+} \rightarrow J/\psi K^{+} \pi^{+} \pi^{-}$, and
$(5.7 \pm 1.2)\%$ for
$B^{+} \rightarrow \psi^{\prime} K^{+} \pi^{+} \pi^{-}$.
The observed branching fractions are corrected for this effect, and the
associated uncertainty
is included as a systematic error.

To determine the sideband normalization factor $f_{B}$ in 
Eq.~\ref{inclusive:eq_nsignal}, the data $\Delta E$ distribution is
fitted as described in Sec.~\ref{events:section_signal_sideband}.
In this fit, the background under the signal is parametrized as a
first-order polynomial.  To estimate the error introduced by this
assumption, a second fit is performed, parametrizing the background 
as a second-order polynomial.  The fractional change in the signal yield
is taken as a systematic error.

Since the signal and sideband regions are 
defined based on the results of 
fitting the data $\Delta E$ distribution,
$\mu_{\Delta E}$ and $\sigma_{\Delta E}$ 
in Eqs.~\ref{events:eq_signal_region} 
and \ref{events:eq_sideband_region}
are varied 
within the fit errors.

In the MC sample used for determining the efficiency, 
$J/\psi$'s from the signal $B$ are forced to decay to  
$e^{+} e^{-}$ or $\mu^{+} \mu^{-}$,
and $\psi^{\prime}$'s from the signal $B$ are forced to decay to  
$e^{+} e^{-}$, $\mu^{+} \mu^{-}$, or $J/\psi \pi^{+} \pi^{-}$.
Thus, 
to obtain branching fractions for 
$B^{+} \rightarrow J/\psi K^{+} \pi^{+} \pi^{-}$
and 
$B^{+} \rightarrow \psi^{\prime} K^{+} \pi^{+} \pi^{-}$,
the branching fractions measured using 
Eq.~\ref{inclusive:eq_branching_fraction} 
are divided by previously-measured values~\cite{pdg08} 
of these $J/\psi$ and $\psi^{\prime}$ decay rates.
The uncertainties of these previous measurements are included as a
systematic error.

Finally, the error in $N_{B}$ is
$1.3\%$.
Table~\ref{inclusive:table_summary} lists the 
components of the systematic errors. 

\begin{table*}[htbp]
\caption{Components of the systematic error in the
   branching-fraction measurements, expressed
   as a percentage of the branching fraction.}
\label{inclusive:table_summary}
\begin{ruledtabular}
\begin{tabular*}{17.8cm}
  {l@{\hspace{1.5cm}}c@{\hspace{2.2cm}}c@{\hspace{2.2cm}}c} 
  Component
  & {$B^{+} \rightarrow J/\psi K^{+} \pi^{+} \pi^{-}$}
  & {$B^{+} \rightarrow \psi^{\prime} K^{+} \pi^{+} \pi^{-}$}
  & {$B^{+} \rightarrow \psi^{\prime} K^{+}$} \\
\colrule
MC statistics                                  & $0.18$ & $0.19$ & $1.12$ \\
Tracking efficiency                            & $6.0$  & $8.7$  & $6.0$  \\
Lepton-ID efficiency                           & $5.1$  & $5.1$  & $5.0$  \\
Kaon-ID efficiency                             & $1.0$  & $2.0$  & $1.0$  \\
Binning                                        & $0.13$ & $0.19$ & $0.26$ \\
Oversubtraction                                & $0.95$ & $1.3$  & $0.95$ \\
Background shape                               & $1.6$  & $1.3$  & $0.19$ \\
Signal/Sideband regions                        & $0.35$ & $4.4$  & $0.27$ \\
$J/\psi$ or $\psi^{\prime}$ branching fraction & $0.71$ & $1.9$  & $0.67$ \\
$N_{B}$                                        & $1.3$  & $1.3$  & $1.3$  \\ 
\end{tabular*}
\end{ruledtabular}
\end{table*}

\subsection{Results}
\label{inclusive:section_results}

The measured branching fractions are 
\begin{eqnarray*}
{\mathcal{B}}(B^{+} \rightarrow J/\psi K^{+} \pi^{+} \pi^{-}) 
 &=&  (7.16 \pm 0.10 \pm 0.60) \times 10^{-4}, \\
{\mathcal{B}}(B^{+} \rightarrow \psi^{\prime} K^{+} \pi^{+} \pi^{-}) 
 &=&  (4.31 \pm 0.20 \pm 0.50) \times 10^{-4}.
\end{eqnarray*}
As a cross-check, we also measure a branching fraction for 
$B^{+} \rightarrow \psi^{\prime} K^{+}$,
using a similar method but reversing
the $\psi^{\prime} \rightarrow J/\psi \pi^{+} \pi^{-}$ veto in
the reconstruction of $B^{+} \rightarrow J/\psi K^{+} \pi^{+} \pi^{-}$.
This branching fraction is
\begin{equation*}
{\mathcal{B}}(B^{+} \rightarrow \psi^{\prime} K^{+}) 
= (6.65 \pm 0.17 \pm 0.55) \times 10^{-4}
,
\end{equation*}
which is consistent with the previously-measured 
value of $(6.48 \pm 0.35) \times 10^{-4}$~\cite{pdg08}.

Our 
$B^{+} \rightarrow J/\psi K^{+} \pi^{+} \pi^{-}$
branching-fraction measurement
represents a significant improvement over 
previous measurements~\cite{pdg08}.
It 
is consistent 
with Ref.~\onlinecite{acosta:2002}
but inconsistent 
with Ref.~\onlinecite{aubert:2005}
at the $3.4$-$\sigma$ level.
Our 
$B^{+} \rightarrow \psi^{\prime} K^{+} \pi^{+} \pi^{-}$
branching-fraction measurement
is also a significant improvement over the 
previous measurement~\cite{albrecht:1990}.

%%%%% Amplitude

\section{Amplitude analyses}
\label{section_amplitude}

To study the resonant structure of the $K^{+}\pi^{+}\pi^{-}$ 
final state 
in
$B^{+} \rightarrow J/\psi K^{+} \pi^{+} \pi^{-}$
and
$B^{+} \rightarrow \psi^{\prime} K^{+} \pi^{+} \pi^{-}$,
we perform amplitude analyses.
Using an unbinned maximum-likelihood method,
we simultaneously fit the data in the three dimensions
$M^{2}(K\pi\pi)$, $M^{2}(K\pi)$, and $M^{2}(\pi\pi)$.

\subsection{Fitting technique}
\label{amplitude:section_fitting_technique}

Signal-region data are fitted by 
maximizing{\footnote{Standalone MINUIT~\cite{minuit}
is used for all maximizations in this section.}}
the log-likelihood function, which is given by
\begin{equation}
\label{amplitude:eq_likelihood}
\ell (\vec a)
= \sum_{i} \ln p (\vec x_{i}; \vec a)
,
\end{equation}
where the sum is over the events in the signal region, 
$\vec x_{i}$ is the vector of coordinates for a given event (i.e., 
$\vec x \equiv [M^{2}(K\pi\pi), M^{2}(K\pi), M^{2}(\pi\pi)]$),
$\vec a$ is the vector of parameters 
with respect to which $\ell$ is maximized, 
and
$p$ is the probability-density function (PDF) 
that is used to model the observed distribution.

The distribution of events in the signal region is modeled as
\begin{equation}
\label{amplitude:eq_pdf}
p(\vec x; \vec a) = n_{B} \, \frac{p_{B}(\vec x)}
                                  {\int p_{B}(\vec x) \, d^{3}x}
                  + n_{S} \, \frac{p_{S}(\vec x; \vec a)}
                                  {\int p_{S}(\vec x; \vec a) \, d^{3}x}
,
\end{equation}
where $p_{B}$ and $p_{S}$ describe the observed shapes of the 
background and signal, respectively.
The constants $n_{B}$ and $n_{S}$ are the background and signal
fractions in the signal region;
the former is given by Eq.~\ref{events:eq_n_b},
and the latter is $1 - n_{B}$.{\footnote{The background
fraction $n_{B}$ is corrected for the oversubtraction effect described
in Sec.~\ref{inclusive:section_systematics}.}} 

The observed signal distribution $p_{S}$ is expressed as
\begin{equation}
\label{amplitude:eq_signal}
p_{S}(\vec x; \vec a) = \varepsilon(\vec x) 
                        \phi(\vec x) 
                        s(\vec x; \vec a)
,
\end{equation}
where $\varepsilon$ is the detector efficiency,
$\phi$ is the phase-space density,
and $s$ is the raw signal function.

Using nonresonant MC simulation, 
we have measured the detector resolution 
to be approximately $3$-$4~{\mathrm{MeV}/c^{2}}$ 
in each of the three coordinates
$M(K\pi\pi)$, $M(K\pi)$, and $M(\pi\pi)$.
Since this is smaller than the width of 
any resonance included in the fits, 
we neglect the effect of detector resolution on line shapes.

The following five sections 
describe the methods followed 
in performing
the integrals of Eq.~\ref{amplitude:eq_pdf}
and in obtaining the functions
$p_{B}(\vec x)$,
$\varepsilon(\vec x)$, 
$\phi(\vec x)$, 
and $s(\vec x; \vec a)$
in Eqs.~\ref{amplitude:eq_pdf}
and \ref{amplitude:eq_signal}.

\subsection{Normalization procedure}
\label{amplitude:section_normalization}

\begin{figure}[hbtp]
\centerline{
\scalebox{0.35}{
\rotatebox{270}{
\includegraphics*[58,37][566,564]
  {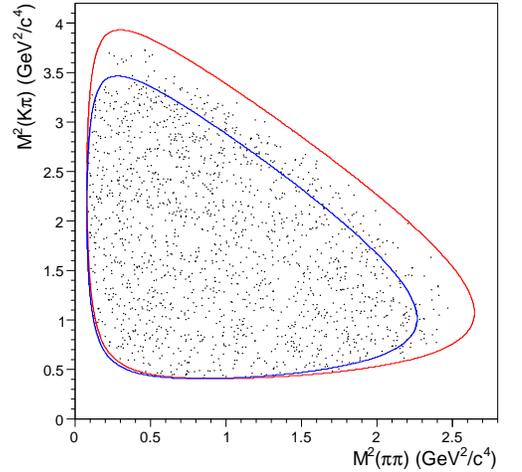}
}}}
\caption{Scatterplot of 
  $M^{2}(K\pi)$ versus $M^{2}(\pi\pi)$ for 
  $M^{2}(K\pi\pi)$ between
  $4.0$~GeV$^{2}/c^{4}$ and $4.5$~GeV$^{2}/c^{4}$
  in sideband data for
  $B^{+} \rightarrow J/\psi K^{+} \pi^{+} \pi^{-}$.
  Blue and red curves show the calculated boundaries corresponding
  to the low and high edge, respectively,
  of the plotted $M^{2}(K\pi\pi)$ region.}
\label{amplitude:fig_boundary}
\end{figure}

The integrations of Eq.~\ref{amplitude:eq_pdf} 
are performed
numerically, using Simpson's rule.
A step size of 
$0.010~{\mathrm{GeV}^{2}/c^{4}}$ 
for $B^{+} \rightarrow J/\psi K^{+} \pi^{+} \pi^{-}$ and
$0.005~{\mathrm{GeV}^{2}/c^{4}}$ 
for $B^{+} \rightarrow \psi^{\prime} K^{+} \pi^{+} \pi^{-}$
is used in each dimension.{\footnote{The larger step size is necessary 
for $B^{+} \rightarrow J/\psi K^{+} \pi^{+} \pi^{-}$ because of the 
larger phase space, which significantly increases 
the CPU time required for the integration.}}
The three-dimensional region of integration can be determined by noting
that the minimum and maximum values of $M^{2}(K\pi\pi)$ are given by
\begin{eqnarray}
M^{2}(K\pi\pi)_{\mathrm{min}} &=& (M_{K} + 2 M_{\pi})^{2}, \\
M^{2}(K\pi\pi)_{\mathrm{max}} &=& (M_{B} - M_{\psi})^{2},
\end{eqnarray}
where $M_{B}$, $M_{K}$, $M_{\pi}$, and $M_{\psi}$ are the nominal 
values of the subscripted particles.
For a given value of $M^{2}(K\pi\pi)$, the minimum and maximum values of
$M^{2}(\pi\pi)$ are
\begin{eqnarray}
M^{2}(\pi\pi)_{\mathrm{min}} &=& (2 M_{\pi})^{2}, \\
M^{2}(\pi\pi)_{\mathrm{max}} 
&=& \left( \sqrt{M^{2}(K\pi\pi)} - M_{K} \right)^{2}.
\end{eqnarray}
For given $M^{2}(K\pi\pi)$ and $M^{2}(\pi\pi)$, the minimum
and maximum values of $M^{2}(K\pi)$ are
\begin{eqnarray}
\label{amplitude:eq_boundary}
M^{2}(K\pi)_{\mathrm{min}}^{\mathrm{max}} 
& = & \frac{1}{2} \Biggl [  M^{2}(K\pi\pi)
                          + M_{K}^{2}
                          + 2 M_{\pi}^{2}
                          - M^{2}(\pi\pi)
                          \Biggr .
 \nonumber\\
{} & \pm & 
 \sqrt{ 1 - \left( 2 M_{\pi} / M(\pi\pi) \right)^{2} }
 \nonumber\\
{} & \times &
 \sqrt{ M^{2}(K\pi\pi) - (M_{K} + M(\pi\pi) )^{2} }
 \nonumber\\
{} & \times & 
 \sqrt{ M^{2}(K\pi\pi) - (M_{K} - M(\pi\pi) )^{2} }
 \Biggl .
  \Biggr ].
\end{eqnarray}
Figure~\ref{amplitude:fig_boundary} shows 
the calculated kinematic boundaries for
$B^{+} \rightarrow J/\psi K^{+} \pi^{+} \pi^{-}$,
along with the observed distributions of sideband data, 
for a slice in $M^{2}(K\pi\pi)$.

Events that do not fall within the calculated boundaries are excluded
from the fits.
Because of the coordinate transformations of 
Sec.~\ref{section_transformations},
such events are rare:
$36$ of the $12913$ sideband events and
 $3$ of the $10594$ signal-region events 
for $B^{+} \rightarrow J/\psi K^{+} \pi^{+} \pi^{-}$, and
$3$ of the $2230$ sideband events and
none of the $1176$ signal-region events 
for $B^{+} \rightarrow \psi^{\prime} K^{+} \pi^{+} \pi^{-}$
fall outside the boundaries.

\subsection{Background functions}
\label{amplitude:section_bkg_functions}

To determine the three-dimensional shape of the background
in the signal region
(i.e., $p_{B}(\vec x)$ in Eq.~\ref{amplitude:eq_pdf}),
an unbinned maximum-likelihood fit is performed 
on the sideband-region data.
The log-likelihood function to maximize is given in this case by
\begin{equation}
\ell_{B} (\vec a_{B})
= \sum_{j} \ln \frac{p_{B} (\vec x_{j}; \vec a_{B})}
                    {\int p_{B}(\vec x; \vec a_{B}) \, d^{3}x}
,
\end{equation}
where the sum is over the events in the sideband region.  
The maximization is performed by varying the parameters 
$\vec a_{B}$, which are then fixed at their optimal
values in fitting the signal region.

The background is modeled as 
the sum of a combinatorial term and a set of noninterfering resonances.
For 
$B^{+} \rightarrow J/\psi K^{+} \pi^{+} \pi^{-}$,
\begin{eqnarray}
\label{amplitude:eq_bkg_jkpp}
\lefteqn{ p_{B}(\vec x; \vec a_{B})  = 
 \Biggl[ \, \sum_{i = 0}^{5} a_{xi} T_{i}(x) \, \Biggr] } 
 \nonumber \\
&& {} \times
 \Biggl[ \, \sum_{j = 0}^{1} a_{yj} T_{j}(y) \, \Biggr] \times
 \Biggl[ \, \sum_{k = 0}^{2} a_{zk} T_{k}(z) \, \Biggr] 
 \nonumber \\
&& {} + e^{-2 x} 
        \Bigl [ \Bigr . 
           a_{K^{*}(892)}  BW_{K^{*}(892)} (y) 
         + a_{\rho}        BW_{\rho}       (z) 
 \nonumber \\ 
&& {~~~~~~~~~~} 
         + a_{D}           G_{D}           (y) 
         + a_{K_{S}}       G_{K_{S}}       (z) 
        \Bigl . \Bigr ] ,
\end{eqnarray}
and for $B^{+} \rightarrow \psi^{\prime} K^{+} \pi^{+} \pi^{-}$,
\begin{eqnarray}
\label{amplitude:eq_bkg_pkpp}
\lefteqn{ p_{B}(\vec x; \vec a_{B}) =
 \Biggl[ \, \sum_{i = 0}^{3} a_{xi} T_{i}(x) \, \Biggr] }
 \nonumber \\ 
&& {} + e^{-2 x}
        \Bigl [ \Bigr .   
           a_{K^{*}(892)}  BW_{K^{*}(892)} (y)
         + a_{\rho}        BW_{\rho}       (z) 
        \Bigl . \Bigr ]
.
\end{eqnarray}
In Eqs.~\ref{amplitude:eq_bkg_jkpp} and \ref{amplitude:eq_bkg_pkpp},
$T_{n}$ represents an $n$th-order Chebyshev 
polynomial.
The variables $x$, $y$, and $z$ stand for
$M^{2}(K\pi\pi)$, $M^{2}(K\pi)$, and $M^{2}(\pi\pi)$,
respectively, and are defined over the intervals
\begin{align*}
  x_{\mathrm{min}} &= (M_{K} + 2 M_{\pi})^{2}\,, 
& x_{\mathrm{max}} &= (M_{B} - M_{\psi})^{2}\,,        \\
  y_{\mathrm{min}} &= (M_{K} + M_{\pi})^{2}\,, 
& y_{\mathrm{max}} &= (M_{B} - M_{\psi} - M_{\pi})^{2}\,,        \\
  z_{\mathrm{min}} &= (2 M_{\pi})^{2}\,, 
& z_{\mathrm{max}} &= (M_{B} - M_{\psi} - M_{K})^{2}\,.
\end{align*}
The peak functions 
$BW_{K^{*}(892)}$, $G_{D}$, $G_{K_{S}}$, and $BW_{\rho}$
are obtained as described in
Sec.~\ref{section_transformations}.
Each peak function $P(\vec x)$ is normalized 
over the kinematically-allowed phase space 
to satisfy
\begin{equation}
1 = \int e^{-2 x} P(\vec x) d^{3}x .
\end{equation}
The factor of $e^{-2x}$ that modulates the peak functions 
was found empirically to produce a good fit to the sideband 
data.{\footnote{Since there are more low-energy particles 
than high-energy particles, 
the background peaks are more pronounced at low $M^{2}(K\pi\pi)$.  
Combining a $K^{*}(892)$ with a random pion, for example, will
tend to produce a low value for $M^{2}(K\pi\pi)$.}}

\begin{figure*}[hbtp]
\centerline{
\scalebox{0.35}{
\rotatebox{270}{
\includegraphics*[270,37][569,691]
  {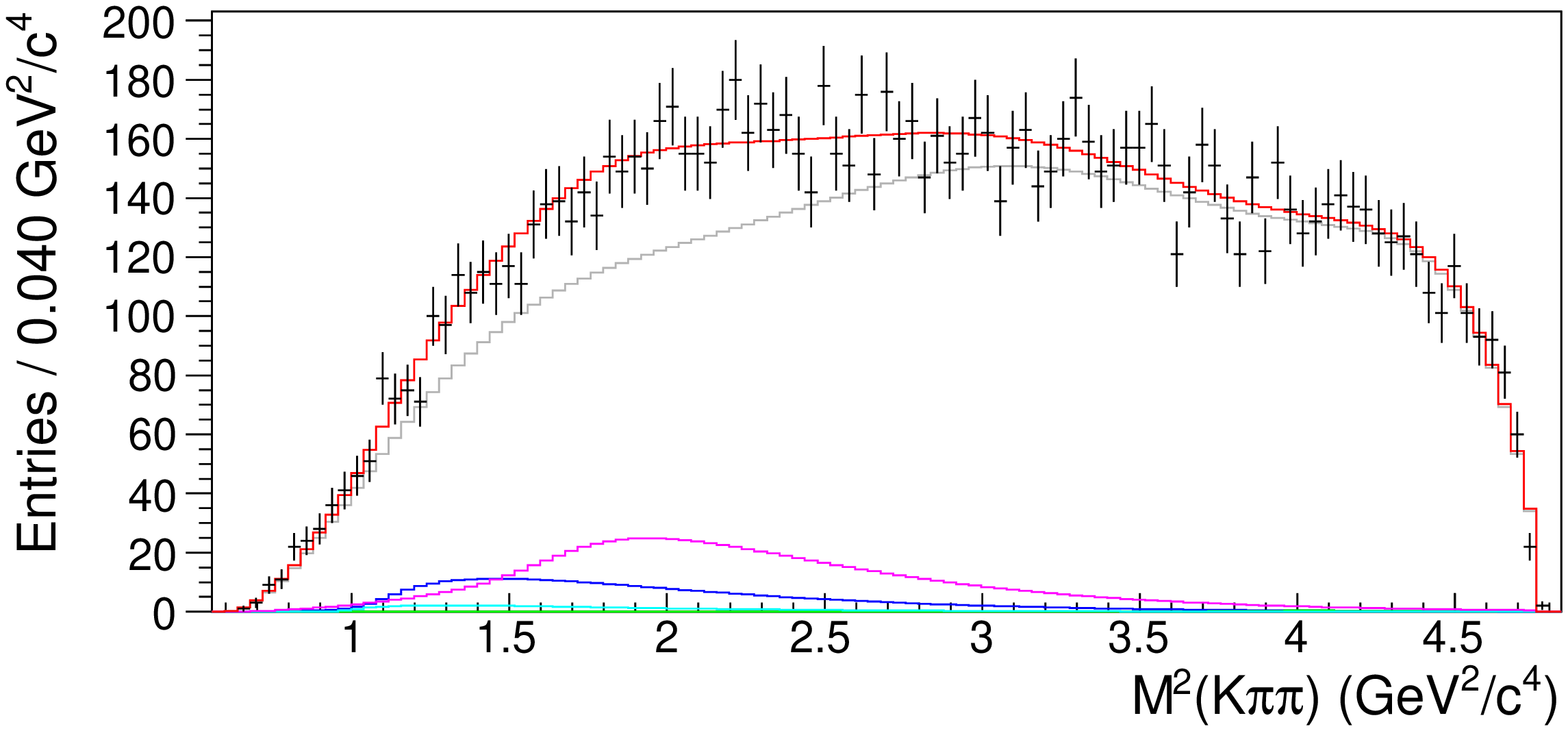}
}}
\hspace{4mm}
\scalebox{0.35}{
\rotatebox{270}{
\includegraphics*[270,37][569,691]
  {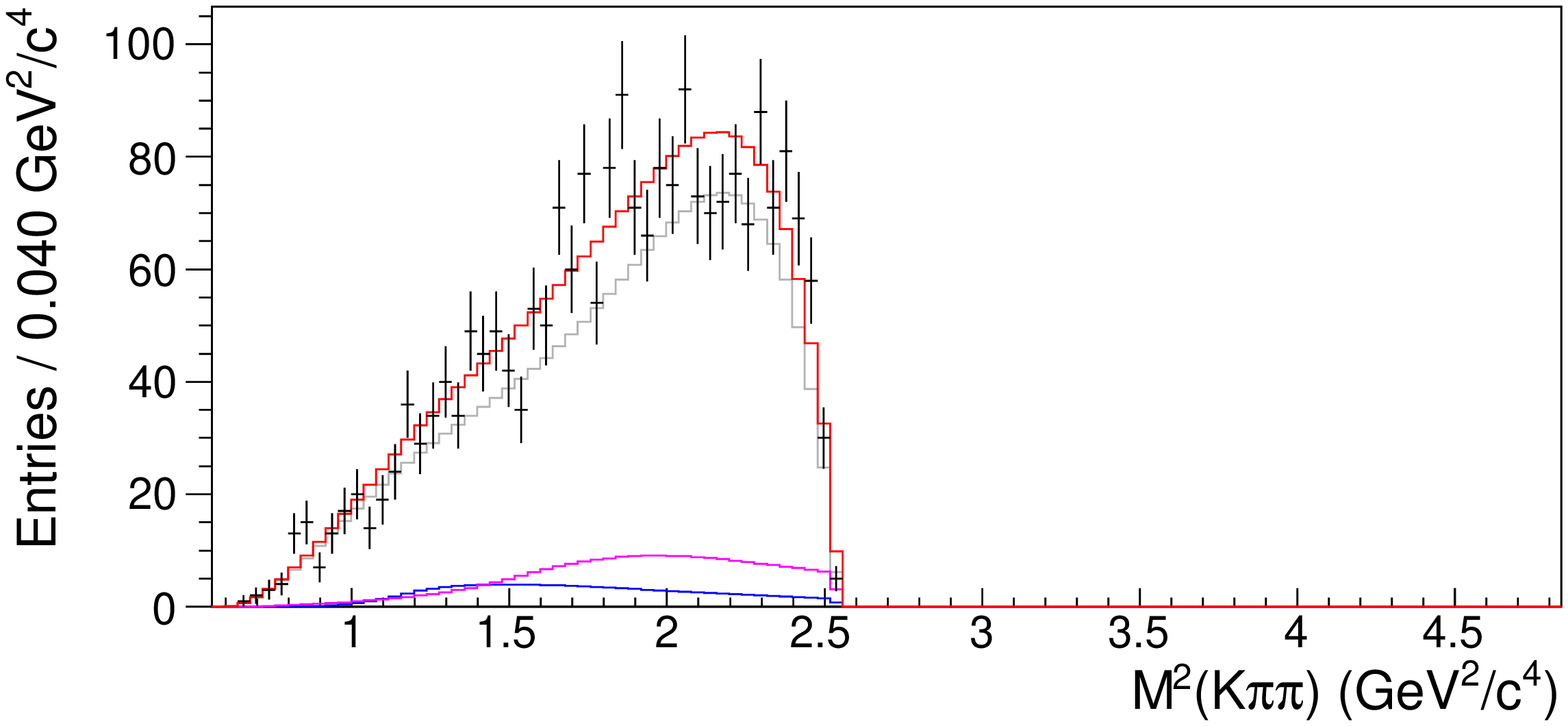}
}}}
\centerline{
\scalebox{0.35}{
\rotatebox{270}{
\includegraphics*[270,37][569,691]
  {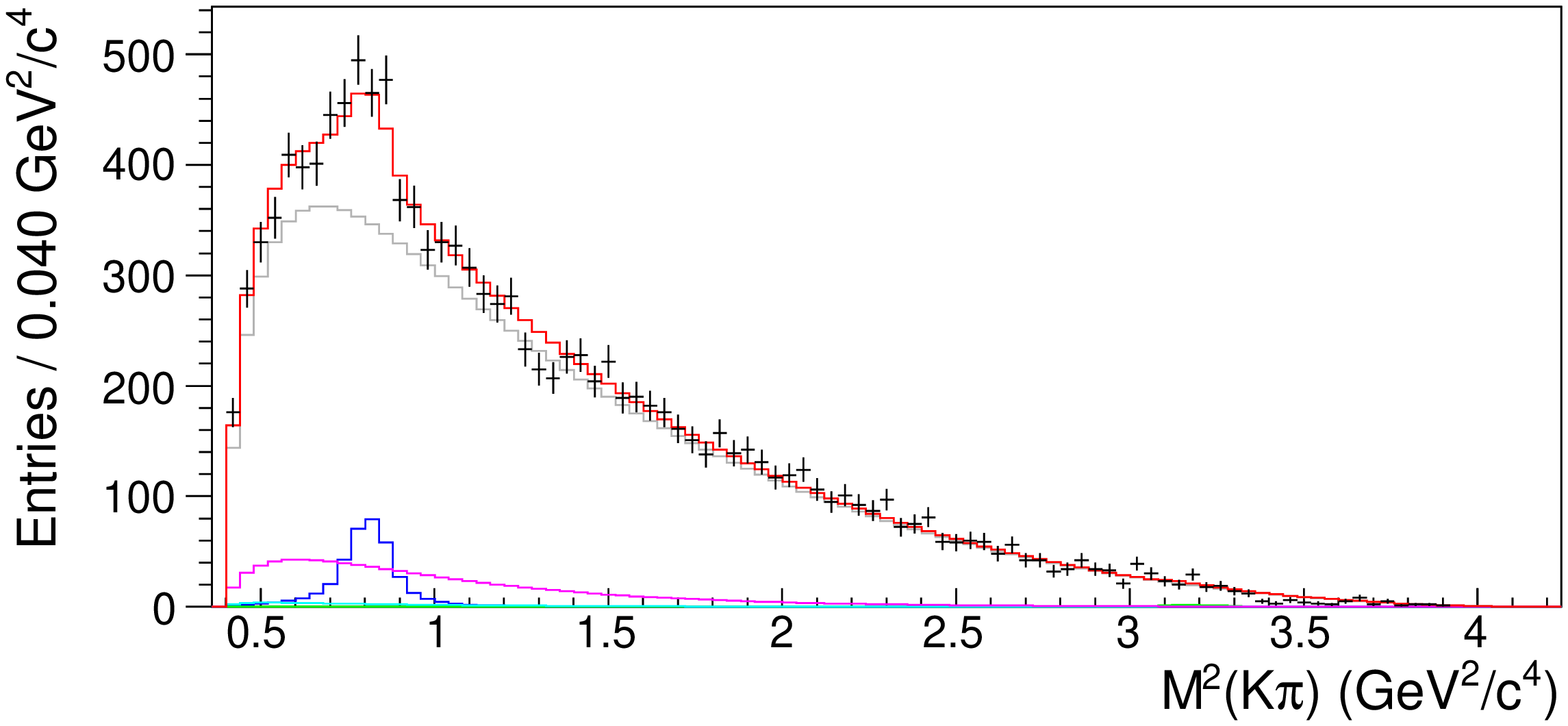}
}}
\hspace{4mm}
\scalebox{0.35}{
\rotatebox{270}{
\includegraphics*[270,37][569,691]
  {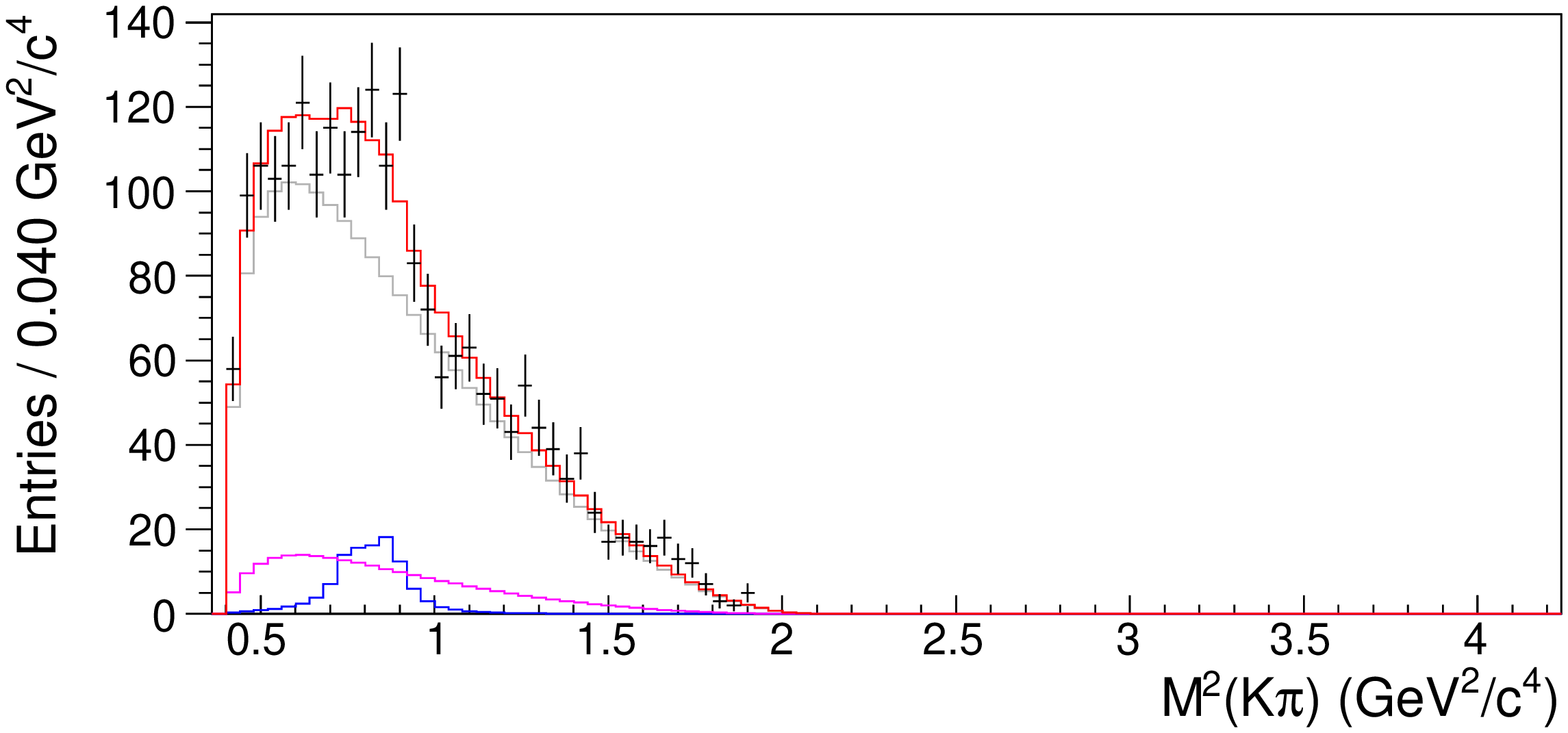}
}}}
\centerline{
\scalebox{0.35}{
\rotatebox{270}{
\includegraphics*[270,37][569,691]
  {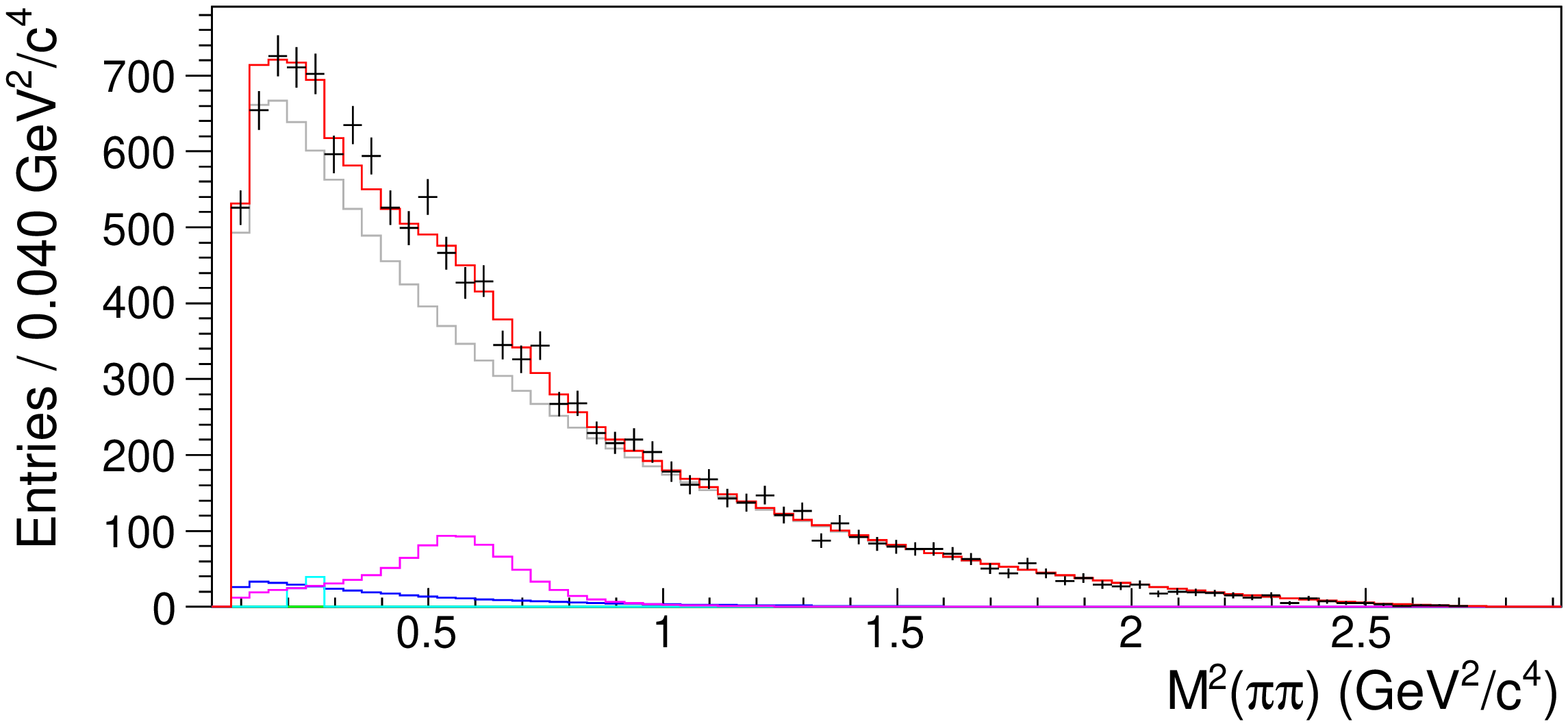}
}}
\hspace{4mm}
\scalebox{0.35}{
\rotatebox{270}{
\includegraphics*[270,37][569,691]
  {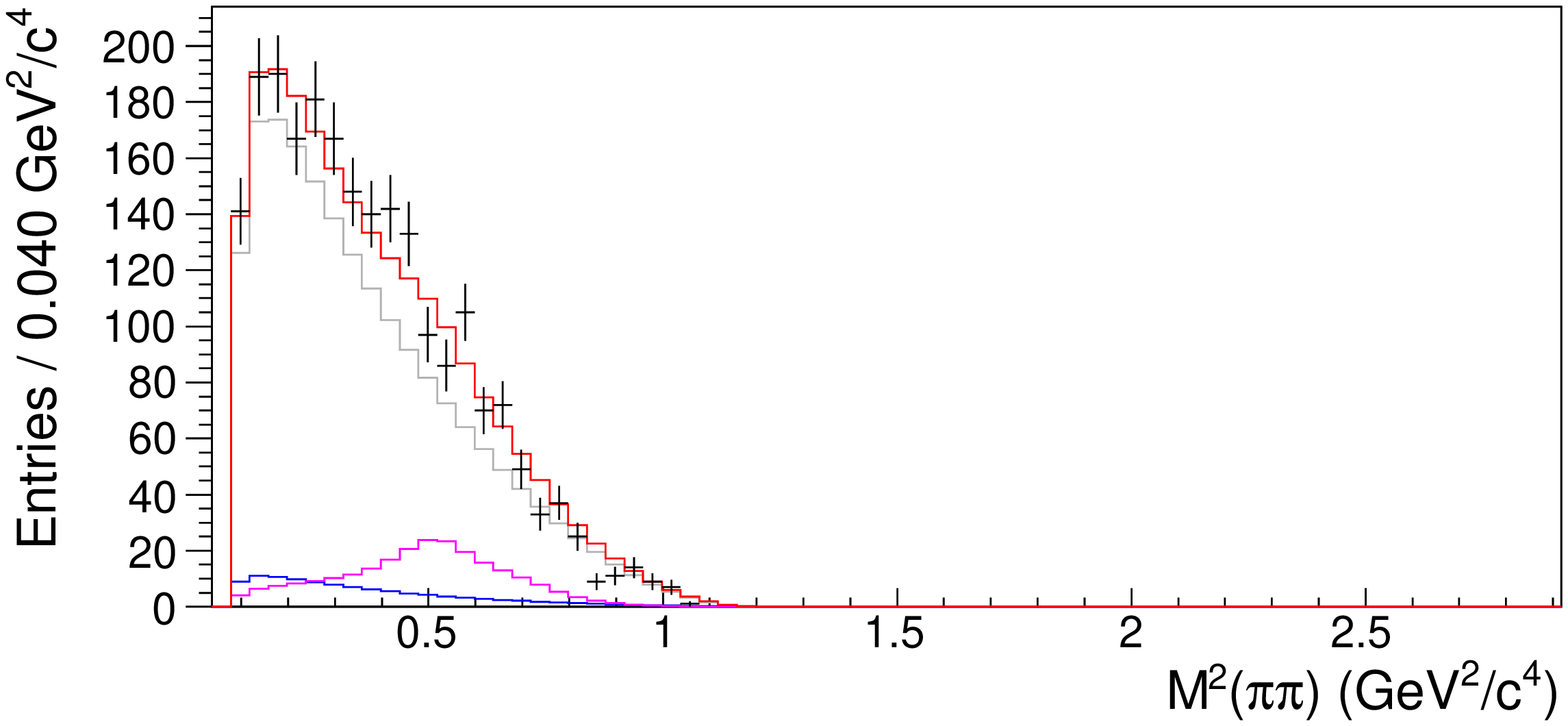}
}}}
\caption{Results of sideband fits for 
  $B^{+} \rightarrow J/\psi K^{+} \pi^{+} \pi^{-}$ (left) and 
  $B^{+} \rightarrow \psi^{\prime} K^{+} \pi^{+} \pi^{-}$ (right).
  Data (points) and fits (histograms) are shown projected onto the 
  three axes.
  The red histograms show the overall background functions.
  The combinatorial components are shown in gray, while
  the $K^{*}(892)$, $\rho$, $K_{S}$ and $D$ backgrounds are
  shown in blue, magenta, cyan, and green, respectively.
  The $K^{*}(892)$ and $\rho$ peaks are broader 
  in
  $B^{+} \rightarrow \psi^{\prime} K^{+} \pi^{+} \pi^{-}$
  than in 
  $B^{+} \rightarrow J/\psi K^{+} \pi^{+} \pi^{-}$ 
  because the distortion 
  shown in Fig.~\ref{transformations:fig_peak_transformation}
  is larger in the former mode.}
\label{amplitude:fig_sidebands}
\end{figure*}

\begin{figure*}[hbtp]
\centerline{
\scalebox{0.35}{
\rotatebox{270}{
\includegraphics*[270,37][569,691]
  {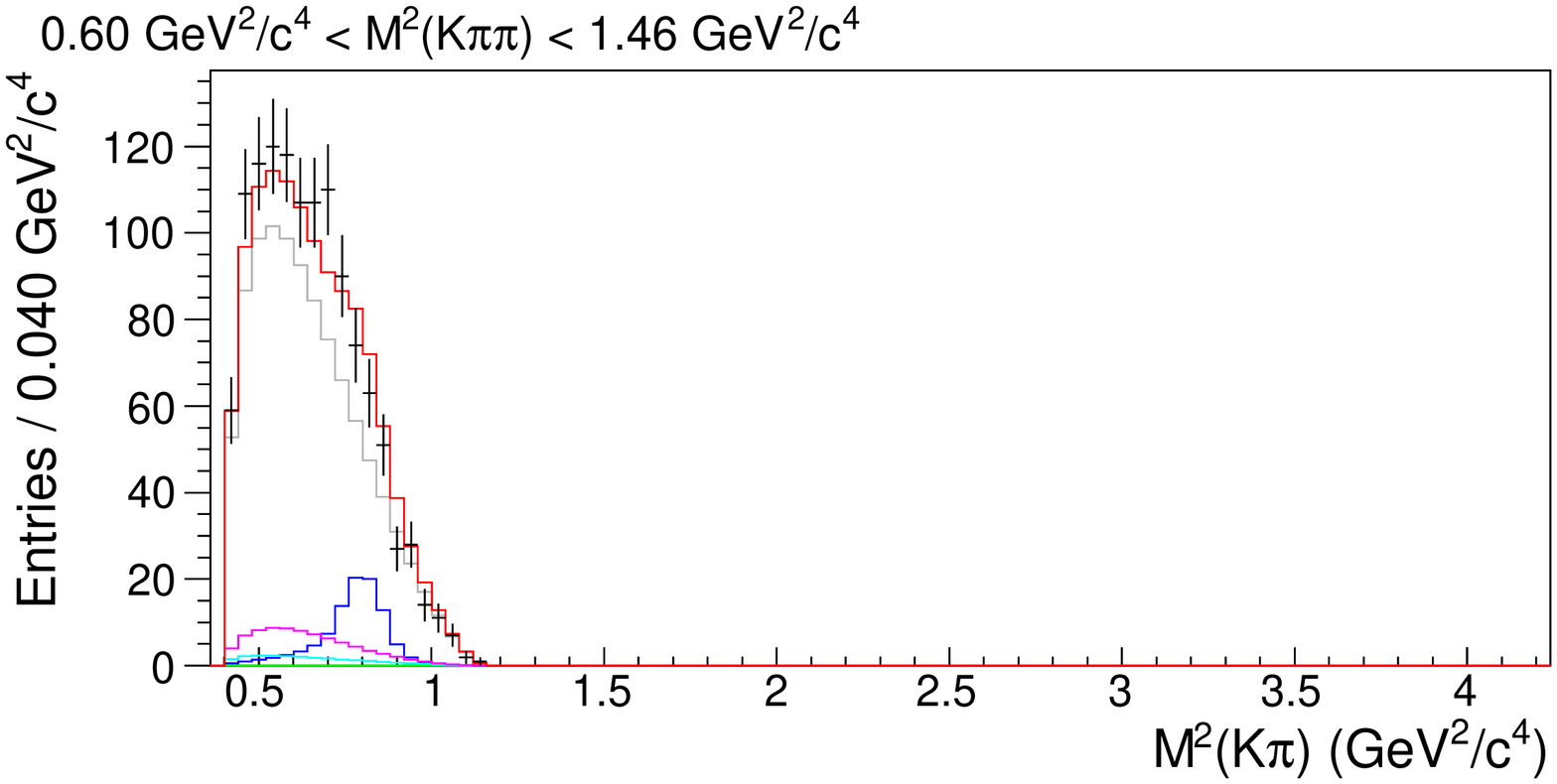}
}}
\hspace{4mm}
\scalebox{0.35}{
\rotatebox{270}{
\includegraphics*[270,37][569,691]
  {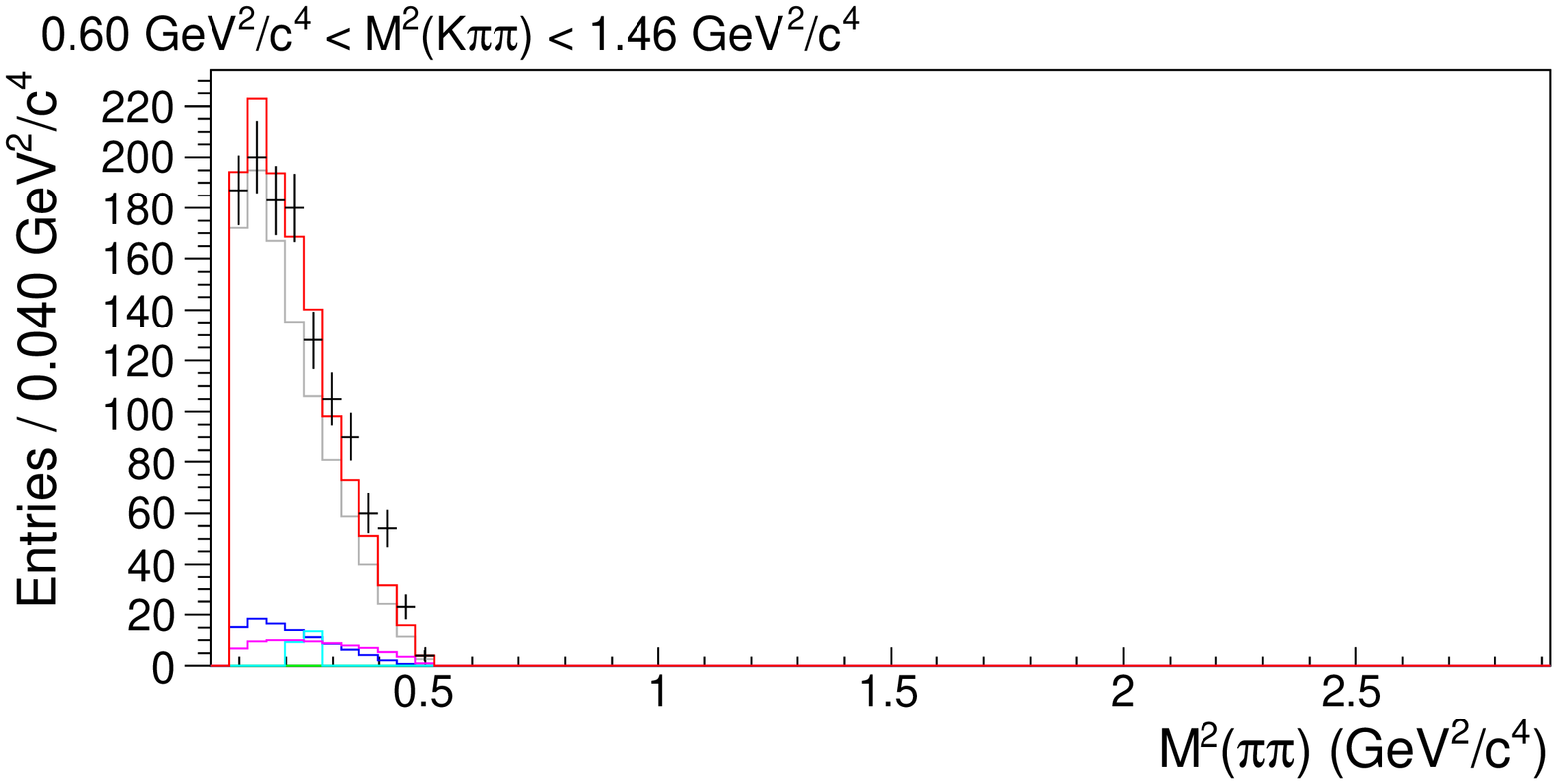}
}}}
\vspace{4mm}
\centerline{
\scalebox{0.35}{
\rotatebox{270}{
\includegraphics*[270,37][569,691]
  {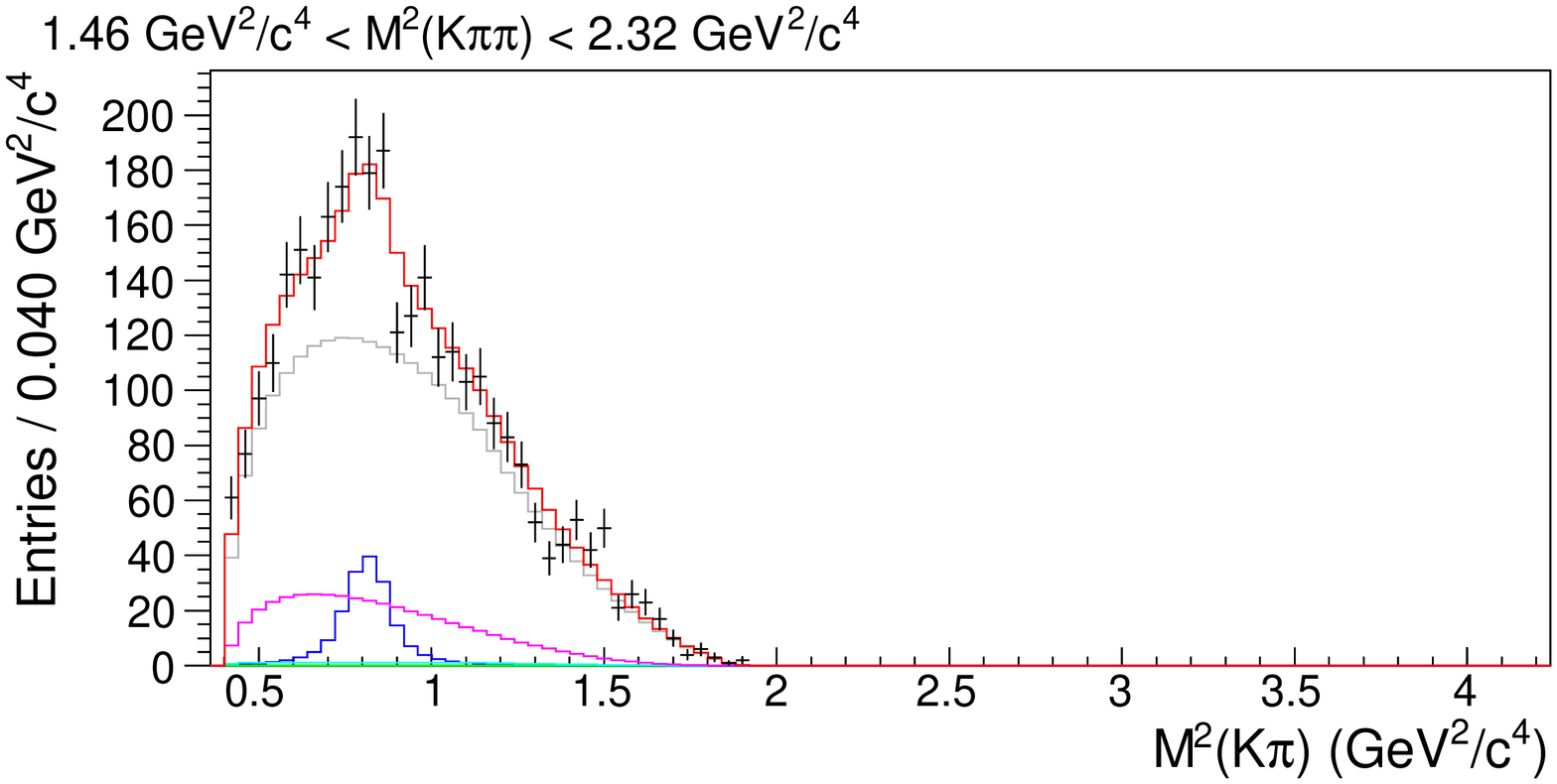}
}}
\hspace{4mm}
\scalebox{0.35}{
\rotatebox{270}{
\includegraphics*[270,37][569,691]
  {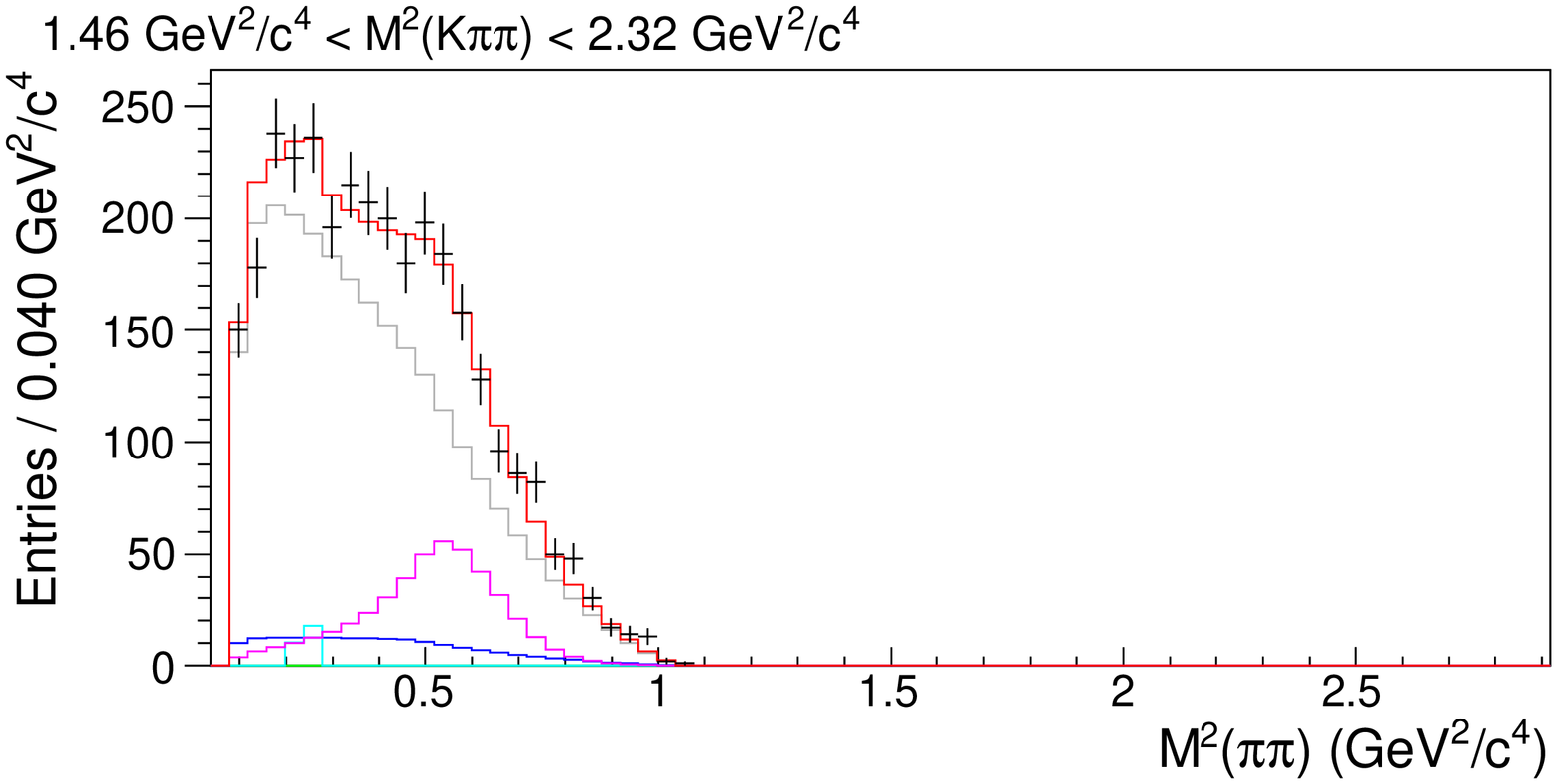}
}}}
\vspace{4mm}
\centerline{
\scalebox{0.35}{
\rotatebox{270}{
\includegraphics*[270,37][569,691]
  {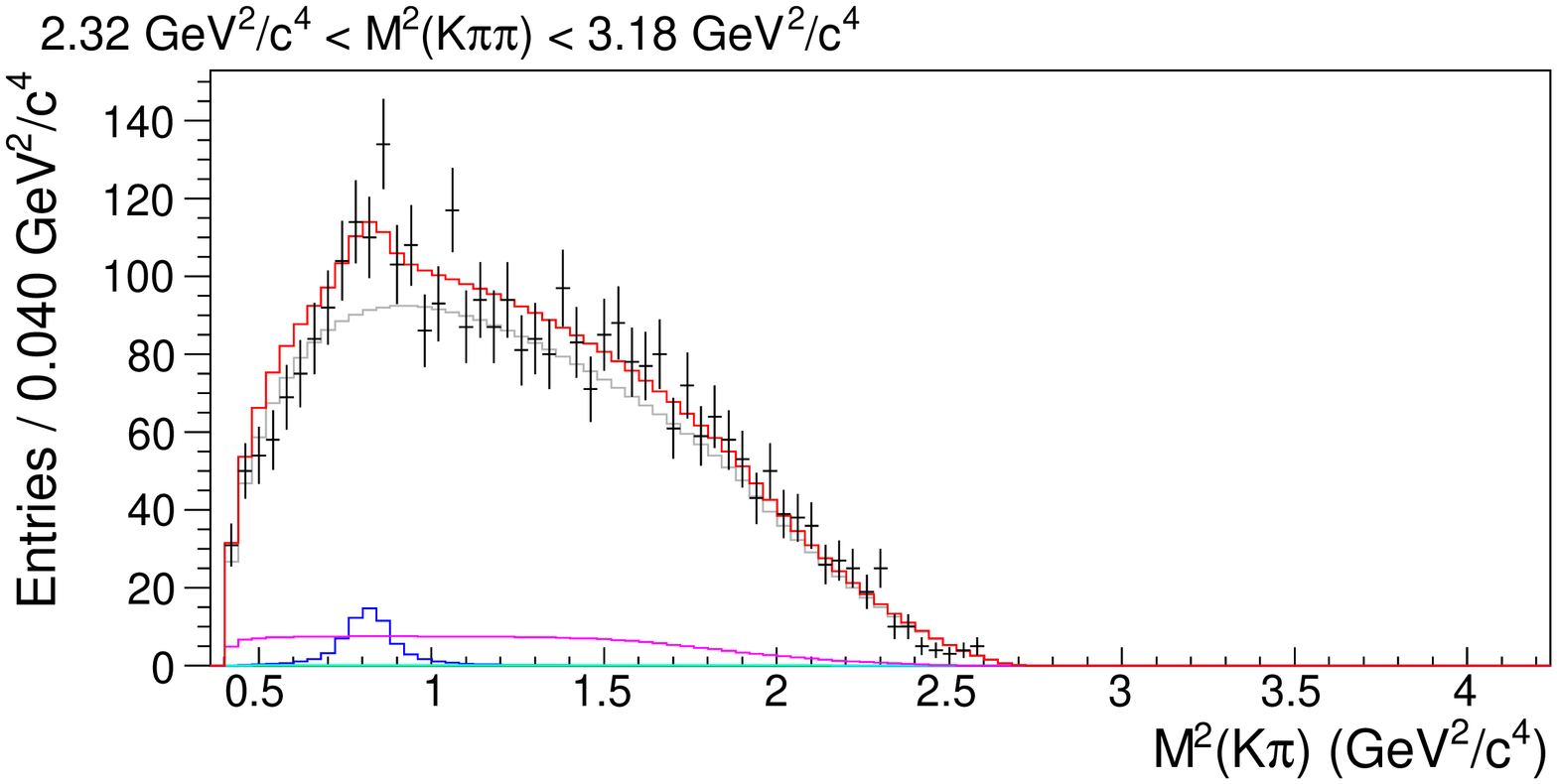}
}}
\hspace{4mm}
\scalebox{0.35}{
\rotatebox{270}{
\includegraphics*[270,37][569,691]
  {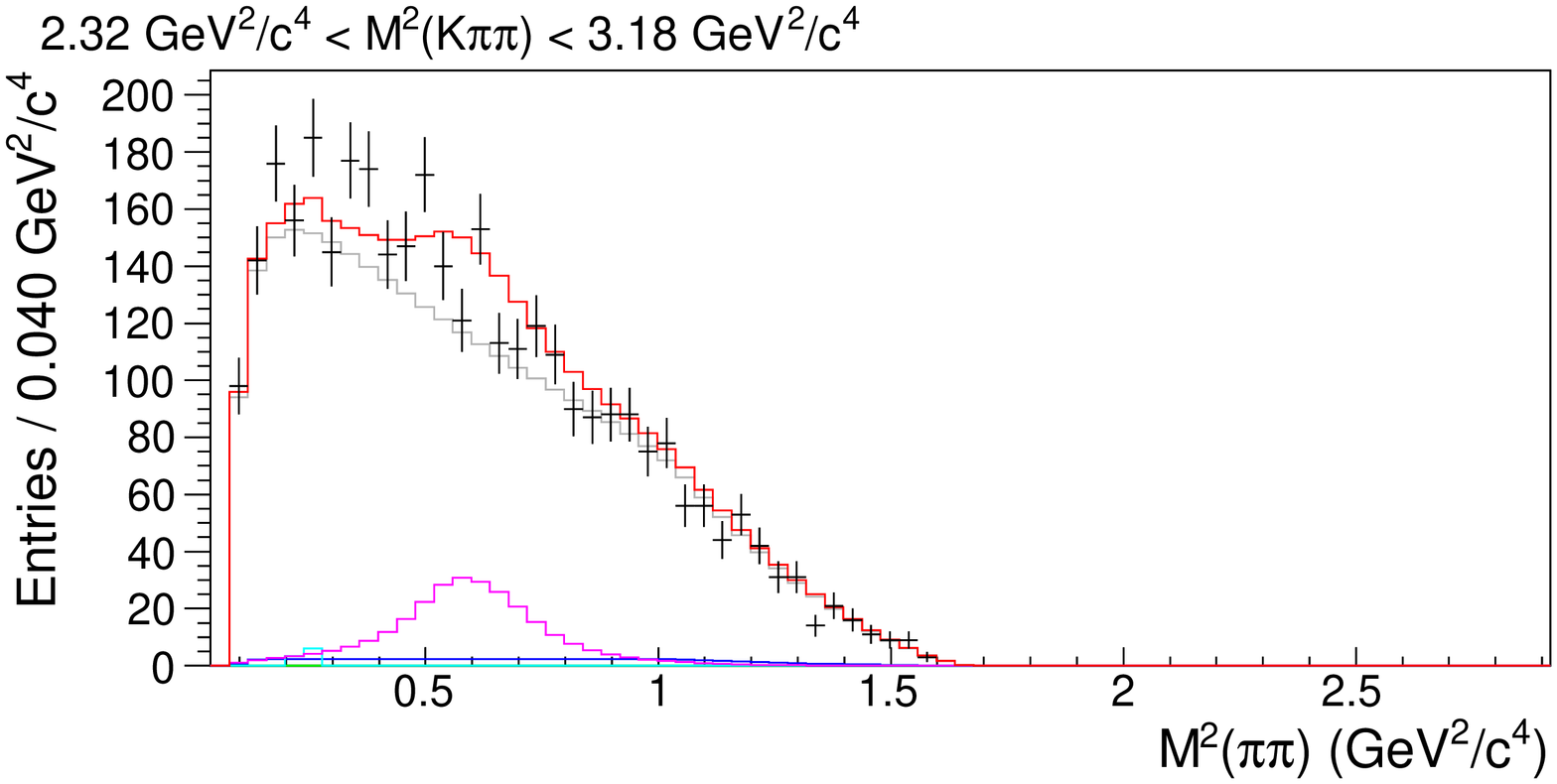}
}}}
\vspace{4mm}
\centerline{
\scalebox{0.35}{
\rotatebox{270}{
\includegraphics*[270,37][569,691]
  {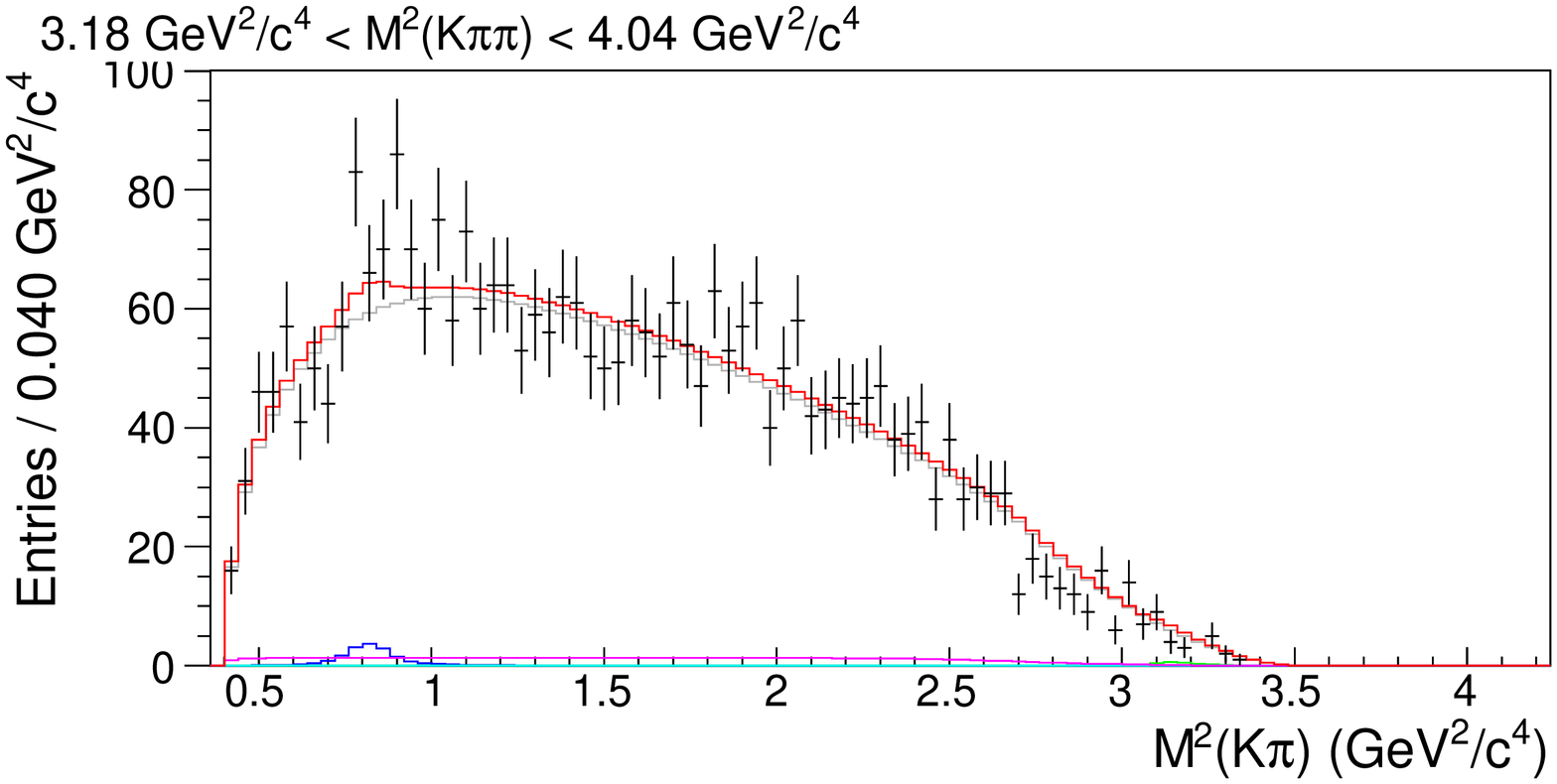}
}}
\hspace{4mm}
\scalebox{0.35}{
\rotatebox{270}{
\includegraphics*[270,37][569,691]
  {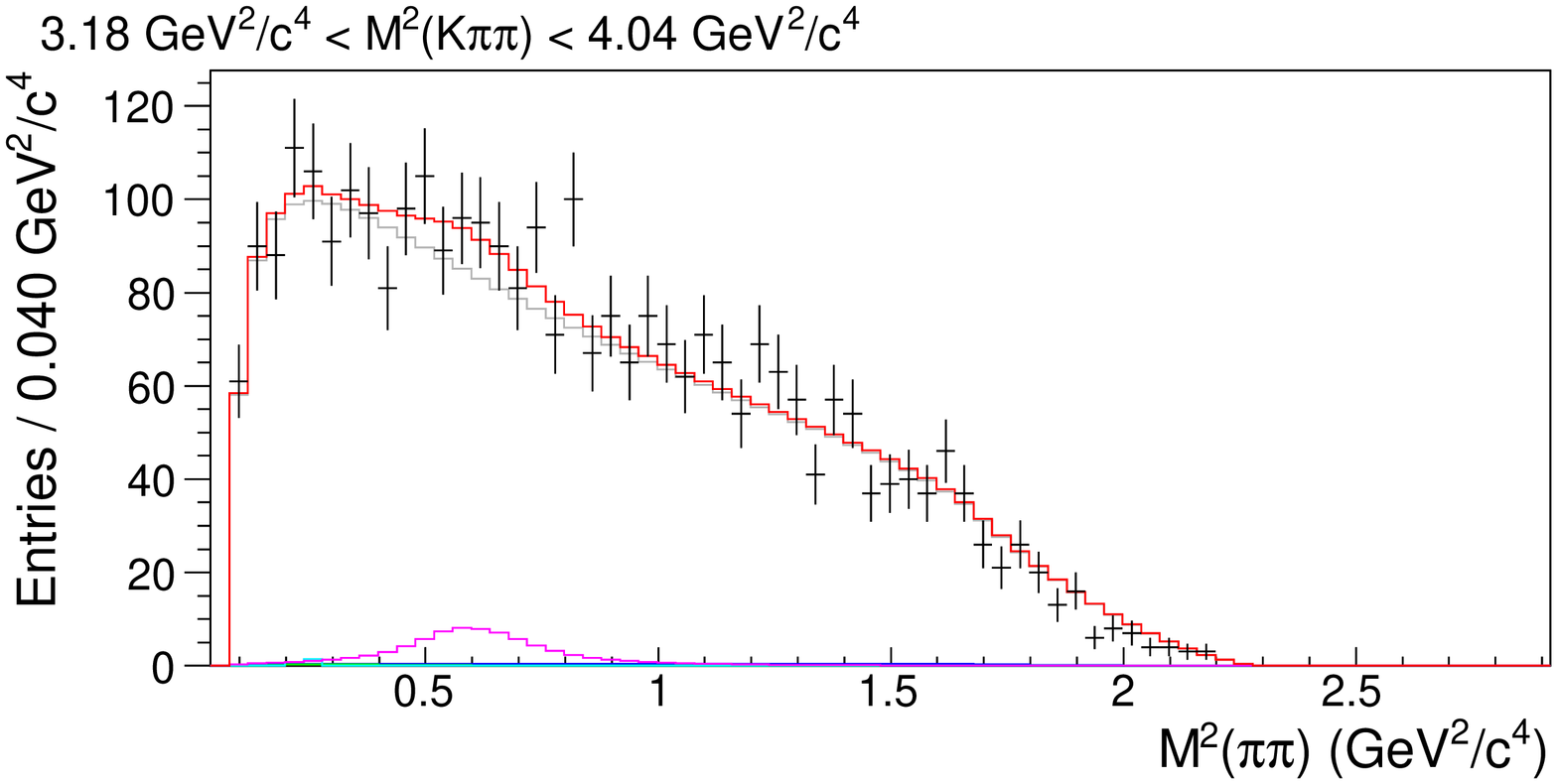}
}}}
\vspace{4mm}
\centerline{
\scalebox{0.35}{
\rotatebox{270}{
\includegraphics*[270,37][569,691]
  {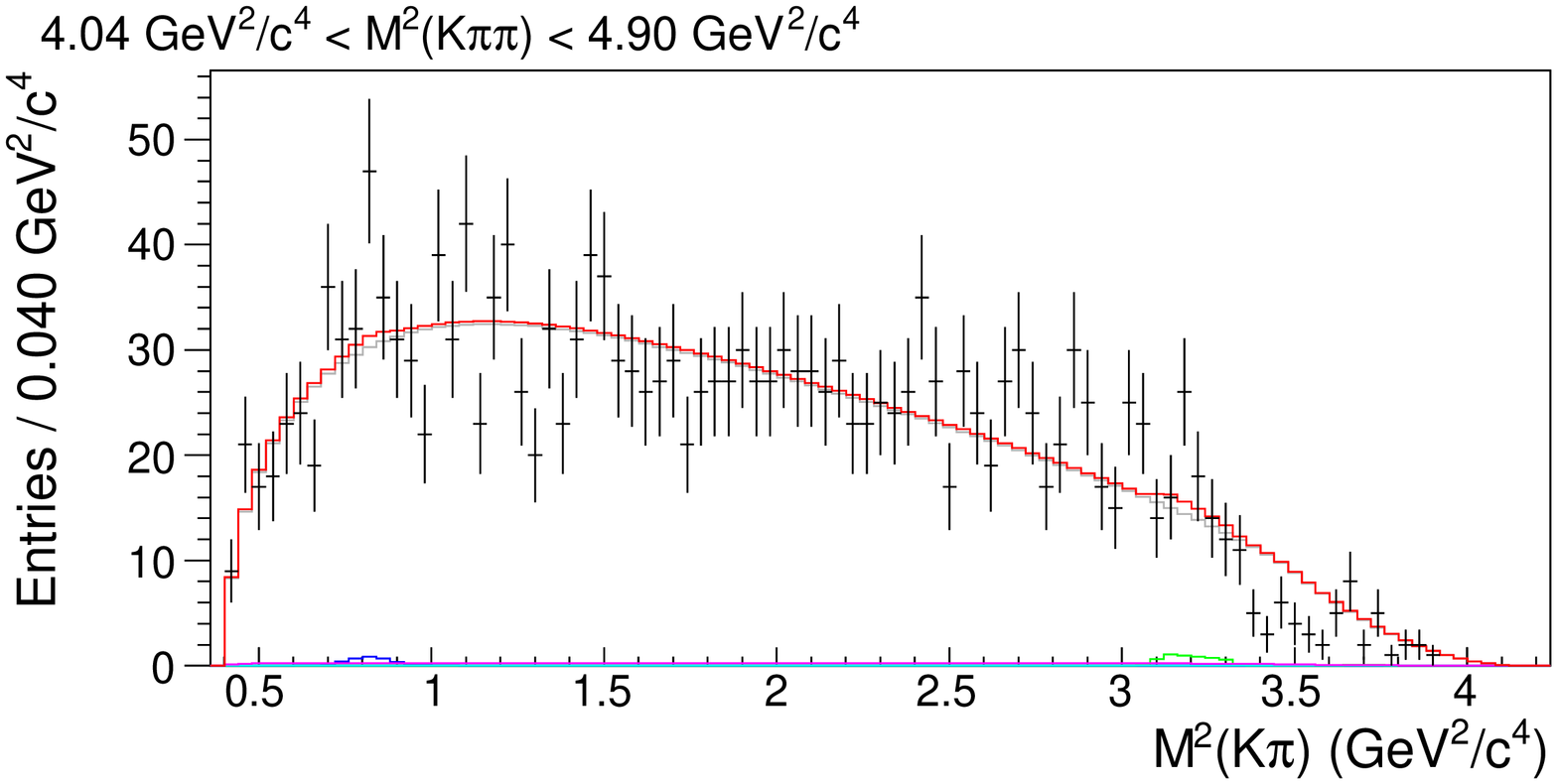}
}}
\hspace{4mm}
\scalebox{0.35}{
\rotatebox{270}{
\includegraphics*[270,37][569,691]
  {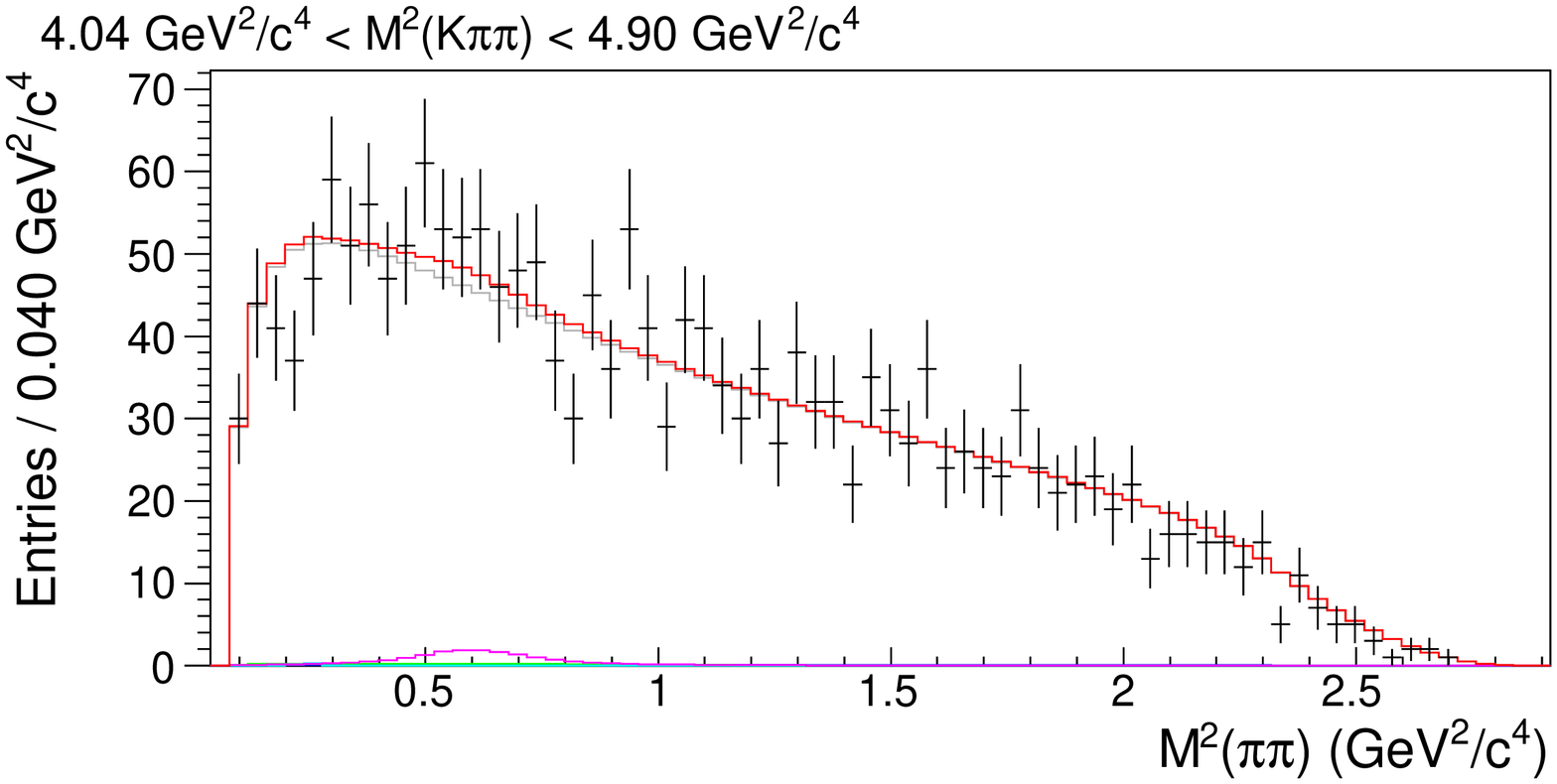}
}}}
\caption{$B^{+} \rightarrow J/\psi K^{+} \pi^{+} \pi^{-}$
  sideband data (points) and fit results (histograms) 
  for slices in $M^{2}(K\pi\pi)$.  The fit components
  are color coded as in Fig.~\ref{amplitude:fig_sidebands}.}
\label{amplitude:fig_sidebands_slices_jkpp}
\end{figure*}

\begin{figure*}[hbtp]
\centerline{
\scalebox{0.35}{
\rotatebox{270}{
\includegraphics*[270,37][569,691]
  {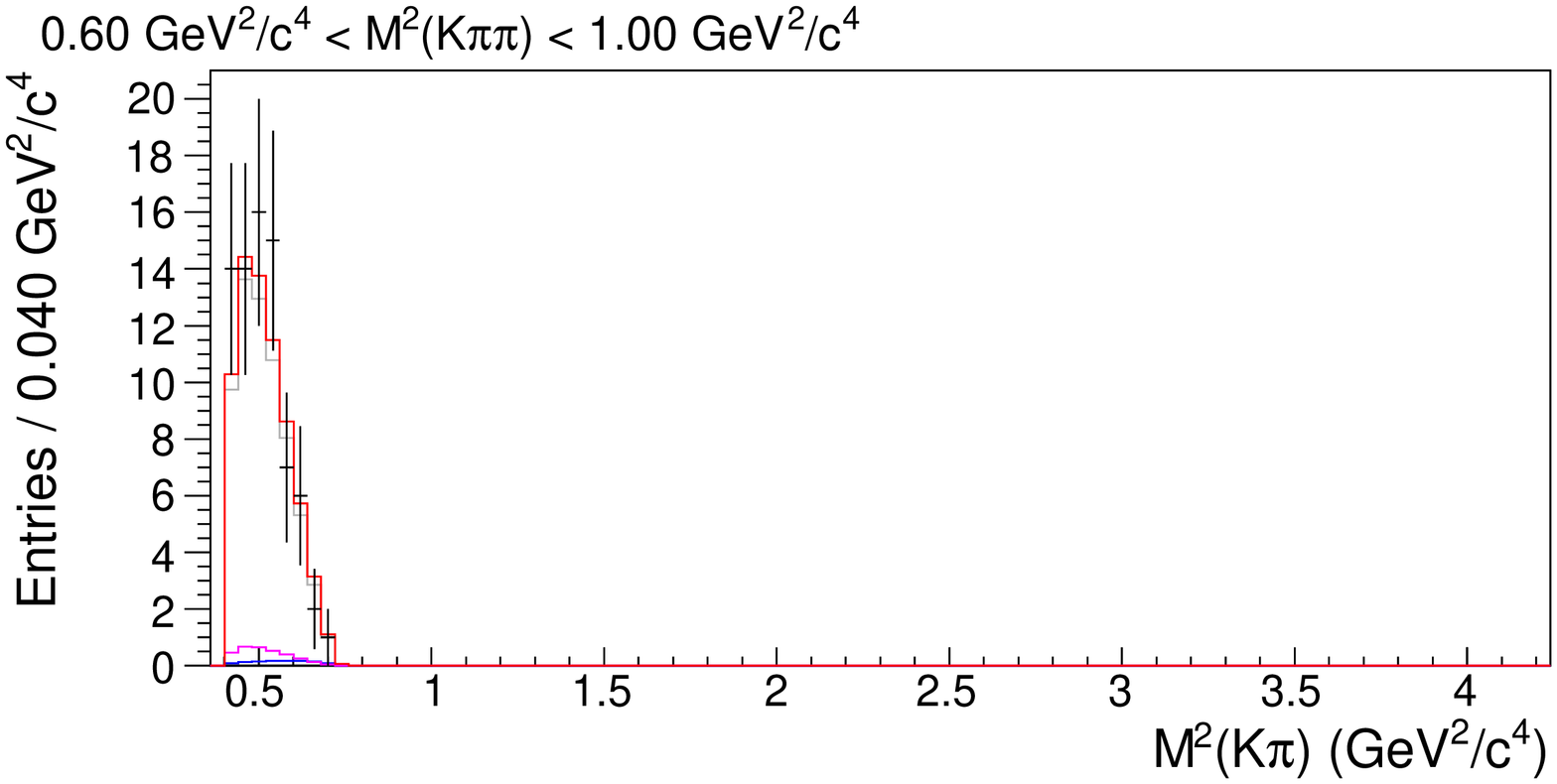}
}}
\hspace{4mm}
\scalebox{0.35}{
\rotatebox{270}{
\includegraphics*[270,37][569,691]
  {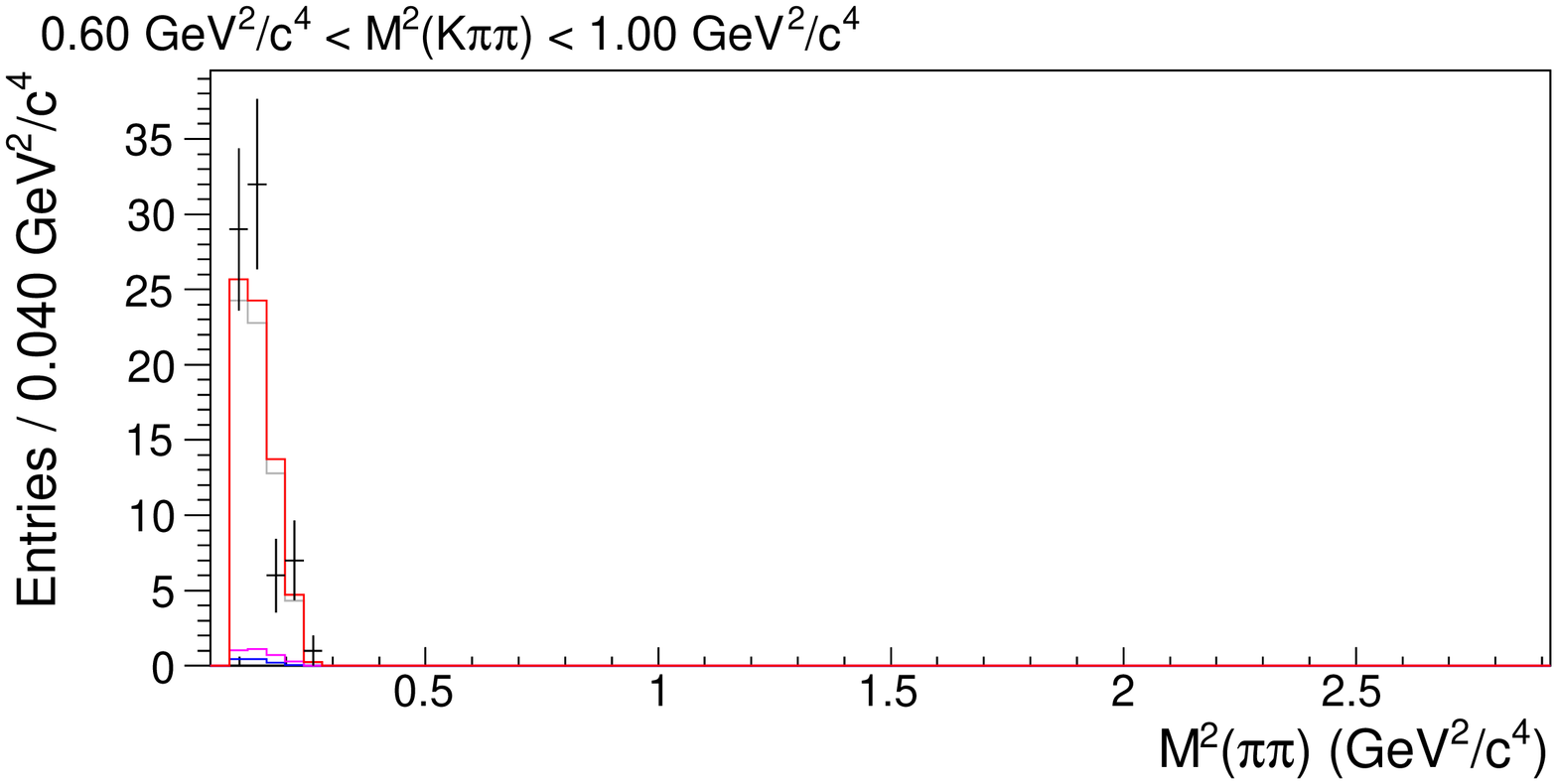}
}}}
\vspace{4mm}
\centerline{
\scalebox{0.35}{
\rotatebox{270}{
\includegraphics*[270,37][569,691]
  {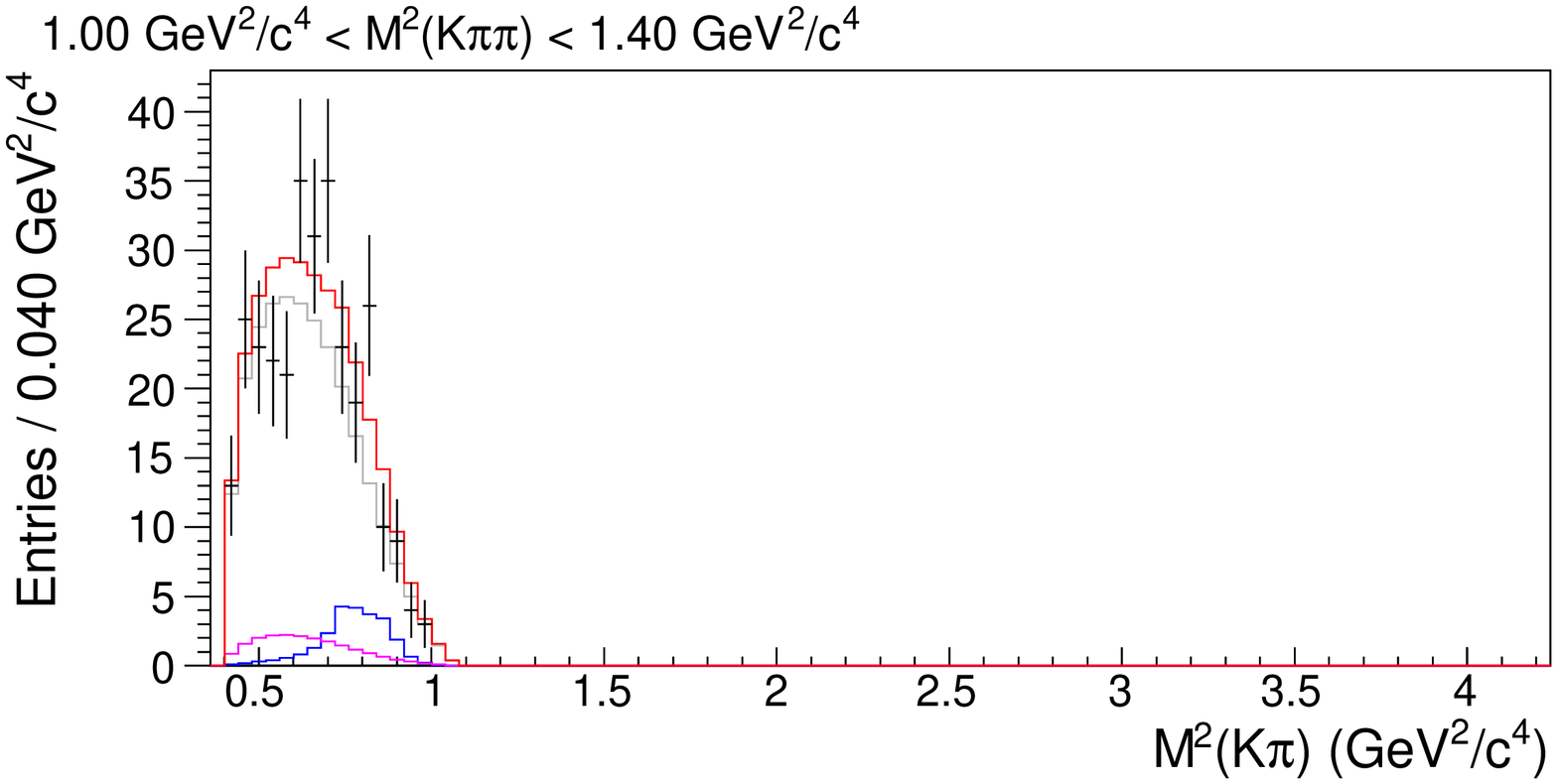}
}}
\hspace{4mm}
\scalebox{0.35}{
\rotatebox{270}{
\includegraphics*[270,37][569,691]
  {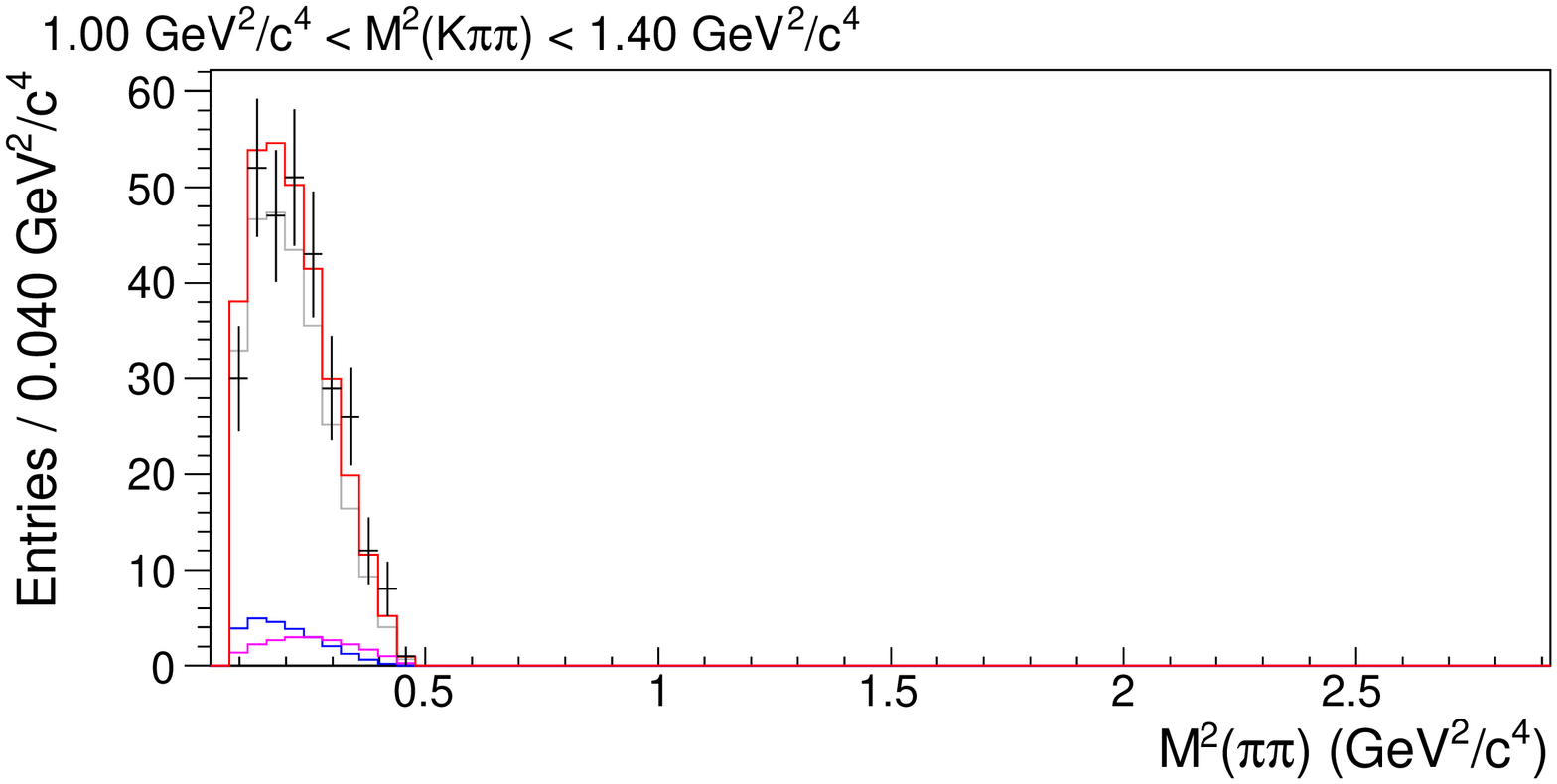}
}}}
\vspace{4mm}
\centerline{
\scalebox{0.35}{
\rotatebox{270}{
\includegraphics*[270,37][569,691]
  {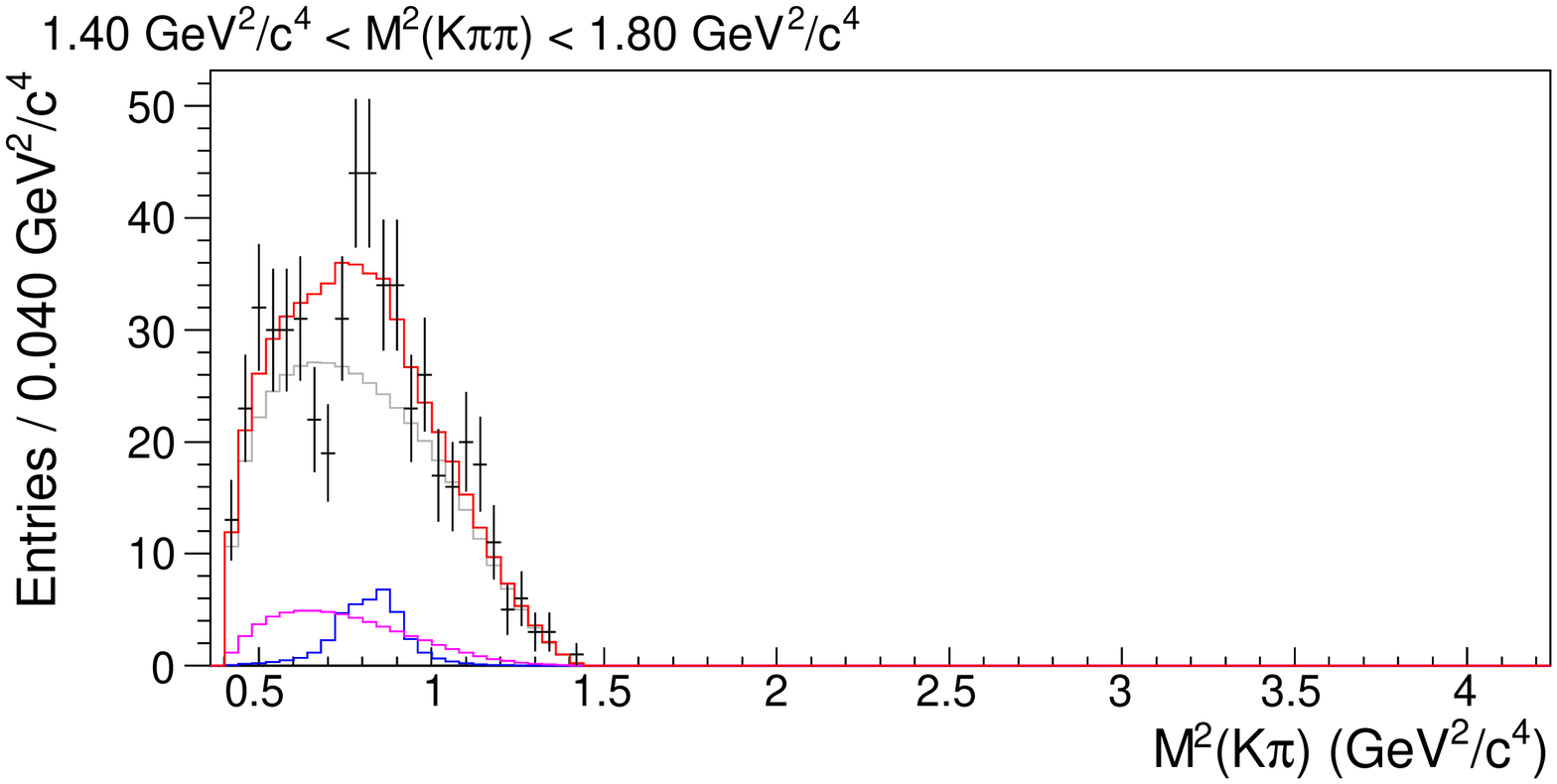}
}}
\hspace{4mm}
\scalebox{0.35}{
\rotatebox{270}{
\includegraphics*[270,37][569,691]
  {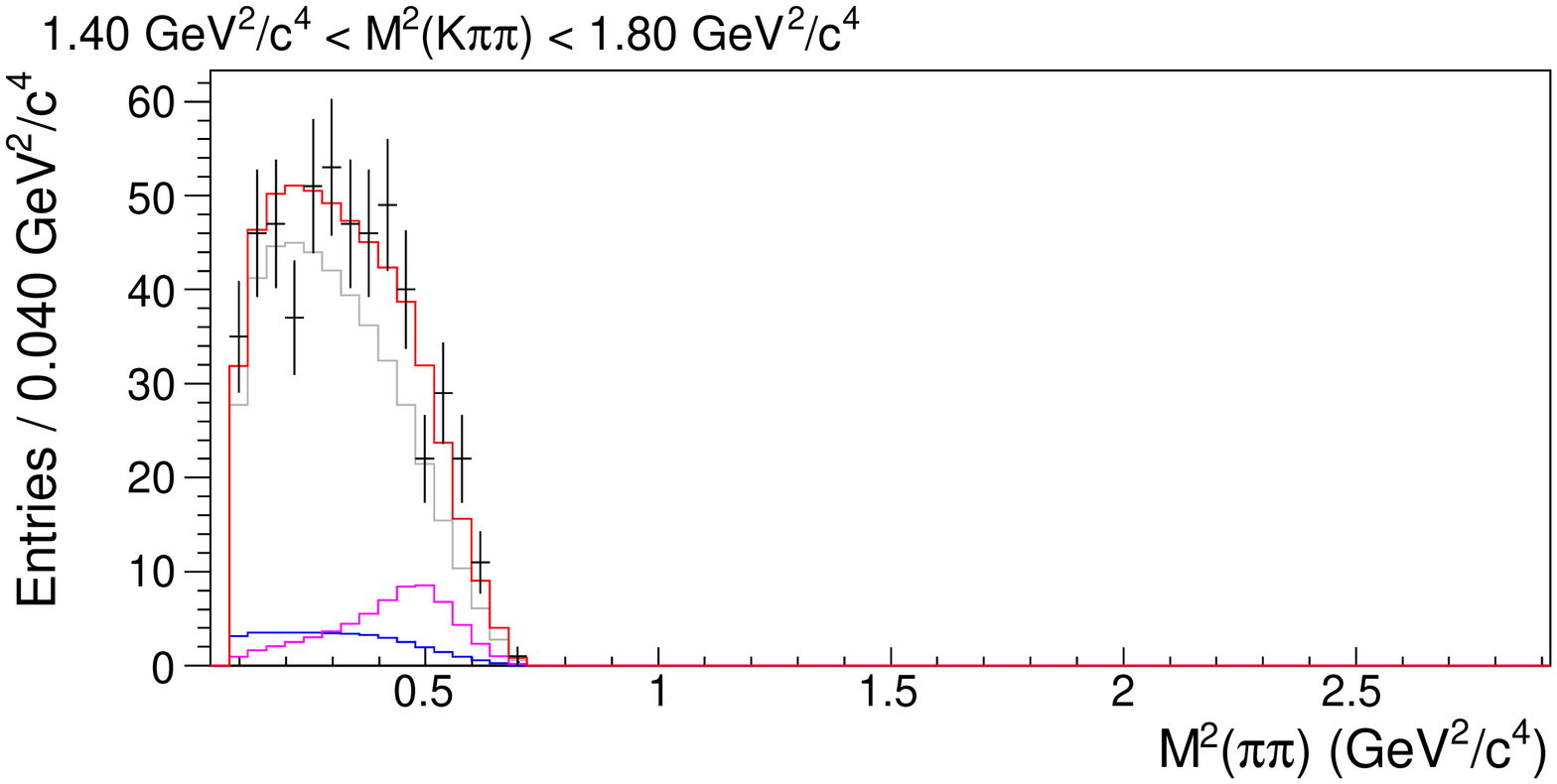}
}}}
\vspace{4mm}
\centerline{
\scalebox{0.35}{
\rotatebox{270}{
\includegraphics*[270,37][569,691]
  {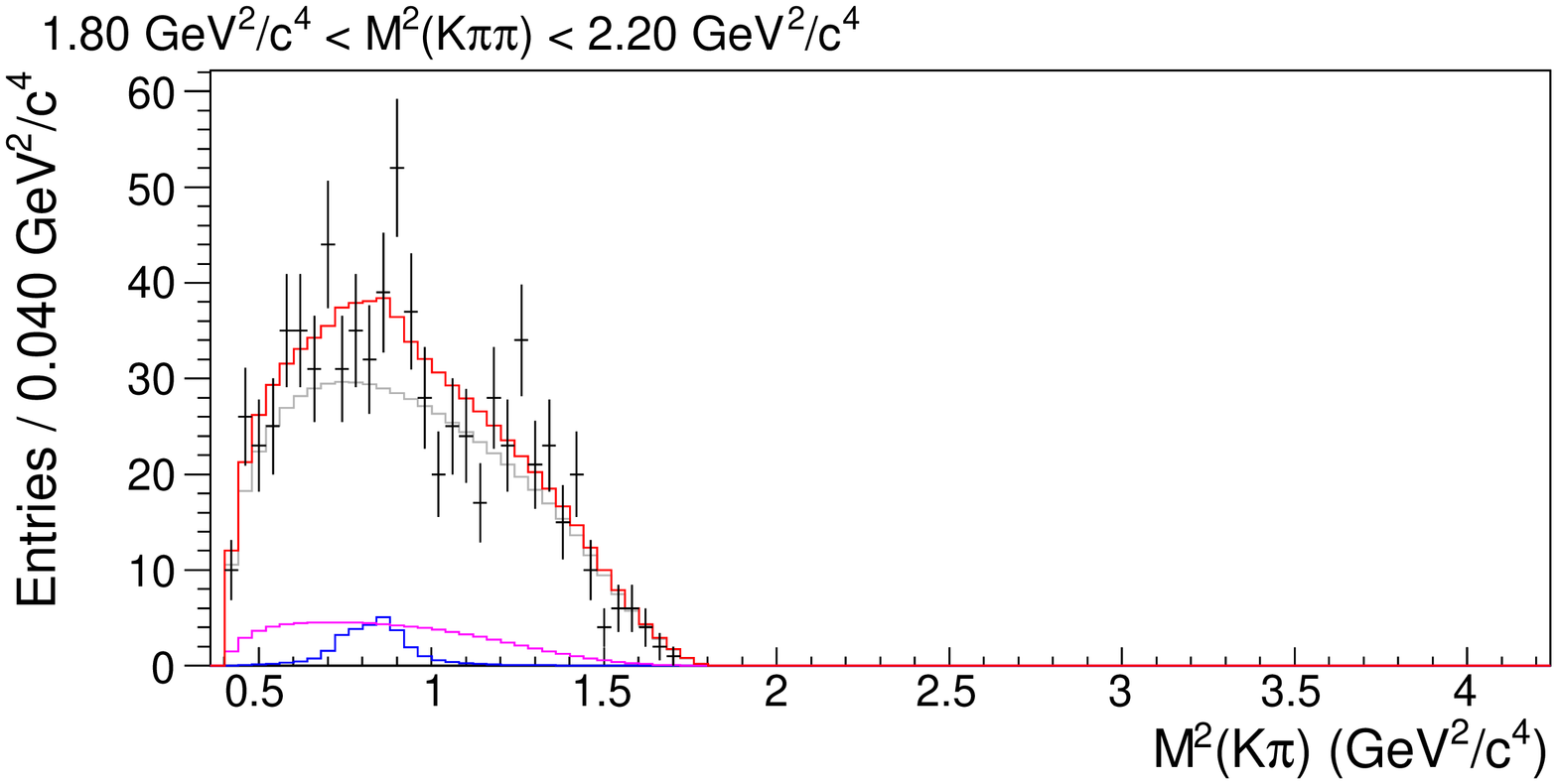}
}}
\hspace{4mm}
\scalebox{0.35}{
\rotatebox{270}{
\includegraphics*[270,37][569,691]
  {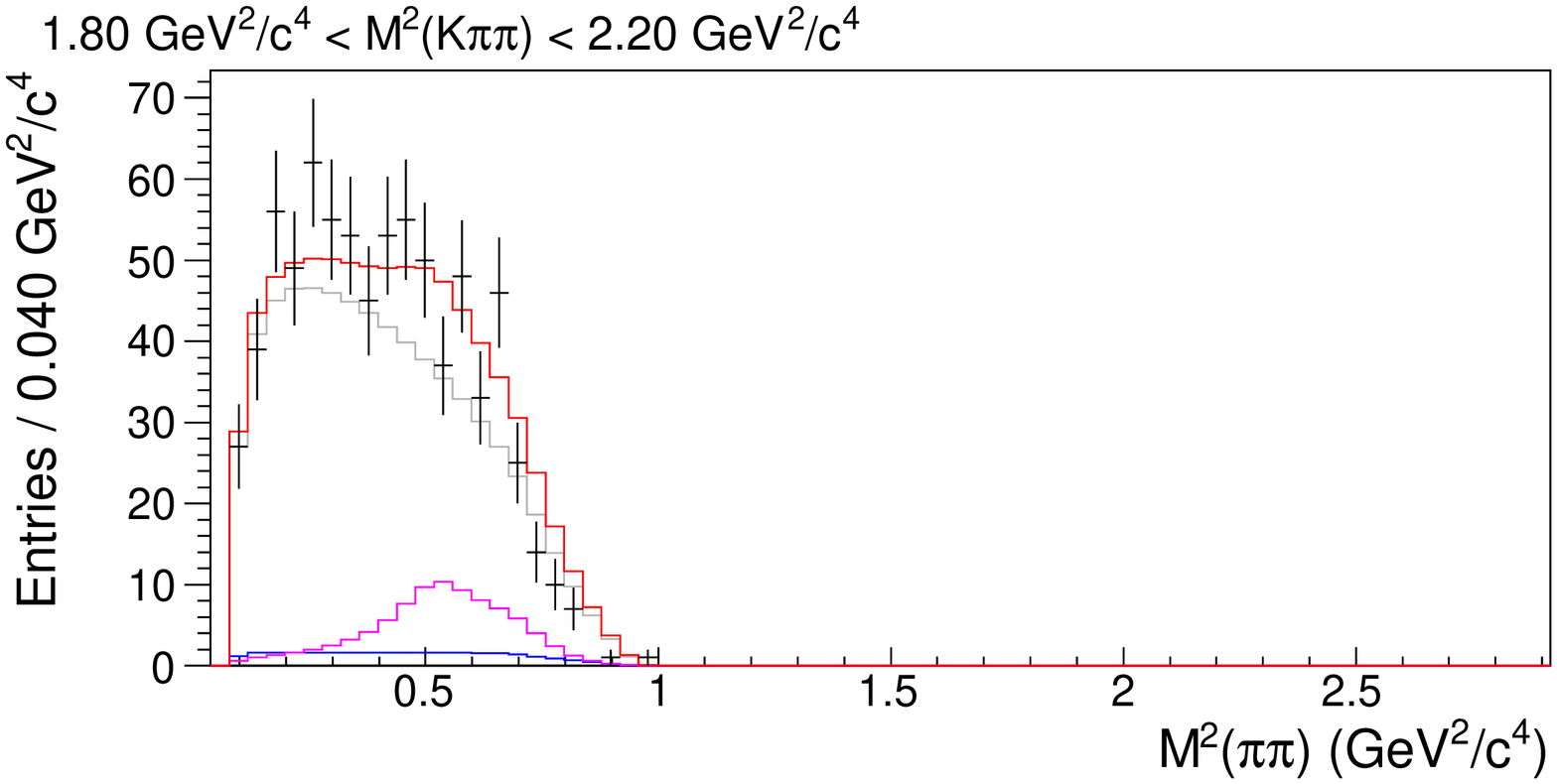}
}}}
\vspace{4mm}
\centerline{
\scalebox{0.35}{
\rotatebox{270}{
\includegraphics*[270,37][569,691]
  {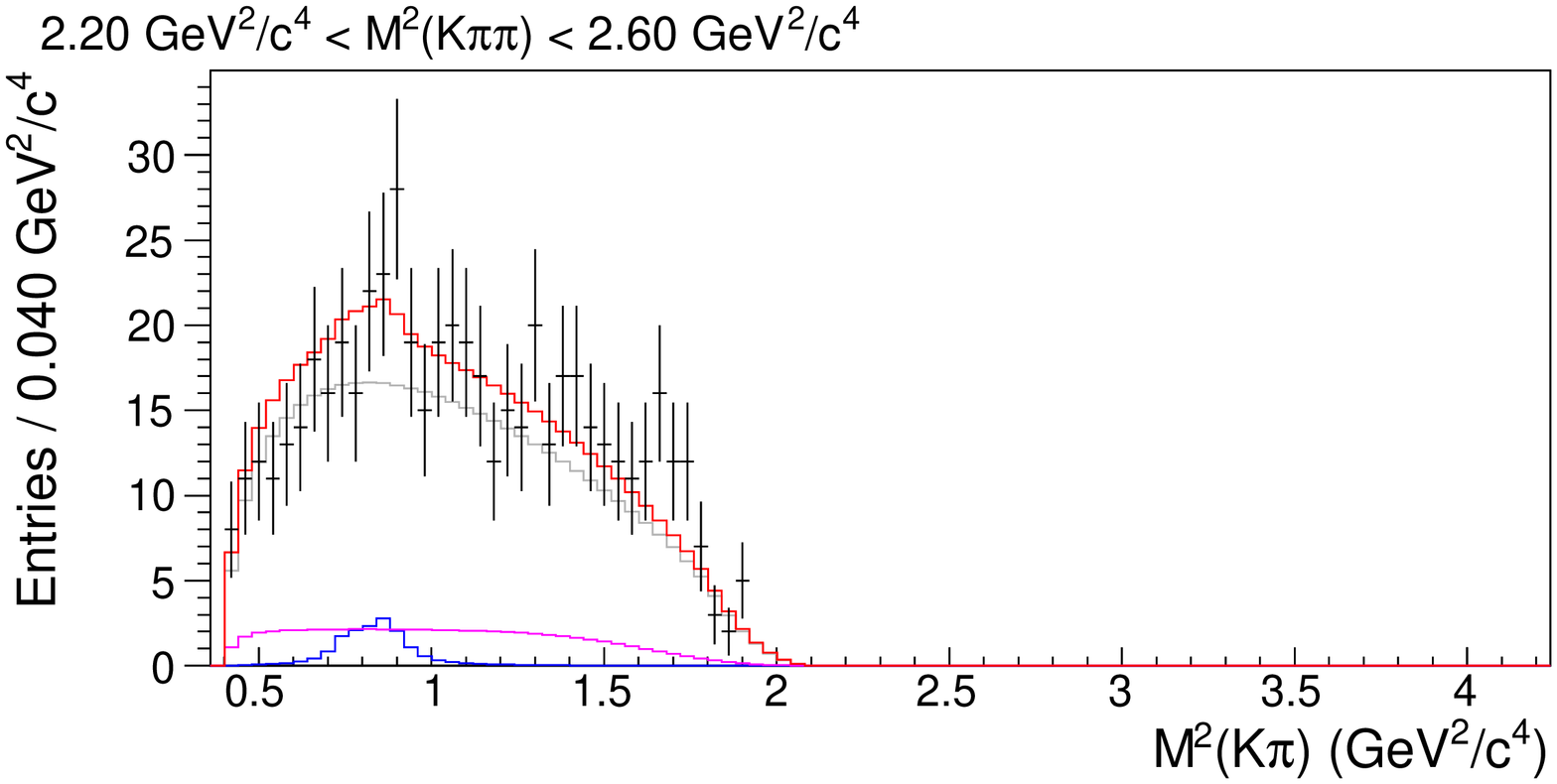}
}}
\hspace{4mm}
\scalebox{0.35}{
\rotatebox{270}{
\includegraphics*[270,37][569,691]
  {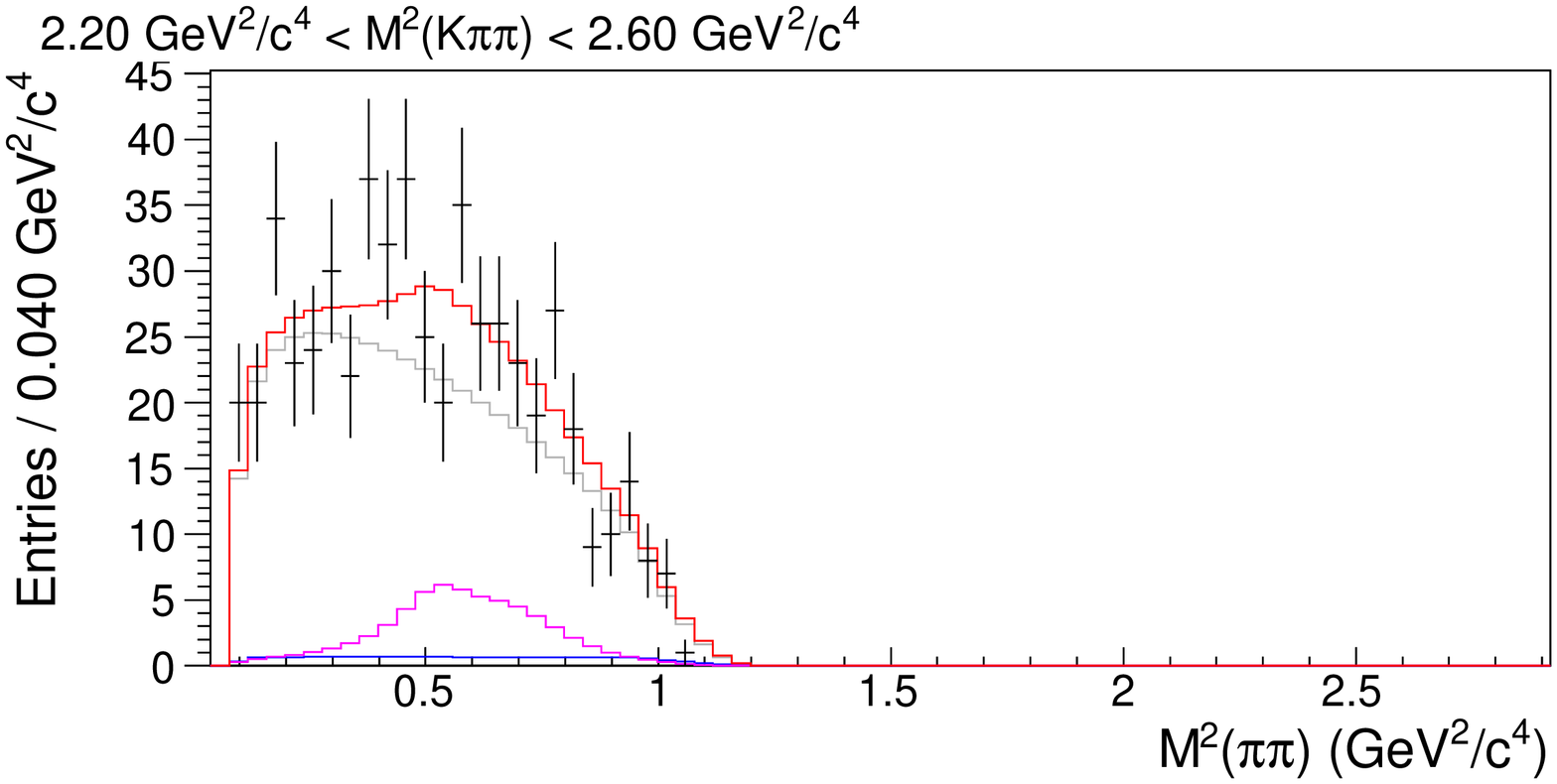}
}}}
\caption{$B^{+} \rightarrow \psi^{\prime} K^{+} \pi^{+} \pi^{-}$ 
  sideband data (points) and fit results (histograms) 
  for slices in $M^{2}(K\pi\pi)$.  The fit components
  are color coded as in Fig.~\ref{amplitude:fig_sidebands}.}
\label{amplitude:fig_sidebands_slices_pkpp}
\end{figure*}

Table~\ref{amplitude:table_sideband_fit_results}
lists the fitted parameters of the background functions.
The statistical error in each parameter is defined as the change 
in that parameter required to reduce the log likelihood by $1/2$.
The fitted functions, 
normalized to the total number of events in the fit,
are shown projected onto the three axes
along with the sideband data 
in Fig.~\ref{amplitude:fig_sidebands}.
Figures \ref{amplitude:fig_sidebands_slices_jkpp} 
and \ref{amplitude:fig_sidebands_slices_pkpp} show 
$M^{2}(K\pi)$ and $M^{2}(\pi\pi)$ projections for 
slices in $M^{2}(K\pi\pi)$.

\begin{table}[htbp]
\caption{Fitted values 
   of the background-function parameters 
   (Eqs.~\ref{amplitude:eq_bkg_jkpp} and~\ref{amplitude:eq_bkg_pkpp}).}
\label{amplitude:table_sideband_fit_results}
\begin{ruledtabular}
\renewcommand{\tabcolsep}{0.0cm}
\begin{tabular*}{8.6cm}
{l@{\hspace{0.7cm}}r@{$~\pm~$}l@{\hspace{0.7cm}}r@{$~\pm~$}l} 
Parameter           
 & \multicolumn{2}{c}{$B^{+} \rightarrow J/\psi K^{+} \pi^{+} \pi^{-}$}
   \hspace{0.7cm}
 & \multicolumn{2}{c}{$B^{+} \rightarrow \psi^{\prime} K^{+} \pi^{+} \pi^{-}$} \\
\colrule
$a_{x0}$            & \multicolumn{2}{c}{$1.0$ (fixed)} \hspace{0.7cm} 
                    & \multicolumn{2}{c}{$1.0$ (fixed)}  \\
$a_{x1}$            & $-1.5901$ & $0.0048$              
                    & $-1.238$  & $0.036$                \\
$a_{x2}$            &  $0.9423$ & $0.0086$              
                    &  $0.480$  & $0.045$                \\
$a_{x3}$            & $-0.4737$ & $0.0087$              
                    & $-0.221$  & $0.021$                \\
$a_{x4}$            &  $0.1778$ & $0.0067$               
                    & \multicolumn{2}{c}{}               \\
$a_{x5}$            & $-0.0488$ & $0.0033$  
                    & \multicolumn{2}{c}{}               \\
$a_{y0}$            & \multicolumn{2}{c}{$1.0$ (fixed)} \hspace{0.7cm} 
                    & \multicolumn{2}{c}{}               \\
$a_{y1}$            &  $0.088$  & $0.021$               
                    & \multicolumn{2}{c}{}               \\
$a_{z0}$            & \multicolumn{2}{c}{$1.0$ (fixed)} \hspace{0.7cm} 
                    & \multicolumn{2}{c}{}               \\
$a_{z1}$            & $-0.022$  & $0.022$              
                    & \multicolumn{2}{c}{}               \\
$a_{z2}$            &  $0.129$  & $0.018$              
                    & \multicolumn{2}{c}{}               \\
$a_{K^{*}(892)}$    & $0.0353$  & $0.0065$              
                    & $0.0161$  & $0.0062$               \\
$a_{D}$             & $0.0007$  & $0.0011$            
                    & \multicolumn{2}{c}{}               \\
$a_{K_{S}}$         & $0.0061$  & $0.0023$           
                    & \multicolumn{2}{c}{}               \\
$a_{\rho}$          & $0.086$   & $0.012$                
                    & $0.0352$  & $0.0099$               \\
\end{tabular*}
\end{ruledtabular}
\end{table}

As a measure of goodness of fit,
a $\chi^{2}$ variable is calculated by
distributing the data into cubic bins that are
$0.1~{\mathrm{GeV}}/c^{2}$ wide on each side.
The normalized PDF, with the parameters set to their best-fit values,
is integrated over each bin and multiplied 
by the total number of events in the fit
to determine the number of events expected in the bin.
Adjacent bins are combined until each bin has at least $6$ data events.
A $\chi^{2}$ variable for the multinomial distribution is then 
calculated as~\cite{baker:1984}
\begin{equation}
\chi^{2} = 2 \sum_{i}^{N_{\mathrm{bins}}} 
             n_{i} \ln \left( \frac{n_{i}}{p_{i}} \right) ,
\end{equation}
where $N_{\mathrm{bins}}$ is the total number of bins used, 
$n_{i}$ is the number of observed events in a given bin, and 
$p_{i}$ is the number expected in that bin based on the PDF.

If the expected distribution $p_{i}$ 
were obtained by a binned maximum-likelihood fit 
of the data distribution $n_{i}$, the
number of degrees of freedom associated with this $\chi^{2}$ 
would be reduced by the number
of fit parameters $N_{\mathrm{par}}$ and would be given by
$N_{\mathrm{DOF}} = N_{\mathrm{bins}} - N_{\mathrm{par}} - 1$.
If, on the other hand, the two distributions were not correlated by a 
fit, the number of degrees of freedom would be
$N_{\mathrm{DOF}} = N_{\mathrm{bins}} - 1$.  
Since, in this case, the distributions are related by an 
unbinned maximum-likelihood fit, the true $N_{\mathrm{DOF}}$ 
can be expected to lie between these extremes~\cite{kopp:2001}.

For the $B^{+} \rightarrow J/\psi K^{+} \pi^{+} \pi^{-}$
sideband-data fit,
$\ell_{B} = -21484.6$,
while
$\chi^{2} = 1709.5$
with
$N_{\mathrm{bins}} = 1707$ and
$N_{\mathrm{par}}  = 12$.
For the $B^{+} \rightarrow \psi^{\prime} K^{+} \pi^{+} \pi^{-}$
sideband-data fit,
$\ell_{B} = 822.7$,
while
$\chi^{2} = 286.8$
with
$N_{\mathrm{bins}} = 294$ and
$N_{\mathrm{par}} = 5$.

\subsection{Efficiency functions}

The dependence of the detector efficiency on the kinematic variables
(i.e., $\varepsilon(\vec x)$ in Eq.~\ref{amplitude:eq_signal})
is obtained for three-dimensional bins, 
$0.15~{\mathrm{GeV}}^{2}/c^{4}$-wide on each side,
using nonresonant signal-MC simulation as described in
Sec.~\ref{section_inclusive} 
and illustrated in 
Fig.~\ref{inclusive:fig_efficiency}.
The function is implemented as a lookup table:
the efficiency for a given data point is 
the efficiency in the corresponding bin.

\subsection{Phase-space densities}

\begin{figure*}[hbtp]
\centerline{
\scalebox{0.308}{
\rotatebox{270}{
\includegraphics*[270,37][569,691]
  {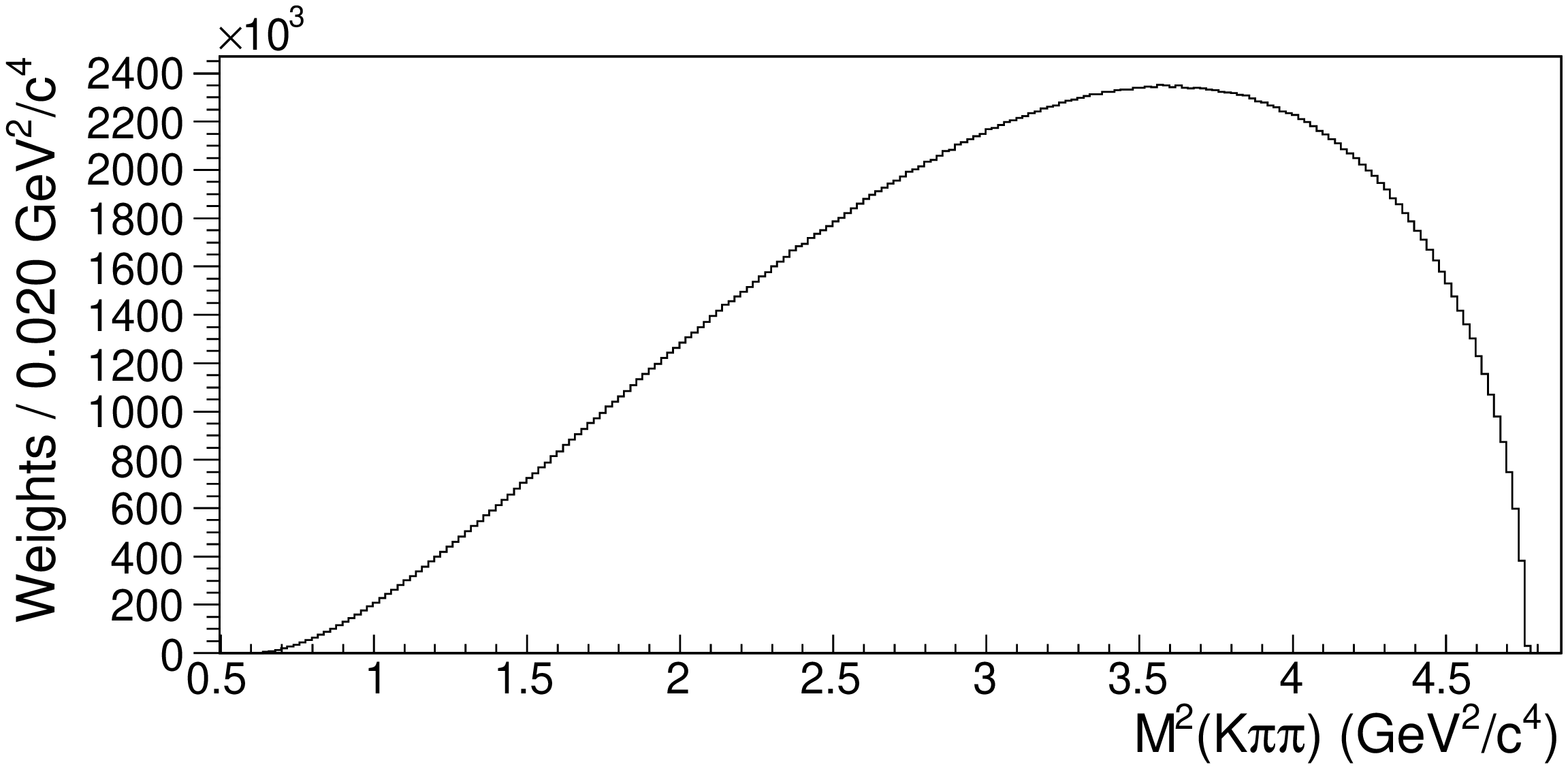}
}}
\hspace{4mm}
\scalebox{0.308}{
\rotatebox{270}{
\includegraphics*[270,37][569,691]
  {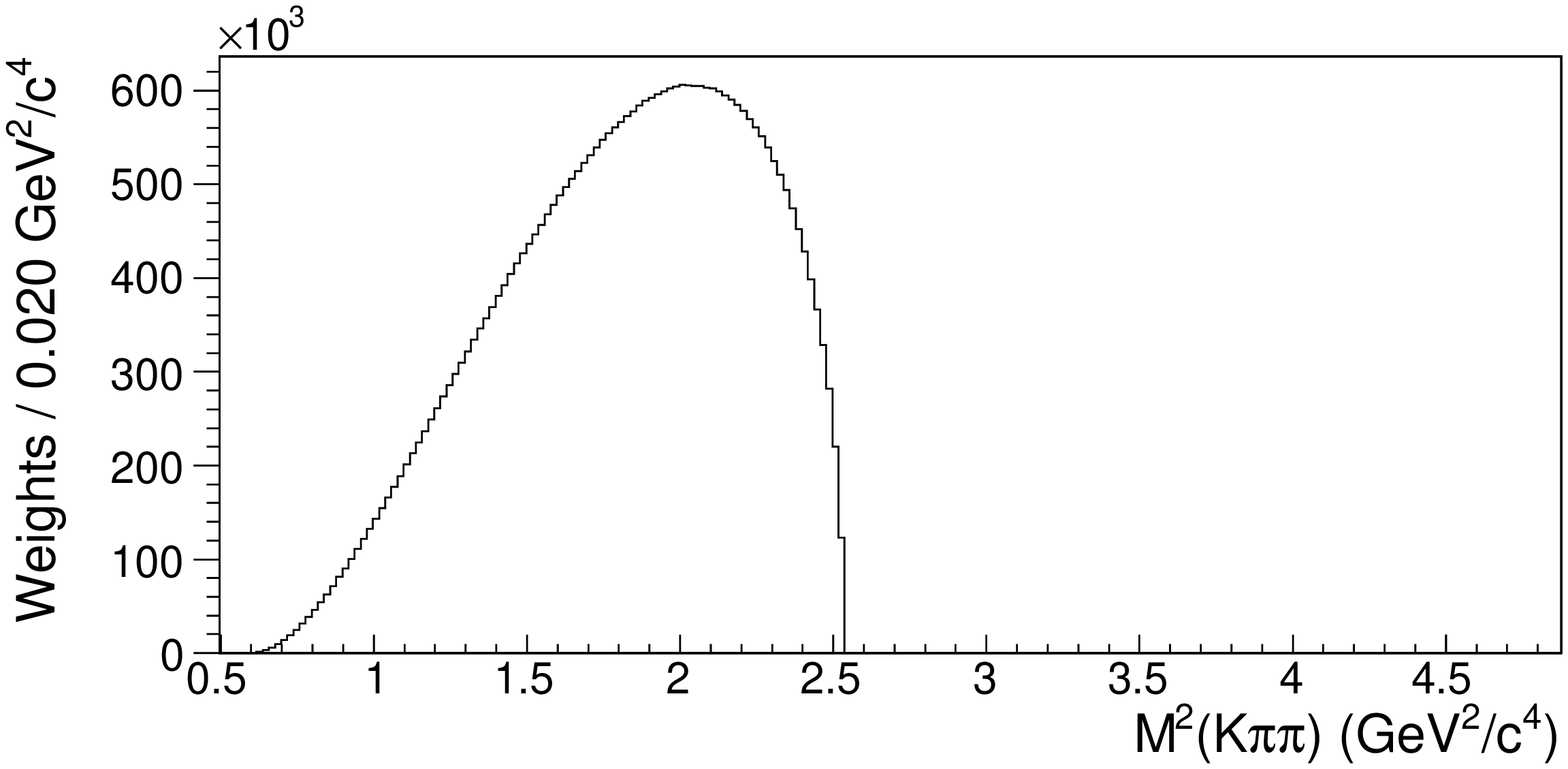}
}}}
\centerline{
\scalebox{0.308}{
\rotatebox{270}{
\includegraphics*[270,37][569,691]
  {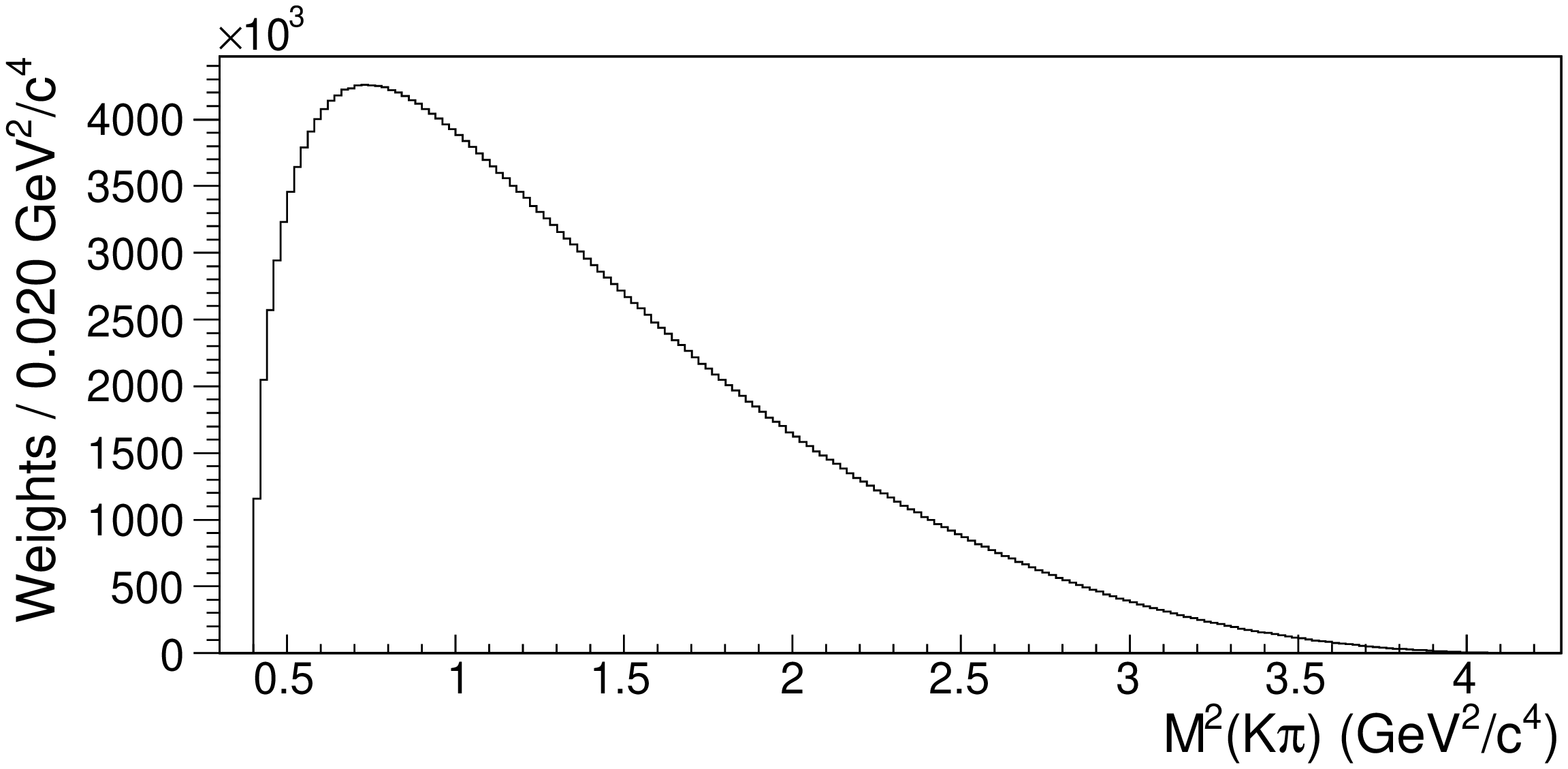}
}}
\hspace{4mm}
\scalebox{0.308}{
\rotatebox{270}{
\includegraphics*[270,37][569,691]
 {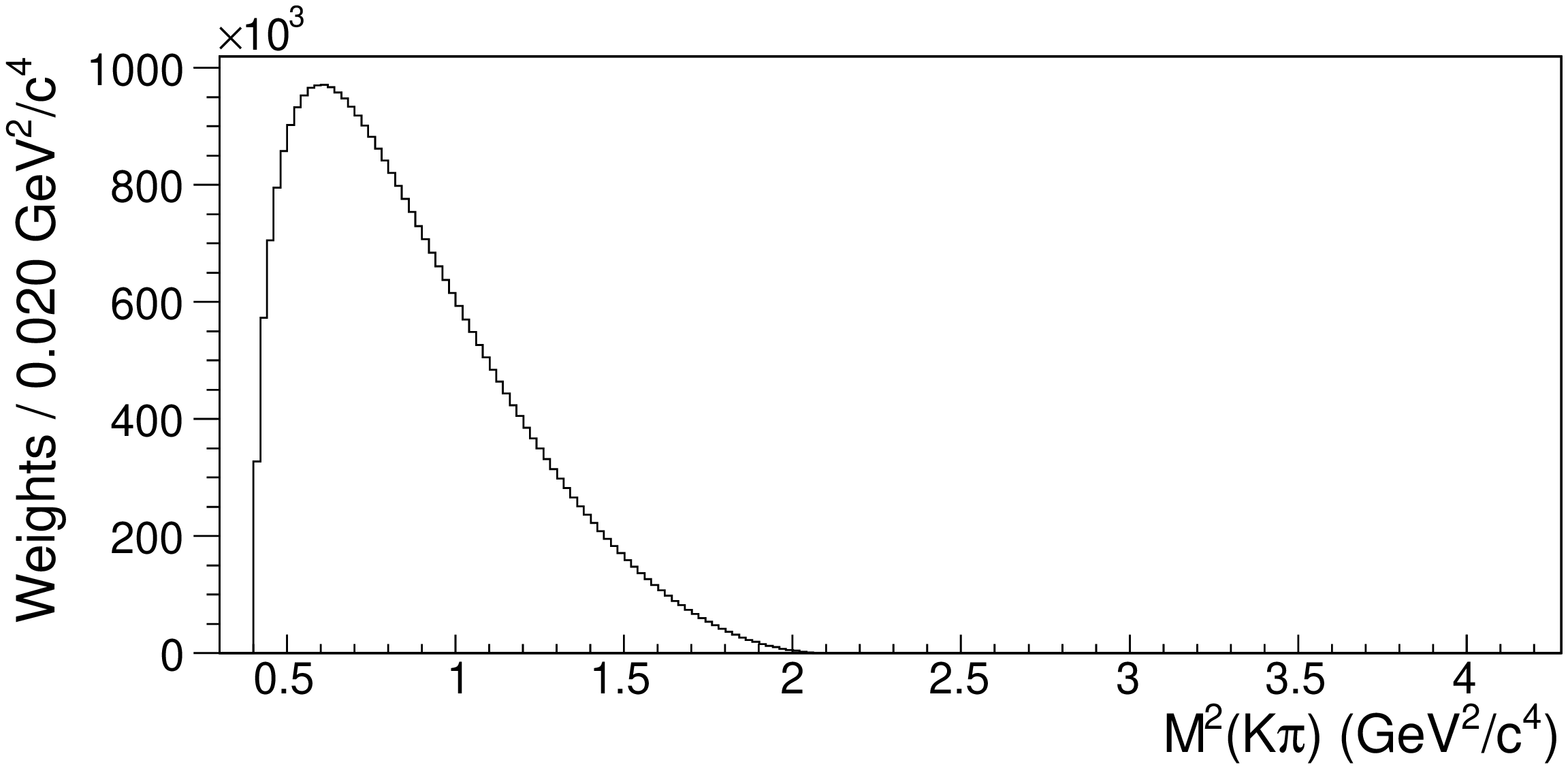}
}}}
\centerline{
\scalebox{0.308}{
\rotatebox{270}{
\includegraphics*[270,37][569,691]
  {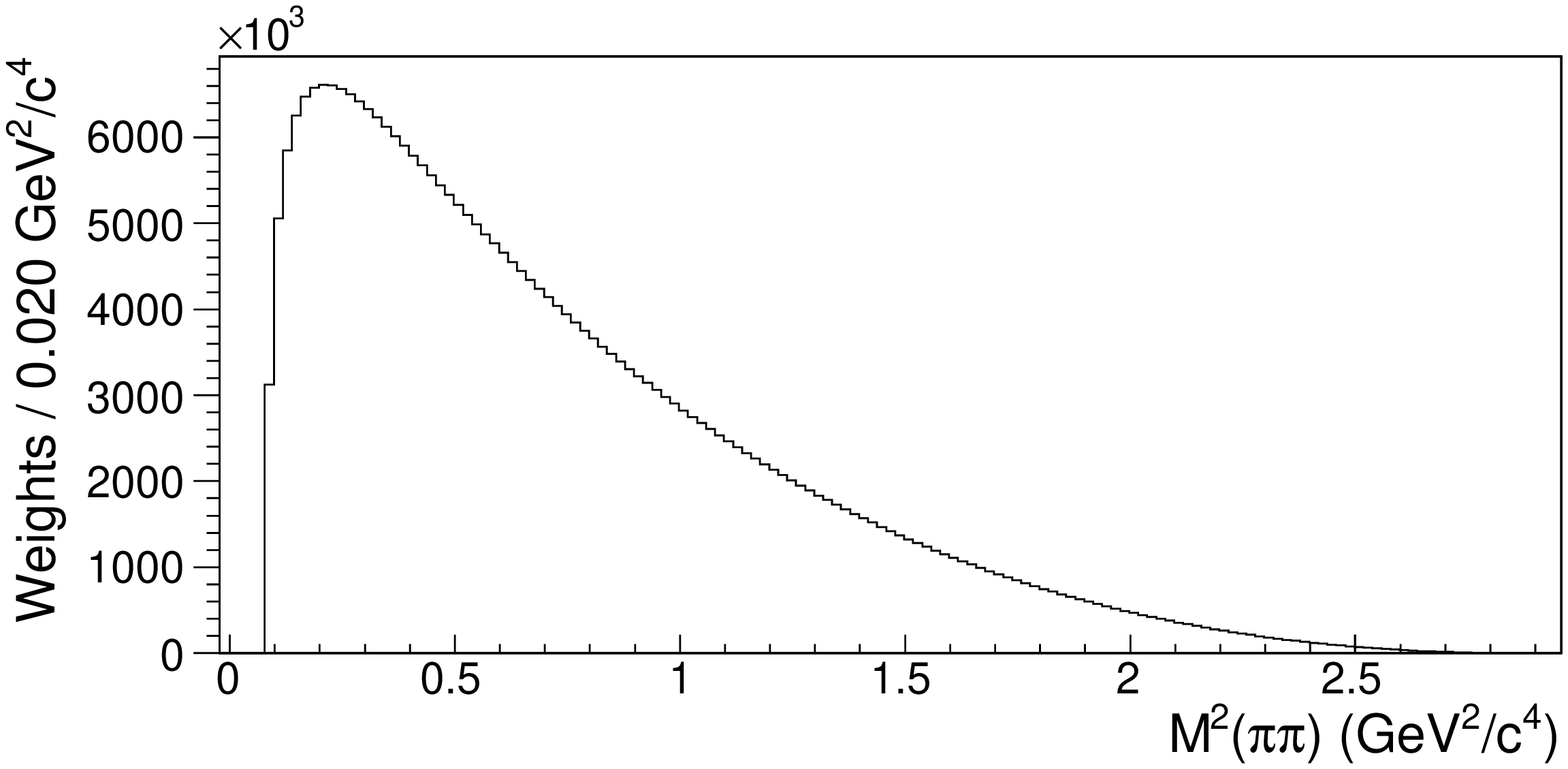}
}}
\hspace{4mm}
\scalebox{0.308}{
\rotatebox{270}{
\includegraphics*[270,37][569,691]
  {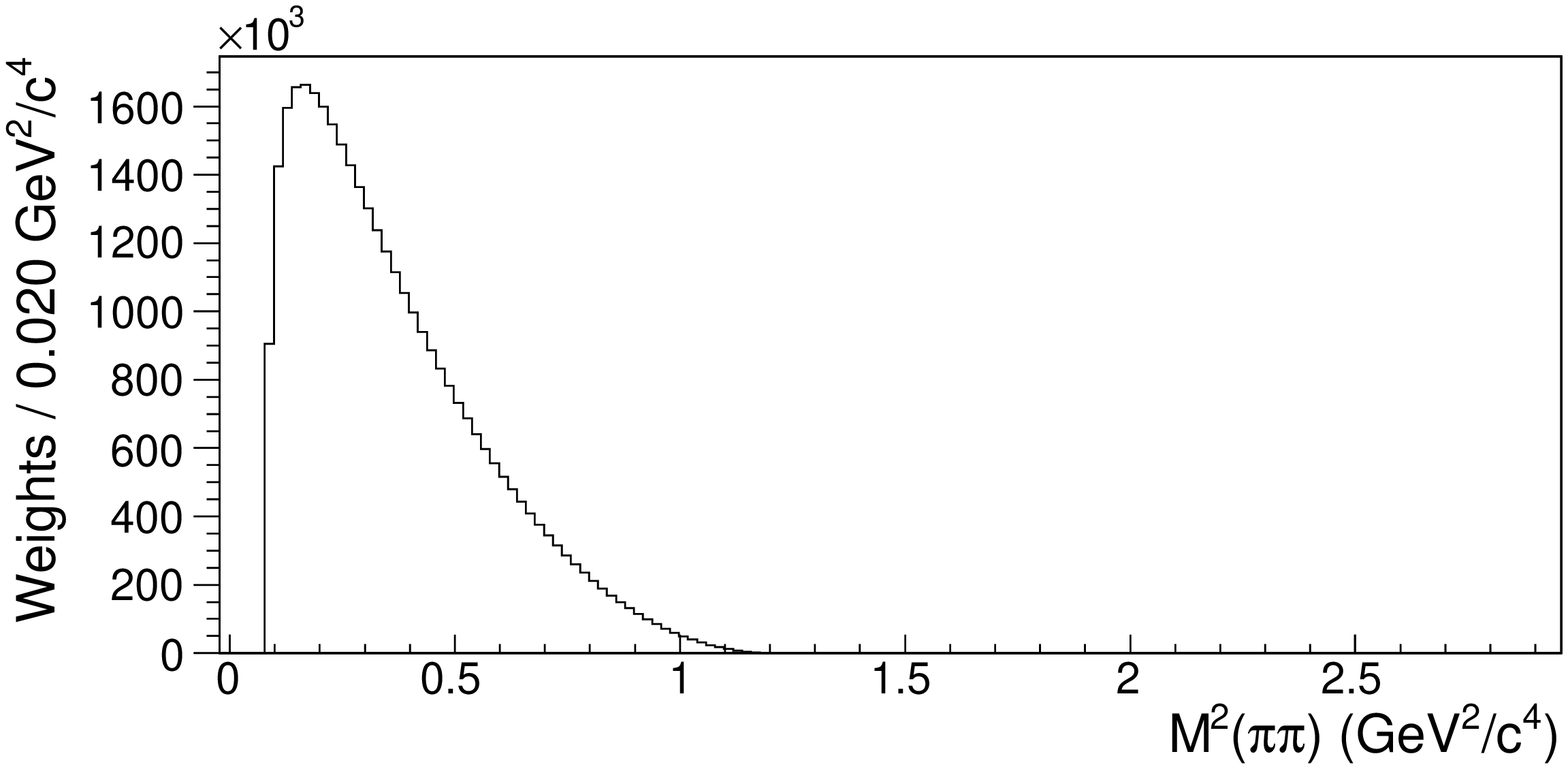}
}}}
\caption{Projections of the three-dimensional histogram of 
  phase-space weights onto the three axes
  for
  $B^{+} \rightarrow J/\psi K^{+} \pi^{+} \pi^{-}$ (left) and 
  $B^{+} \rightarrow \psi^{\prime} K^{+} \pi^{+} \pi^{-}$ (right).}
\label{amplitude:fig_phsp}
\end{figure*}

Four-body phase-space densities 
(i.e., $\phi(\vec x)$ in Eq.~\ref{amplitude:eq_signal})
for
$B^{+} \rightarrow J/\psi K^{+} \pi^{+} \pi^{-}$ and 
$B^{+} \rightarrow \psi^{\prime} K^{+} \pi^{+} \pi^{-}$
are obtained by using 
GENBOD~\cite{genbod} 
to generate final-state-particle four-momenta
that are weighted by the density of states in 
phase space~\cite{james:1968u}.
For each decay mode, $10^{8}$ events are generated.
Event phase-space weights 
are distributed 
into cubic bins in 
$M^{2}(K\pi\pi)$, $M^{2}(K\pi)$, and $M^{2}(\pi\pi)$,
with a bin width of 
$0.02~{\mathrm{GeV}}^{2}/c^{4}$.
The phase-space density is implemented as a lookup table:
the value of $\phi(\vec x)$ 
for a given data point is the 
total 
phase-space weight in the 
corresponding bin.{\footnote{Boundary effects
are
insignificant.}}
In Fig.~\ref{amplitude:fig_phsp},
the three-dimensional histogram
of phase-space weights
is projected onto the three axes,
showing 
the distribution 
that signal events would have 
in the absence of resonant effects.

\begin{figure*}[hbtp]
\centerline{
\scalebox{0.308}{
\rotatebox{270}{
\includegraphics*[245,45][572,691]
  {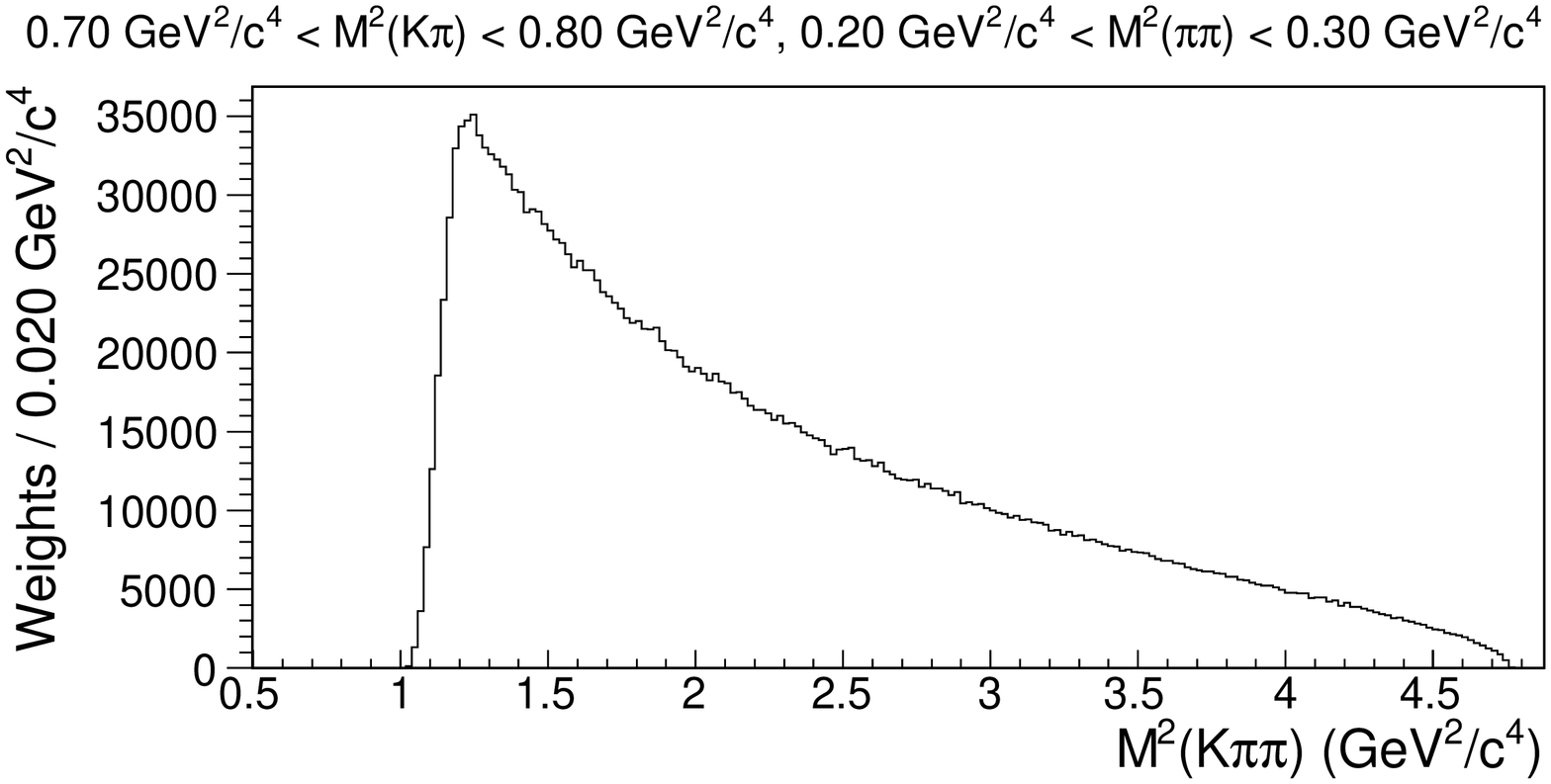}
}}
\hspace{4mm}
\scalebox{0.308}{
\rotatebox{270}{
\includegraphics*[245,45][572,691]
  {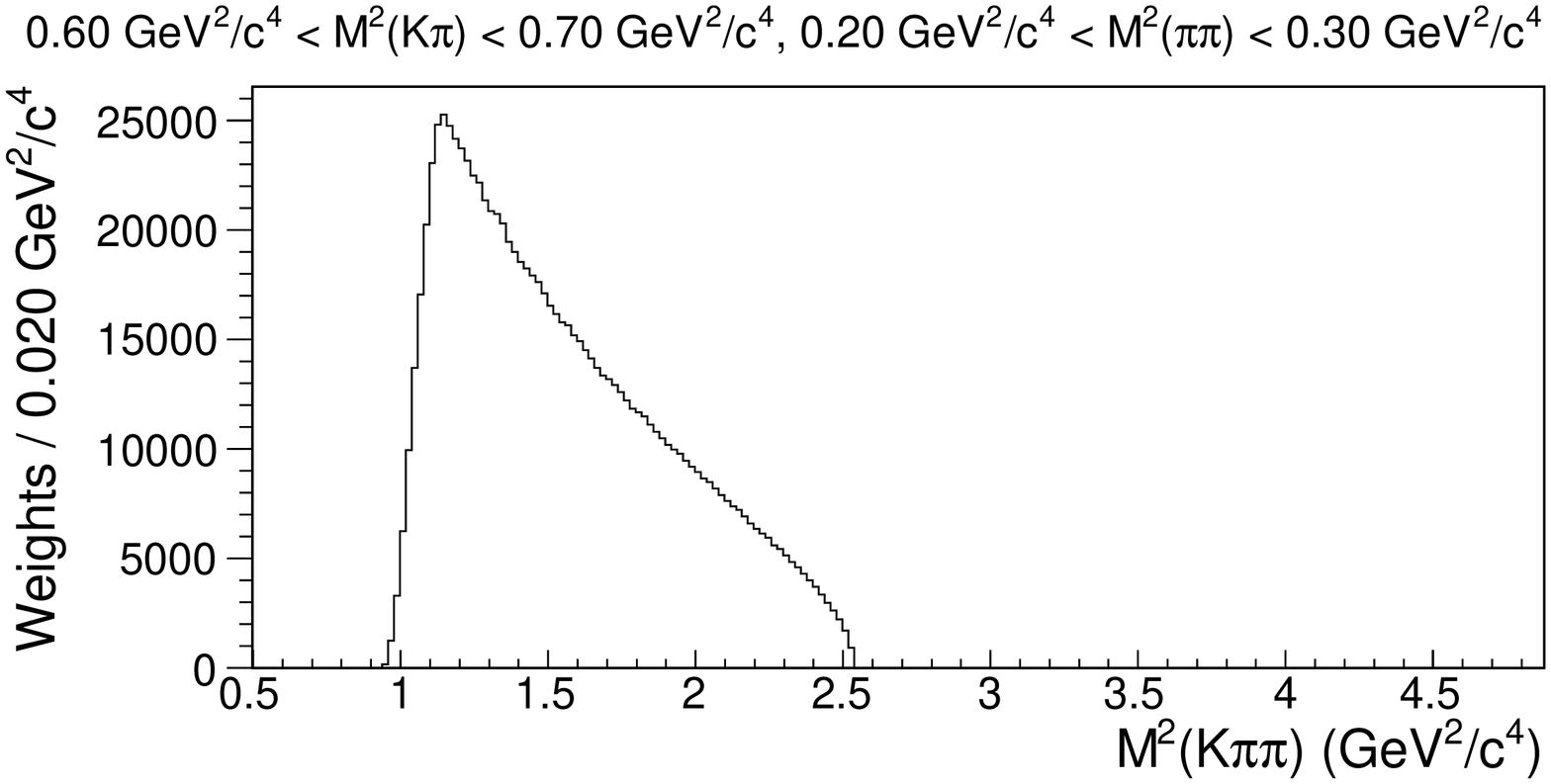}
}}}
\centerline{
\scalebox{0.308}{
\rotatebox{270}{
\includegraphics*[245,45][572,691]
  {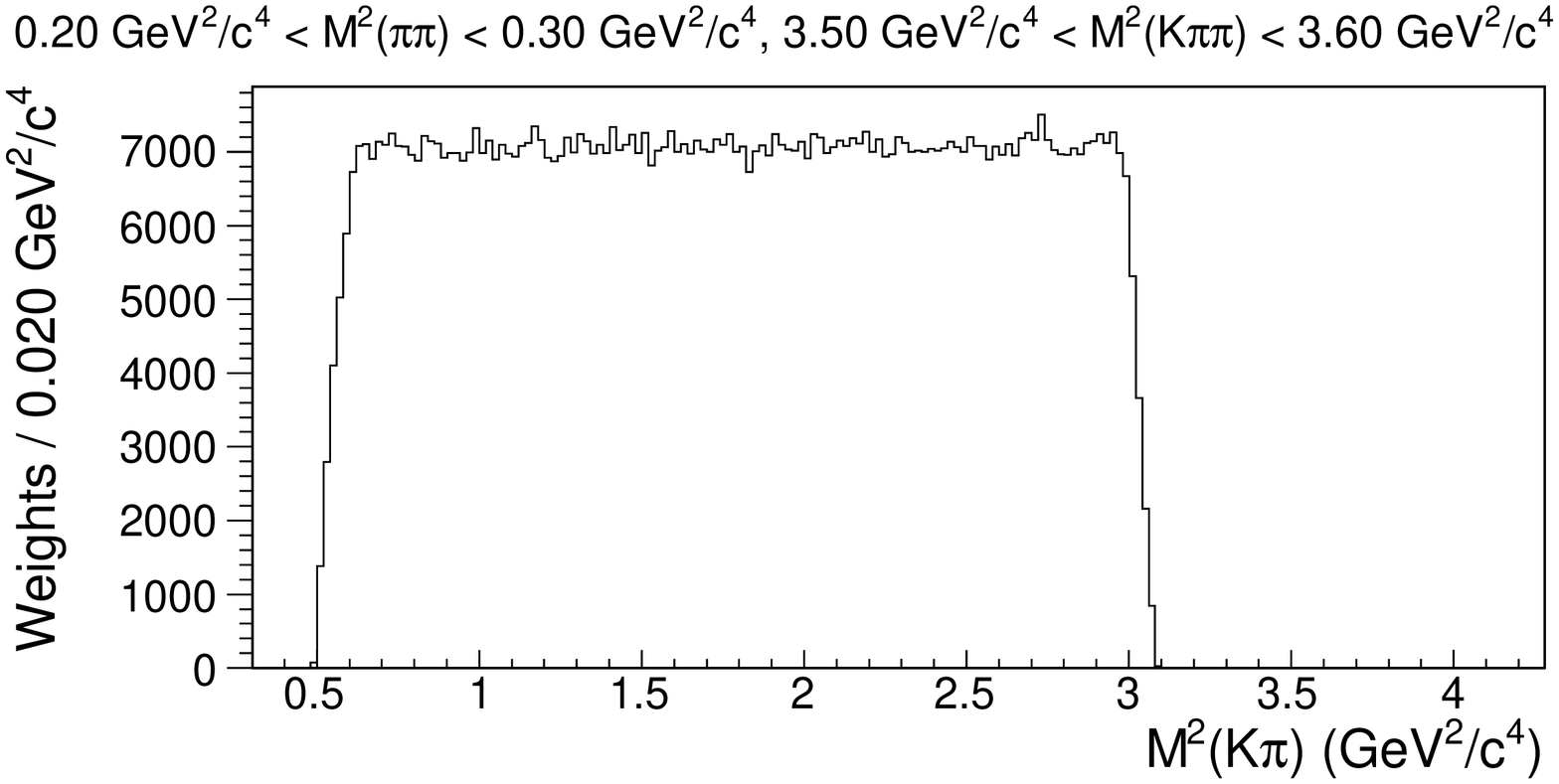}
}}
\hspace{4mm}
\scalebox{0.308}{
\rotatebox{270}{
\includegraphics*[245,45][572,691]
 {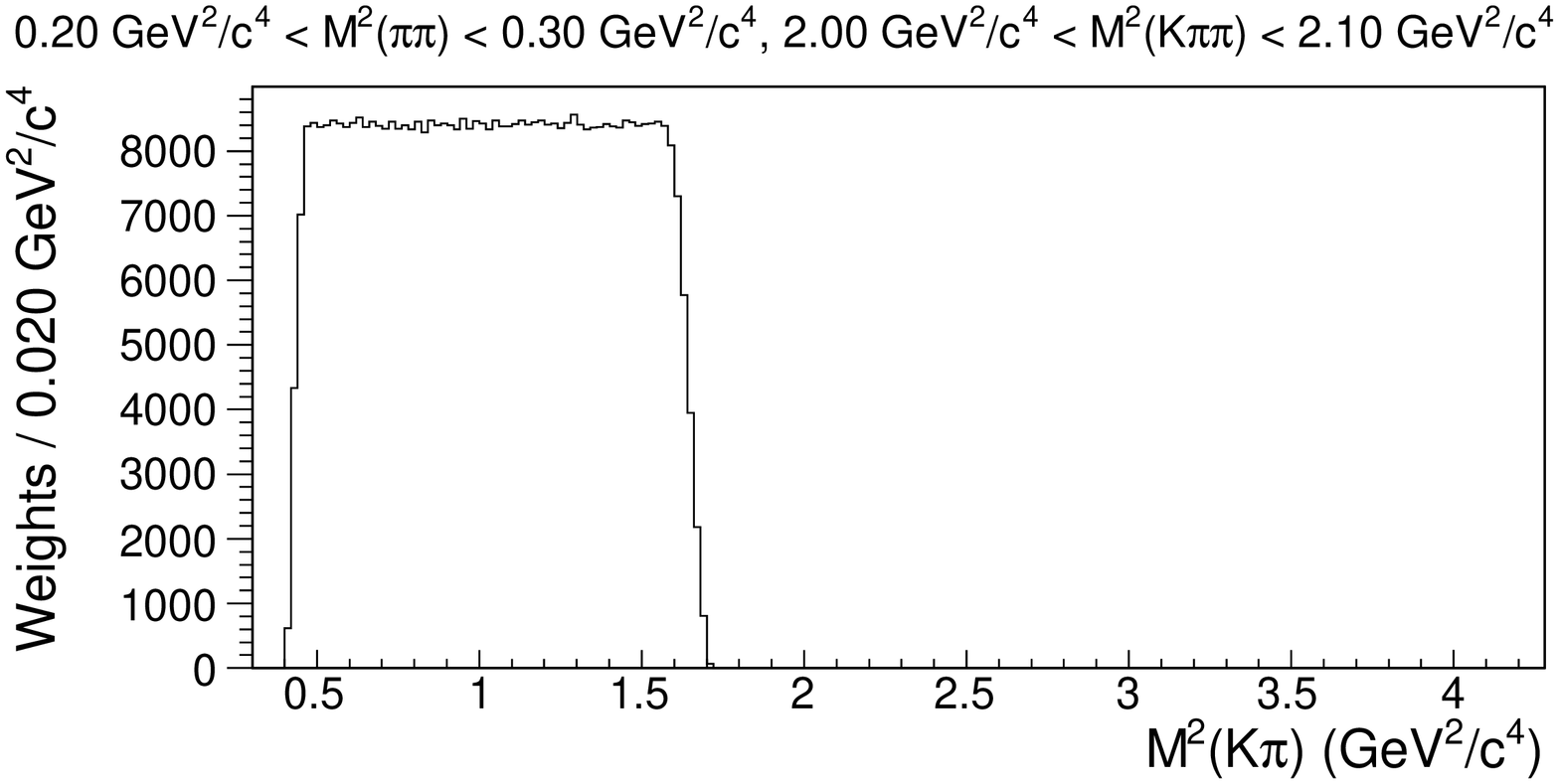}
}}}
\centerline{
\scalebox{0.308}{
\rotatebox{270}{
\includegraphics*[245,45][572,691]
  {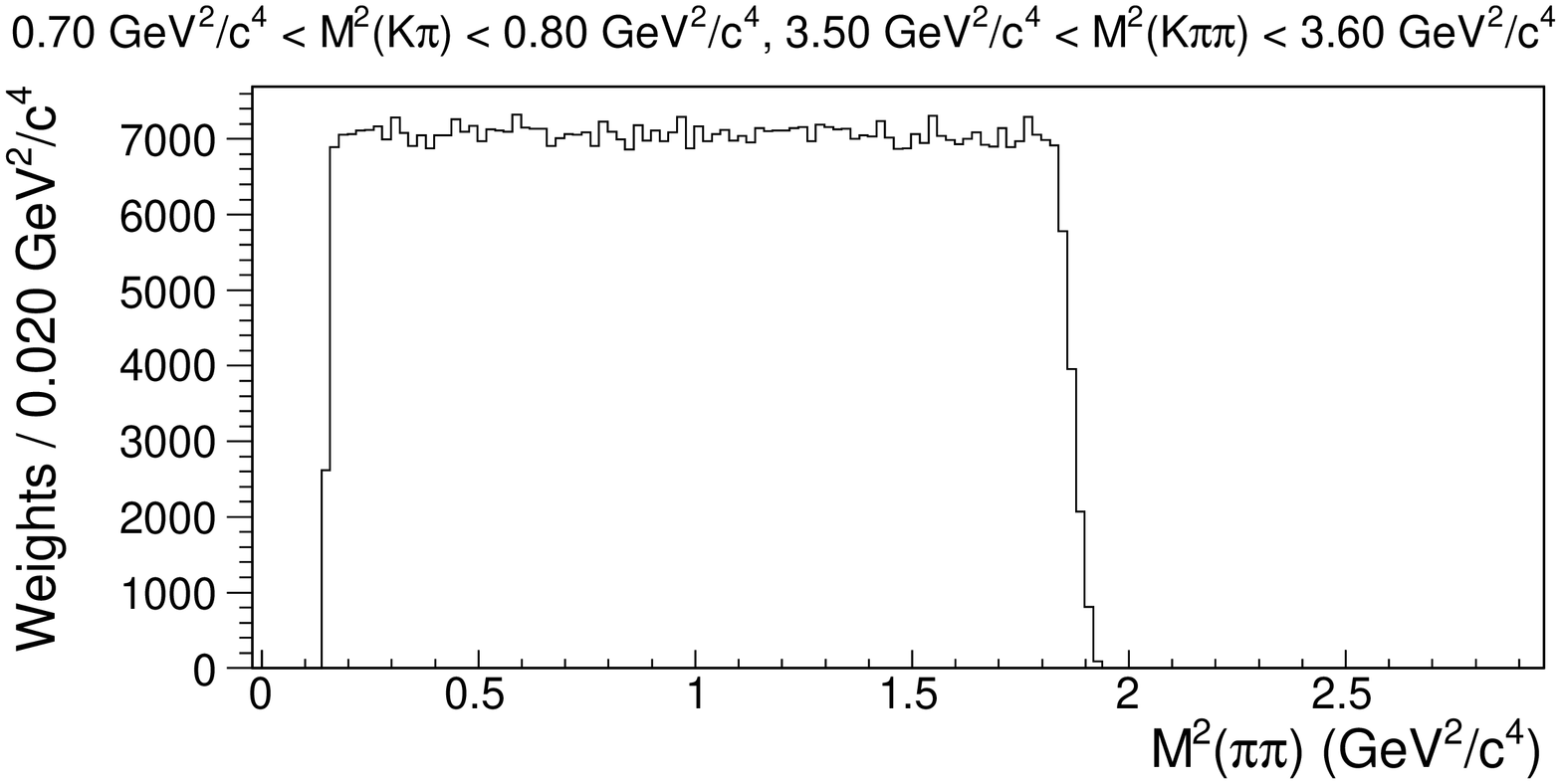}
}}
\hspace{4mm}
\scalebox{0.308}{
\rotatebox{270}{
\includegraphics*[245,45][572,691]
  {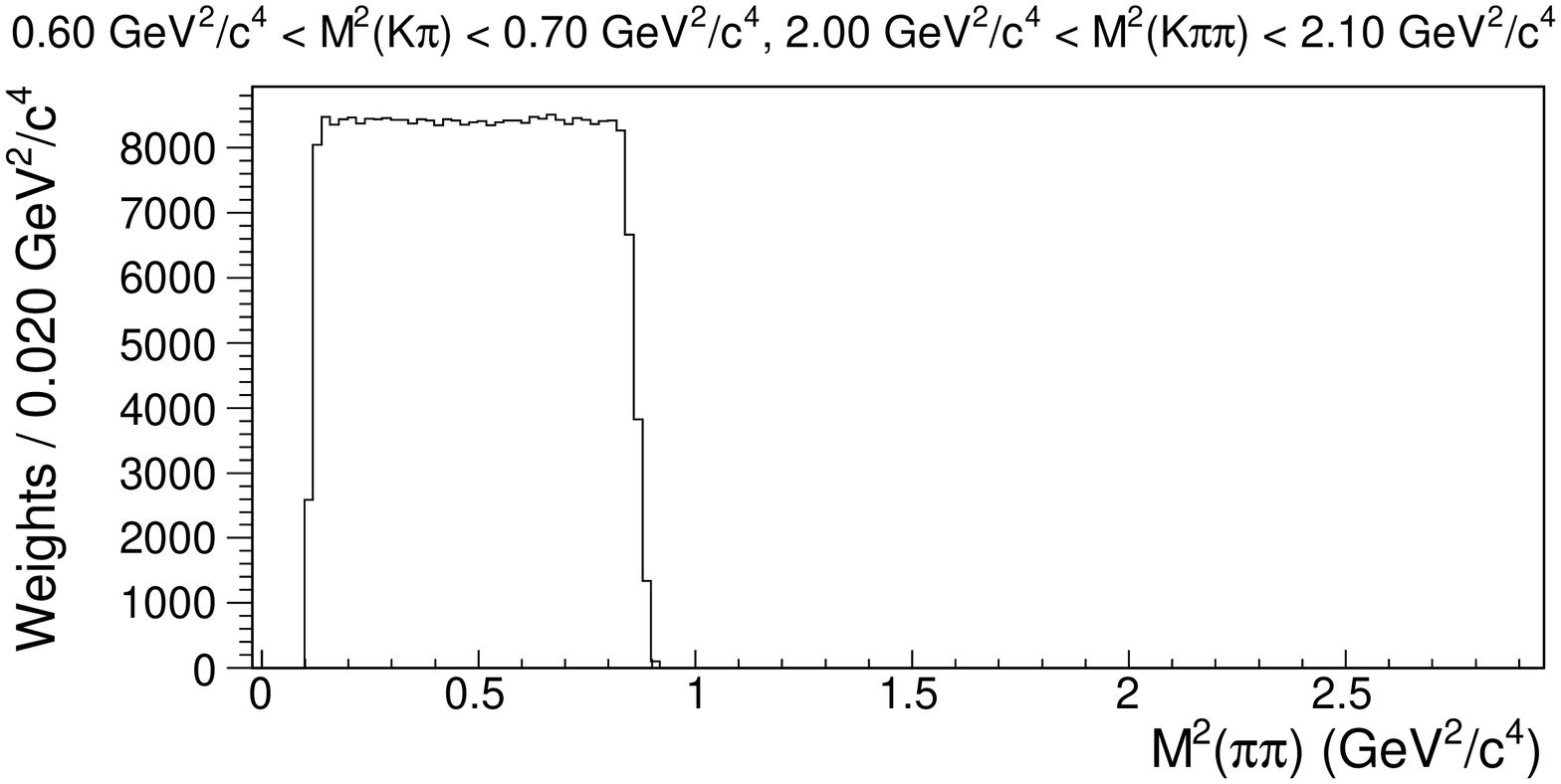}
}}}
\caption{Phase-space densities 
  for selected regions,
  as a function of 
  $M^{2}(K\pi\pi)$, $M^{2}(K\pi)$, and $M^{2}(\pi\pi)$, for
  $B^{+} \rightarrow J/\psi K^{+} \pi^{+} \pi^{-}$ (left) and 
  $B^{+} \rightarrow \psi^{\prime} K^{+} \pi^{+} \pi^{-}$ (right).
  In each case, the region selected 
  is indicated above the plot.}
\label{amplitude:fig_phsp_slices}
\end{figure*}

Figure~\ref{amplitude:fig_phsp} does not indicate the
functional form of $\phi(\vec x)$,
since 
the projection onto a single dimension
effectively integrates 
over the other two dimensions,
and the region of integration is the complicated 
one 
described in
Sec.~\ref{amplitude:section_normalization}.
In Fig.~\ref{amplitude:fig_phsp_slices},
the same projections are performed 
over 
a narrow slice in each of the other two dimensions,
to illustrate the
dependence of the function 
$\phi(\vec x)$
on each variable.

\subsection{Signal functions}

The $K^{+}\pi^{+}\pi^{-}$ final state is modeled as 
a nonresonant signal
plus a superposition of 
initial-state resonances $R_{1}$.
The latter are assumed to decay through 
intermediate-state resonances $R_{2}$
as $R_{1} \rightarrow a R_{2}$, $R_{2} \rightarrow b c$, 
where $a$, $b$, and $c$ are the final-state particles.
Specifically,
the function $s(\vec x; \vec a)$ of Eq.~\ref{amplitude:eq_signal} 
is expressed as
\begin{eqnarray}
\label{amplitude:eq_raw_signal}
\lefteqn{
s(\vec x; \vec a) 
  \equiv s(\vec x; a_{k}) } \nonumber \\
  & = & \left | a_{nr} A_{nr}(\vec x)\right |^{2} +
    \sum_{J_{1}} 
    \left | 
    \sum_{J_{2}} 
      a_{J_{1}J_{2}} A_{J_{1}J_{2}}(\vec x)
    \right |^{2} 
.
\end{eqnarray}
Here, $J_{1}$ and $J_{2}$ stand for the spin-parity ($J^{P}$)
of $R_{1}$ and $R_{2}$, respectively.
Resonances with different $J_{1}$ are added
incoherently, while those with the same $J_{1}$
are added coherently.
The parameters varied in the fit are 
the complex coefficients $a_{nr}$ and $a_{J_{1}J_{2}}$,
collectively referred to as $a_{k}$.
While the nonresonant signal is assumed to be constant 
over the phase space,
\begin{equation}
A_{nr}(\vec x) = 1 ,
\end{equation}
the resonant decay amplitudes $A_{J_{1}J_{2}}$ are expressed as
\begin{eqnarray}
\label{amplitude:eq_amplitude}
A_{J_{1}J_{2}}(\vec x) 
 & = & \alpha_{J_{1}J_{2}} 
 \nonumber\\
 & \times &
 \frac{\sqrt{M_{R_{1}} \Gamma_{R_{1}}}}
      {M_{R_{1}}^{2} - m_{abc}^{2} - i M_{R_{1}} \Gamma_{R_{1}}}
 \nonumber\\
 & \times &
 \frac{\sqrt{M_{R_{2}} \Gamma_{R_{2}}(m_{bc})}}
      {M_{R_{2}}^{2} - m_{bc}^{2} - i M_{R_{2}} \Gamma_{R_{2}}(m_{bc})}
,
\end{eqnarray}
where $\Gamma_{R_{2}}(m_{bc})$ is the mass-dependent width
\begin{equation}
\Gamma_{R_{2}}(m_{bc}) = \Gamma_{R_{2}}
\left( \frac{q}{q_{0}} \right)^{2J_{2}+1}
\left( \frac{M_{R_{2}}}{m_{bc}} \right)
F_{R}^{2}
,
\end{equation}
and $F_{R}$ is the Blatt-Weisskopf barrier factor
\begin{alignat}{2}
\label{amplitude:eq_blatt-weisskopf}
F_{R} & = 1   
            && {\text{ for }} J_{2} = 0,  \nonumber \\ 
      & = \frac {\sqrt{1 + R^{2}q_{0}^{2}}} 
                {\sqrt{1 + R^{2}q^{2}}}   
            && {\text{ for }} J_{2} = 1,            \\
      & = \frac {\sqrt{9 + 3R^{2}q_{0}^{2} + R^{4}q_{0}^{4}}} 
                {\sqrt{9 + 3R^{2}q^{2} + R^{4}q^{4}}}  
            && {\text{ for }} J_{2} = 2   \nonumber
.
\end{alignat}
The meson radial parameter $R$ is set to 
$1.5~({\mathrm{GeV}}/c)^{-1}$.
The function $\alpha_{J_{1}J_{2}}$ describes the 
spin-dependent angular distribution of the final state
and is shown for various combinations of $J_{1}$ and $J_{2}$ in 
Table~\ref{amplitude:table_angular_formulas}.
Resonances with spin greater than two are not included in the fitting
model.  In cases where there is more than one covariant spin amplitude,
only the lowest spin is included.

\begin{table}[hbtp]
\caption{Angular distribution of the 
   $K^{+}\pi^{+}\pi^{-}$ final state
   for various combinations of initial and intermediate-state
   spin parities.  See Ref.~\onlinecite{filippini:1995} for 
   derivation and conditions of applicability.}
\label{amplitude:table_angular_formulas}
\renewcommand{\arraystretch}{1.2}
\begin{ruledtabular}
\begin{tabular*}{8.6cm}{ccc}
$J_{1}$  & $J_{2}$ & $\alpha_{J_{1}J_{2}}$         \\ \colrule
         & $0^{+}$ &                               \\ 
\raisebox{1.9ex}[0pt]{Any} & $0^{-}$ & 
  \raisebox{1.9ex}[0pt]{$1$}   \\ \colrule
$0^{+}$  & $1^{+}$ &                               \\ 
$0^{-}$  & $1^{-}$ & 
  \raisebox{1.9ex}[0pt]{$(1+z^{2})\cos^{2}\theta$} \\ \colrule
$1^{+}$  & $1^{-}$ &                               \\ 
$1^{-}$  & $1^{+}$ & 
  \raisebox{1.9ex}[0pt]{$1+z^{2}\cos^{2}\theta$}   \\ \colrule
$1^{+}$  & $1^{+}$ &                               \\ 
$1^{-}$  & $1^{-}$ & 
  \raisebox{1.9ex}[0pt]{$1-\cos^{2}\theta$}        \\ \colrule
$1^{+}$  & $2^{+}$ & {$(1+z^{2})$}                 \\ 
$1^{-}$  & $2^{-}$ & 
  {${} \times \left[ 1 + 3 \cos^{2}\theta +
  9z^{2} ( \cos^{2} \theta - 1/3 ) ^{2} \right]$} 
                                                   \\ \colrule
$2^{+}$  & $1^{+}$ &                               \\ 
$2^{-}$  & $1^{-}$ & 
  \raisebox{1.9ex}[0pt]{$3 + (1 + 4z^{2})\cos^{2}\theta$} \\ \colrule
$2^{+}$  & $1^{-}$ &                               \\ 
$2^{-}$  & $1^{+}$ & 
  \raisebox{1.9ex}[0pt]{$1-\cos^{2}\theta$}        \\ \colrule
$2^{+}$  & $2^{+}$ &                              
  {$1 + z^{2} / 9 + ( z^{2} / 3 - 1 ) \cos^{2}\theta$} \\ 
$2^{-}$  & $2^{-}$ & 
  {${} - z^{2} ( \cos^{2}\theta - 1/3)^{2} $} \\ \colrule
$2^{+}$  & $2^{-}$ &                           \\   
$2^{-}$  & $2^{+}$ & 
  \raisebox{1.9ex}[0pt]{$1 + z^{2}/3 + z^{2}\cos^{2}\theta + z^{4} ( \cos^{2}\theta - 1/3 )^{2}$} \\ 
\end{tabular*}
\end{ruledtabular}
\end{table}

In Eqs.~\ref{amplitude:eq_amplitude}-\ref{amplitude:eq_blatt-weisskopf} 
and Table \ref{amplitude:table_angular_formulas},
the nominal masses of the resonances $R_{1}$ and $R_{2}$
are denoted by $M_{R_{1}}$ and $M_{R_{2}}$, 
and the nominal widths by $\Gamma_{R_{1}}$ and $\Gamma_{R_{2}}$. 
The angle $\theta$ is 
between $a$ and $b$ in the $bc$ rest frame and can be expressed as
\begin{eqnarray}
\lefteqn{ \cos \theta 
 = 
\frac{m_{bc}}{4pqm_{abc}} }
 \nonumber\\
& \times &
\left[
    m_{ac}^{2} 
  - m_{ab}^{2} 
  + \frac { (m_{abc}^{2}-m_{a}^{2}) (m_{b}^{2}-m_{c}^{2}) } 
          { m_{bc}^{2} }
\right]
.
\end{eqnarray}
The variable $z$ is given by
\begin{equation}
z = p / m_{abc} .
\end{equation}
The breakup momentum $p$ is 
the momentum of $a$ or $bc$ in the $abc$ rest frame:
\begin{equation}
p^{2} = 
\frac{(m_{abc}^{2} - (m_{a} + m_{bc})^{2})
      (m_{abc}^{2} - (m_{a} - m_{bc})^{2})}
     {4m_{abc}^{2}}
,
\end{equation}
while $q$ is the momentum of $b$ or $c$ in the $bc$ rest frame:
\begin{equation}
q^{2} = 
\frac{(m_{bc}^{2} - (m_{b} + m_{c})^{2})
      (m_{bc}^{2} - (m_{b} - m_{c})^{2})}
     {4m_{bc}^{2}}
,
\end{equation}
where the constant $q_{0}$ is the value of $q$ 
evaluated at $m_{bc} = M_{R_{2}}$.

Since the components of the signal function
are not individually normalized,
it is not meaningful to compare
the moduli of the
complex coefficients $a_{k}$
in Eq.~\ref{amplitude:eq_raw_signal}.
A decay fraction is therefore calculated for each component
by integrating the
component
over the kinematically-allowed region
and dividing by the integral of the full signal function
\begin{equation}
\label{amplitude:eq_decay_fraction}
f_{k} = \frac{ \int \phi(\vec x) |a_{k} A_{k}(\vec x)|^{2} d^{3} x }
             { \int \phi(\vec x) s(\vec x; \vec a) d^{3}x }
.
\end{equation}
The integrations in Eq.~\ref{amplitude:eq_decay_fraction} 
are performed
as described in Sec.~\ref{amplitude:section_normalization}.
Because of interference effects, 
decay fractions for a given final state
will not, in general, add up to unity.

\subsection{Statistical errors}

As with the sideband-region fits, the statistical uncertainties 
in the fit parameters (i.e., moduli and phases)
are determined by the fitter: the error in a given parameter is 
the change in that parameter that reduces the log likelihood by $1/2$.
The statistical uncertainties in the decay fractions, 
on the other hand, are more complicated.  
Since a given decay fraction involves the integral of the full signal
function, the error in a single decay fraction
incorporates the errors in all of the parameters.
To determine the statistical errors in the decay fractions,
$1000$ sets of 
correlated signal-function parameters 
are drawn 
from
Gaussian distributions using the fitted parameter values and the
error matrix.{\footnote{Correlated Gaussian distributions are 
generated using CORSET and CORGEN~\cite{corset}.}}
Decay fractions are calculated 
for each set of generated parameters.
The rms of the resulting distribution provides
an estimate of the statistical error in the decay fraction.

\subsection{Systematic errors}
\label{amplitude:section_systematics}

Several sources of systematic error are considered, as described below.
They are added in quadrature to obtain the 
systematic errors
reported in Sec.~\ref{amplitude:section_results}.

\subsubsection{Background parametrization}

A possible source of systematic error in the fits is
the fixed
background fraction $n_{B}$ in Eq.~\ref{amplitude:eq_pdf}.
While the error in $n_{B}$ is small,
the correction for the oversubtraction,
described in Sec.~\ref{inclusive:section_systematics},
lowers $n_{B}$
by $10.8\%$
for $B^{+} \rightarrow J/\psi K^{+} \pi^{+} \pi^{-}$
and by $11.4\%$ 
for $B^{+} \rightarrow \psi^{\prime} K^{+} \pi^{+} \pi^{-}$.
The systematic error associated with this correction is estimated
conservatively as the change in each parameter
when the fits are performed with the uncorrected values of $n_{B}$.

There may be an additional systematic error 
if the background in the signal region is not correctly parametrized
by the shape determined by fitting the sidebands.
As noted in Sec.~\ref{section_transformations}, generic-MC studies 
suggest that not enough of the $K_{S}^{0}$ and $\rho$ 
background peaks are removed by the sideband subtraction.
To estimate this error,
a fit is performed in which the coefficients of the background peaks 
in Eqs.~\ref{amplitude:eq_bkg_jkpp} and \ref{amplitude:eq_bkg_pkpp}
are doubled.

\subsubsection{Efficiency}

To estimate the error introduced by binning the efficiency information,
the fits are repeated 
using bin sizes of 
$0.10~{{\mathrm{GeV}}^{2}/c^{4}}$ and
$0.20~{{\mathrm{GeV}}^{2}/c^{4}}$ 
for the efficiency.
The average absolute change in each parameter is 
the estimate of the error.

Another possible source of error is that
the MC simulation may not faithfully reproduce the detector efficiency
for low-momentum particles.
To test for such an effect, 
two additional fits are performed.
In the first fit, 
only charged particles with a momentum greater than
$200~{\mathrm{MeV}}/c$
are included.
In the second fit,
the $|dr|$ and $|dz|$ requirements described in Sec.~\ref{section_events}
are loosened
from $0.4~{\mathrm{cm}}$ to $0.8~{\mathrm{cm}}$, and 
from $1.5~{\mathrm{cm}}$ to $3.0~{\mathrm{cm}}$, respectively.
The changes in each parameter observed in these two fits
are added in quadrature to 
obtain an estimate of the error
due to inaccuracies in the efficiency estimation.

Using only the three variables
$M^{2}(K\pi\pi)$, $M^{2}(K\pi)$, and $M^{2}(\pi\pi)$
in this analysis
is equivalent to integrating over
variables that describe the relative momentum
of the $J/\psi$ or $\psi^{\prime}$
with respect to the
$K^{+} \pi^{+} \pi^{-}$ system.
In this integration,
the terms corresponding to $K^{+} \pi^{+} \pi^{-}$
states with different initial-state spin-parity cancel out,
producing Eq.~\ref{amplitude:eq_raw_signal}.
This cancellation, however, is exact
only if the
detector efficiency is flat
over the extra variables.
To determine the effect of neglecting these variables,
an additional set of fits is performed,
in which the efficiency in Eq.~\ref{amplitude:eq_signal}
is calculated as a function of the two angles between 
the $J/\psi$ or $\psi^{\prime}$ and the $K^{+} \pi^{+} \pi^{-}$ system,
rather than $M^{2}(K\pi\pi)$, $M^{2}(K\pi)$, and $M^{2}(\pi\pi)$.
The resulting fitted parameters are compared
to those obtained by a fit in which the efficiency is held  
constant.{\footnote{If the efficiency is calculated as a function
of all five dimensions,
the accuracy of the result becomes dominated by the 
MC statistics.}}
The absolute change in each parameter is 
found to be 
small (less than $15\%$ of the statistical error)
and is included in the systematic error.

\subsubsection{Integration step size}

To estimate the error introduced by the finite step size used in
the numerical integrals
of Secs.~\ref{section_transformations}
and~\ref{amplitude:section_normalization},
the fits are repeated, using a step size of
$0.005~{\mathrm{GeV}^{2}/c^{4}}$ 
for $B^{+} \rightarrow J/\psi K^{+} \pi^{+} \pi^{-}$ and
$0.010~{\mathrm{GeV}^{2}/c^{4}}$ 
for $B^{+} \rightarrow \psi^{\prime} K^{+} \pi^{+} \pi^{-}$.
The change in each parameter is an estimate of the 
uncertainty associated with the numerical integration.

\subsubsection{Modeling of the signal}

The masses and widths of the resonances included in the fits 
are listed in Table~\ref{amplitude:table_masses_and_widths}.
To estimate the systematic error associated with
the uncertainties in these quantities,
the fits are repeated,
varying each fixed quantity within its errors.
For each mass or width,
the average absolute change in each parameter is recorded.
These average changes are then added in quadrature.

\begin{table}[btp]
\caption{Masses, widths, and spin-parity values
   of the resonances included in the fits.
   With the exception of the $K(1600)$ parameters
   (discussed in Sec.~\ref{amplitude:section_l_region}),
   all values are from~\cite{pdg08}.}
\label{amplitude:table_masses_and_widths}
\renewcommand{\tabcolsep}{0.3cm}
\renewcommand{\arraystretch}{1.2}
\begin{ruledtabular}
\begin{tabular*}{8.6cm}{cr@{$~\pm~$}lr@{$~\pm~$}lc} 
  & \multicolumn{2}{c}{Mass} & \multicolumn{2}{c}{Width} & \\[-0.5ex]
\raisebox{1.5ex}[0pt]{Resonance}
  & \multicolumn{2}{c}{(${\mathrm{MeV}}/c^{2}$)}
  & \multicolumn{2}{c}{(${\mathrm{MeV}}/c^{2}$)}  
  & \raisebox{1.5ex}[0pt]{$J^{P}$}                             \\
\colrule
$\rho^{0}$            &  $775.49$ &   $0.34$
                      &  $146.2$  &   $0.7$          & $1^{-}$ \\
$\omega$              &  $782.65$ &   $0.12$ 
                      &    $8.49$ &   $0.08$         & $1^{-}$ \\
$f_{0}(980)^{0}$      &  $980$    &   $10$
                      &  $50$     & $^{50}_{10}$     & $0^{+}$ \\
$f_{2}(1270)^{0}$     & $1275.1$  &   $1.2$ 
                      &  $185.1$  & $^{2.9}_{2.4}$   & $0^{+}$ \\
$K^{*}(892)^{0}$      &  $896.00$ &   $0.25$ 
                      &   $50.3$  &   $0.6$          & $1^{-}$ \\
$K_{1}(1270)^{+}$     & $1272$    &   $7$    
                      &   $90$    &  $20$            & $1^{+}$ \\
$K_{1}(1400)^{+}$     & $1403$    &   $7$    
                      &  $174$    &  $13$            & $1^{+}$ \\
$K^{*}(1410)^{+}$     & $1414$    &  $15$    
                      &  $232$    &  $21$            & $1^{-}$ \\
$K_{0}^{*}(1430)^{+}$ & $1425$    &  $50$    
                      &  $270$    &  $80$            & $0^{+}$ \\
$K_{2}^{*}(1430)^{+}$ & $1425.6$  &   $1.5$  
                      &   $98.5$  &   $2.7$          & $2^{+}$ \\
$K_{2}^{*}(1430)^{0}$ & $1432.4$  &   $1.3$  
                      &  $109$    &   $5$            & $2^{+}$ \\
$K(1600)^{+}$         & $1605$    &  $15$
                      &  $115$    &  $15$            & $2^{-}$ \\
$K^{*}(1680)^{+}$     & $1717$    &  $27$    
                      &  $322$    & $110$            & $1^{-}$ \\
$K_{2}(1770)^{+}$     & $1773$    &   $8$    
                      &  $186$    &  $14$            & $2^{-}$ \\
$K_{2}^{*}(1980)^{+}$ & $1973$    &  $26$    
                      &  $373$    &  $68$            & $2^{+}$ \\ 
\end{tabular*}
\end{ruledtabular}
\end{table}

In fitting the 
$B^{+} \rightarrow J/\psi K^{+} \pi^{+} \pi^{-}$ data,
the modulus for
$K_{2}^{*}(1430) \rightarrow K^{*}(892) \pi$ 
is allowed to float.
Relative to this modulus,
the moduli{\footnote{The phases of the three submodes are 
allowed to float.}}
for
$K_{2}^{*}(1430) \rightarrow K \rho$ 
and
$K_{2}^{*}(1430) \rightarrow K \omega$ 
are fixed based on 
previously-measured relative branching
fractions~\cite{pdg08}.
To estimate the associated systematic error,
additional fits are performed,
varying these branching fractions within their uncertainties.

\subsection{Results}
\label{amplitude:section_results}

\begin{figure}[btp]
\centerline{
\scalebox{0.35}{
\rotatebox{270}{
\includegraphics*[270,37][569,691]
  {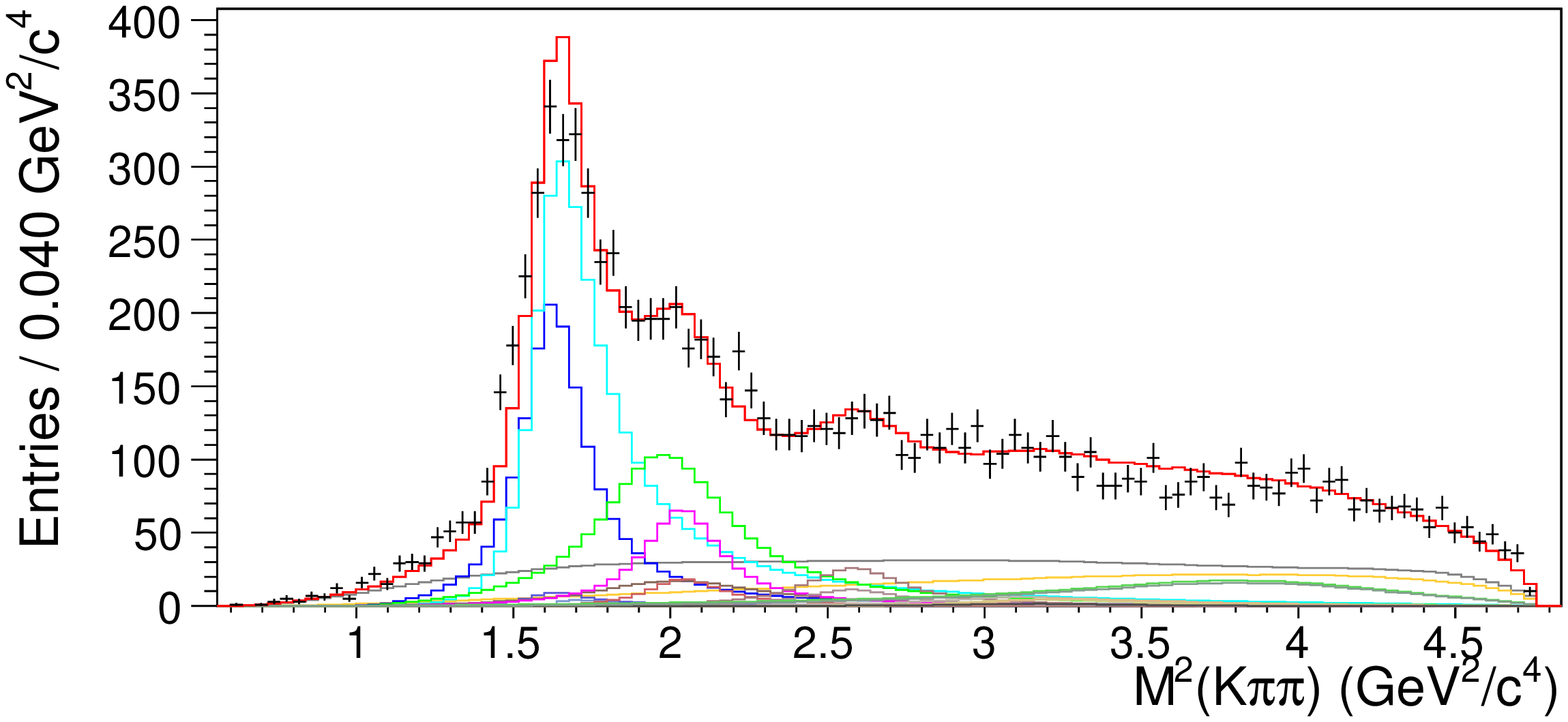}
}}}
\vspace{2mm}
\centerline{
\scalebox{0.35}{
\rotatebox{270}{
\includegraphics*[270,37][569,691]
  {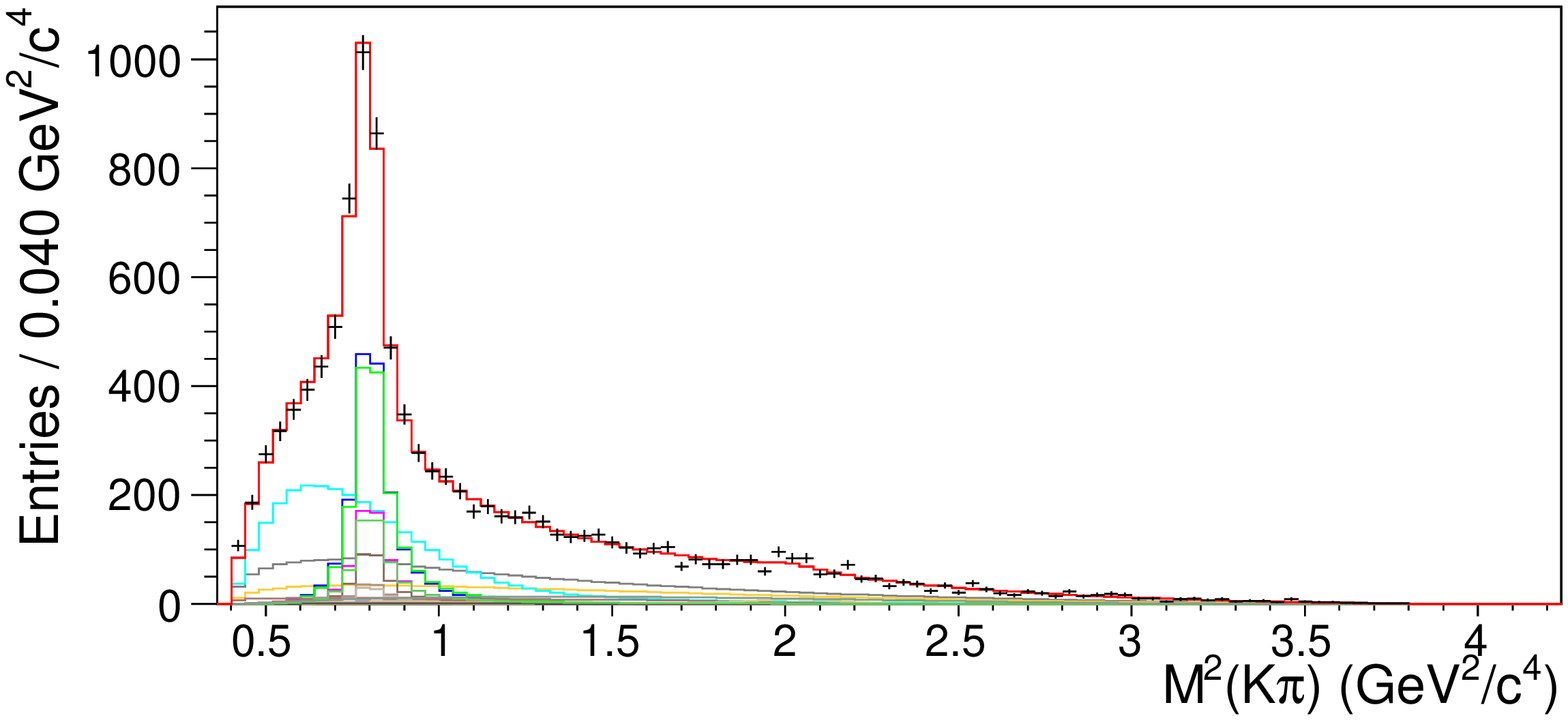}
}}}
\vspace{2mm}
\centerline{
\scalebox{0.35}{
\rotatebox{270}{
\includegraphics*[270,37][569,691]
  {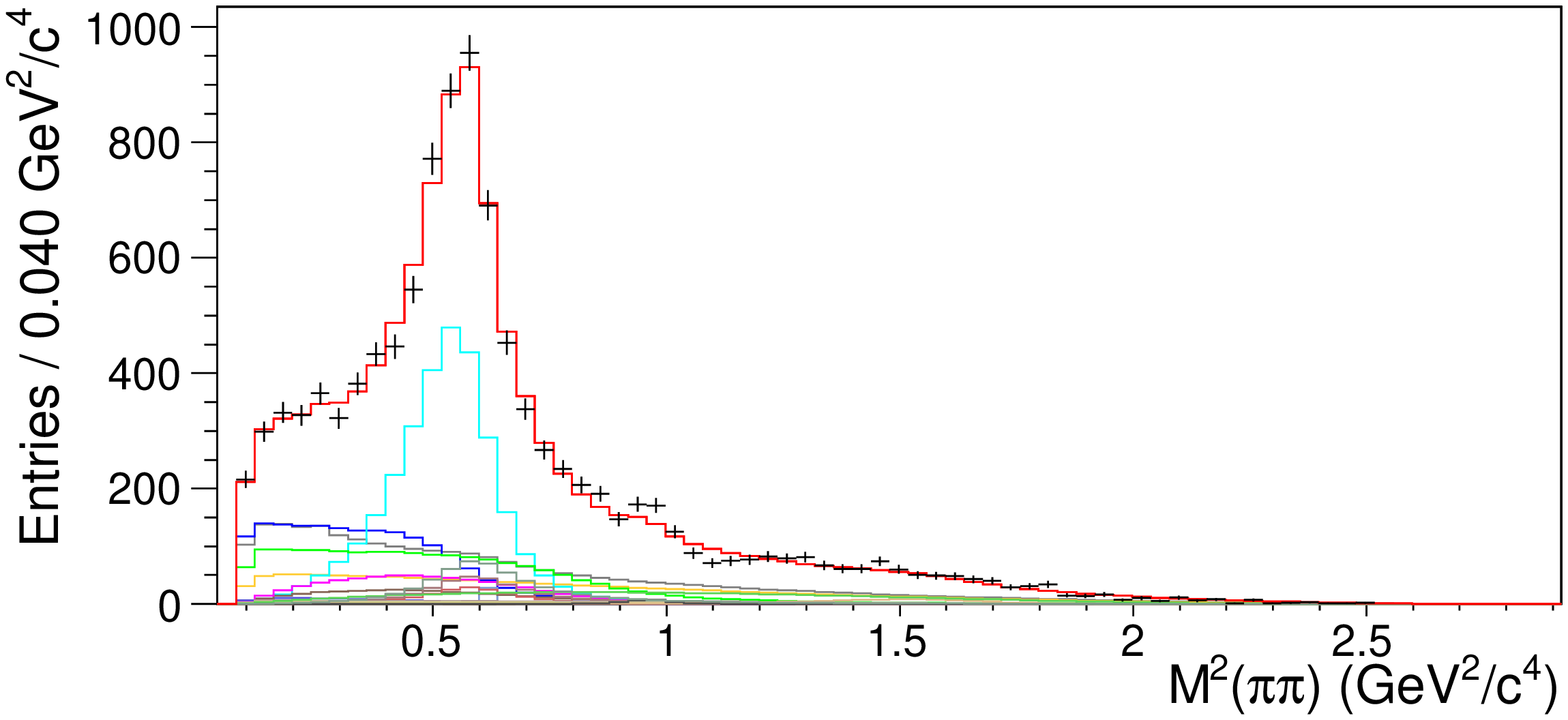}
}}}
\caption{Results of signal-region fits for 
  $B^{+} \rightarrow J/\psi K^{+} \pi^{+} \pi^{-}$.
  Data (points) and fits (histograms) are shown projected onto the 
  three axes.
  The fit components are color coded as shown in
  Fig.~\ref{amplitude:fig_signal_legend}.}
\label{amplitude:fig_signal_pdg_jkpp}
\end{figure}

\begin{figure}[btp]
\centerline{
\scalebox{0.32}{
\rotatebox{270}{
\includegraphics*[53,41][549,738]
  {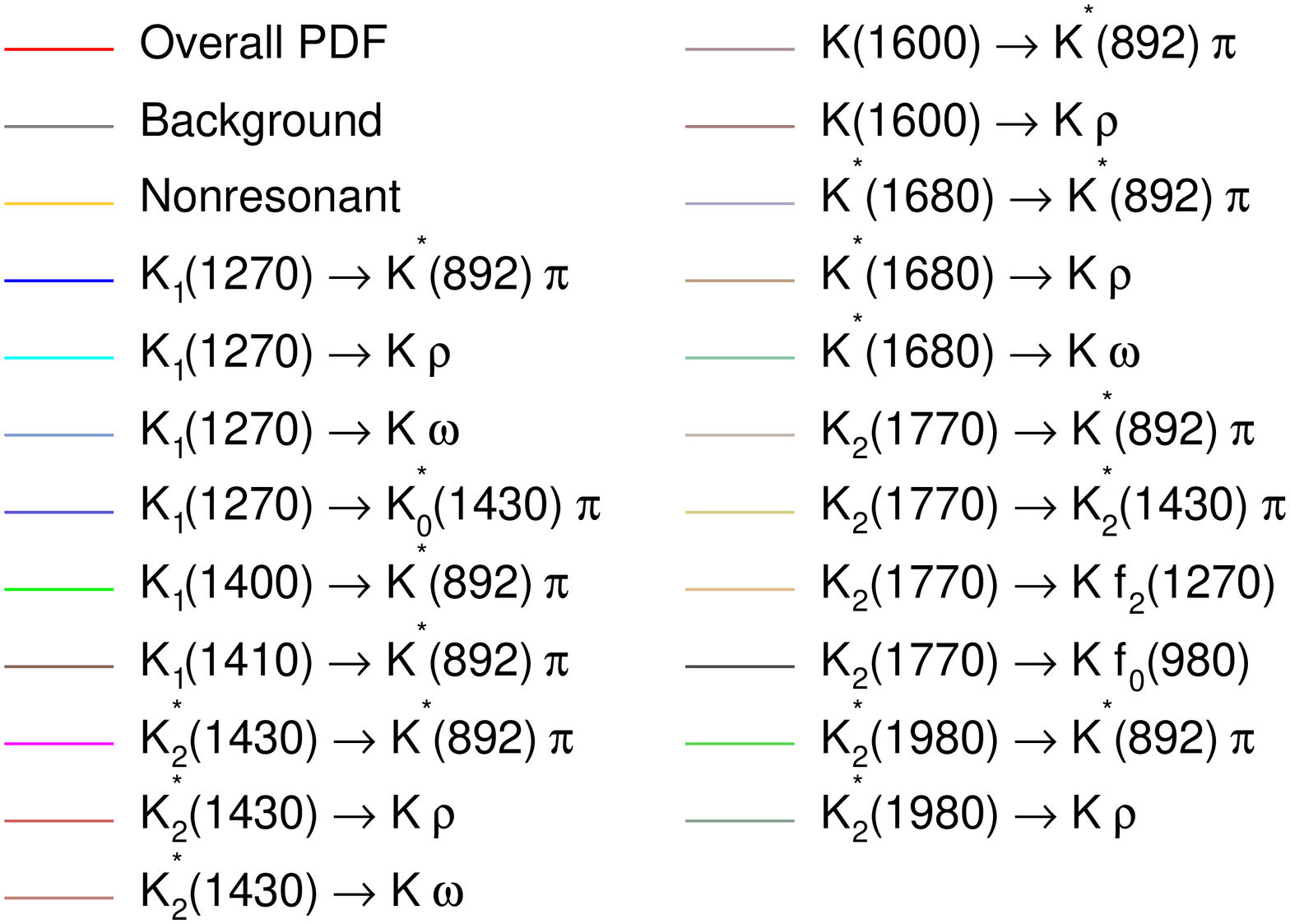}
}}}
\caption{Legend for
  Figs.~\ref{amplitude:fig_signal_pdg_jkpp},
        \ref{amplitude:fig_signal_pdg_slices_jkpp},
        \ref{amplitude:fig_signal_pdg_pkpp},
        \ref{amplitude:fig_signal_pdg_slices_pkpp},
        \ref{amplitude:fig_signal_float_jkpp}, and
        \ref{amplitude:fig_signal_float_slices_jkpp}.}
\label{amplitude:fig_signal_legend}
\end{figure}

\begin{figure*}[hbtp]
\centerline{
\scalebox{0.35}{
\rotatebox{270}{
\includegraphics*[270,37][569,691]
  {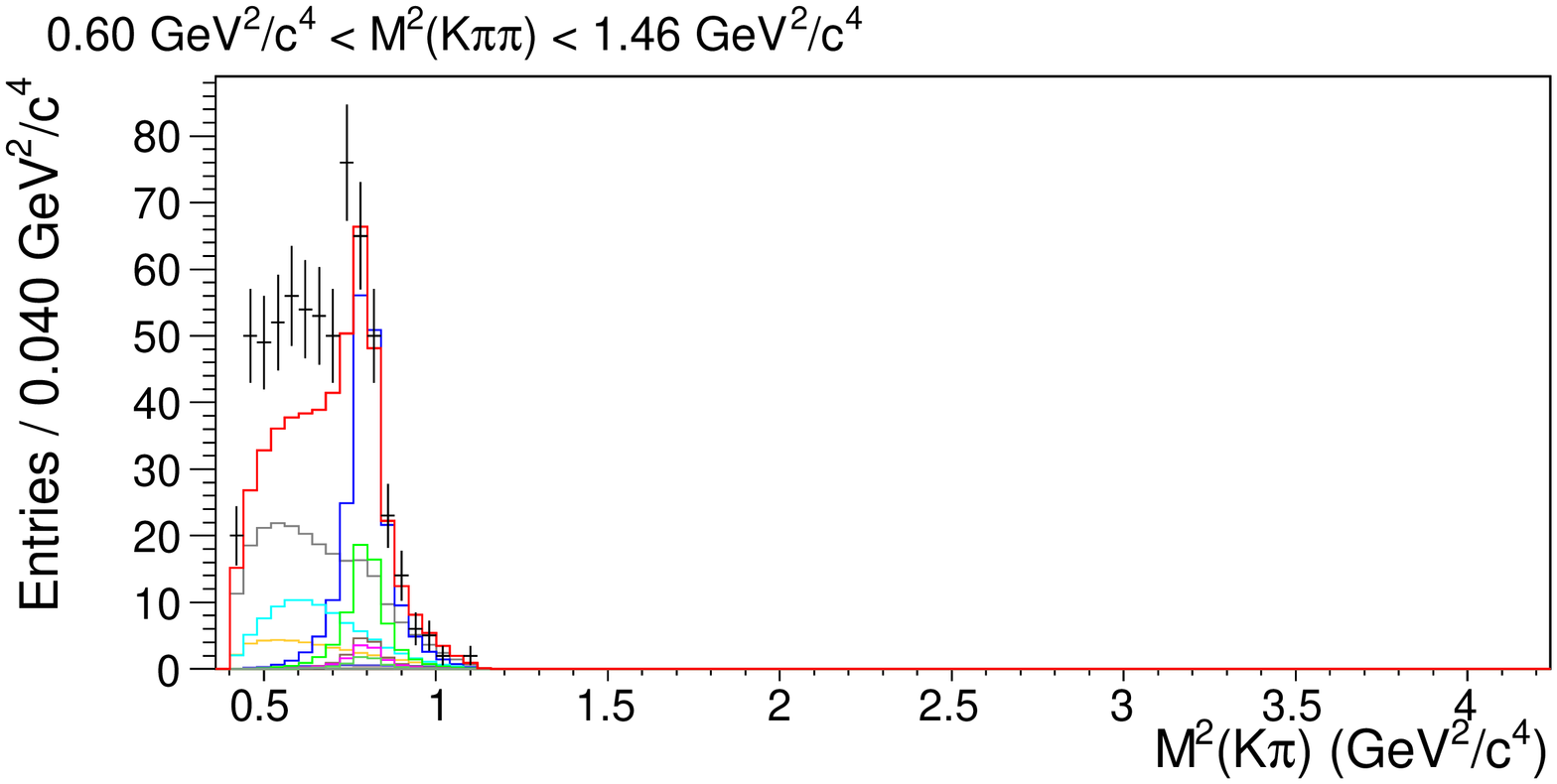}
}}
\hspace{4mm}
\scalebox{0.35}{
\rotatebox{270}{
\includegraphics*[270,37][569,691]
  {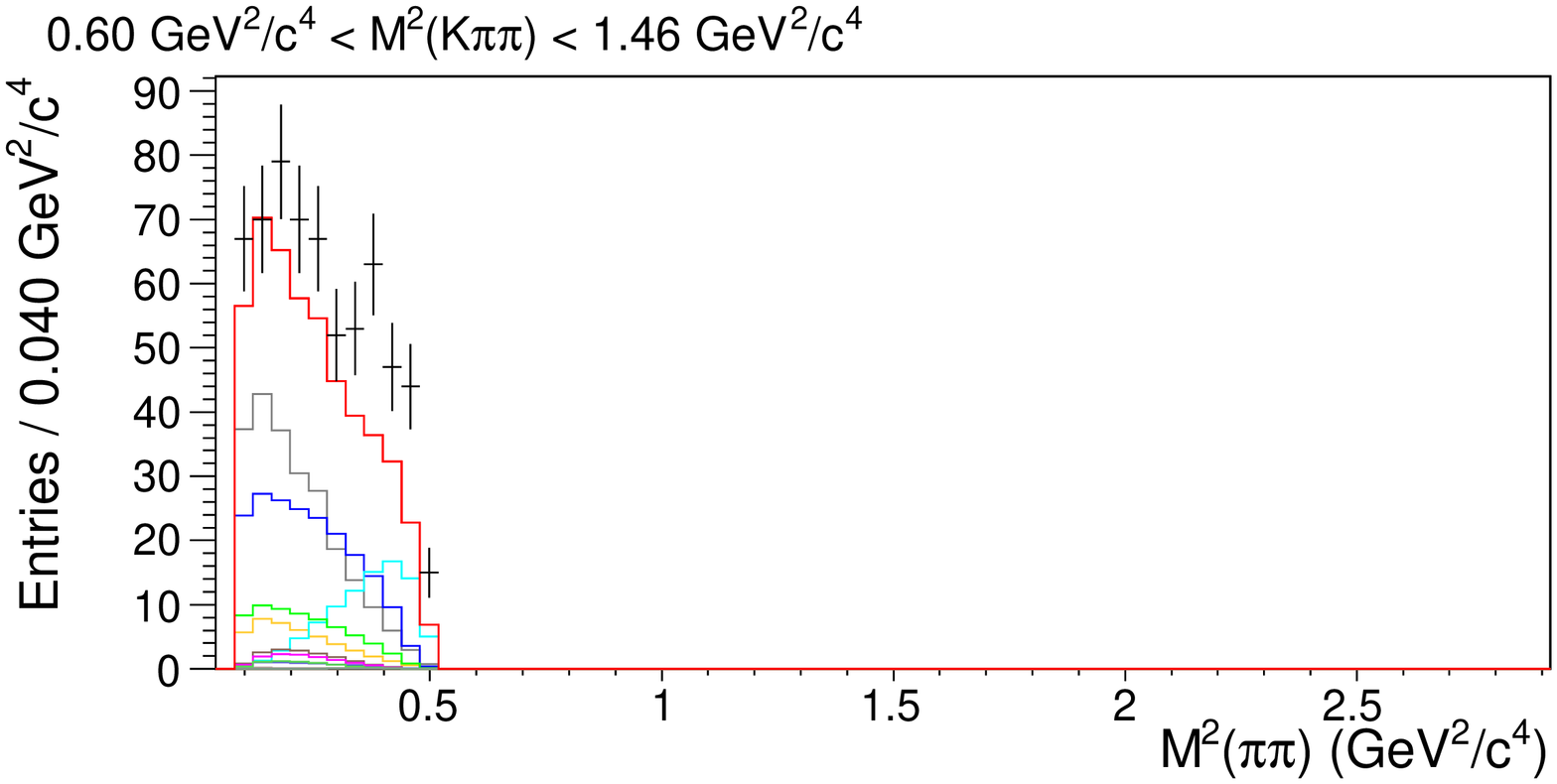}
}}}
\vspace{4mm}
\centerline{
\scalebox{0.35}{
\rotatebox{270}{
\includegraphics*[270,37][569,691]
  {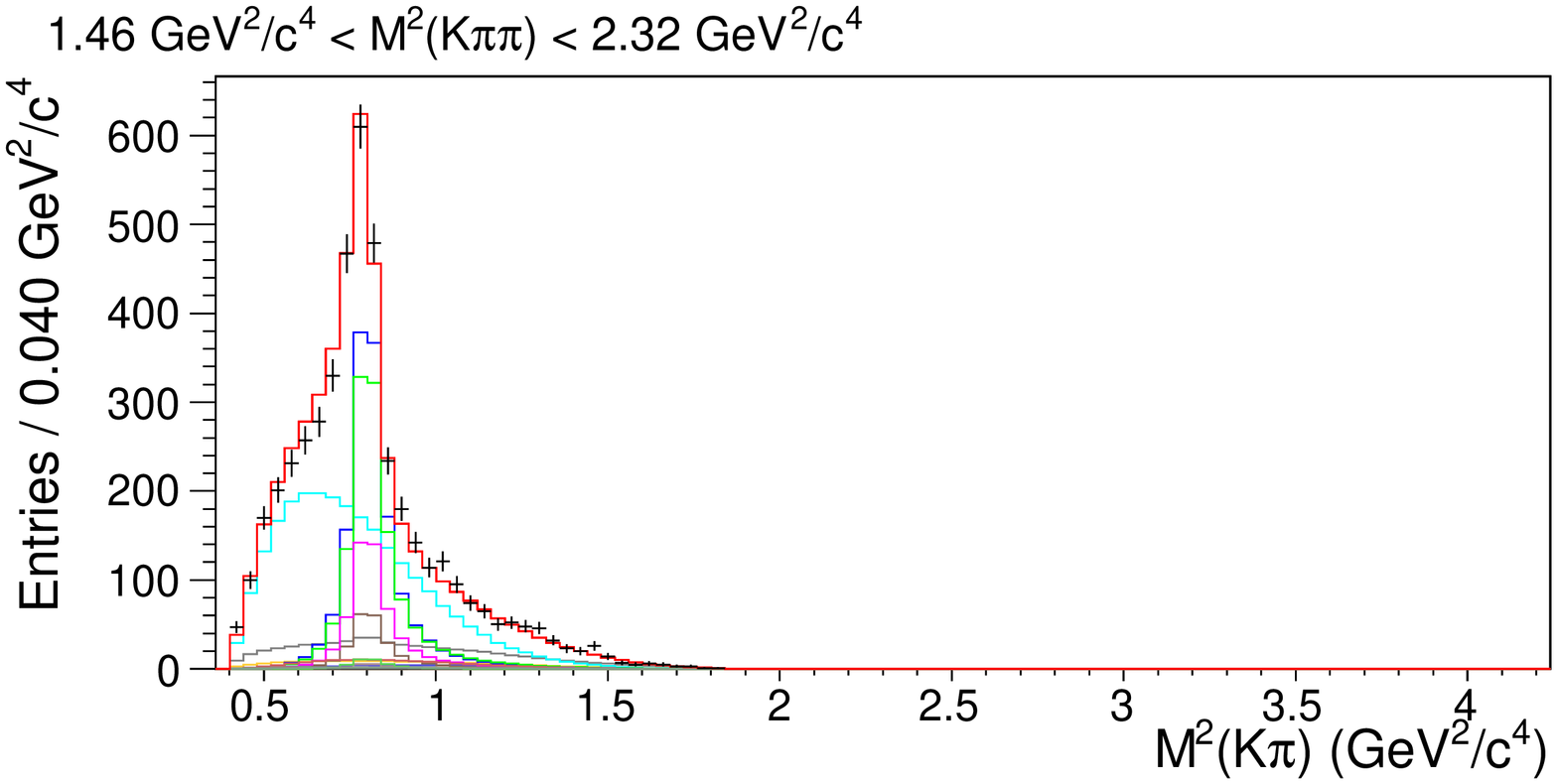}
}}
\hspace{4mm}
\scalebox{0.35}{
\rotatebox{270}{
\includegraphics*[270,37][569,691]
  {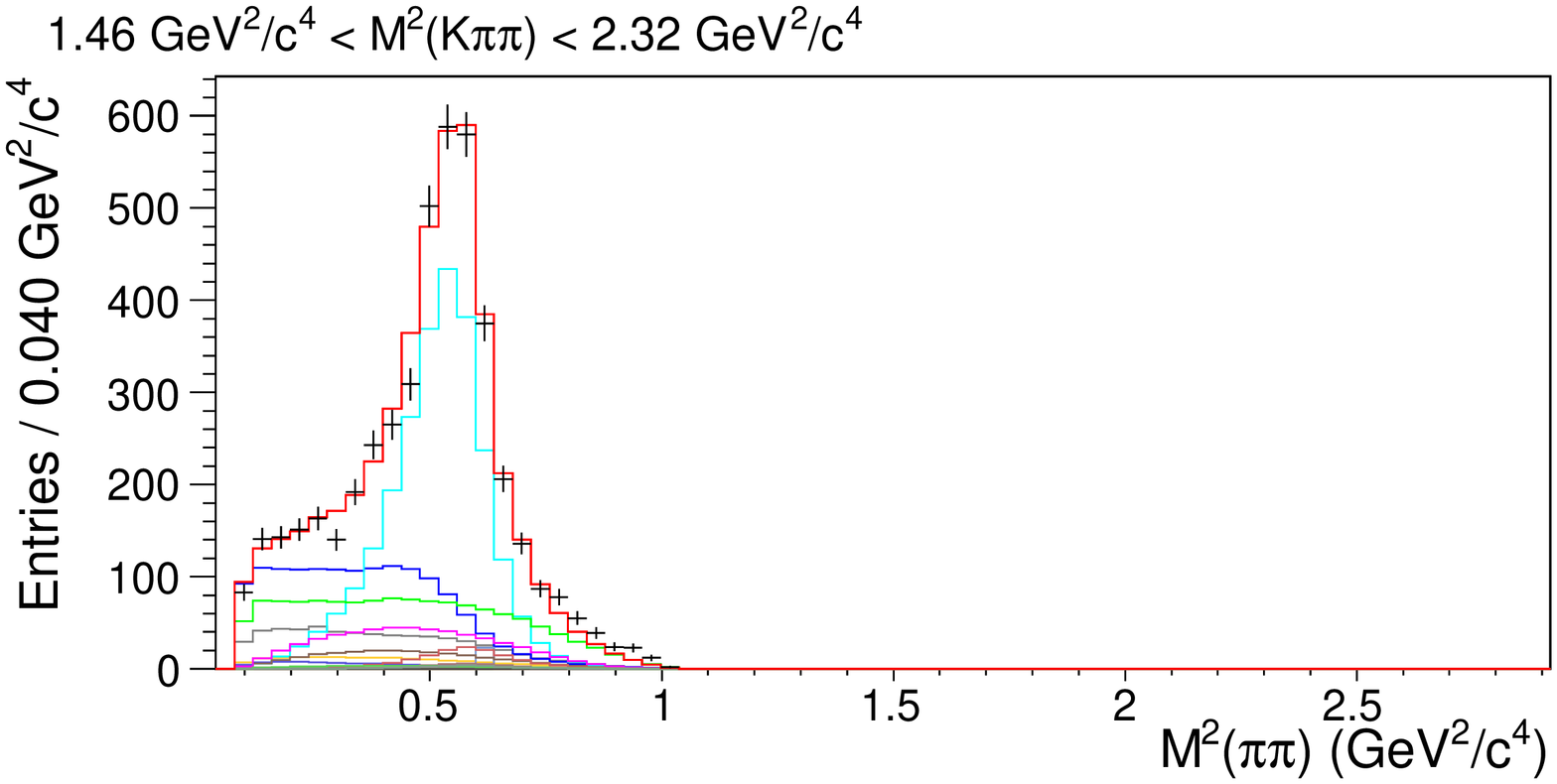}
}}}
\vspace{4mm}
\centerline{
\scalebox{0.35}{
\rotatebox{270}{
\includegraphics*[270,37][569,691]
  {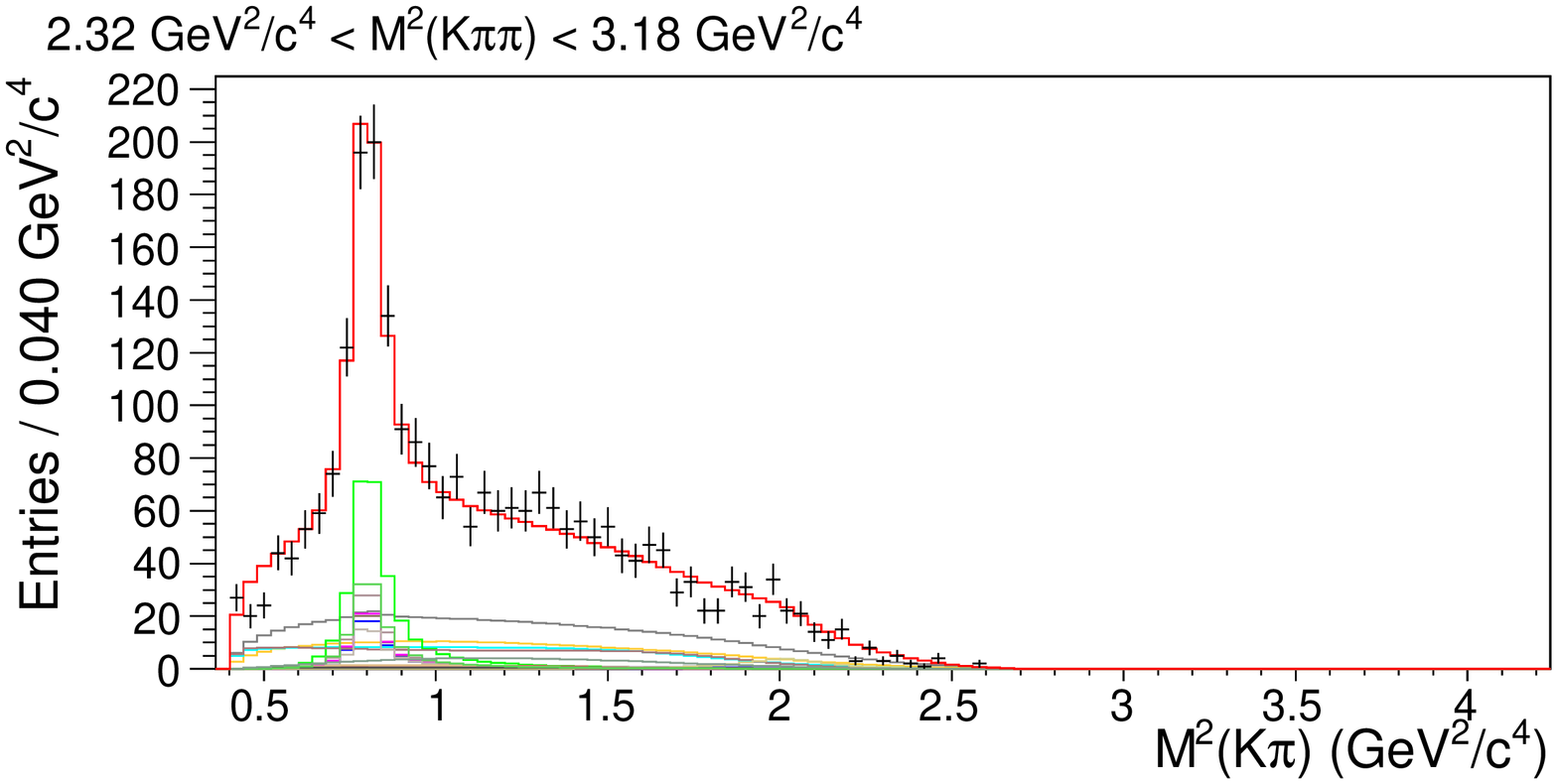}
}}
\hspace{4mm}
\scalebox{0.35}{
\rotatebox{270}{
\includegraphics*[270,37][569,691]
  {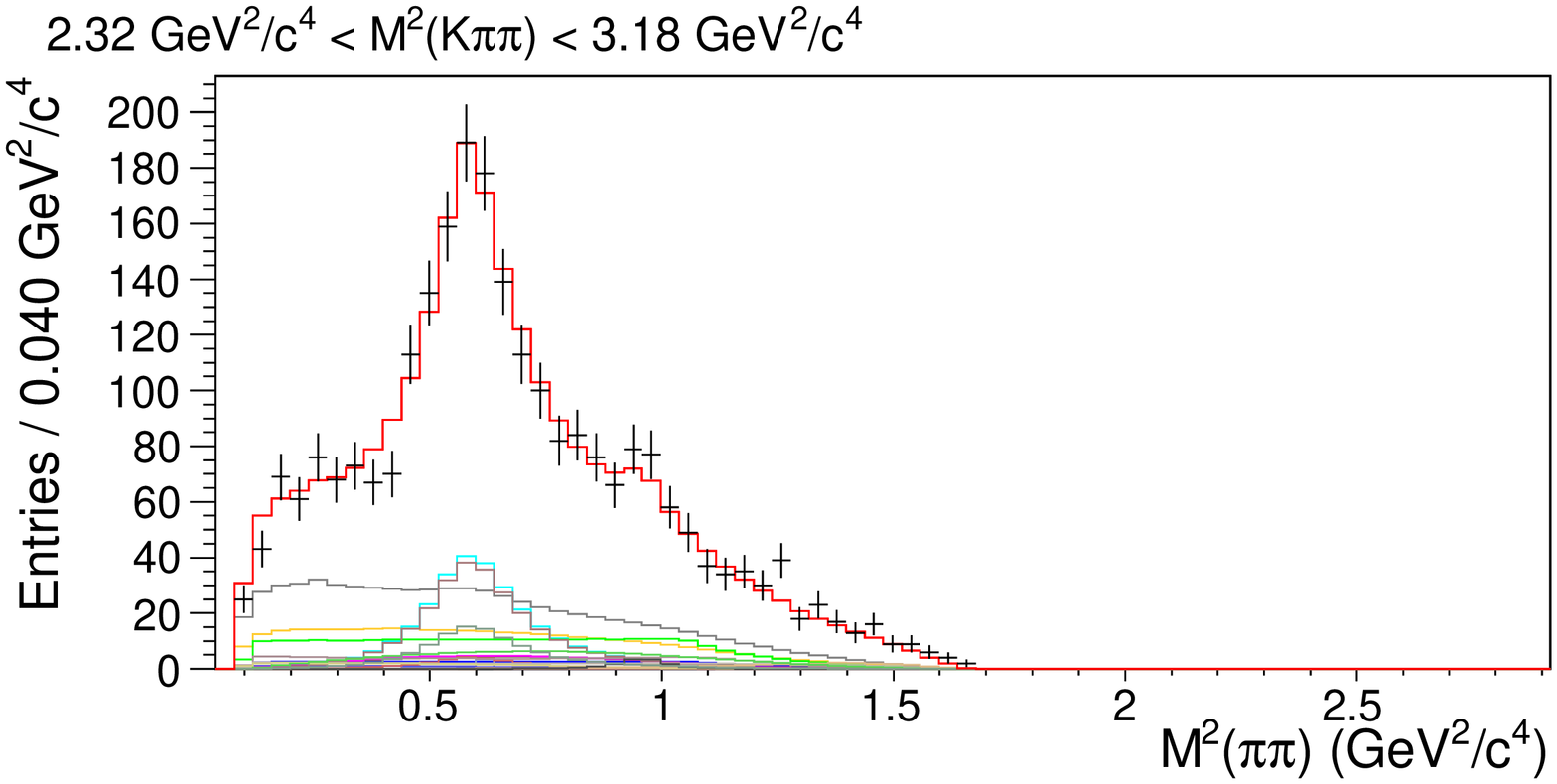}
}}}
\vspace{4mm}
\centerline{
\scalebox{0.35}{
\rotatebox{270}{
\includegraphics*[270,37][569,691]
  {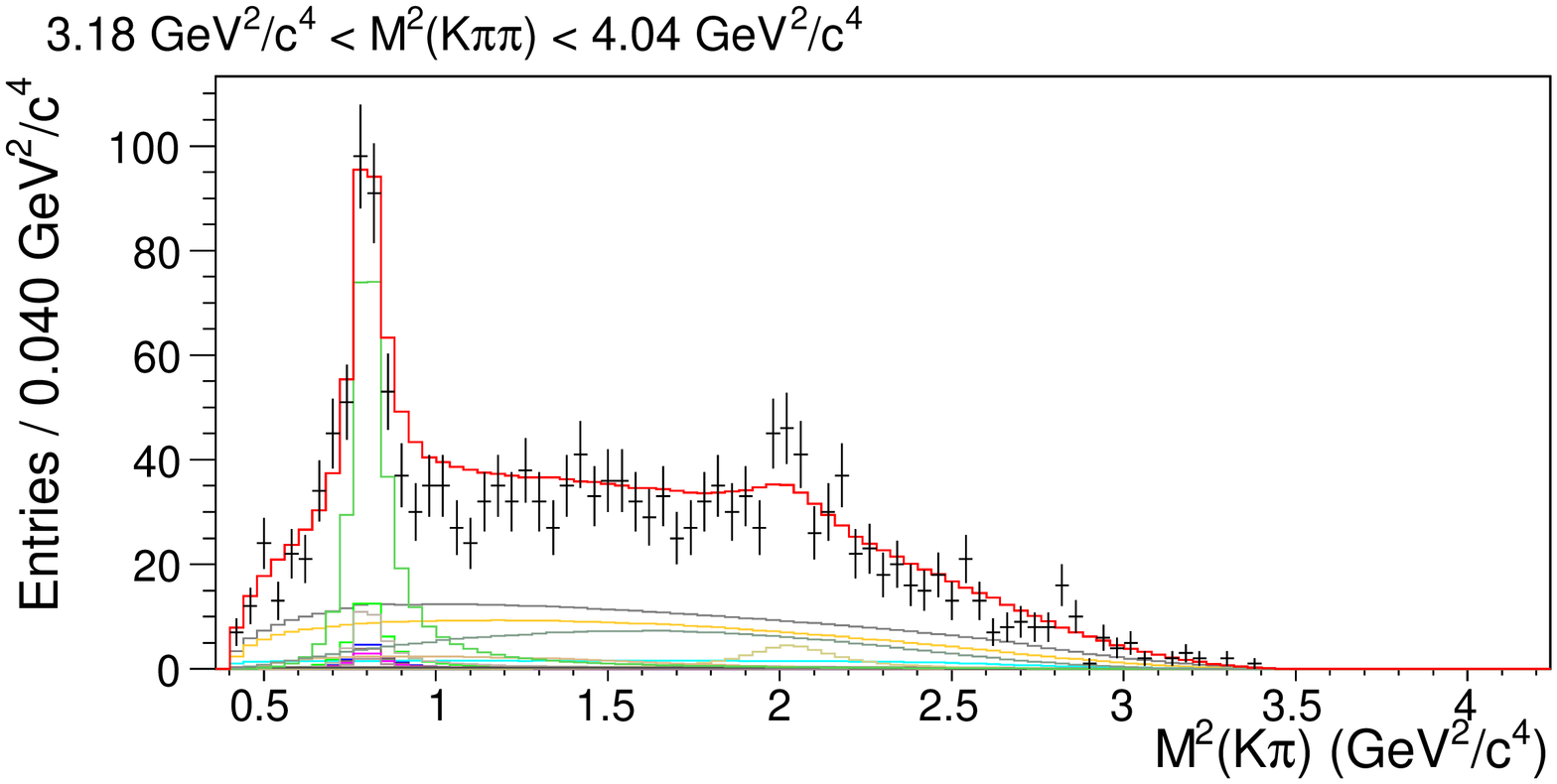}
}}
\hspace{4mm}
\scalebox{0.35}{
\rotatebox{270}{
\includegraphics*[270,37][569,691]
  {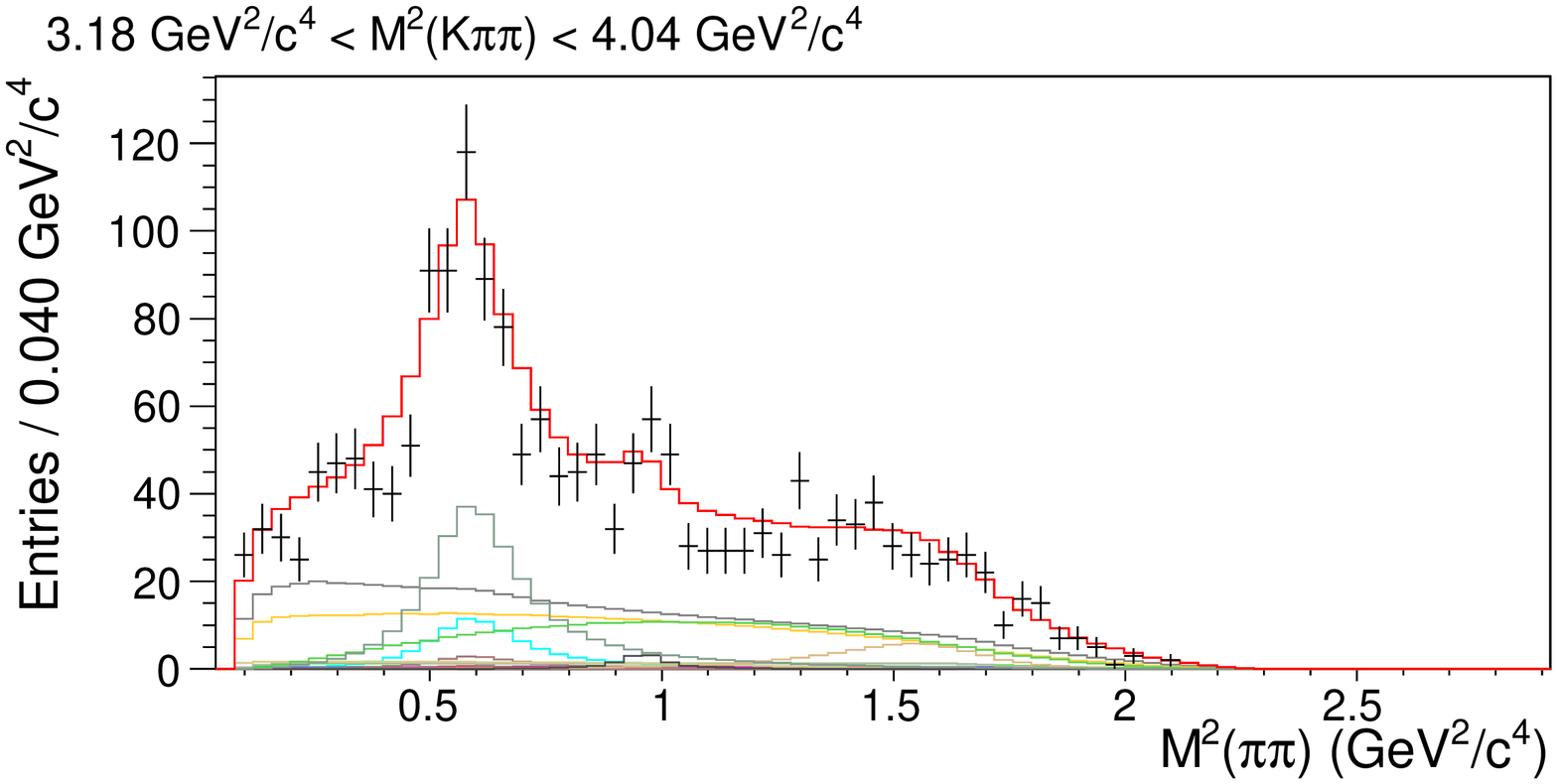}
}}}
\vspace{4mm}
\centerline{
\scalebox{0.35}{
\rotatebox{270}{
\includegraphics*[270,37][569,691]
  {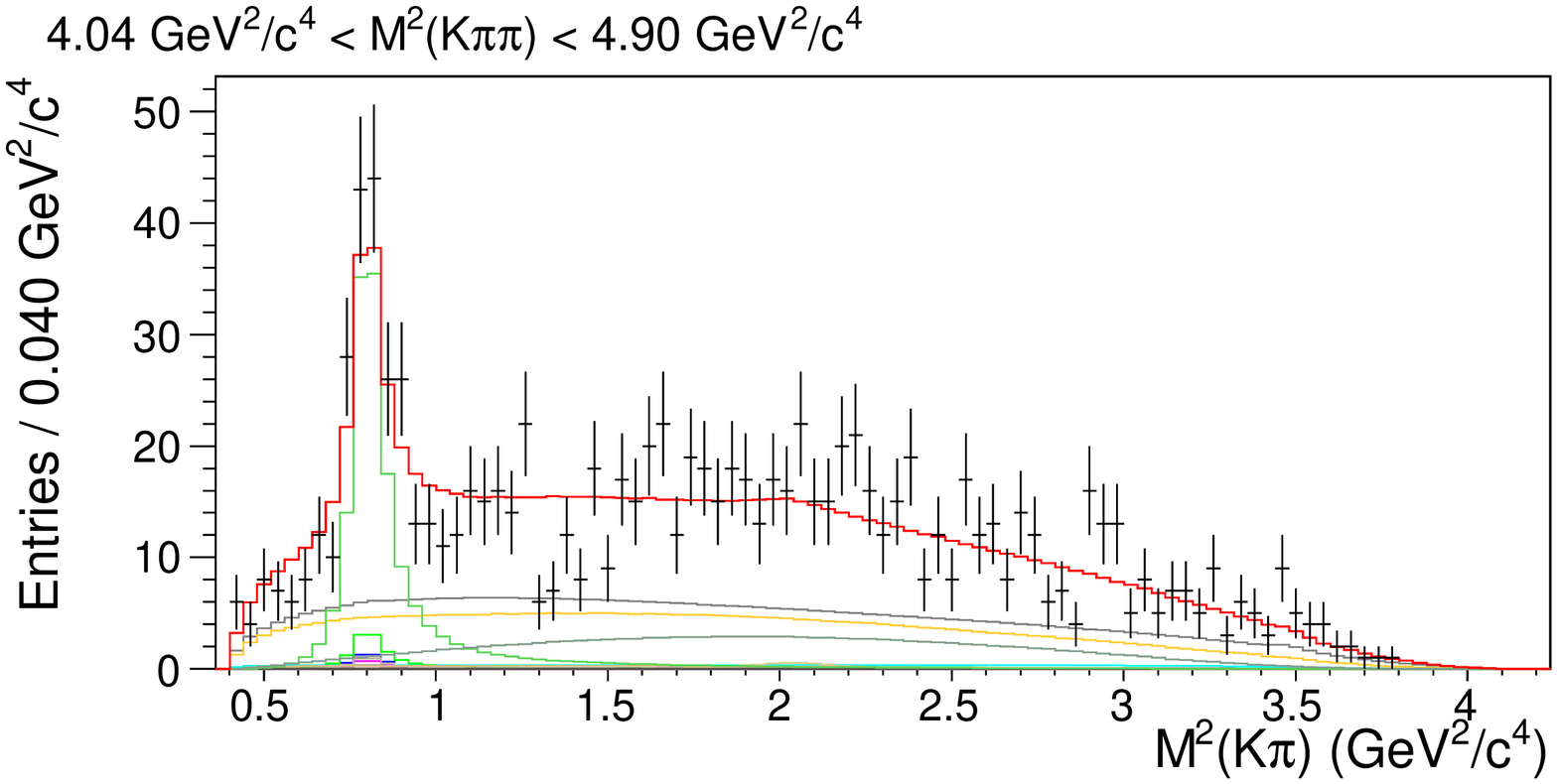}
}}
\hspace{4mm}
\scalebox{0.35}{
\rotatebox{270}{
\includegraphics*[270,37][569,691]
  {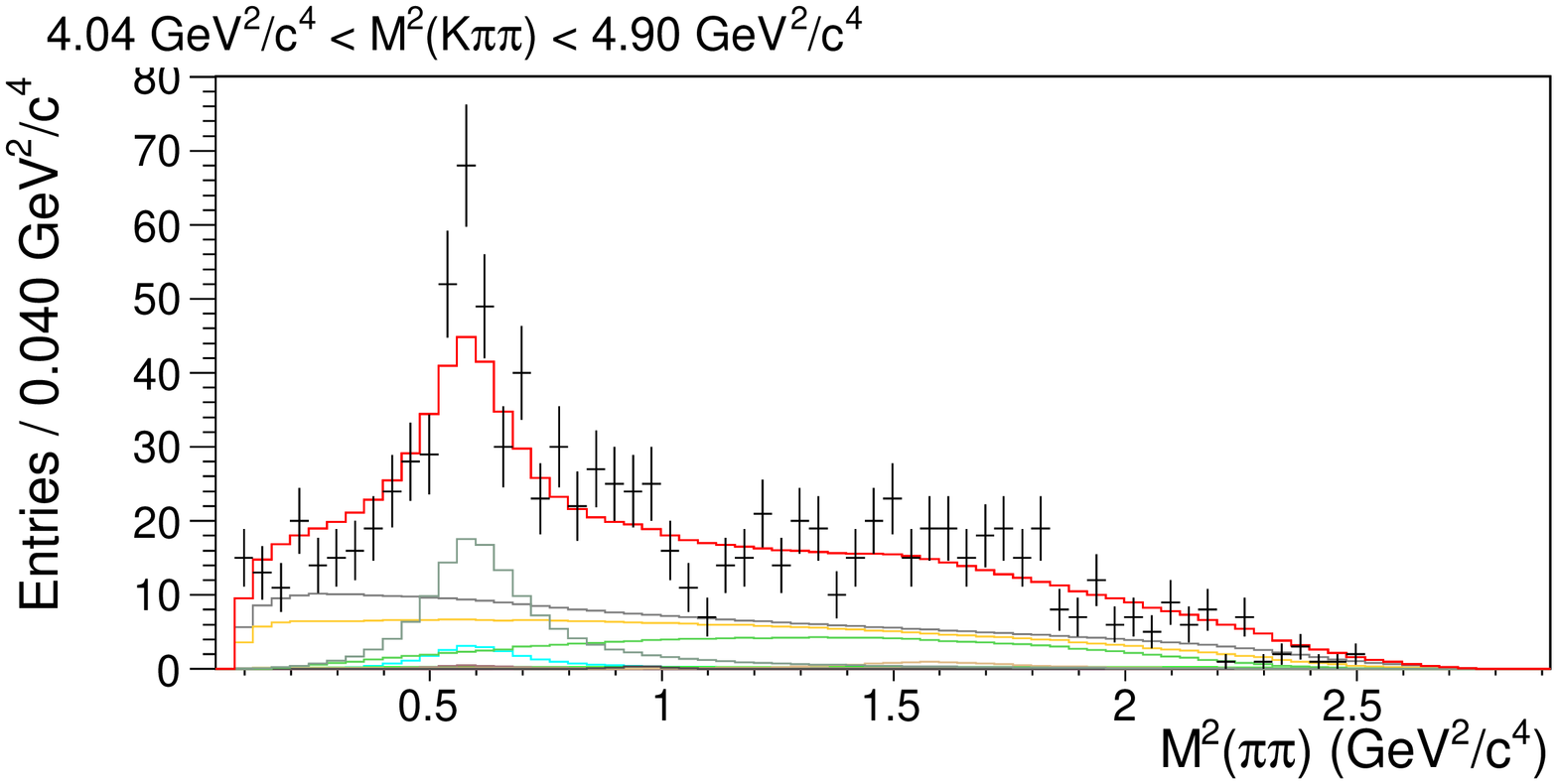}
}}}
\caption{$B^{+} \rightarrow J/\psi K^{+} \pi^{+} \pi^{-}$
  signal data (points) and fit results (histograms) 
  for slices in $M^{2}(K\pi\pi)$.  
  The fit components are color coded as shown in
  Fig.~\ref{amplitude:fig_signal_legend}.}
\label{amplitude:fig_signal_pdg_slices_jkpp}
\end{figure*}

Table~\ref{amplitude:table_fit_pdg_jkpp}
lists the values of the
moduli and phases of the complex coefficients $a_{k}$
of Eq.~\ref{amplitude:eq_raw_signal}
obtained by fitting signal-region data for
$B^{+} \rightarrow J/\psi K^{+} \pi^{+} \pi^{-}$, 
as well as the corresponding values of the
decay fractions,
given by
Eq.~\ref{amplitude:eq_decay_fraction}.
The fitted PDF is shown projected onto the three axes,
along with the data, in
Fig.~\ref{amplitude:fig_signal_pdg_jkpp}.
Figure~\ref{amplitude:fig_signal_pdg_slices_jkpp}
shows $M^{2}(K\pi)$ and $M^{2}(\pi\pi)$ projections
for slices in $M^{2}(K\pi\pi)$.
The legend is presented in Fig.~\ref{amplitude:fig_signal_legend}.
In this fit,
$\ell = -10575.4$,
while
$\chi^{2} = 1475.9$
with
$N_{\mathrm{bins}} = 1202$
and
$N_{\mathrm{par}} = 28$.

\begin{figure}[hbtp]
\centerline{
\scalebox{0.35}{
\rotatebox{270}{
\includegraphics*[270,37][569,691]
  {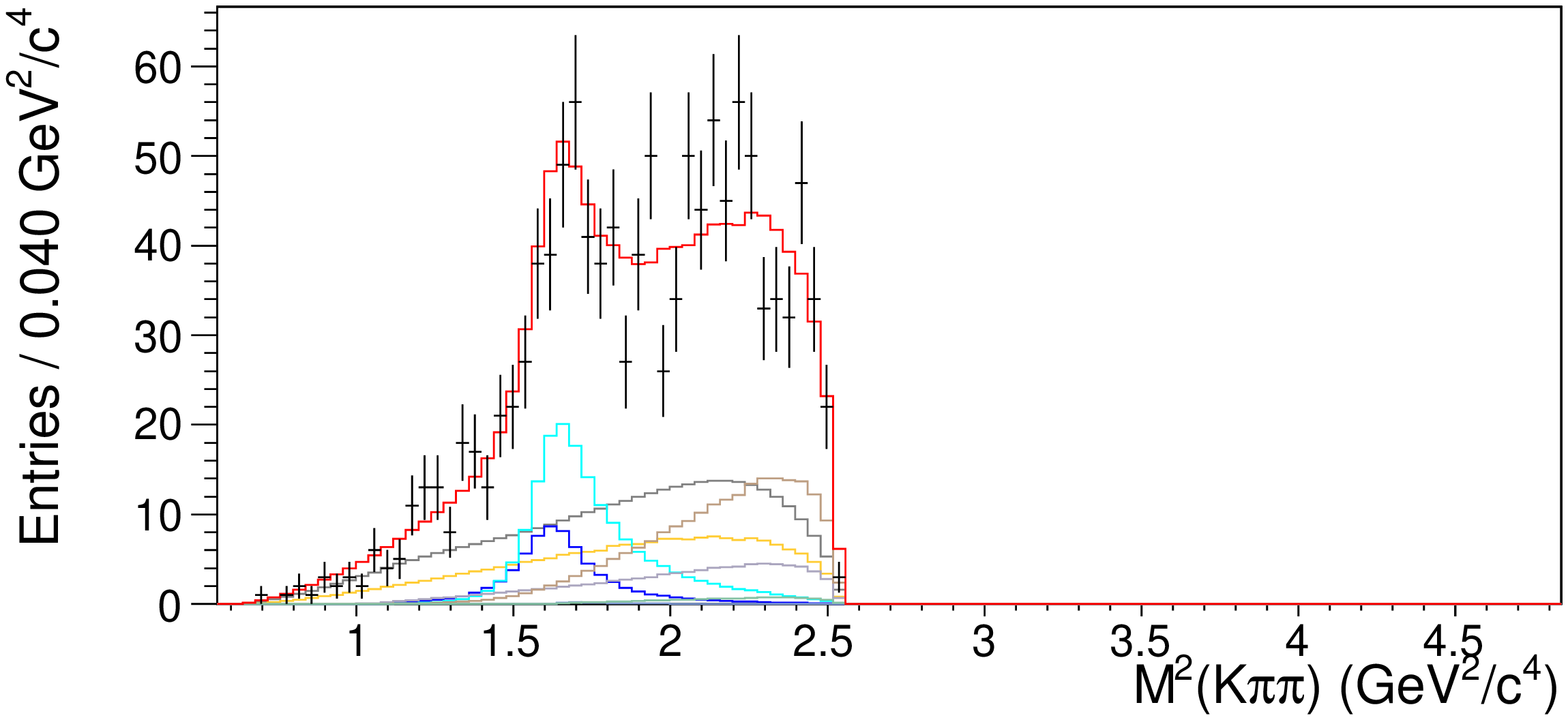}
}}}
\vspace{2mm}
\centerline{
\scalebox{0.35}{
\rotatebox{270}{
\includegraphics*[270,37][569,691]
  {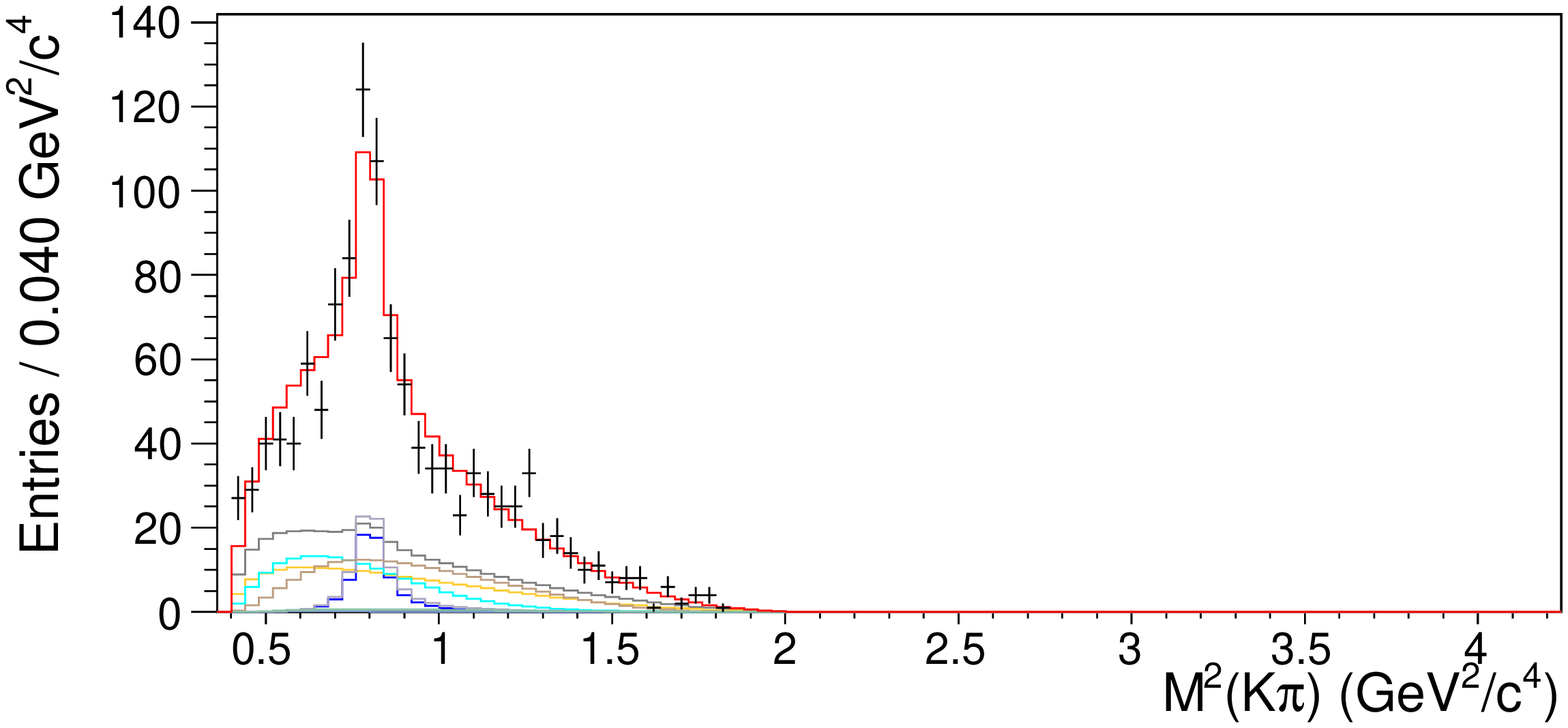}
}}}
\vspace{2mm}
\centerline{
\scalebox{0.35}{
\rotatebox{270}{
\includegraphics*[270,37][569,691]
  {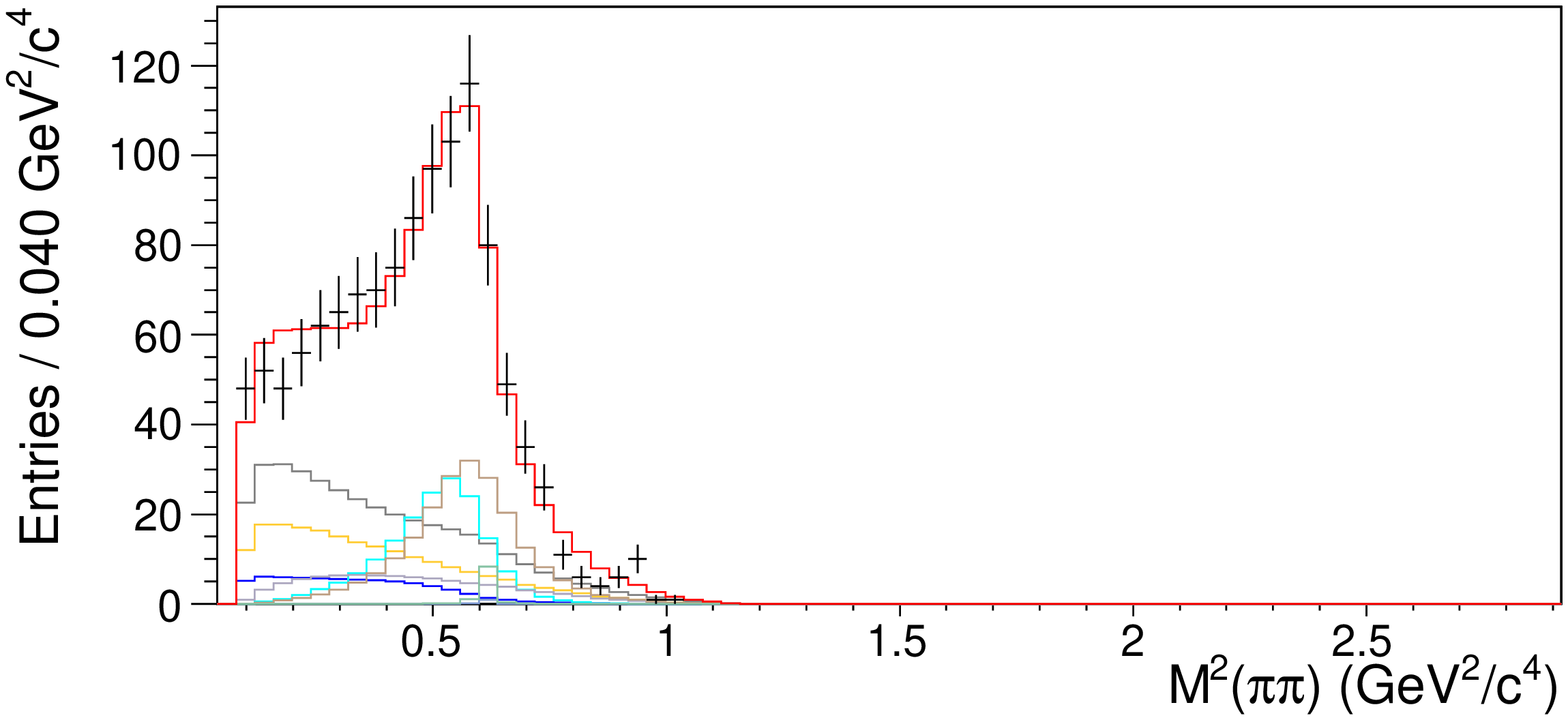}
}}}
\caption{Results of signal-region fits for 
  $B^{+} \rightarrow \psi^{\prime} K^{+} \pi^{+} \pi^{-}$.
  Data (points) and fits (histograms) are shown projected onto the 
  three axes.
  The fit components are color coded as shown in
  Fig.~\ref{amplitude:fig_signal_legend}.}
\label{amplitude:fig_signal_pdg_pkpp}
\end{figure}

\begin{figure*}[hbtp]
\centerline{
\scalebox{0.35}{
\rotatebox{270}{
\includegraphics*[270,37][569,691]
  {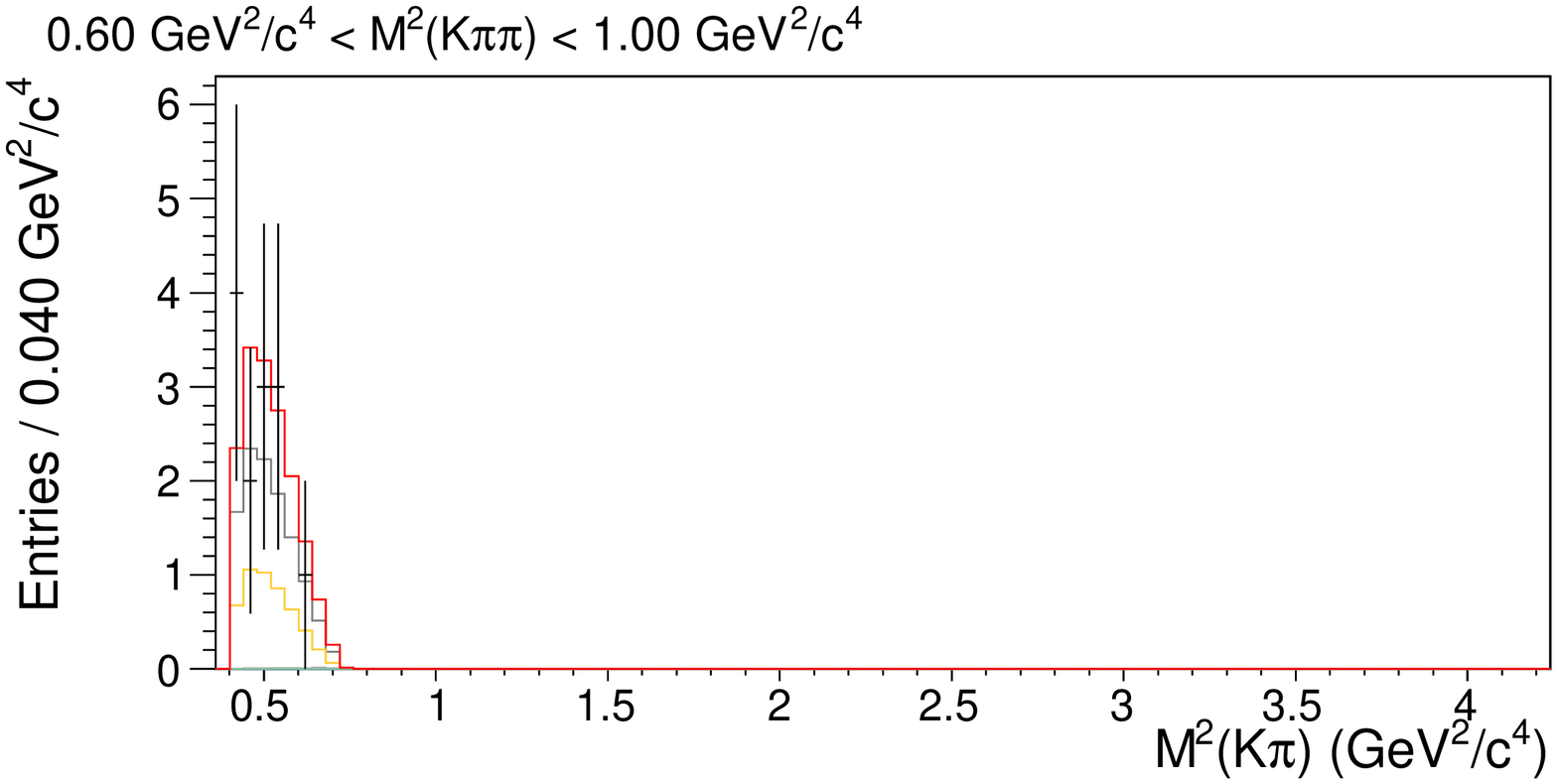}
}}
\hspace{4mm}
\scalebox{0.35}{
\rotatebox{270}{
\includegraphics*[270,37][569,691]
  {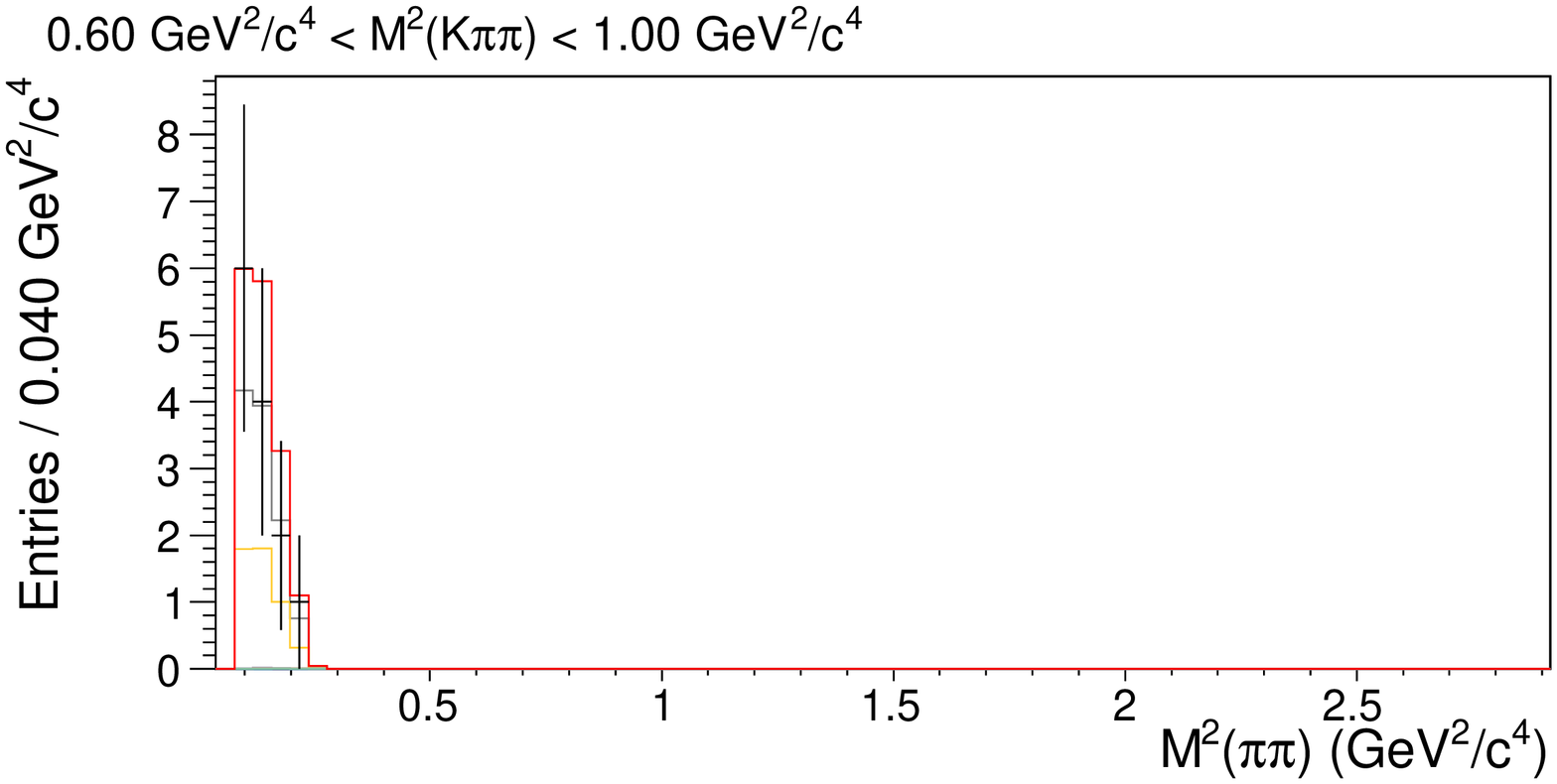}
}}}
\vspace{4mm}
\centerline{
\scalebox{0.35}{
\rotatebox{270}{
\includegraphics*[270,37][569,691]
  {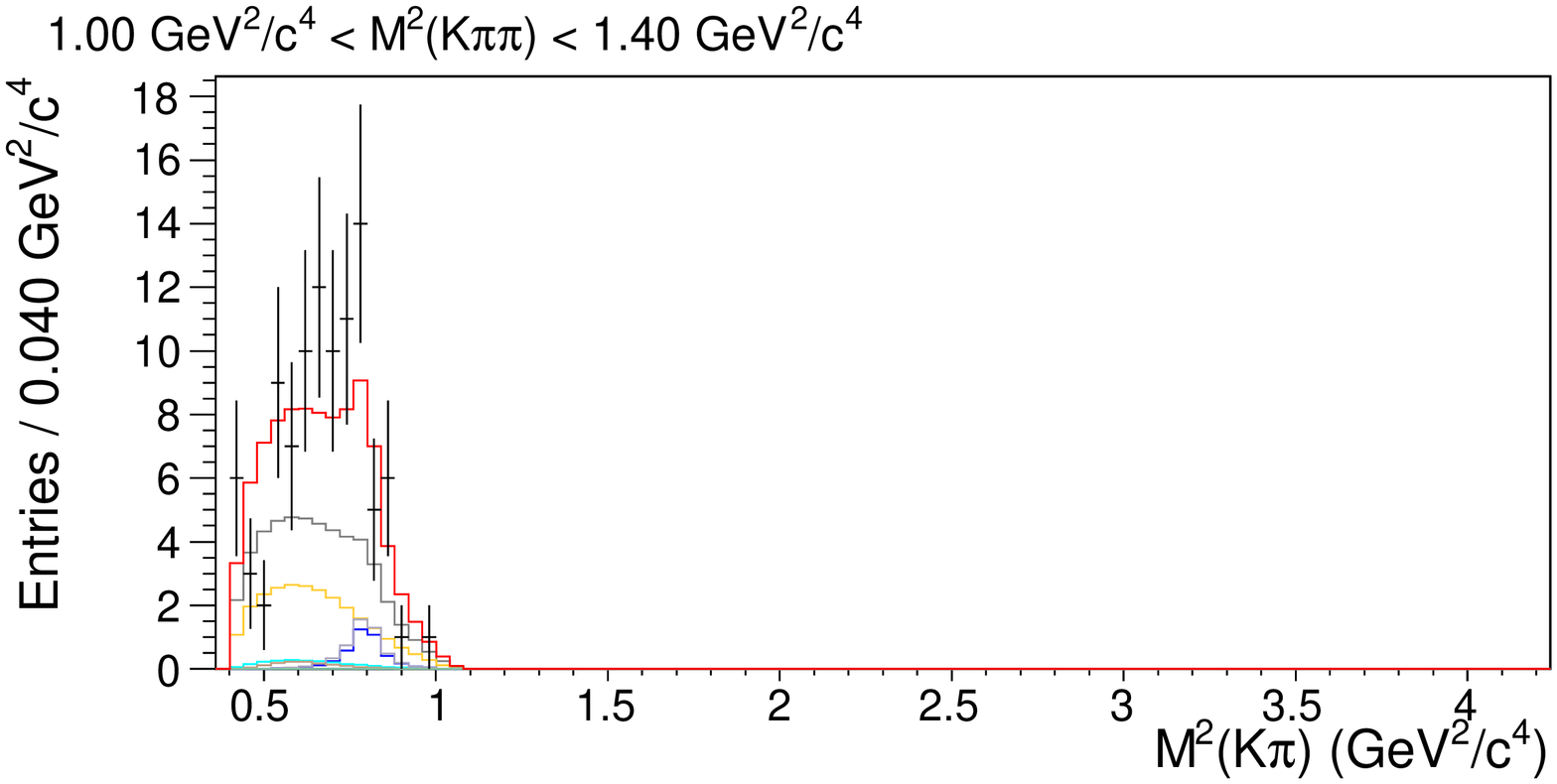}
}}
\hspace{4mm}
\scalebox{0.35}{
\rotatebox{270}{
\includegraphics*[270,37][569,691]
  {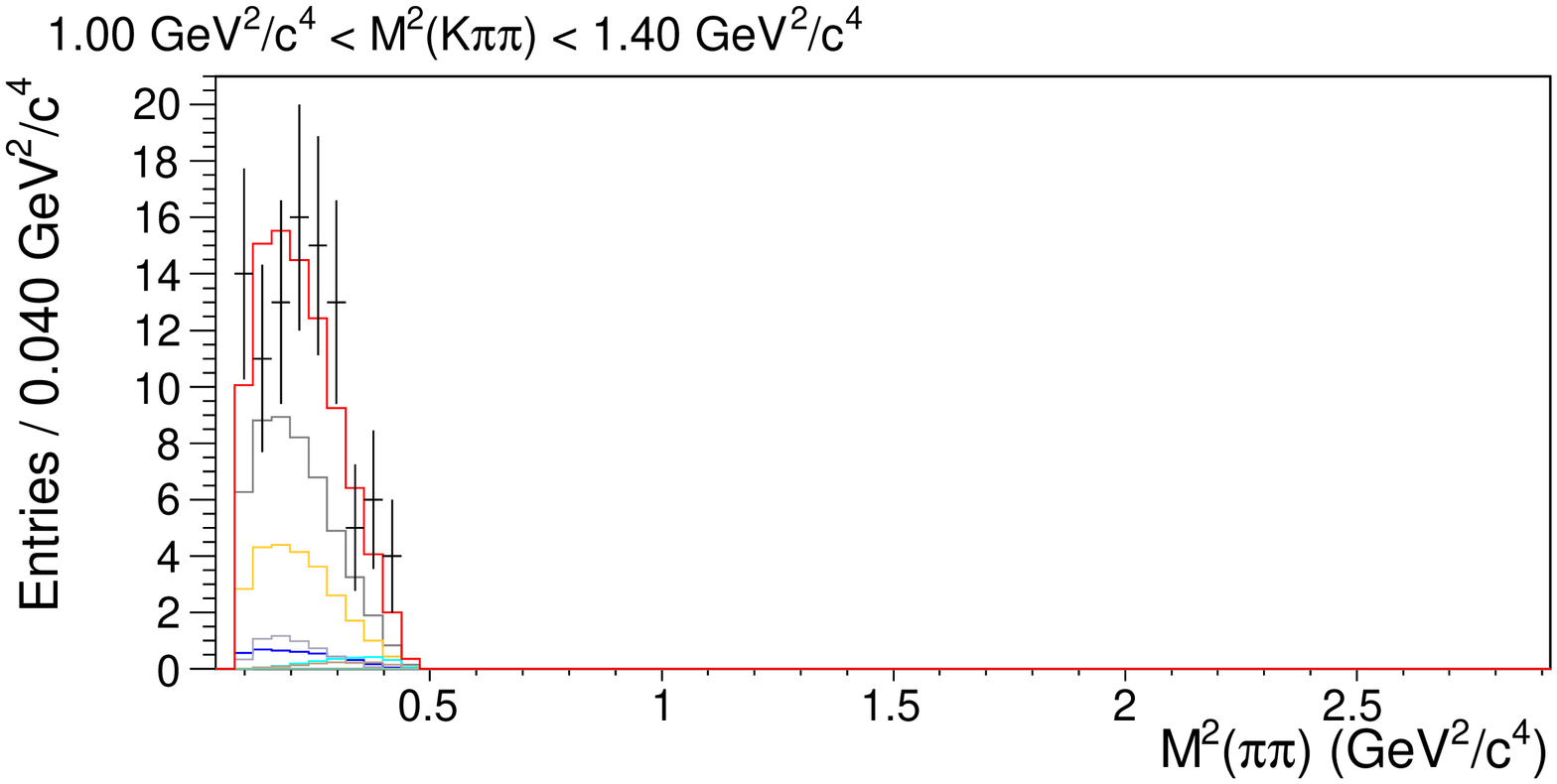}
}}}
\vspace{4mm}
\centerline{
\scalebox{0.35}{
\rotatebox{270}{
\includegraphics*[270,37][569,691]
  {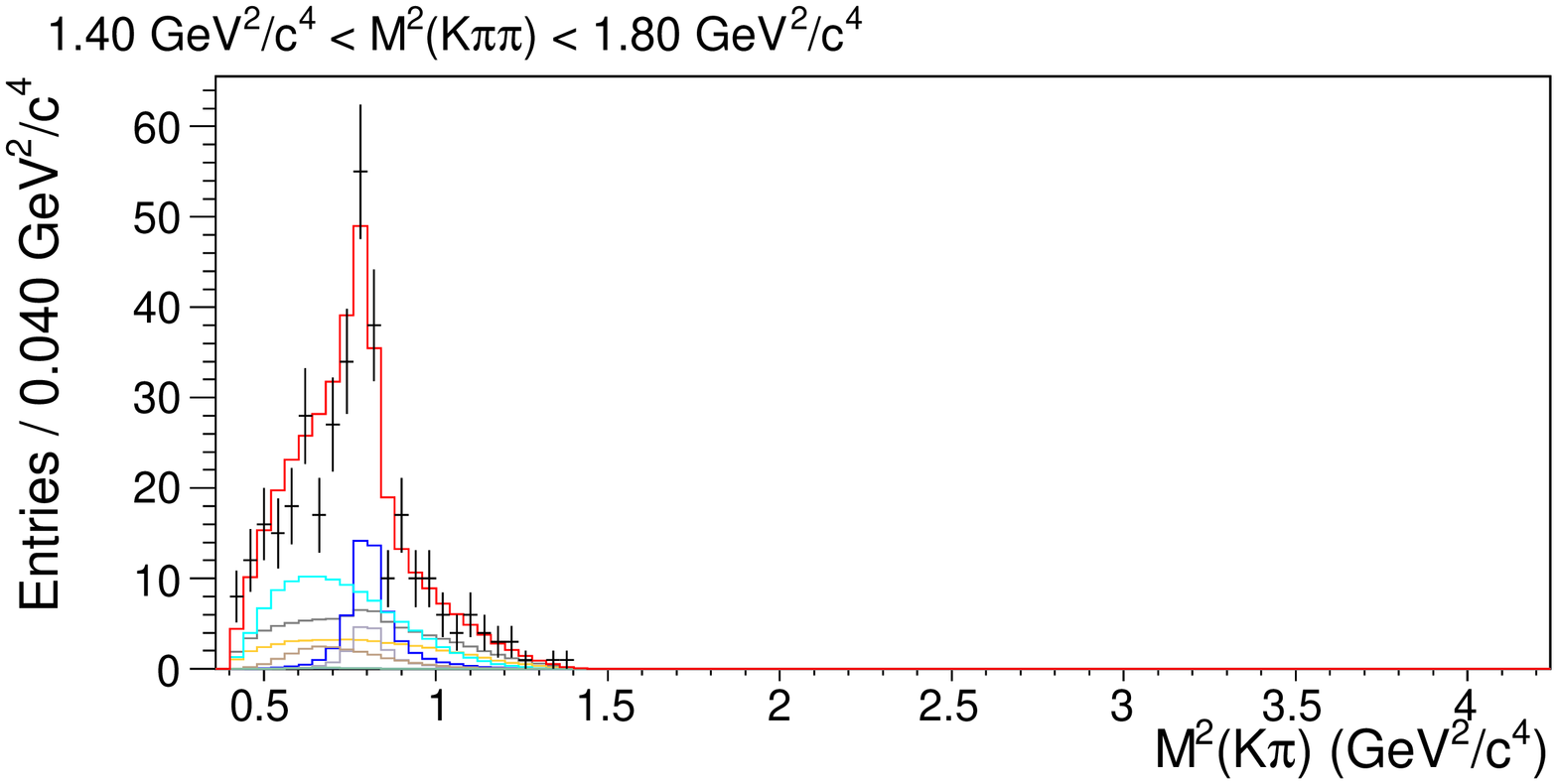}
}}
\hspace{4mm}
\scalebox{0.35}{
\rotatebox{270}{
\includegraphics*[270,37][569,691]
  {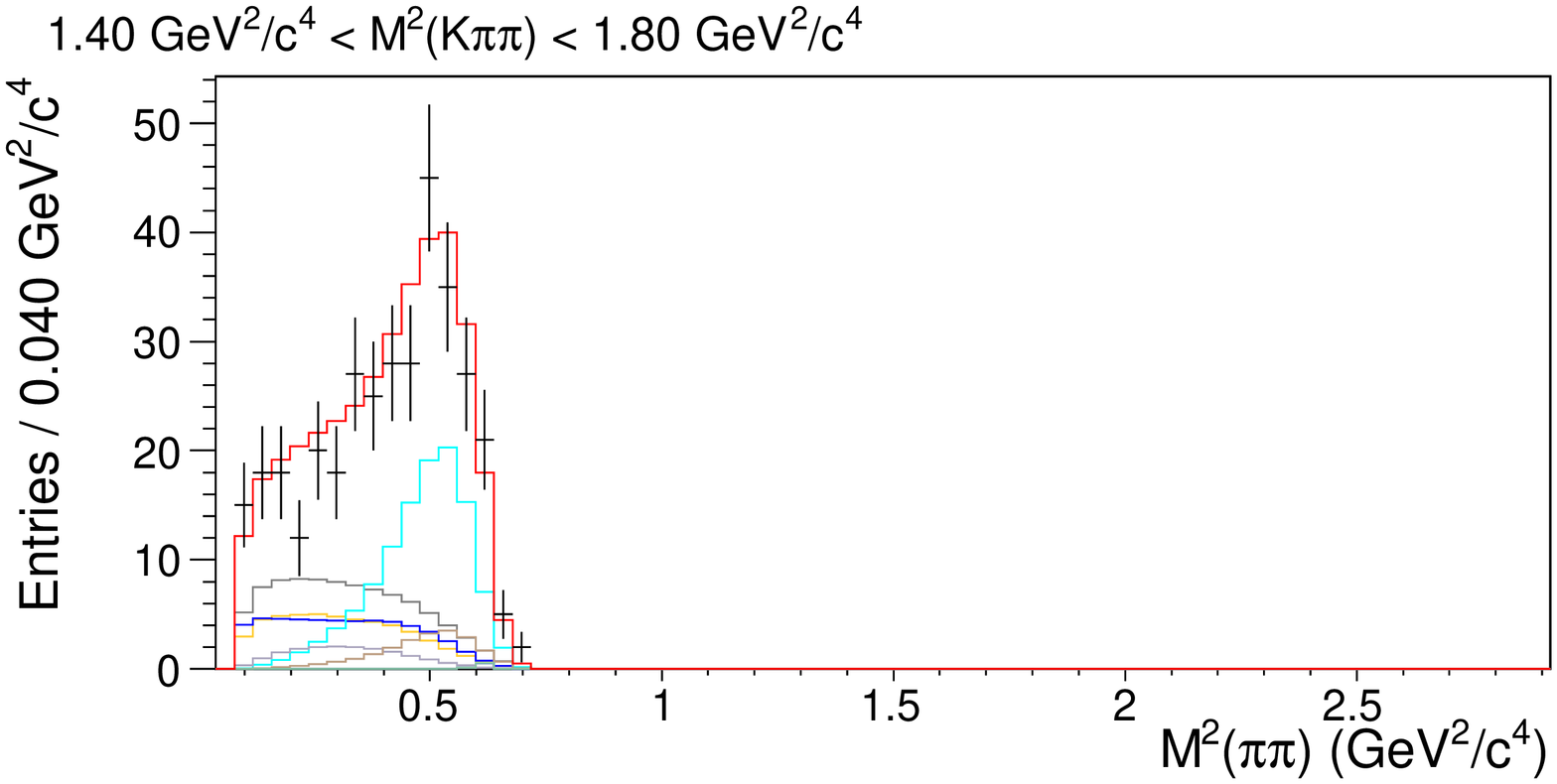}
}}}
\vspace{4mm}
\centerline{
\scalebox{0.35}{
\rotatebox{270}{
\includegraphics*[270,37][569,691]
  {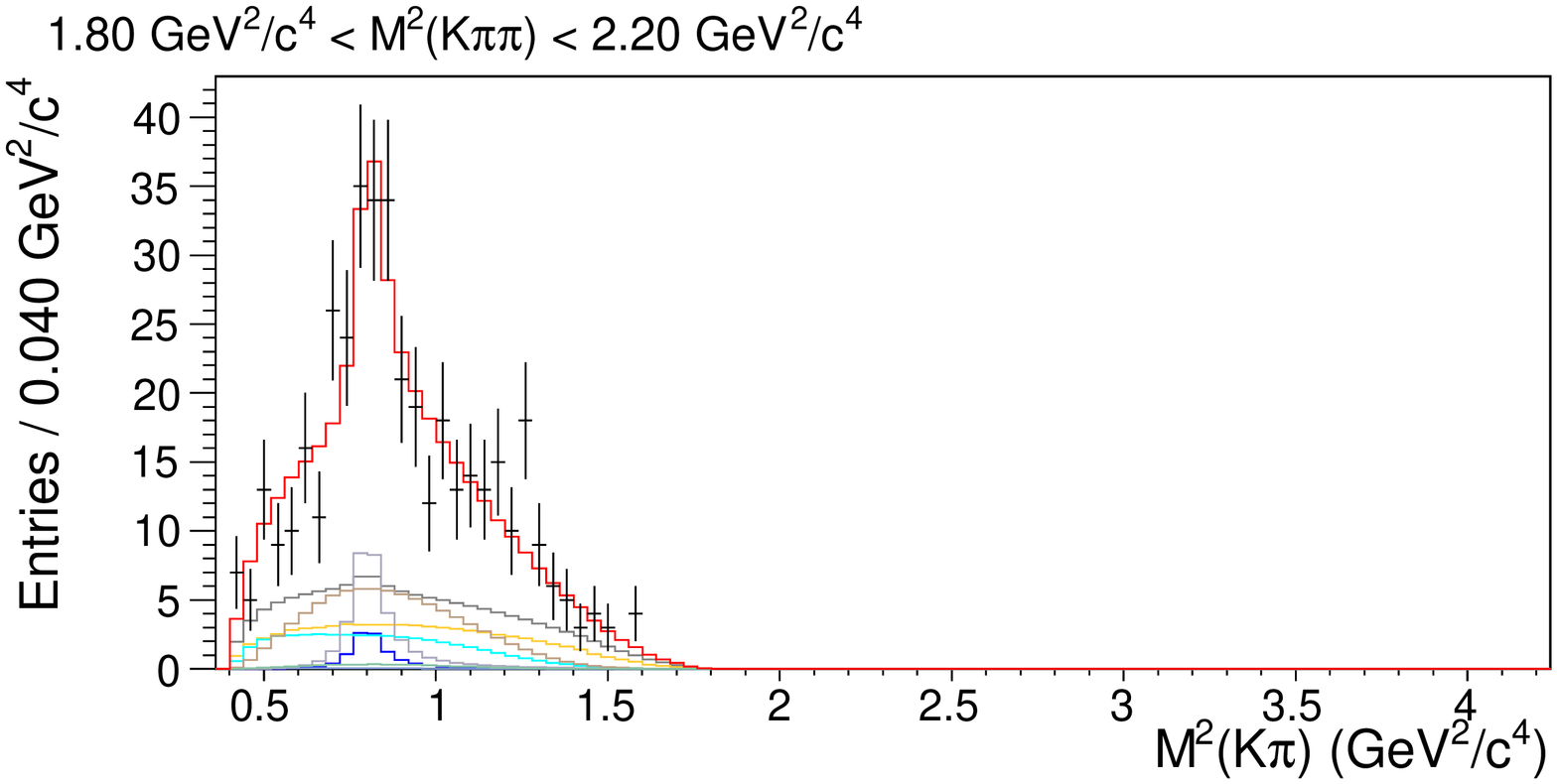}
}}
\hspace{4mm}
\scalebox{0.35}{
\rotatebox{270}{
\includegraphics*[270,37][569,691]
  {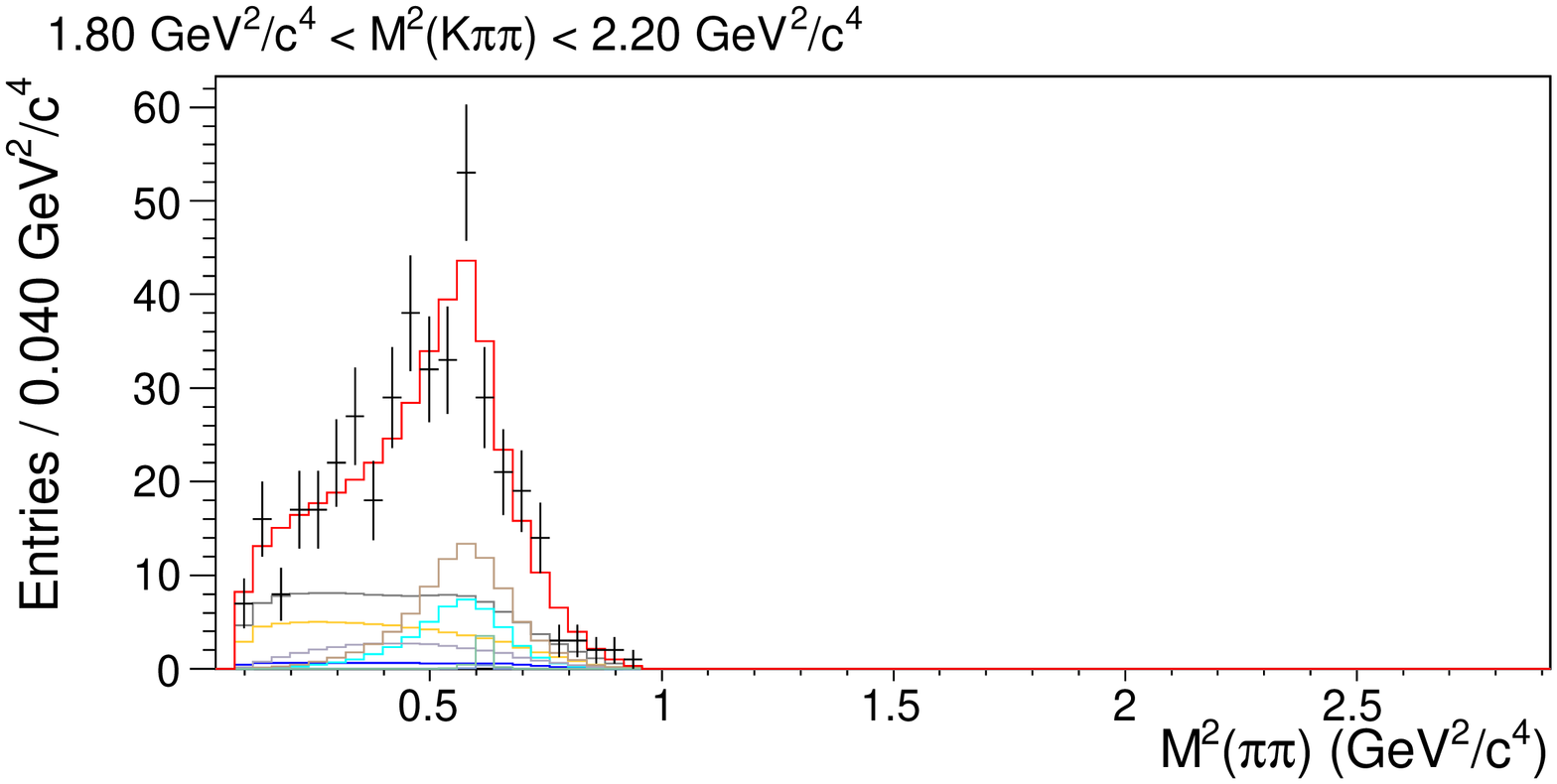}
}}}
\vspace{4mm}
\centerline{
\scalebox{0.35}{
\rotatebox{270}{
\includegraphics*[270,37][569,691]
  {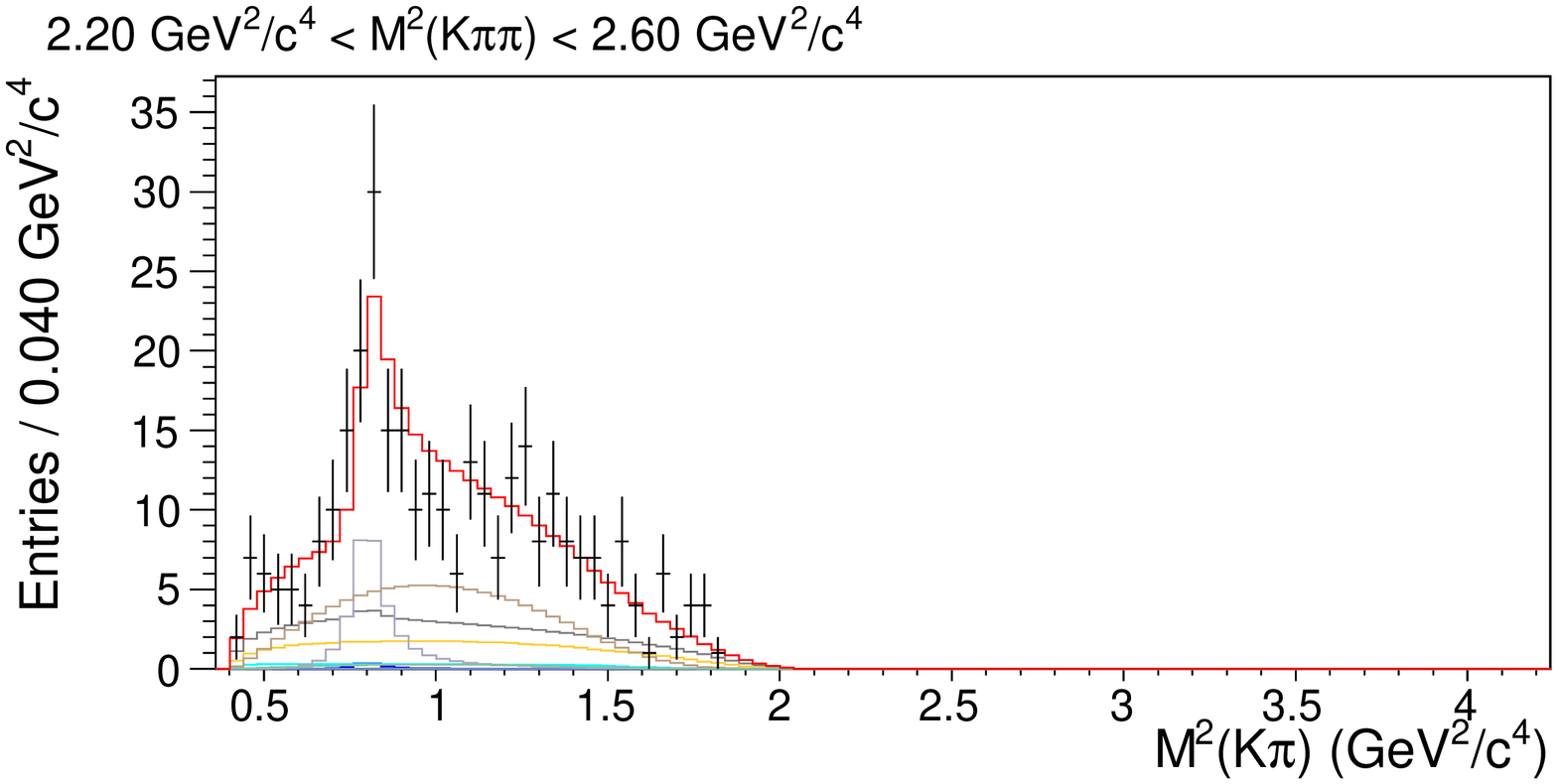}
}}
\hspace{4mm}
\scalebox{0.35}{
\rotatebox{270}{
\includegraphics*[270,37][569,691]
  {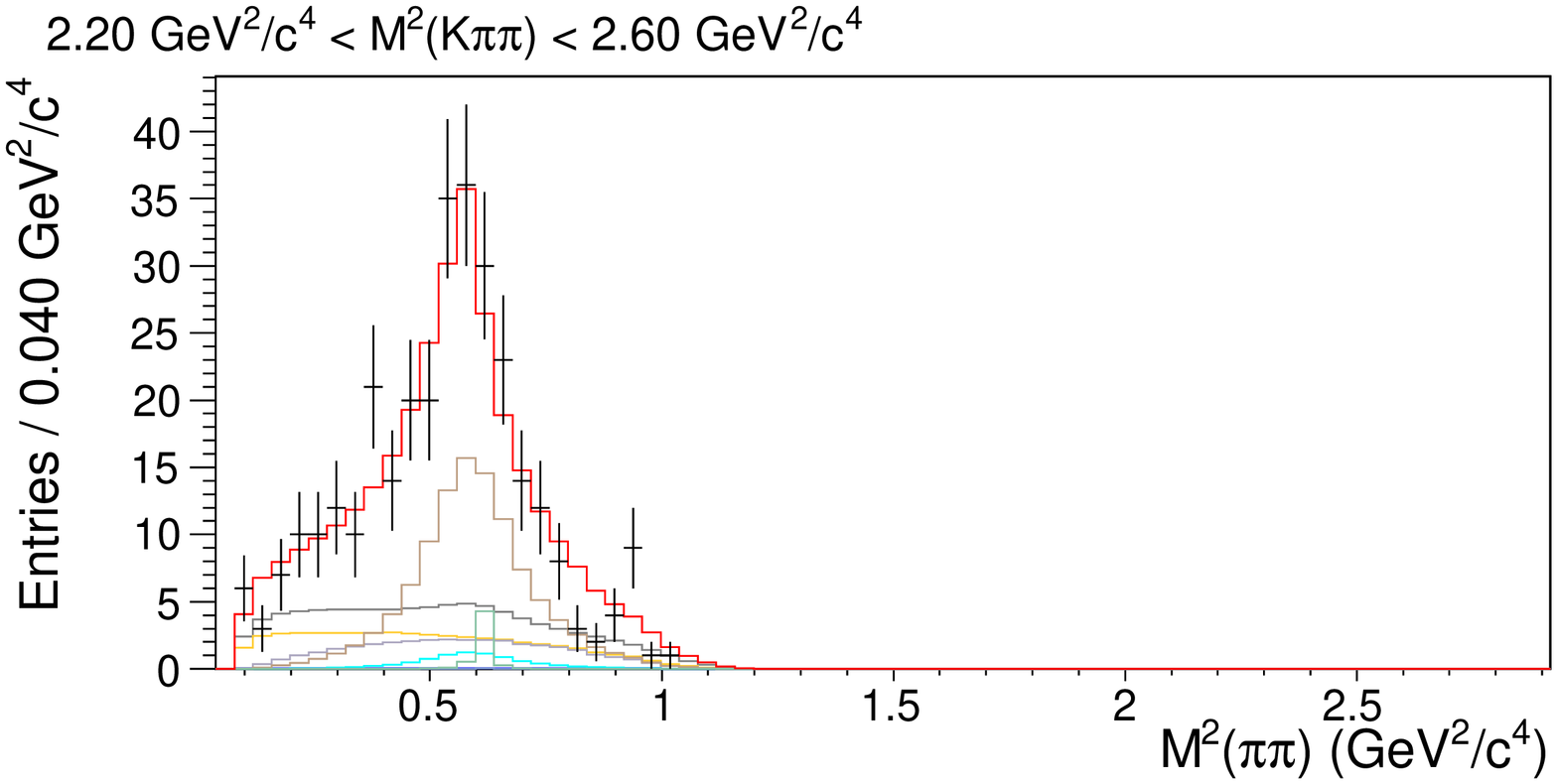}
}}}
\caption{$B^{+} \rightarrow \psi^{\prime} K^{+} \pi^{+} \pi^{-}$ 
  signal data (points) and fit results (histograms) 
  for slices in $M^{2}(K\pi\pi)$.  
  The fit components are color coded as shown in
  Fig.~\ref{amplitude:fig_signal_legend}.}
\label{amplitude:fig_signal_pdg_slices_pkpp}
\end{figure*}

Similarly, Table~\ref{amplitude:table_fit_pdg_pkpp}
shows the fitted parameters
for
$B^{+} \rightarrow \psi^{\prime} K^{+} \pi^{+} \pi^{-}$
signal-region data,
as well as the corresponding decay fractions.
Figure~\ref{amplitude:fig_signal_pdg_pkpp}
shows the fitted PDF and data
projected onto the three axes,
while
Fig.~\ref{amplitude:fig_signal_pdg_slices_pkpp}
shows $M^{2}(K\pi)$ and $M^{2}(\pi\pi)$ projections
for slices in $M^{2}(K\pi\pi)$.
In this fit,
$\ell = 638.3$,
while
$\chi^{2} = 180.1$
with
$N_{\mathrm{bins}} = 168$
and
$N_{\mathrm{par}} = 10$.

\begin{figure}[hbtp]
\centerline{
\scalebox{0.35}{
\rotatebox{270}{
\includegraphics*[270,37][569,691]
  {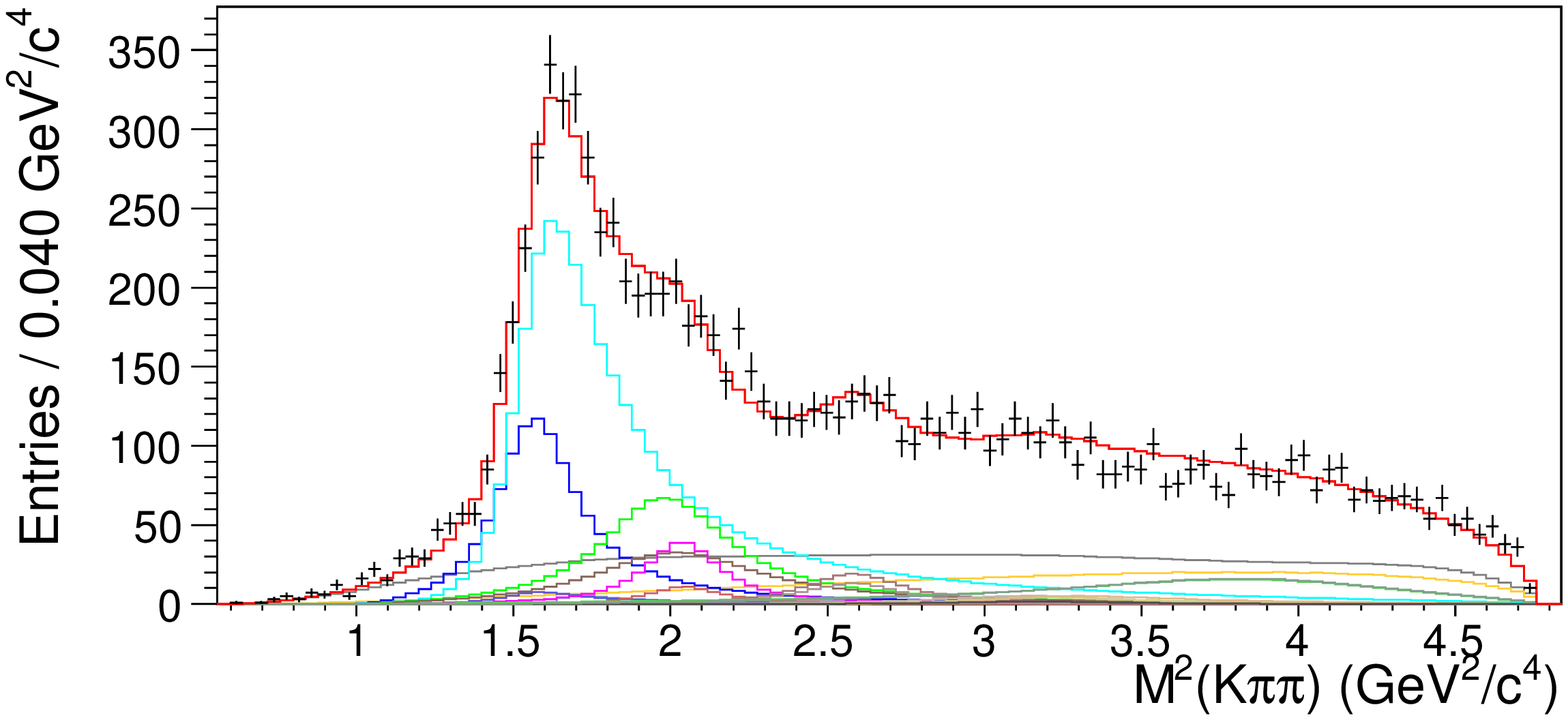}
}}}
\vspace{2mm}
\centerline{
\scalebox{0.35}{
\rotatebox{270}{
\includegraphics*[270,37][569,691]
  {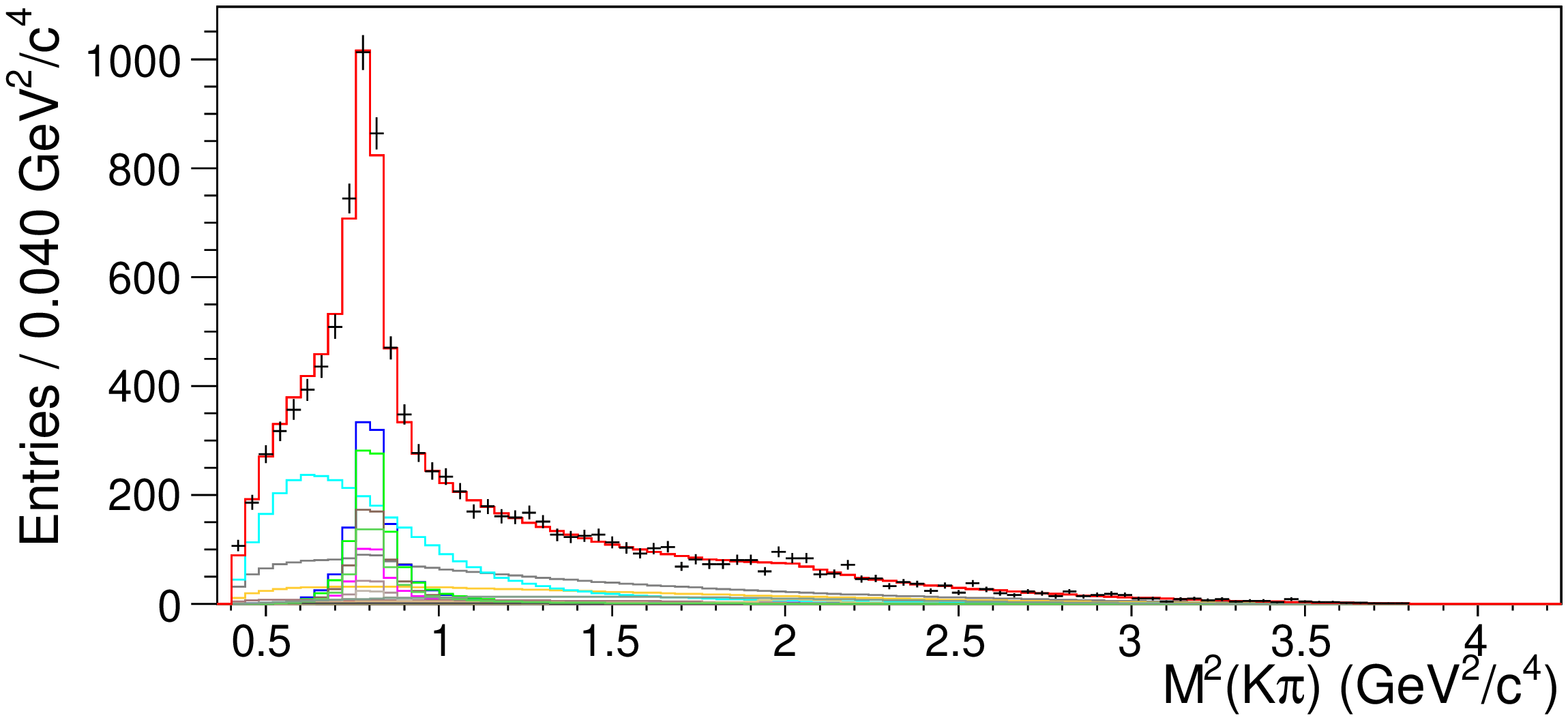}
}}}
\vspace{2mm}
\centerline{
\scalebox{0.35}{
\rotatebox{270}{
\includegraphics*[270,37][569,691]
  {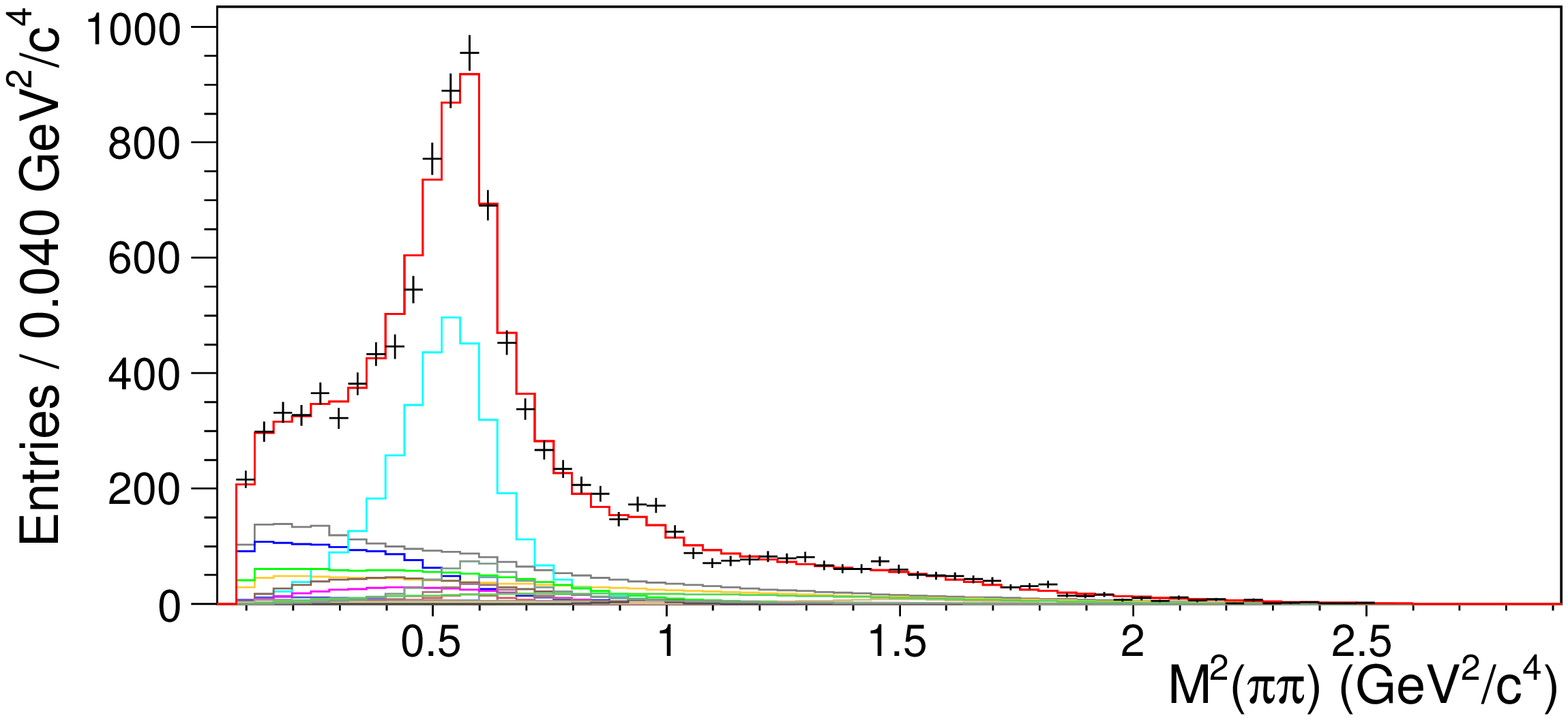}
}}}
\caption{Results of signal-region fits for 
  $B^{+} \rightarrow J/\psi K^{+} \pi^{+} \pi^{-}$,
  with the mass and width of the $K_{1}(1270)$ floated.
  Data (points) and fits (histograms) are shown projected onto the 
  three axes.
  The fit components are color coded as shown in
  Fig.~\ref{amplitude:fig_signal_legend}.}
\label{amplitude:fig_signal_float_jkpp}
\end{figure}

\begin{figure*}[hbtp]
\centerline{
\scalebox{0.35}{
\rotatebox{270}{
\includegraphics*[270,37][569,691]
  {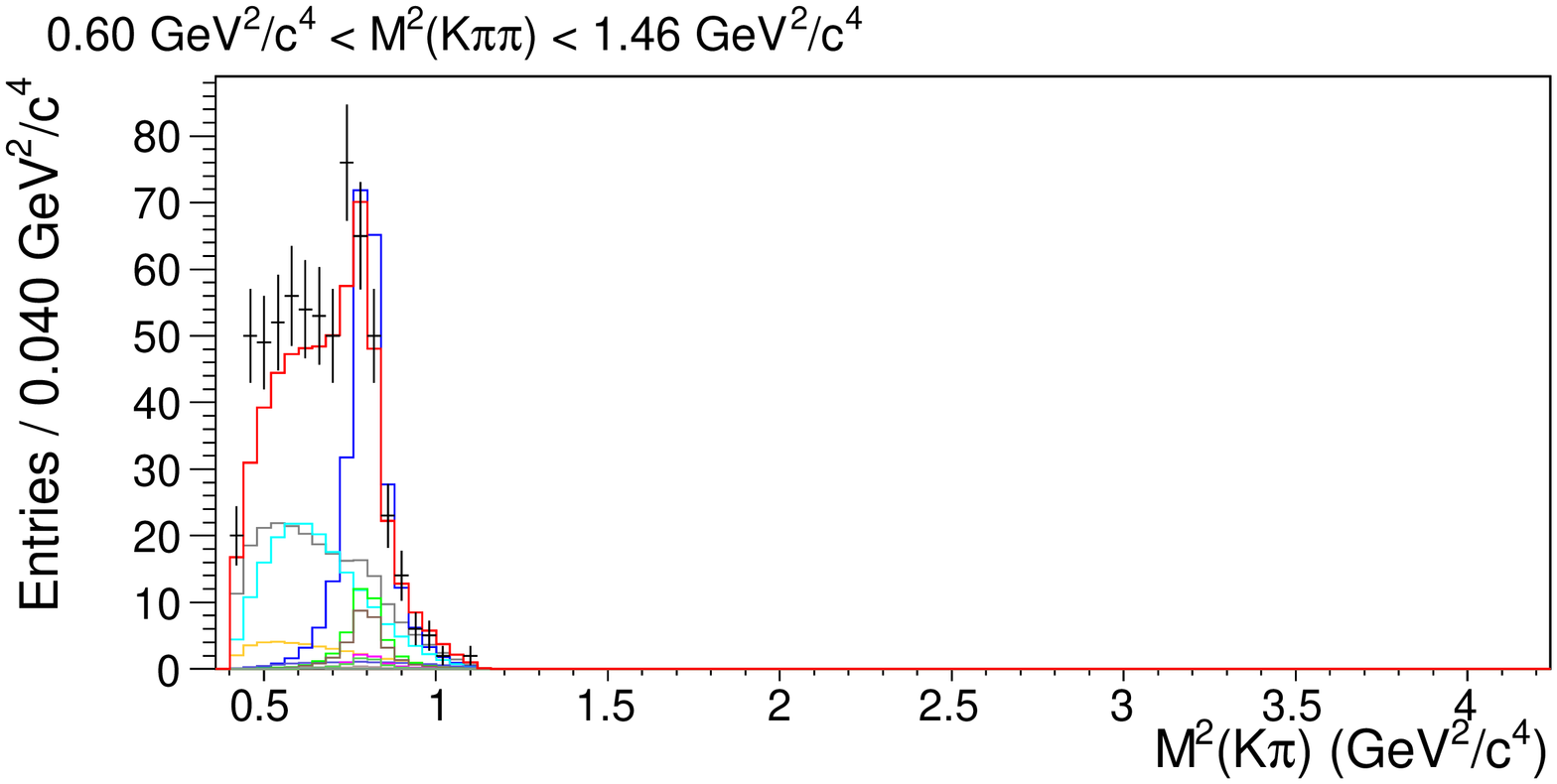}
}}
\hspace{4mm}
\scalebox{0.35}{
\rotatebox{270}{
\includegraphics*[270,37][569,691]
  {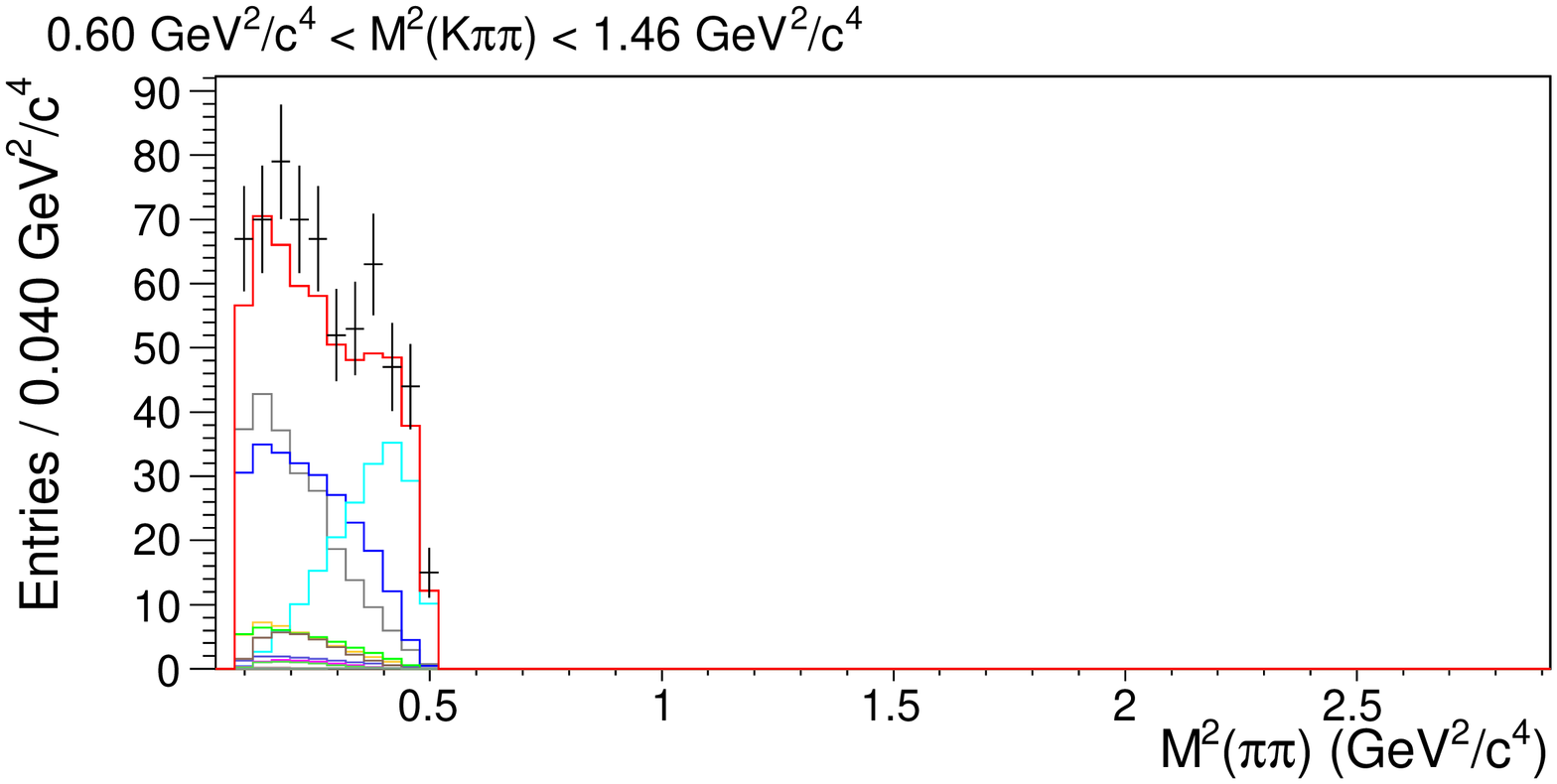}
}}}
\vspace{4mm}
\centerline{
\scalebox{0.35}{
\rotatebox{270}{
\includegraphics*[270,37][569,691]
  {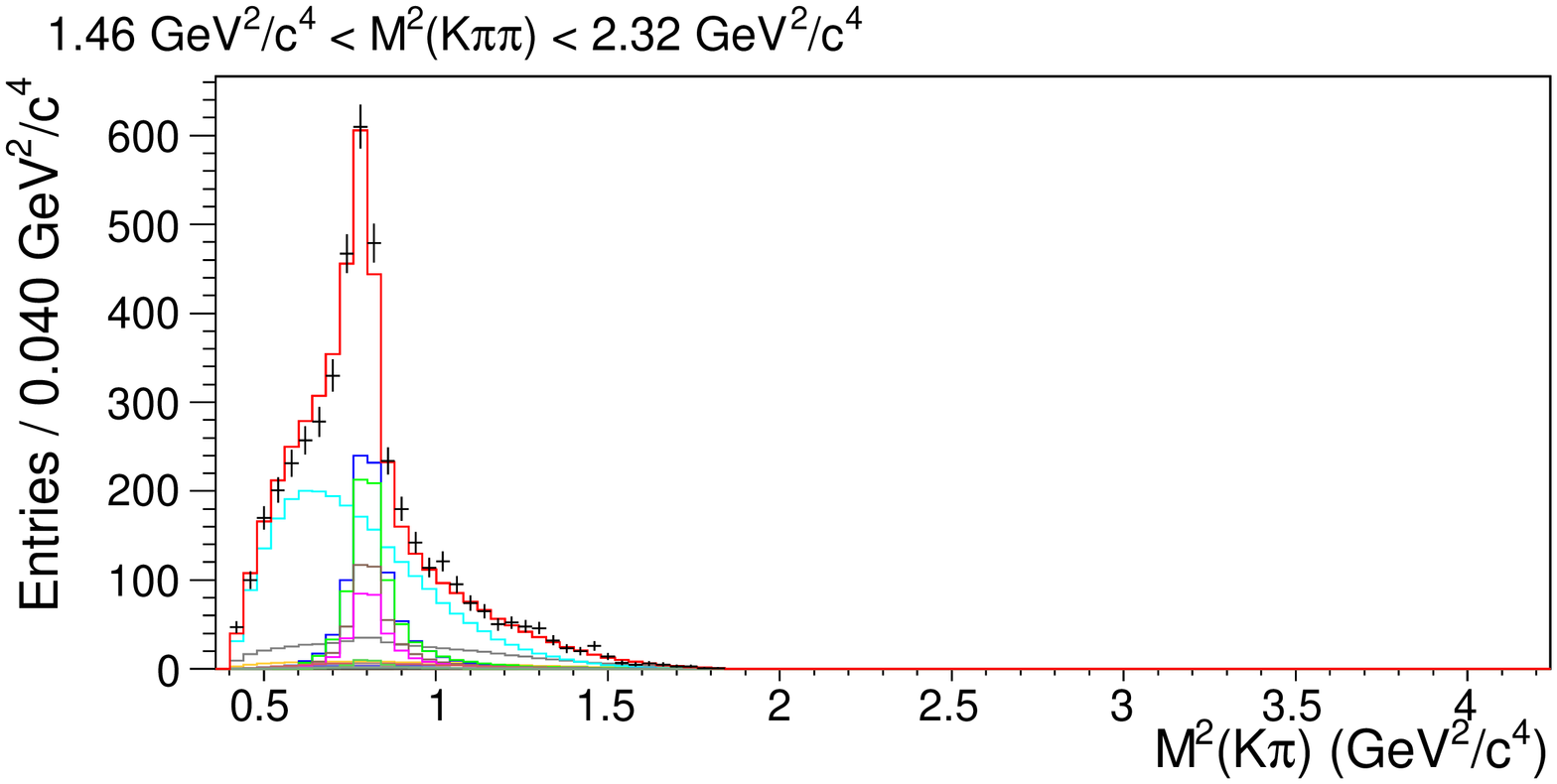}
}}
\hspace{4mm}
\scalebox{0.35}{
\rotatebox{270}{
\includegraphics*[270,37][569,691]
  {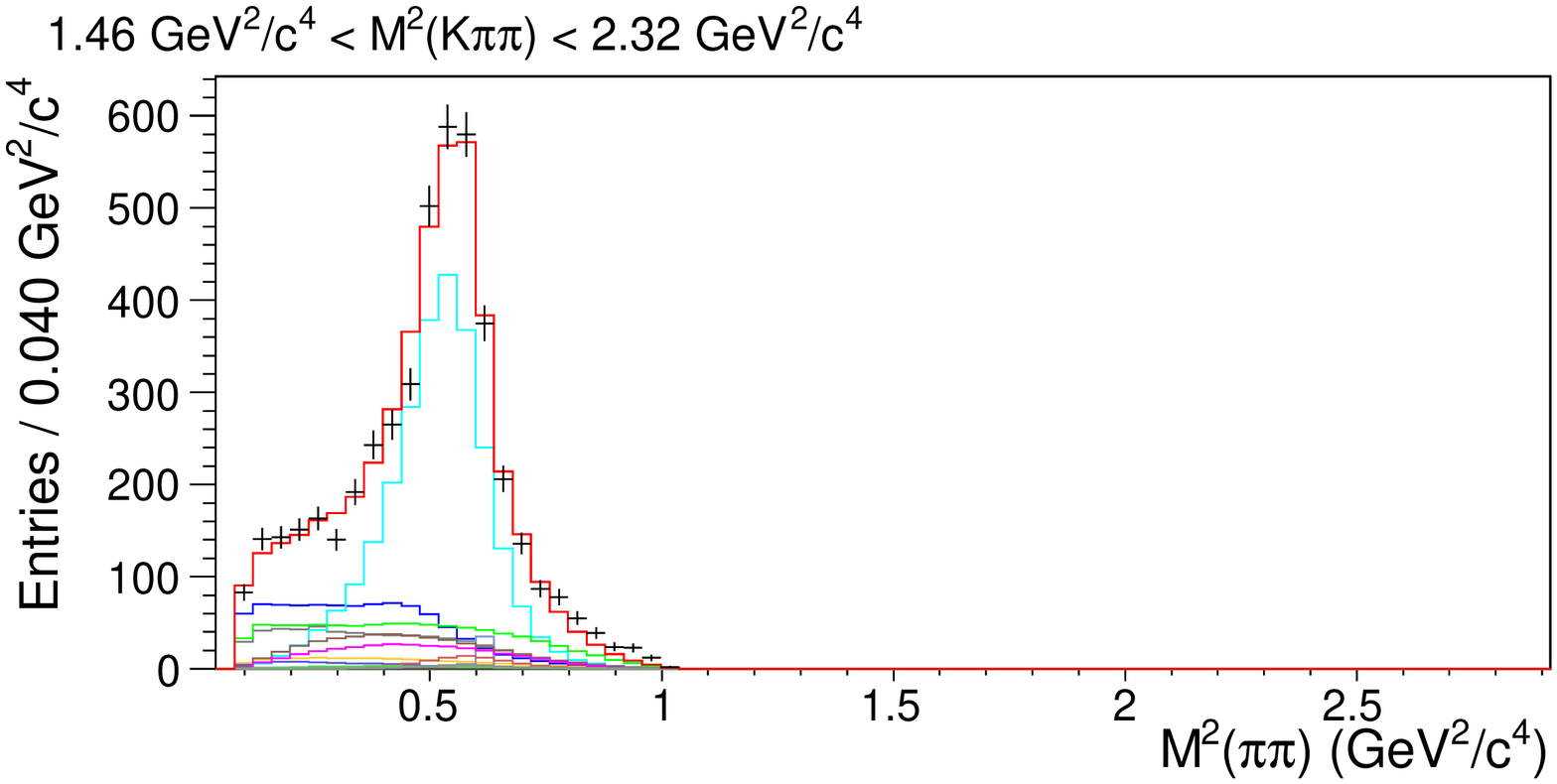}
}}}
\vspace{4mm}
\centerline{
\scalebox{0.35}{
\rotatebox{270}{
\includegraphics*[270,37][569,691]
  {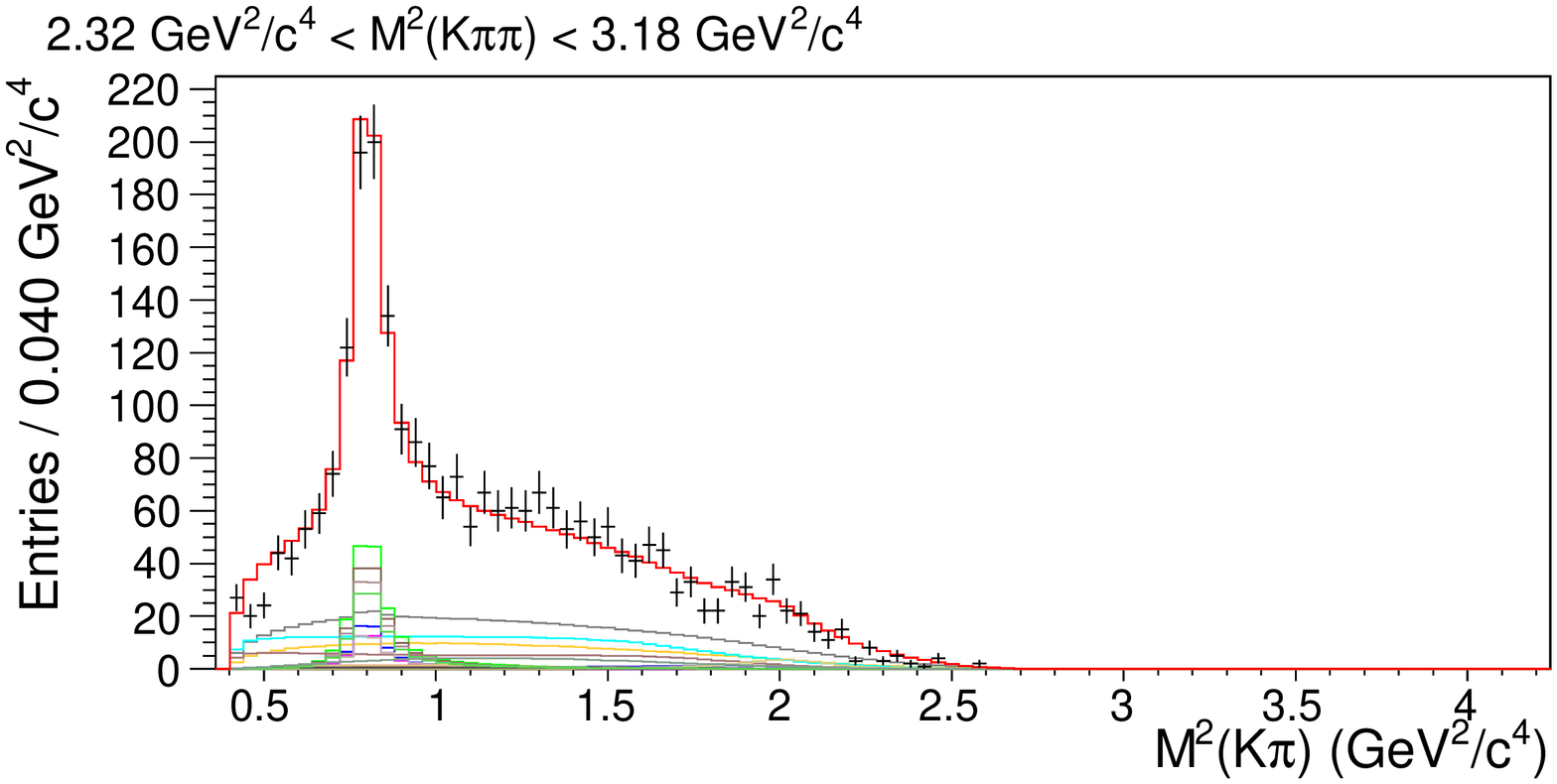}
}}
\hspace{4mm}
\scalebox{0.35}{
\rotatebox{270}{
\includegraphics*[270,37][569,691]
  {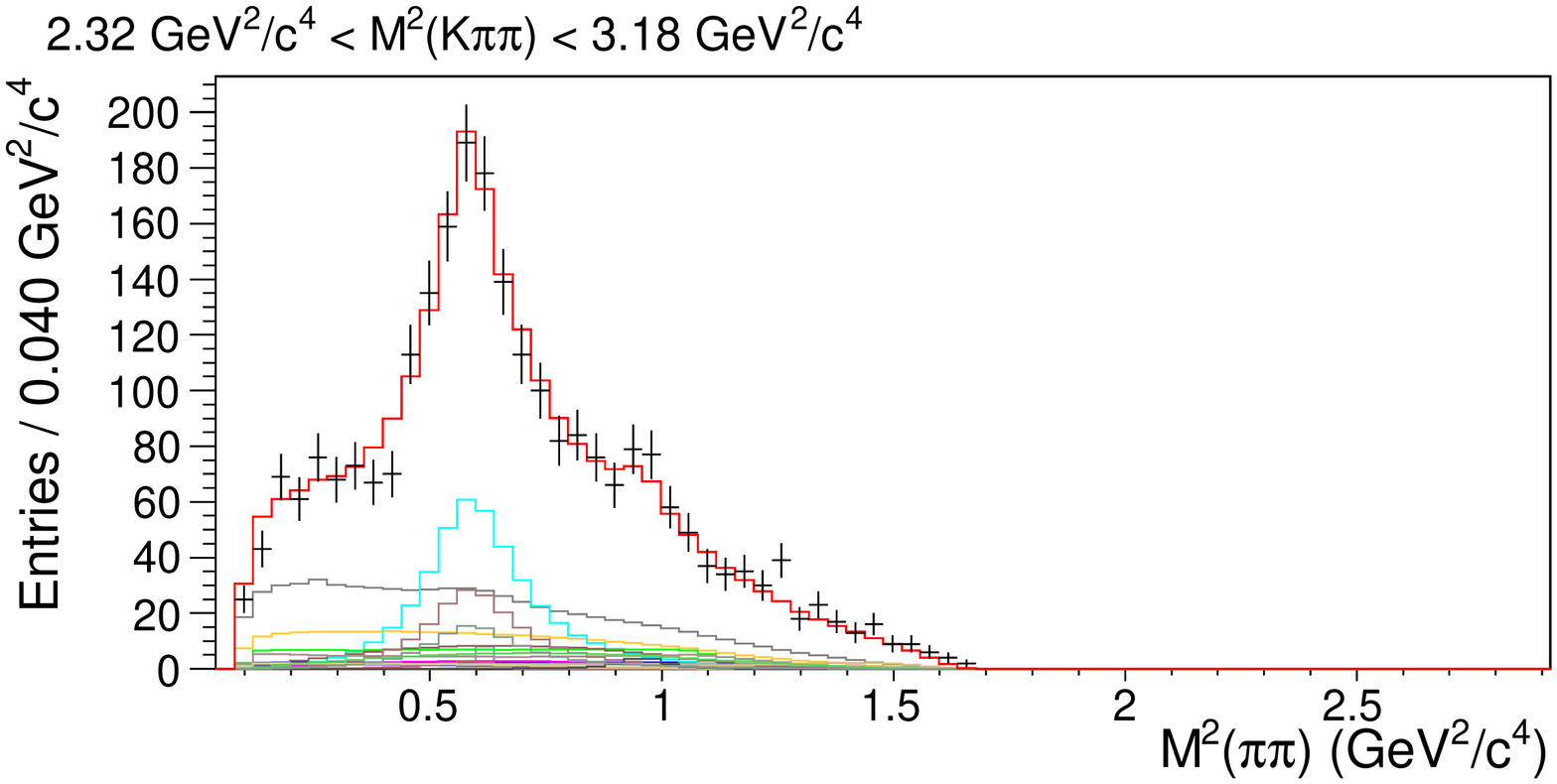}
}}}
\vspace{4mm}
\centerline{
\scalebox{0.35}{
\rotatebox{270}{
\includegraphics*[270,37][569,691]
  {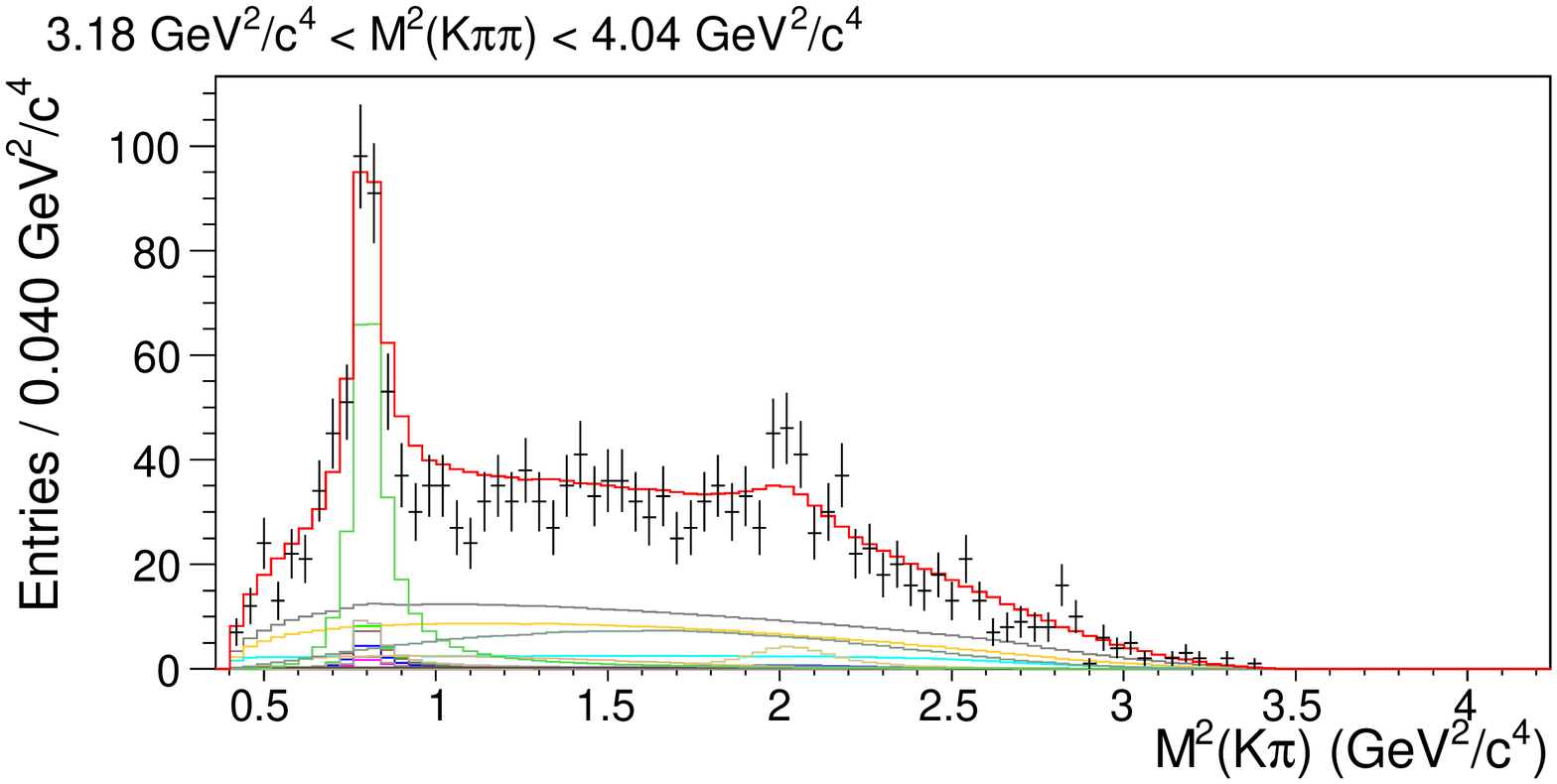}
}}
\hspace{4mm}
\scalebox{0.35}{
\rotatebox{270}{
\includegraphics*[270,37][569,691]
  {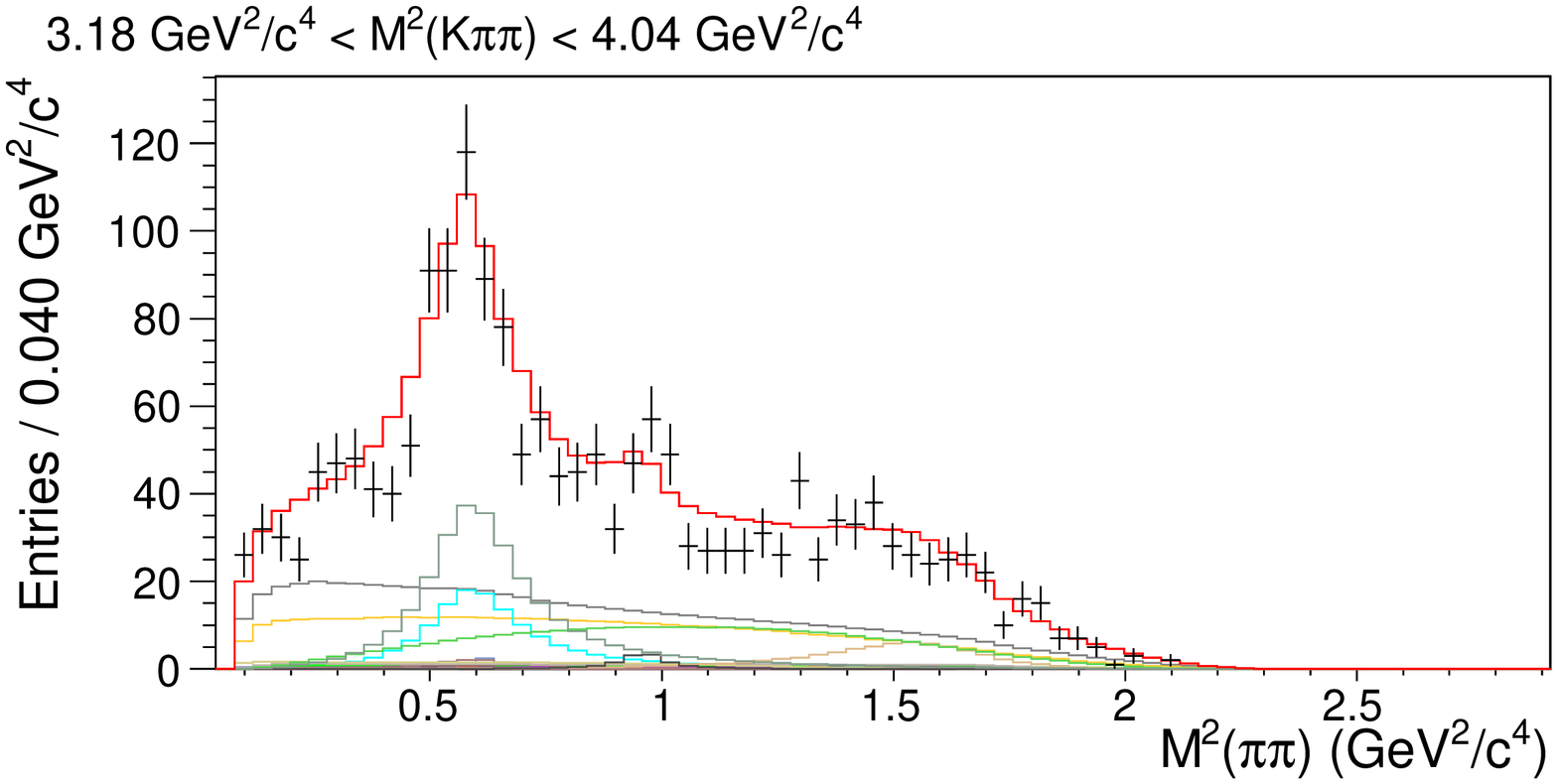}
}}}
\vspace{4mm}
\centerline{
\scalebox{0.35}{
\rotatebox{270}{
\includegraphics*[270,37][569,691]
  {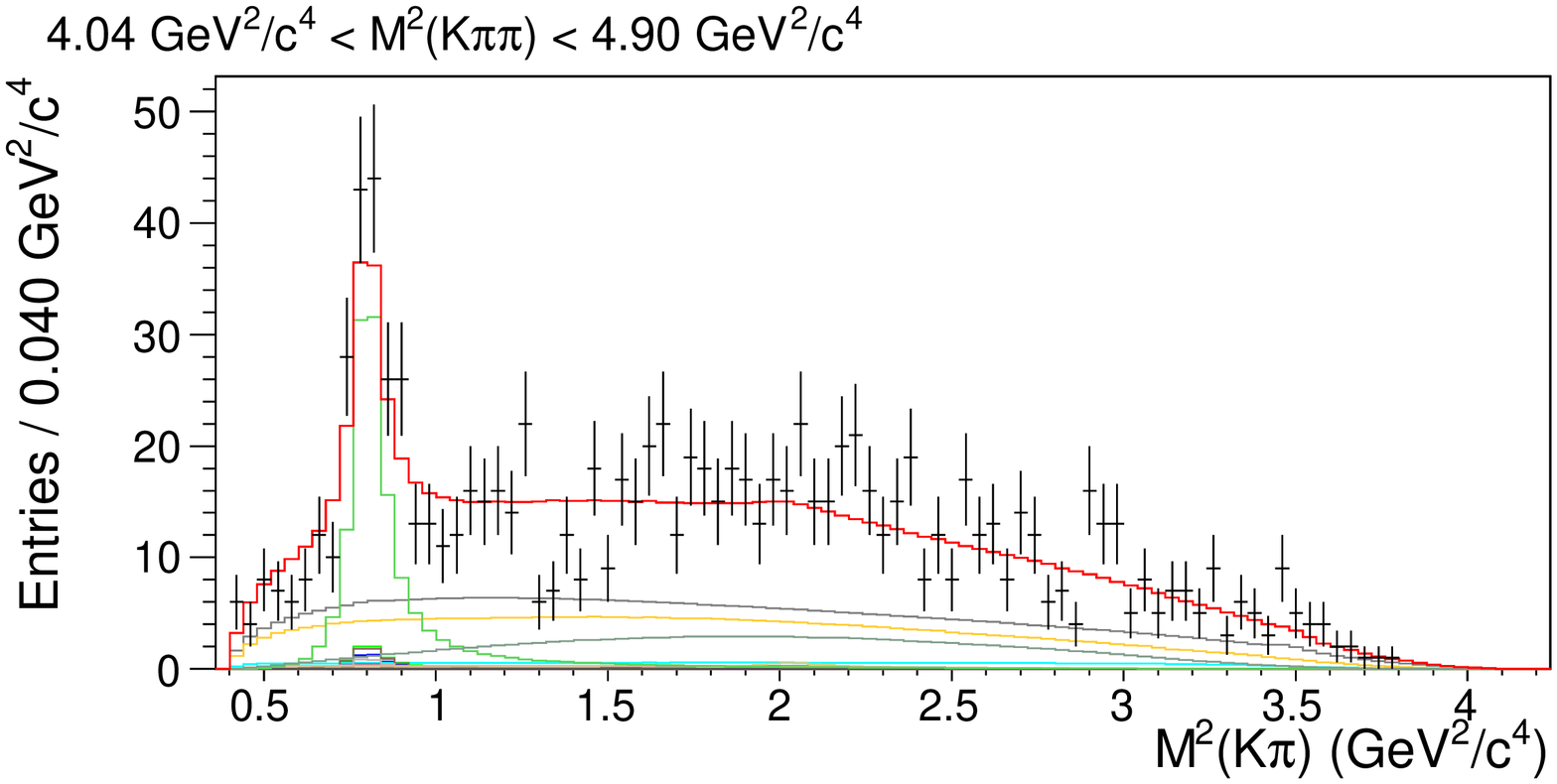}
}}
\hspace{4mm}
\scalebox{0.35}{
\rotatebox{270}{
\includegraphics*[270,37][569,691]
  {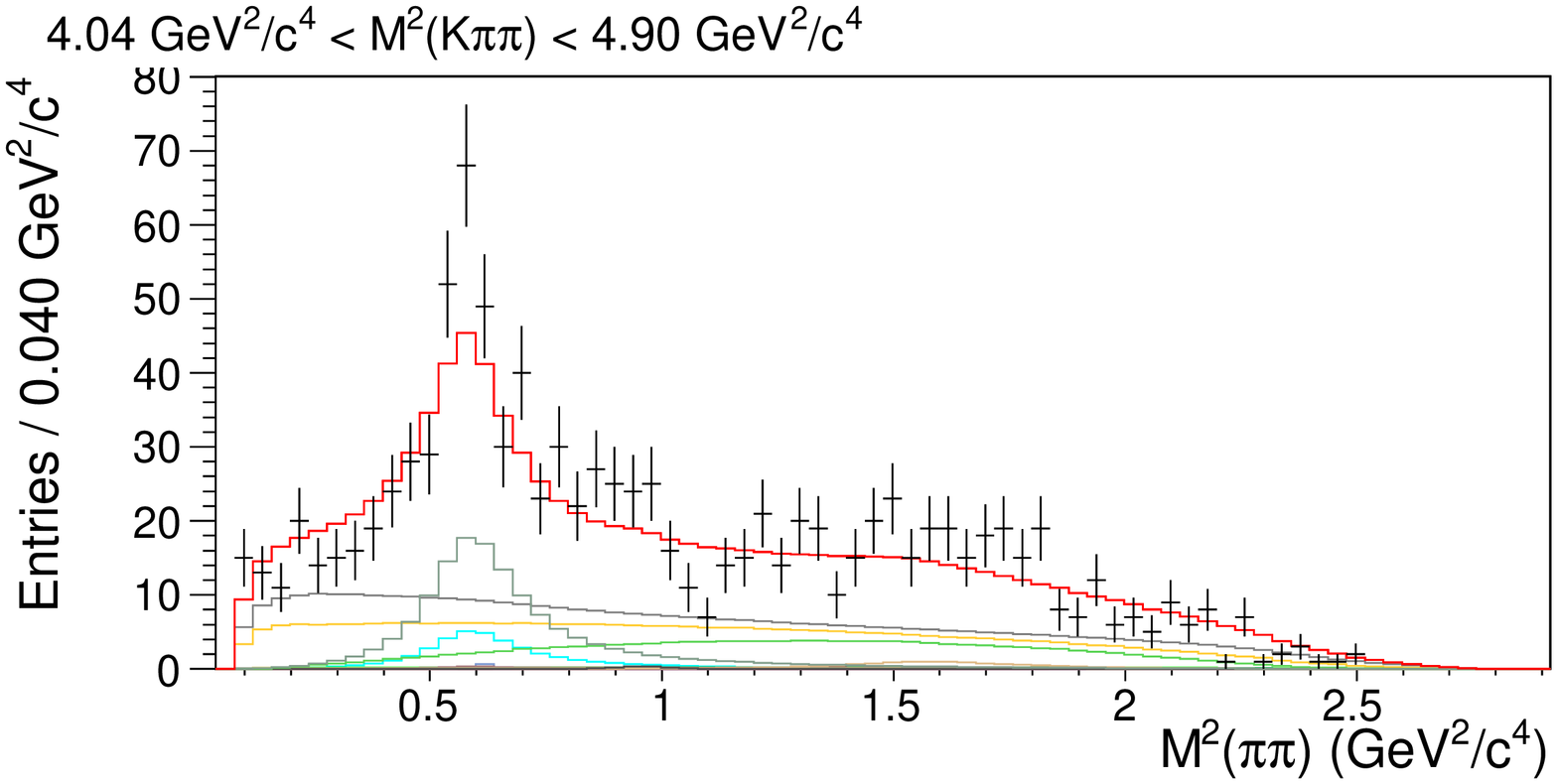}
}}}
\caption{$B^{+} \rightarrow J/\psi K^{+} \pi^{+} \pi^{-}$
  signal data (points) and fit results (histograms) 
  for slices in $M^{2}(K\pi\pi)$. 
  The fit components are color coded as shown in
  Fig.~\ref{amplitude:fig_signal_legend}.
  The mass and width of the $K_{1}(1270)$ floated in the fit.}
\label{amplitude:fig_signal_float_slices_jkpp}
\end{figure*}

Finally, the
$B^{+} \rightarrow J/\psi K^{+} \pi^{+} \pi^{-}$
signal-region data
are fitted
again,
this time 
floating the mass and width of the $K_{1}(1270)$.
The fitted mass and width are
\begin{eqnarray}
\label{amplitude:eq_k1270_mass}
M_{K_{1}(1270)}      
 & = & (1248.1 \pm 3.3 \pm 1.4)~{\mathrm{MeV}}/c^{2},  \\
\label{amplitude:eq_k1270_width}
\Gamma_{K_{1}(1270)} 
 & = & (119.5 \pm 5.2 \pm 6.7)~{\mathrm{MeV}}/c^{2}. 
\end{eqnarray}
Table~\ref{amplitude:table_fit_float_jkpp}
shows the fitted parameters,
along with the corresponding decay fractions.
Figure~\ref{amplitude:fig_signal_float_jkpp}
shows the fitted PDF and data
projected onto the three axes,
while
Fig.~\ref{amplitude:fig_signal_float_slices_jkpp}
shows $M^{2}(K\pi)$ and $M^{2}(\pi\pi)$ projections
for slices in $M^{2}(K\pi\pi)$.
In this fit,
$\ell = -10525.3$,
while
$\chi^{2} = 1404.4$
with
$N_{\mathrm{bins}} = 1202$
and
$N_{\mathrm{par}} = 30$.

A comparison of 
Tables~\ref{amplitude:table_fit_pdg_jkpp}
and~\ref{amplitude:table_fit_float_jkpp}
reveals that in many cases, 
the effect of floating the mass and width of the $K_{1}(1270)$ 
results in a substantial decrease of the systematic error, 
which is somewhat offset by an  increase in the corresponding statistical error.
In particular, the 
$K_{1}(1270) \rightarrow K^{*}(892) \pi$ 
decay fraction 
is especially sensitive to the $K_{1}(1270)$ 
mass and width.

In any fit involving many floating parameters, 
local likelihood maxima can present a problem.
To ensure that the fit results are 
global maxima,
$100$
additional fits were performed for each of the three cases,
selecting random starting values for the parameters.  
None of these fits yielded better likelihoods than those
presented above.
The local maxima encountered in the course of this test
are discussed in the Appendix.

\begin{table*}[htbp]
\renewcommand{\arraystretch}{1.1}
\caption{Fitted parameters 
   of the signal function
   for $B^{+} \rightarrow J/\psi K^{+} \pi^{+} \pi^{-}$, 
   along with the corresponding decay fractions.}
\label{amplitude:table_fit_pdg_jkpp}
\begin{ruledtabular}
\begin{tabular*}{17.8cm}
  {@{\hspace{0.6cm}}c@{\hspace{0.6cm}}l@{\hspace{0.8cm}}l@{$\,\pm\,$}l@{$\,\pm\,$}l@{\hspace{0.8cm}}r@{$\,\pm\,$}l@{$\,\pm\,$}l@{\hspace{0.8cm}}l@{$~\pm~$}l@{$~\pm~$}l}
$J_{1}$
  & Submode
    & \multicolumn{3}{c}{Modulus}         \hspace{0.8cm} 
    & \multicolumn{3}{c}{Phase (radians)} \hspace{0.8cm}
    & \multicolumn{3}{c}{Decay fraction}  
  \\ 
\colrule
  & Nonresonant $K^{+}\pi^{+}\pi^{-}$
    & \multicolumn{3}{c}{$1.0$ (fixed)}   \hspace{0.8cm} 
    & \multicolumn{3}{c}{$0$ (fixed)}     \hspace{0.8cm}
    & $0.152$   &  $0.013$  &  $0.028$
  \\
\colrule
  & $K_{1}(1270) \rightarrow K^{*}(892) \pi$ 
    &  $0.962$  &  $0.058$  &  $0.176$
    & \multicolumn{3}{c}{$0$ (fixed)}     \hspace{0.8cm}  
    & $0.232$   &  $0.017$  &  $0.058$
  \\
  & $K_{1}(1270) \rightarrow K \rho$ 
    &  $1.813$  &  $0.090$  &  $0.243$
    & $-0.764$  &  $0.069$  &  $0.127$                   
    & $0.383$   &  $0.016$  &  $0.036$
  \\
$1^{+}$
  & $K_{1}(1270) \rightarrow K \omega$ 
    &  $0.198$  &  $0.036$  &  $0.041$
    &  $1.09$   &  $0.18$   &  $0.18$                    
    & $0.0045$  &  $0.0017$ &  $0.0014$
  \\
  & $K_{1}(1270) \rightarrow K_{0}^{*}(1430) \pi$ 
    &  $0.95$   &  $0.16$   &  $0.24$
    &  $2.83$   &  $0.18$   &  $0.18$                    
    & $0.0157$  &  $0.0052$ &  $0.0049$
  \\
  & $K_{1}(1400) \rightarrow K^{*}(892) \pi$ 
    &  $0.894$  &  $0.066$  &  $0.125$
    & $-2.300$  &  $0.044$  &  $0.078$                   
    & $0.223$   &  $0.026$  &  $0.036$
  \\
\colrule
$1^{-}$
  & $K^{*}(1410) \rightarrow K^{*}(892) \pi$ 
    &  $0.516$  &  $0.090$  &  $0.103$
    & \multicolumn{3}{c}{$0$ (fixed)}     \hspace{0.8cm}             
    & $0.047$   &  $0.016$  &  $0.015$
  \\
\colrule
  & $K_{2}^{*}(1430) \rightarrow K^{*}(892) \pi$ 
    &  $0.663$  &  $0.051$  &  $0.085$
    & \multicolumn{3}{c}{$0$ (fixed)}     \hspace{0.8cm}
    & $0.088$   &  $0.011$  &  $0.011$
  \\
  & $K_{2}^{*}(1430) \rightarrow K \rho$ 
    & \multicolumn{3}{c}{$0.371$ (fixed)} \hspace{0.8cm}
    & $-1.12$   &  $0.22$   &  $0.29$                    
    & \multicolumn{3}{c}{$0.0233$ (fixed)}
  \\
$2^{+}$
  & $K_{2}^{*}(1430) \rightarrow K \omega$ 
    & \multicolumn{3}{c}{$0.040$ (fixed)} \hspace{0.8cm}
    &  $0.58$   &  $0.51$   &  $0.27$                    
    & \multicolumn{3}{c}{$0.00036$ (fixed)}
  \\
  & $K_{2}^{*}(1980) \rightarrow K^{*}(892) \pi$ 
    &  $0.775$  &  $0.054$  &  $0.118$
    & $-1.59$   &  $0.15$   &  $0.14$                    
    & $0.0739$  &  $0.0073$ &  $0.0095$
  \\
  & $K_{2}^{*}(1980) \rightarrow K \rho$ 
    &  $0.660$  &  $0.048$  &  $0.101$
    &  $0.86$   &  $0.22$   &  $0.21$                    
    & $0.0613$  &  $0.0058$ &  $0.0059$
  \\ 
\colrule
  & $K(1600) \rightarrow K^{*}(892) \pi$ 
    &  $0.131$  &  $0.021$  &  $0.024$
    & \multicolumn{3}{c}{$0$ (fixed)}     \hspace{0.8cm}
    & $0.0187$  &  $0.0058$ &  $0.0050$
  \\
  & $K(1600) \rightarrow K \rho$ 
    &  $0.193$  &  $0.017$  &  $0.029$
    & $-0.27$   &  $0.27$   &  $0.18$                    
    & $0.0424$  &  $0.0062$ &  $0.0110$
  \\
  & $K_{2}(1770) \rightarrow K^{*}(892) \pi$ 
    &  $0.122$  &  $0.021$  &  $0.026$
    &  $2.22$   &  $0.49$   &  $0.37$                    
    & $0.0164$  &  $0.0055$ &  $0.0061$
  \\
\raisebox{1.5ex}[0pt]{$2^{-}$}
  & $K_{2}(1770) \rightarrow K_{2}^{*}(1430) \pi$ 
    &  $0.286$  &  $0.043$  &  $0.044$
    &  $1.78$   &  $0.39$   &  $0.24$                    
    & $0.0100$  &  $0.0028$ &  $0.0020$
  \\
  & $K_{2}(1770) \rightarrow K f_{2}(1270)$ 
    &  $0.444$  &  $0.069$  &  $0.077$
    &  $2.30$   &  $0.37$   &  $0.32$                    
    & $0.0124$  &  $0.0033$ &  $0.0022$
  \\
  & $K_{2}(1770) \rightarrow K f_{0}(980)$ 
    &  $0.113$  &  $0.029$  &  $0.024$
    &  $1.83$   &  $0.45$   &  $0.53$                    
    & $0.0034$  &  $0.0017$ &  $0.0011$
  \\
\end{tabular*}
\end{ruledtabular}
\end{table*}

\begin{table*}[htbp]
\caption{Fitted parameters 
   of the signal function
   for $B^{+} \rightarrow \psi^{\prime} K^{+} \pi^{+} \pi^{-}$,
   along with the corresponding decay fractions.}
\label{amplitude:table_fit_pdg_pkpp}
\renewcommand{\arraystretch}{1.1}
\begin{ruledtabular}
\begin{tabular*}{17.8cm}
  {@{\hspace{0.6cm}}c@{\hspace{0.6cm}}l@{\hspace{0.9cm}}l@{$\,\pm\,$}l@{$\,\pm\,$}l@{\hspace{0.9cm}}r@{$\,\pm\,$}l@{$\,\pm\,$}l@{\hspace{0.9cm}}l@{$~\pm~$}l@{$~\pm~$}l}
$J_{1}$
  & Submode
    & \multicolumn{3}{c}{Modulus}         \hspace{0.9cm} 
    & \multicolumn{3}{c}{Phase (radians)} \hspace{0.9cm}
    & \multicolumn{3}{c}{Decay Fraction}  
  \\ 
\colrule
  & Nonresonant $K^{+}\pi^{+}\pi^{-}$
    & \multicolumn{3}{c}{$1.0$ (fixed)}   \hspace{0.9cm}
    & \multicolumn{3}{c}{$0$ (fixed)}     \hspace{0.9cm}
    &  $0.253$  &  $0.045$  &  $0.102$
  \\ 
\colrule
  & $K_{1}(1270) \rightarrow K^{*}(892) \pi$ 
    &  $0.213$  &  $0.037$  &  $0.049$
    & \multicolumn{3}{c}{$0$ (fixed)}     \hspace{0.9cm}
    &  $0.090$  &  $0.024$  &  $0.013$
  \\ 
$1^{+}$
  & $K_{1}(1270) \rightarrow K \rho$ 
    &  $0.513$  &  $0.070$  &  $0.141$
    & $-0.66$   &  $0.26$   &  $0.11$
    &  $0.215$  &  $0.038$  &  $0.045$
  \\ 
  & $K_{1}(1270) \rightarrow K \omega$ 
    &  $0.048$  &  $0.041$  &  $0.022$
    & $-0.37$   &  $1.21$   &  $0.52$
    &  $0.0017$ &  $0.0033$ &  $0.0013$
  \\ 
\colrule
  & $K^{*}(1680) \rightarrow K^{*}(892) \pi$
    &  $0.67$   &  $0.12$   &  $0.15$
    & \multicolumn{3}{c}{$0$ (fixed)}     \hspace{0.9cm}
    &  $0.106$  &  $0.031$  &  $0.017$
  \\ 
$1^{-}$
  & $K^{*}(1680) \rightarrow K \rho$
    &  $1.17$   &  $0.16$   &  $0.32$
    &  $1.27$   &  $0.24$   &  $0.15$
    &  $0.241$  &  $0.047$  &  $0.050$
  \\ 
  & $K^{*}(1680) \rightarrow K \omega$
    &  $0.233$  &  $0.097$  &  $0.047$
    & $-3.06$   &  $0.43$   &  $0.45$
    &  $0.0119$ &  $0.0106$ &  $0.0061$
  \\ 
\end{tabular*}
\end{ruledtabular}
\end{table*}

\begin{table*}[htbp]
\caption{Fitted parameters 
   of the signal function
   for $B^{+} \rightarrow J/\psi K^{+} \pi^{+} \pi^{-}$
   when the $K_{1}(1270)$ mass and width are floated,
   along with the corresponding decay fractions.}
\label{amplitude:table_fit_float_jkpp}
\renewcommand{\arraystretch}{1.1}
\begin{ruledtabular}
\begin{tabular*}{17.8cm}
  {@{\hspace{0.6cm}}c@{\hspace{0.6cm}}l@{\hspace{0.8cm}}l@{$\,\pm\,$}l@{$\,\pm\,$}l@{\hspace{0.6cm}}r@{$\,\pm\,$}l@{$\,\pm\,$}l@{\hspace{0.6cm}}l@{$~\pm~$}l@{$~\pm~$}l}
$J_{1}$
  & Submode
    & \multicolumn{3}{c}{Modulus}         \hspace{0.8cm} 
    & \multicolumn{3}{c}{Phase (radians)} \hspace{0.6cm}
    & \multicolumn{3}{c}{Decay Fraction}  
  \\ 
\colrule
  & Nonresonant $K^{+}\pi^{+}\pi^{-}$
    & \multicolumn{3}{c}{$1.0$ (fixed)}   \hspace{0.8cm} 
    & \multicolumn{3}{c}{$0$ (fixed)}     \hspace{0.6cm}
    & $0.142$   &  $0.013$   &  $0.026$
  \\ 
\colrule
  & $K_{1}(1270) \rightarrow K^{*}(892) \pi$ 
    &  $0.882$  &  $0.076$   &  $0.090$
    & \multicolumn{3}{c}{$0$ (fixed)}     \hspace{0.6cm}
    & $0.168$   &  $0.023$   &  $0.012$
  \\ 
  & $K_{1}(1270) \rightarrow K \rho$ 
    &  $2.14$   &  $0.12$    &  $0.27$
    & $-0.588$  &  $0.084$   &  $0.110$ 
    & $0.430$   &  $0.018$   &  $0.027$
  \\ 
$1^{+}$
  & $K_{1}(1270) \rightarrow K \omega$ 
    &  $0.289$  &  $0.043$   &  $0.040$
    &  $1.25$   &  $0.16$    &  $0.14$
    & $0.00758$ &  $0.00216$ &  $0.00076$
  \\ 
  & $K_{1}(1270) \rightarrow K_{0}^{*}(1430) \pi$ 
    &  $1.09$   &  $0.18$    &  $0.24$
    &  $2.93$   &  $0.18$    &  $0.16$
    & $0.0184$  &  $0.0055$  &  $0.0046$
  \\ 
  & $K_{1}(1400) \rightarrow K^{*}(892) \pi$ 
    &  $0.746$  &  $0.085$   &  $0.089$
    & $-2.585$  &  $0.100$   &  $0.076$
    & $0.145$   &  $0.029$   &  $0.017$
  \\ 
\colrule
$1^{-}$
  & $K^{*}(1410) \rightarrow K^{*}(892) \pi$ 
    &  $0.736$  &  $0.084$   &  $0.098$
    & \multicolumn{3}{c}{$0$ (fixed)}     \hspace{0.6cm}
    & $0.089$   &  $0.019$   &  $0.010$
  \\ 
\colrule
  & $K_{2}^{*}(1430) \rightarrow K^{*}(892) \pi$ 
    &  $0.529$  &  $0.064$   &  $0.070$
    & \multicolumn{3}{c}{$0$ (fixed)}     \hspace{0.6cm}
    & $0.0525$  &  $0.0120$  &  $0.0070$
  \\ 
  & $K_{2}^{*}(1430) \rightarrow K \rho$ 
    & \multicolumn{3}{c}{$0.296$ (fixed)} \hspace{0.8cm}
    & $-0.61$   &  $0.39$    &  $1.07$
    & \multicolumn{3}{c}{$0.014$ (fixed)}
  \\ 
$2^{+}$
  & $K_{2}^{*}(1430) \rightarrow K \omega$ 
    & \multicolumn{3}{c}{$0.032$ (fixed)} \hspace{0.8cm}
    &  $1.41$   &  $0.80$    &  $0.25$
    & \multicolumn{3}{c}{$0.00021$ (fixed)}
  \\
  & $K_{2}^{*}(1980) \rightarrow K^{*}(892) \pi$ 
    &  $0.756$  &  $0.060$   &  $0.119$
    & $-1.46$   &  $0.23$    &  $0.22$
    & $0.0659$  &  $0.0073$  &  $0.0088$
  \\ 
  & $K_{2}^{*}(1980) \rightarrow K \rho$ 
    &  $0.685$  &  $0.052$   &  $0.106$
    &  $1.15$   &  $0.30$    &  $0.20$
    & $0.0617$  &  $0.0061$  &  $0.0065$
  \\ 
\colrule
  & $K(1600) \rightarrow K^{*}(892) \pi$ 
    &  $0.147$  &  $0.021$   &  $0.026$
    & \multicolumn{3}{c}{$0$ (fixed)}     \hspace{0.6cm}
    & $0.0222$  &  $0.0058$  &  $0.0054$
  \\ 
  & $K(1600) \rightarrow K \rho$ 
    &  $0.171$  &  $0.020$   &  $0.023$
    & $-0.21$   &  $0.29$    &  $0.12$  
    & $0.0312$  &  $0.0065$  &  $0.0040$
  \\ 
  & $K_{2}(1770) \rightarrow K^{*}(892) \pi$ 
    &  $0.116$  &  $0.020$   &  $0.029$
    &  $1.93$   &  $0.52$    &  $0.43$ 
    & $0.0137$  &  $0.0049$  &  $0.0058$
  \\ 
\raisebox{1.5ex}[0pt]{$2^{-}$}
  & $K_{2}(1770) \rightarrow K_{2}^{*}(1430) \pi$ 
    &  $0.288$  &  $0.045$   &  $0.045$
    &  $1.81$   &  $0.40$    &  $0.24$
    & $0.0095$  &  $0.0027$  &  $0.0018$
  \\ 
  & $K_{2}(1770) \rightarrow K f_{2}(1270)$ 
    &  $0.466$  &  $0.073$   &  $0.080$
    &  $2.22$   &  $0.36$    &  $0.33$
    & $0.0128$  &  $0.0035$  &  $0.0021$
  \\ 
  & $K_{2}(1770) \rightarrow K f_{0}(980)$ 
    &  $0.118$  &  $0.030$   &  $0.025$
    &  $1.89$   &  $0.45$    &  $0.53$
    & $0.0035$  &  $0.0017$  &  $0.0011$
  \\ 
\end{tabular*}
\end{ruledtabular}
\end{table*}

\subsection{Discussion}

\subsubsection{Signal components}

In choosing the signal components to be included in the fits,
the data were used as a guide.
As
the $K_{1}(1270)$ signal is prominent
in both $B^{+} \rightarrow J/\psi        K^{+} \pi^{+} \pi^{-}$
and     $B^{+} \rightarrow \psi^{\prime} K^{+} \pi^{+} \pi^{-}$ data,
the initial fits were done with only
$K_{1}(1270) \rightarrow K^{*}(892) \pi$
and 
$K_{1}(1270) \rightarrow K \rho$
on top of the nonresonant component.  
Additional decay channels were added successively
until a reasonable level of agreement 
between fit and data
was obtained.

\begin{figure}[hbtp]
\centerline{
\scalebox{0.36}{
\rotatebox{270}{
\includegraphics*[280,35][561,694]
  {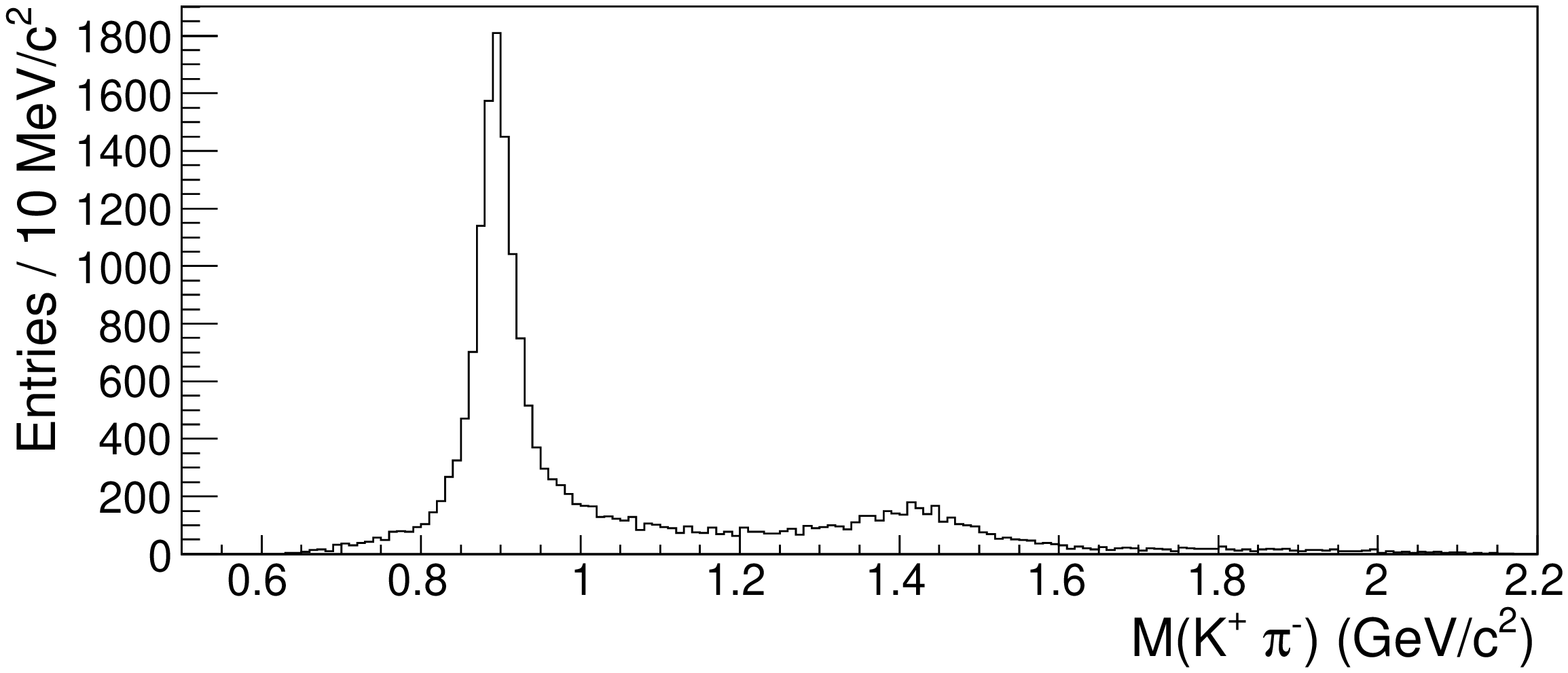}
}}}
\vspace{2mm}
\centerline{
\scalebox{0.36}{
\rotatebox{270}{
\includegraphics*[280,35][561,694]
  {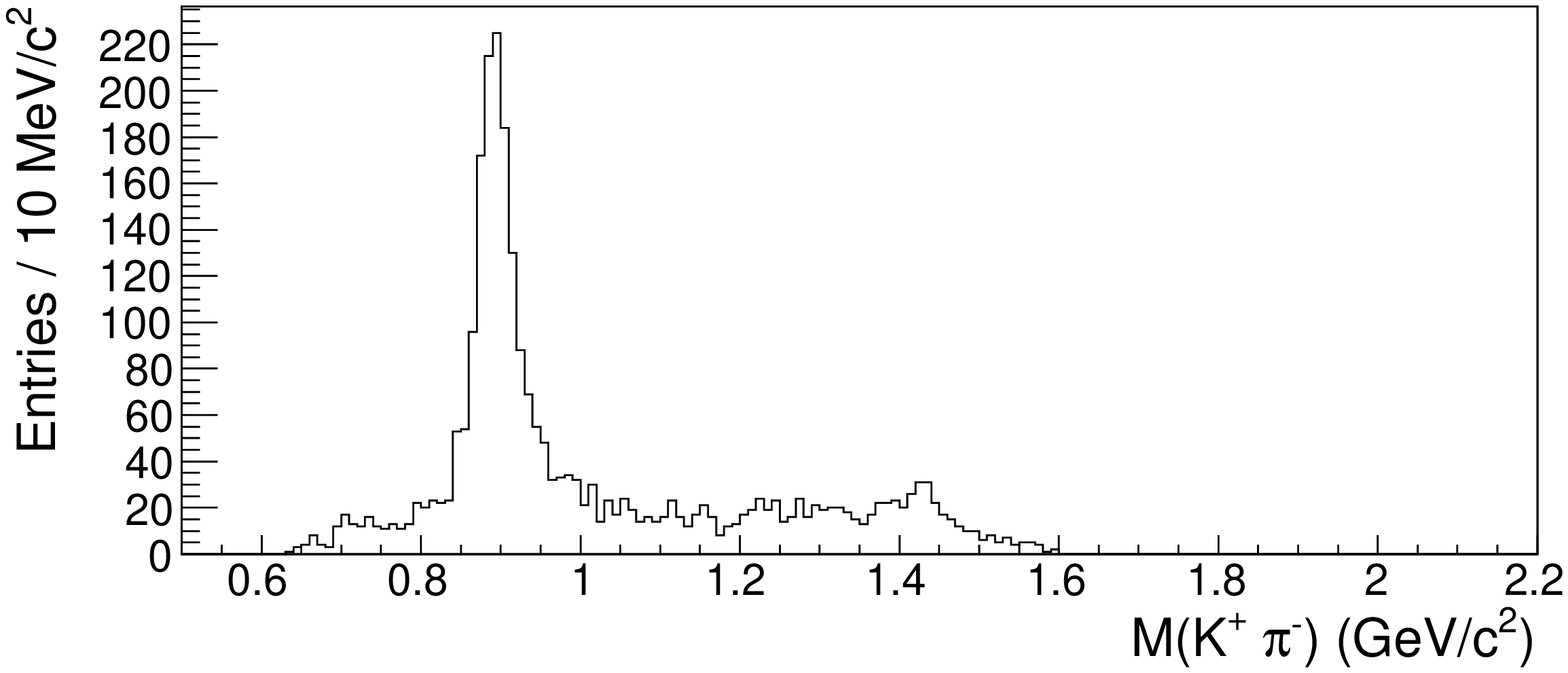}
}}}
\caption{Observed $K \pi$ mass spectra for
  $B^{0} \rightarrow J/\psi K^{+} \pi^{-}$ (top) and
  $B^{0} \rightarrow \psi^{\prime} K^{+} \pi^{-}$ (bottom)
  data.}
\label{amplitude:fig_mkp}
\end{figure}

As a further guide,
the decays
$B^{0} \rightarrow J/\psi K^{+} \pi^{-}$ and
$B^{0} \rightarrow \psi^{\prime} K^{+} \pi^{-}$
were reconstructed.
The observed $K \pi$ mass spectra
are shown in Fig.~\ref{amplitude:fig_mkp}.
Consistent with the $1^{+}$ spin-parity assignment
of the $K_{1}(1270)$, 
no $K_{1}(1270) \rightarrow K \pi$ signal appears in these spectra.
In both modes, 
a small peak can be seen near $1.4~{\mathrm{GeV}}/c^{2}$ in $M(K\pi)$;
this may have contributions from
$K^{*}(1410)$ or $K_{2}^{*}(1430)$, as well as $K_{0}^{*}(1430)$. 
The absence of a $K^{*}(1680)$ peak 
in $B^{0} \rightarrow J/\psi K^{+} \pi^{-}$ 
is noteworthy,
although a precise statement would require an analysis
of the efficiency and phase space for these modes.{\footnote{A 
detailed analysis of 
$B \rightarrow J/\psi K \pi$
and
$B \rightarrow \psi^{\prime} K \pi$
is beyond the scope of this work.
A Dalitz analysis of the latter mode was presented in
Ref.~\onlinecite{mizuk:2009}.}} 
In 
$B^{0} \rightarrow \psi^{\prime} K^{+} \pi^{-}$,
the kinematically-allowed 
$M(K\pi)$ region 
does not allow any conclusions to be drawn
about the presence or absence of 
a low $K^{*}(1680)$ tail.

\subsubsection{Interference effects}

The inclusion of interference among submodes 
sharing the same initial-state spin-parity
is essential to obtaining good fits to the data.
In particular, dramatic interference effects are observed
between $K_{1}(1270) \rightarrow K^{*}(892) \pi$
and     $K_{1}(1270) \rightarrow K \rho$,
as well as between $K_{1}(1270) \rightarrow K \rho$
and         $K_{1}(1270) \rightarrow K \omega$.

\begin{figure*}[hbtp]
\centerline{
\scalebox{0.275}{
\rotatebox{270}{
\includegraphics*[13,29][575,608]
  {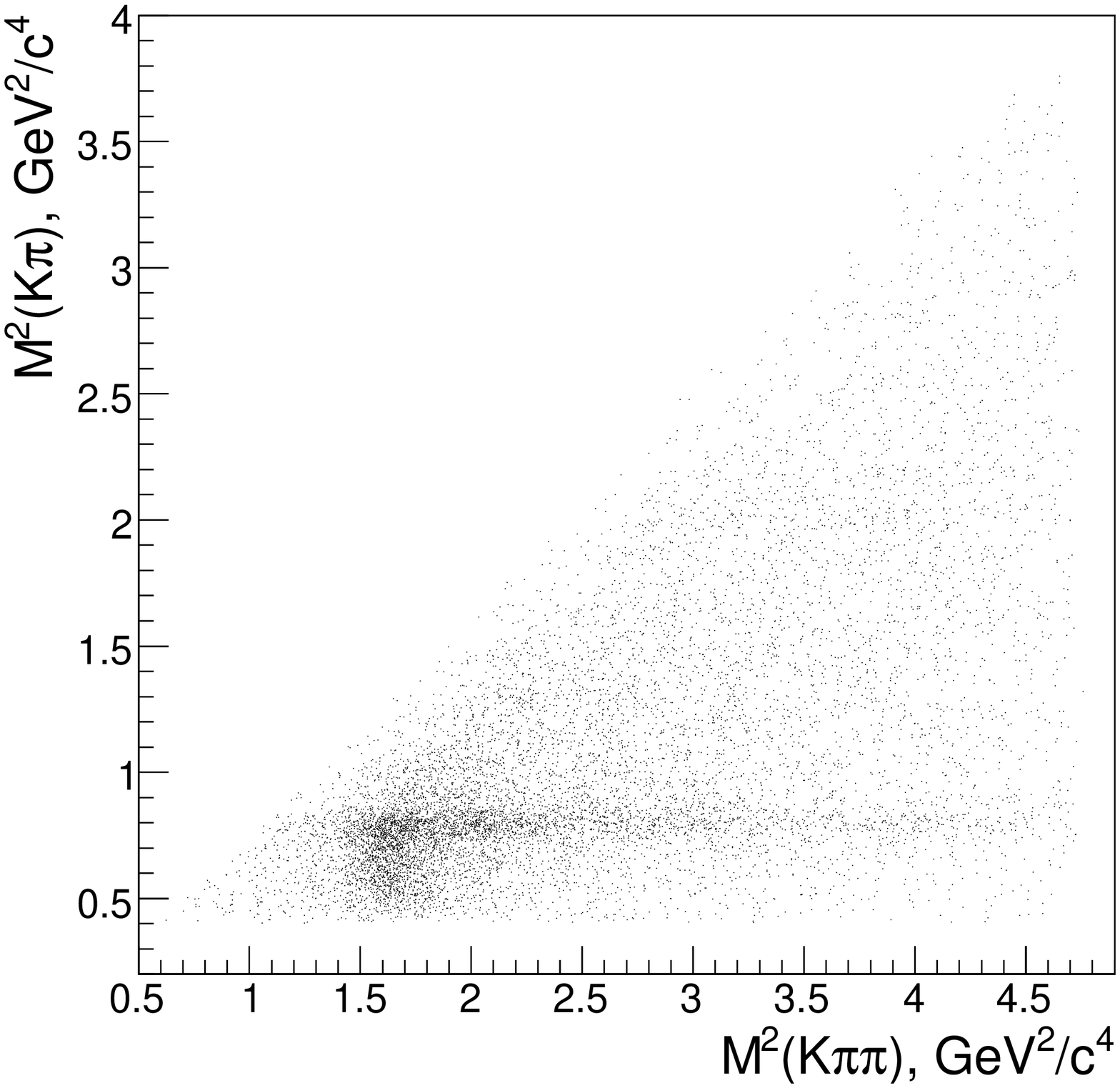}
}}
\hspace{1mm}
\scalebox{0.275}{
\rotatebox{270}{
\includegraphics*[13,29][575,608]
  {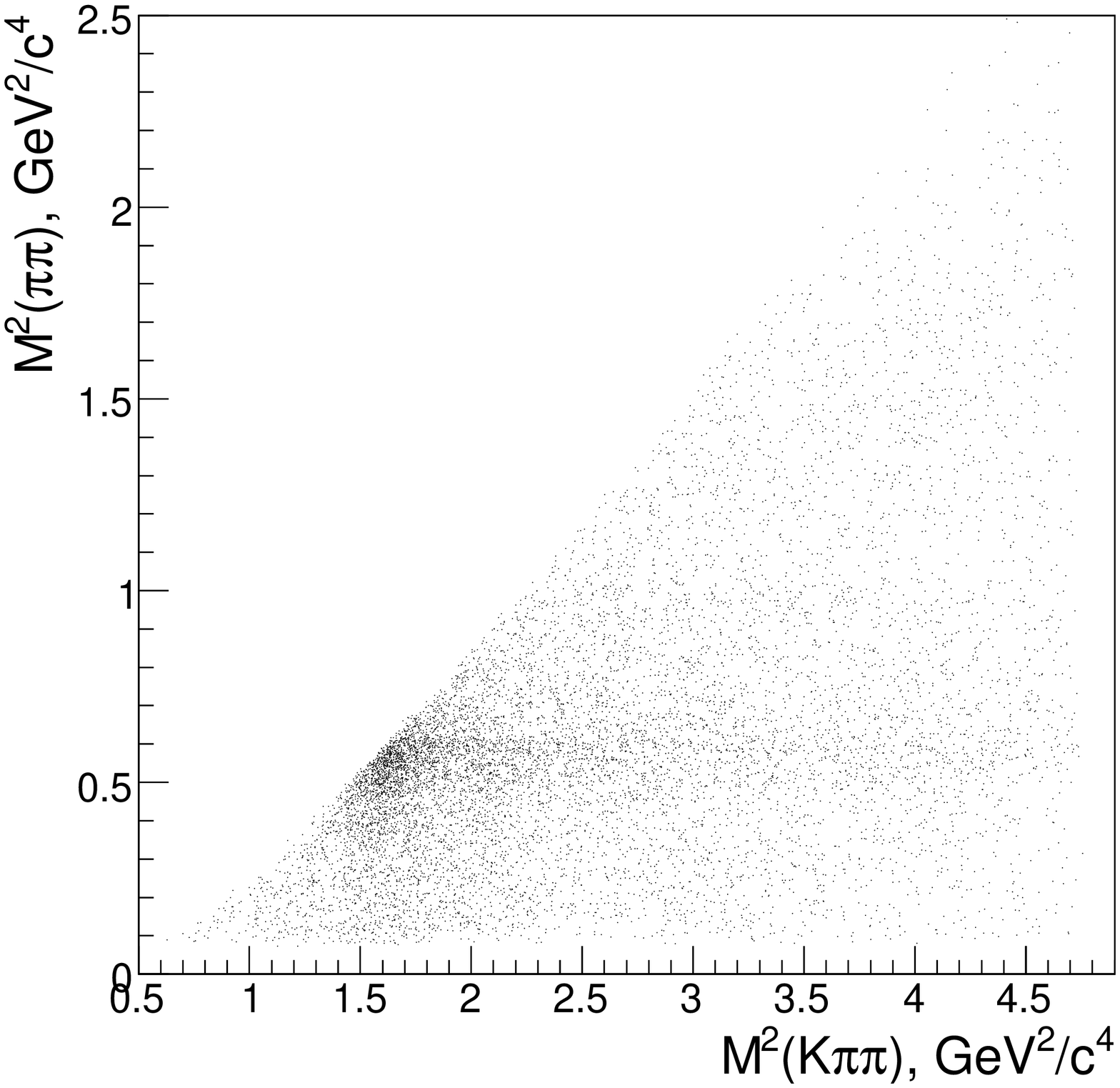}
}}
\hspace{1mm}
\scalebox{0.275}{
\rotatebox{270}{
\includegraphics*[13,29][575,608]
  {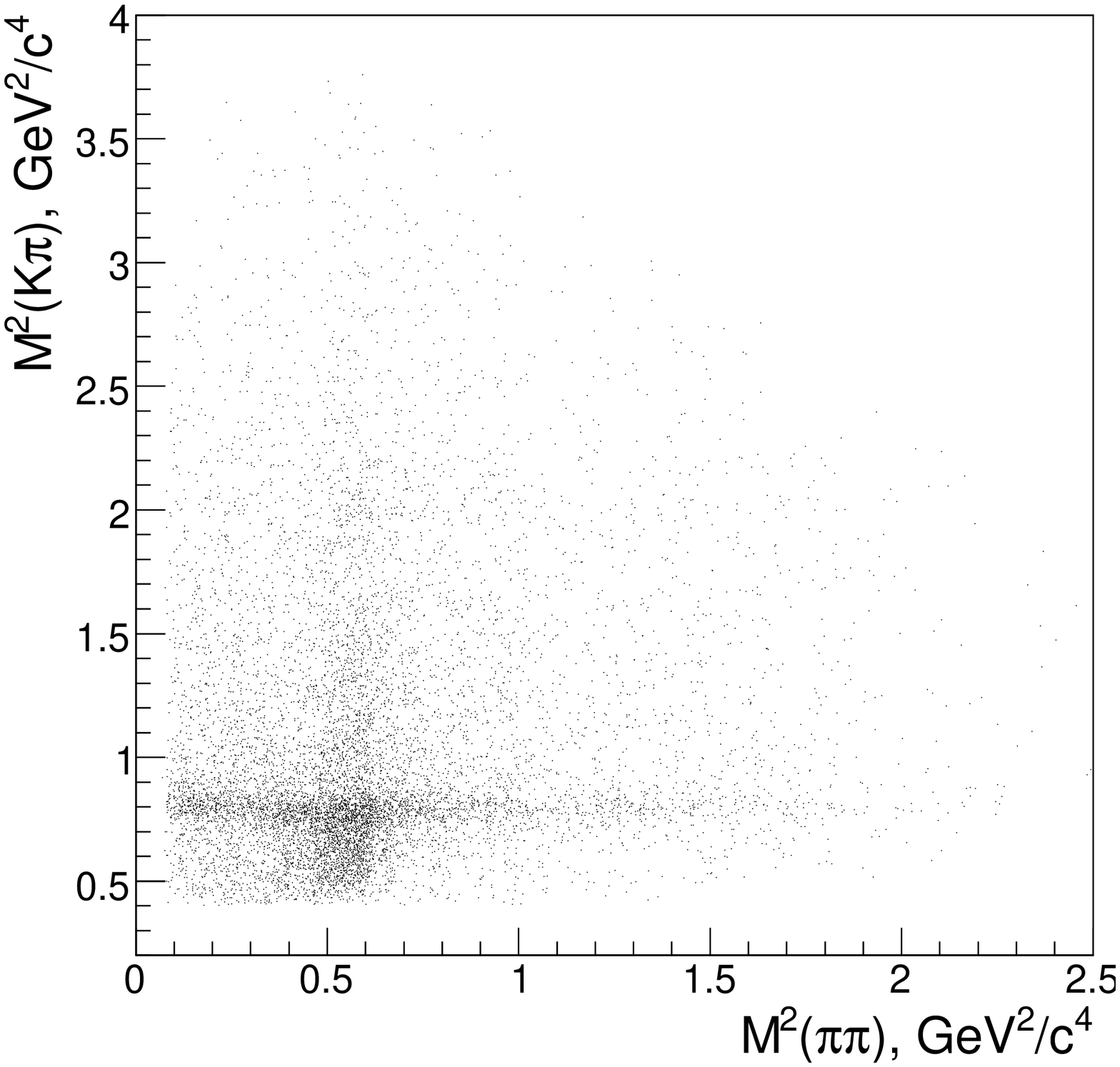}
}}}
\caption{Scatterplots 
  for $B^{+} \rightarrow J/\psi   K^{+} \pi^{+} \pi^{-}$ data,
  showing 
  $M^{2}(K\pi)$   versus $M^{2}(K\pi\pi)$ (left), 
  $M^{2}(\pi\pi)$ versus $M^{2}(K\pi\pi)$ (middle), and
  $M^{2}(K\pi)$   versus $M^{2}(\pi\pi)$  (right).
  Interference between the $K \rho$ and $K^{*}(892) \pi$ submodes 
  of the $K_{1}(1270)$ is responsible for the
  abrupt fading of the $K^{*}(892)$ signal at 
  $M(K\pi) > M_{K^{*}(892)}$.
  The effect is most apparent in the left and right plots.
  Scatterplots 
  for 
  $B^{+} \rightarrow \psi^{\prime} K^{+} \pi^{+} \pi^{-}$ data
  are similar but limited by statistics.}
\label{amplitude:fig_scatterplots}
\end{figure*}

Figure~\ref{amplitude:fig_scatterplots} shows
scatterplots of signal-region 
$B^{+} \rightarrow J/\psi   K^{+} \pi^{+} \pi^{-}$ 
data over the three coordinates.
Interference between
$K_{1}(1270) \rightarrow K^{*}(892) \pi$ and 
$K_{1}(1270) \rightarrow K \rho$ 
is responsible for the weakening of the latter signal 
at $M(K\pi) > M_{K^{*}(892)}$.
Although the four-body phase space
decreases with increasing $M(K\pi)$,
this is not sufficient to account for the abrupt falloff.
To describe the data in this region,
the two modes must be added coherently.

\begin{figure}[hbtp]
\centerline{
\scalebox{0.35}{
\rotatebox{270}{
\includegraphics*[270,37][569,691]
  {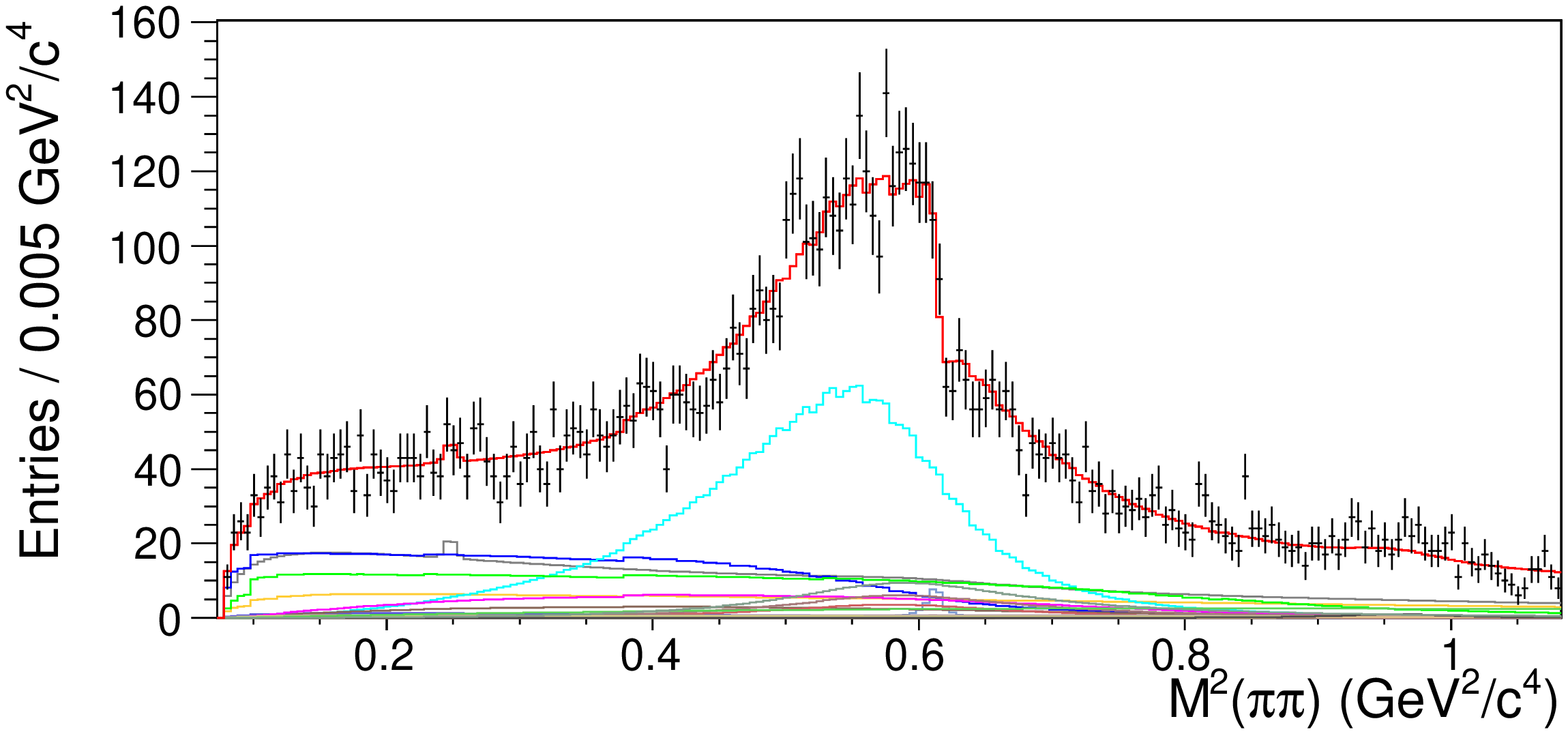}
}}}
\vspace{4mm}
\centerline{
\scalebox{0.35}{
\rotatebox{270}{
\includegraphics*[270,37][569,691]
  {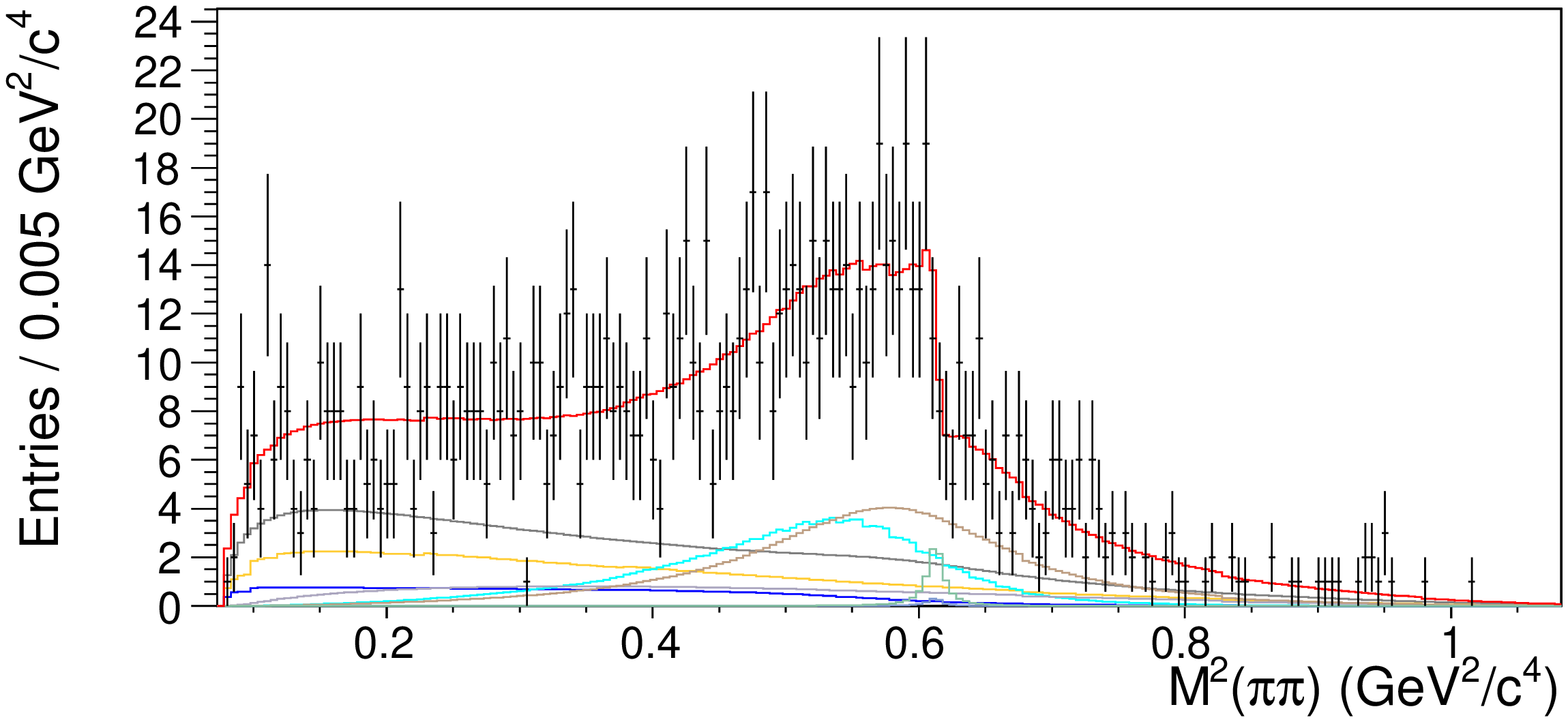}
}}}
\vspace{4mm}
\centerline{
\scalebox{0.35}{
\rotatebox{270}{
\includegraphics*[270,37][569,691]
  {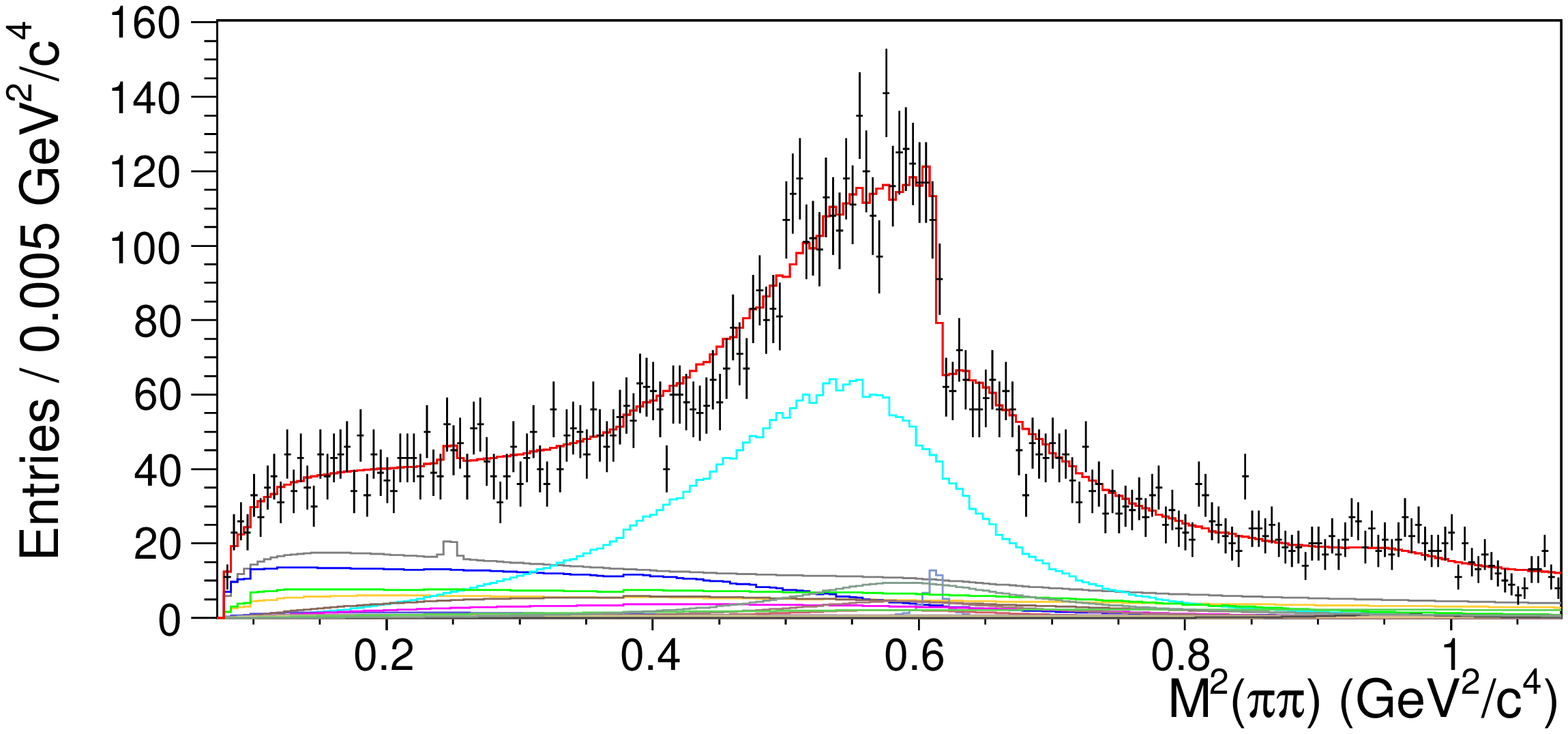}
}}}
\caption{Finely-binned projections 
  onto the $M^{2}(\pi\pi)$ axis
  of
  signal-region data (fits)
  and fit results (histograms) for 
  $B^{+} \rightarrow J/\psi K^{+} \pi^{+} \pi^{-}$ (top),
  $B^{+} \rightarrow \psi^{\prime} K^{+} \pi^{+} \pi^{-}$ (middle),
  and 
  $B^{+} \rightarrow J/\psi K^{+} \pi^{+} \pi^{-}$ with
  the mass and width of the $K_{1}(1270)$ floated (bottom).
  The fit components
  are color coded as shown in Fig.~\ref{amplitude:fig_sidebands}.
  The discontinuity at the $\omega$ mass
  is due to $\rho$-$\omega$ interference.}
\label{amplitude:fig_rho_omega}
\end{figure}

Since the previously-measured~\cite{pdg08} branching fraction for 
$K_{1}(1270) \rightarrow K \omega$ 
is small compared to that for 
$K_{1}(1270) \rightarrow K \rho$,
and since only $1.5\%$ of $\omega$'s decay to 
$\pi^{+}\pi^{-}$,{\footnote{Although $\omega$
decays dominantly to to $\pi^{+}\pi^{-}\pi^{0}$, 
it can also decay to $\pi^{+}\pi^{-}$ 
through $G$-parity violation~\cite{perkins_3},
which causes mixing between $\rho$ and $\omega$.
An $\omega$ component is therefore present 
whenever a particle decays to $\rho$.}}
one might expect 
$K_{1}(1270) \rightarrow K \omega$ 
to play a negligible role in this analysis.
Nonetheless, 
since the $\omega$ is much narrower than the $\rho$,
it significantly distorts the observed $\rho$
line shape through interference~\cite{lafferty:1993}.
In Fig.~\ref{amplitude:fig_rho_omega},
the $M^{2}(\pi\pi)$ projections
of Figs.~\ref{amplitude:fig_signal_pdg_jkpp},
\ref{amplitude:fig_signal_pdg_pkpp}
and \ref{amplitude:fig_signal_float_jkpp}
are finely binned
to demonstrate this interference pattern,
which is accurately modeled by the PDFs.

The peculiar shape
of the observed $\rho$-$\omega$ interference pattern
is caused by
kinematic effects.
The largest contribution to the $\rho$ signal
comes from $K_{1}(1270) \rightarrow K \rho$,
which straddles the edge of phase space,
as can be seen in the middle panel of
Fig.~\ref{amplitude:fig_scatterplots}.
The distortion
that is caused by this kinematic cutoff
is taken into account automatically by integrating
the signal function
only over the kinematically-allowed
region,
as described in
Sec.~\ref{amplitude:section_fitting_technique}.
Modeling the data accurately
requires
including $\rho$-$\omega$ interference,
incorporating the four-body phase space factor into the signal function,
and integrating the signal function over only the kinematically-allowed
phase space.

\subsubsection{The $L$ region}
\label{amplitude:section_l_region}

The $M(K\pi\pi)$ region between $1.5$ and $2.0~{\mathrm{GeV}}/c^{2}$,
historically referred to as the $L$ region,
comprises several wide, 
overlapping resonances~\cite{amaldi:1976,otter:1979,daum:1981}.
The large uncertainties in the 
masses and widths of the known states in this region
make it difficult to characterize 
this region in this analysis.
The model presented here is not necessarily the only one
supported by the data.

To describe the structure observed 
at $2.6~{\mathrm{GeV}}^{2}/c^{4}$ 
in the $M^{2}(K\pi\pi)$ distribution 
of $B^{+} \rightarrow J/\psi K^{+} \pi^{+} \pi^{-}$,
a peak with a mass of $1.605~{\mathrm{GeV}}/c^{2}$ 
and a width of $115~{\mathrm{MeV}}/c^{2}$ is included in the fit,
decaying to $K^{*}(892) \pi$ and $K \rho$.
This peak, which is referred to as $K(1600)$ in this paper, 
may be the $K_{2}(1580)$, 
an as-yet unconfirmed $J^{P} = 2^{-}$ state 
that has previously been observed 
decaying to $K^{*}(892) \pi$~\cite{otter:1979}.

As can be seen in Fig.~\ref{amplitude:fig_signal_pdg_slices_jkpp},
the high end of the
$M^{2}(K\pi\pi)$ spectrum
of $B^{+} \rightarrow J/\psi K^{+} \pi^{+} \pi^{-}$
exhibits $K^{*}$ and $\rho$ signals.
To fit the data in this region, 
we include a $K_{2}^{*}(1980)$ resonance,
which is another state that currently requires confirmation.

Even after including $K(1600)$ and $K_{2}^{*}(1980)$ resonances
in the $B^{+} \rightarrow J/\psi K^{+} \pi^{+} \pi^{-}$ fit,
a slight enhancement remains around
$3~{\mathrm{GeV}}^{2}/c^{4}$ in $M^{2}(K\pi\pi)$.
A $K_{2}(1770)$ signal is therefore also 
included, with its known decays to $K^{*}(892) \pi$, 
$K_{2}^{*}(1430) \pi$, $K f_{0}(980)$, and $K f_{2}(1270)$.

Fitting the $B^{+} \rightarrow \psi^{\prime} K^{+} \pi^{+} \pi^{-}$
data is more difficult still, as there are fewer events to analyze,
and only a small portion of the $L$-region is within the 
kinematic limits of the decay.  
In addition to the $K_{1}(1270)$ signal, the $M^{2}(K\pi\pi)$ spectrum
contains what appears to be the low-mass tail of 
at least one high-mass resonance.
As Fig.~\ref{amplitude:fig_signal_pdg_slices_pkpp} shows, 
there are clear $K^{*}(892)$ and $\rho$ peaks at high $M^{2}(K\pi\pi)$;
these are not reproduced by the PDF if no high-mass resonance is 
included in the model.
If the enhancement is modeled as a single resonance, the data favor 
a mass of roughly $1.7~{\mathrm{GeV}}/c^{2}$ 
and a width of $400$-$500~{\mathrm{MeV}}/c^{2}$.
In this analysis, the enhancement is modeled as the $K^{*}(1680)$.  
The data do not preclude other possibilities,
such as the $K_{2}(1770)$.
Indeed, the hint of $f_{0}(980)$ in the last $M^{2}(K\pi\pi)$ slice
in Fig.~\ref{amplitude:fig_signal_pdg_slices_pkpp}
cannot come from a $1^{-}$ state 
such as the $K^{*}(1680)$, or from a $2^{+}$ state such as the
$K_{2}^{*}(1430)$.

\subsubsection{Comparison with previous measurements}

It is interesting to compare 
the relative decay fractions for $K_{1}(1270)$ submodes
in the 
$B^{+} \rightarrow J/\psi K^{+} \pi^{+} \pi^{-}$ fits
to previous measurements of
$K_{1}(1270)$ branching fractions.
For this purpose, we use the 
decay fractions with phase space 
shown in 
Tables~\ref{amplitude:table_fit_pdg_jkpp} and
\ref{amplitude:table_fit_float_jkpp},
include isospin factors, 
and assume 
branching fractions of 
$(1.53^{+0.11}_{-0.13})\%$ for $\omega \rightarrow \pi^{+} \pi^{-}$,
and 
$(93 \pm 10)\%$ for $K_{0}^{*}(1430) \rightarrow K \pi$~\cite{pdg08}.
The calculation 
neglects the systematic errors in the decay fractions 
and
assumes that the statistical errors among the decay fractions 
are uncorrelated.
Moreover, it assumes
that the $K_{1}(1270)$ decays
only to $K^{*}(892) \pi$, $K \rho$, $K \omega$, and $K_{0}^{*}(1430) \pi$,
and neglects interference among these decay channels.
The comparison is shown in Table~\ref{amplitude:table_k1270_bf}.
While the 
ratios of the $K_{1}(1270)$ branching fractions to
$K^{*}(892) \pi$, $K \rho$, and $K \omega$
are consistent with the previously-measured values,
the branching fraction to $K_{0}^{*}(1430) \pi$ 
is significantly smaller.

\begin{table}[htbp]
\caption{Comparison of branching fractions for 
   $K_{1}(1270)$ decays
   according to the Particle Data Group (PDG)~\cite{pdg08} 
   and based on the results shown in
   Table~\ref{amplitude:table_fit_pdg_jkpp} (fit 1)
   and 
   Table~\ref{amplitude:table_fit_float_jkpp} (fit 2).
   See text for assumptions.}
\label{amplitude:table_k1270_bf}
\renewcommand{\tabcolsep}{2.5mm}
\begin{ruledtabular}
\begin{tabular*}{8.6cm}{l@{\hspace{1.0cm}}r@{$~\pm~$}lr@{$~\pm~$}lr@{$~\pm~$}l} 
\multicolumn{1}{l}{$K_{1}(1270)$}
  & \multicolumn{6}{c}{Branching Fraction (\%)}  \\ 
\multicolumn{1}{l}{Decay mode} 
  & \multicolumn{2}{c}{PDG} 
  & \multicolumn{2}{c}{Fit 1} 
  & \multicolumn{2}{c}{Fit 2} 
\\ \colrule
$K \rho$ &
$42$     &   $6$   &   $57.3$    &   $3.5$        
                   &   $58.4$    &   $4.3$       \\ 
$K_{0}^{*}(1430) \pi$ &
$28$     &   $4$   &    $1.90$   &   $0.66$ 
                   &    $2.01$   &   $0.64$      \\ 
$K^{*}(892) \pi$ &
$16$     &   $5$   &   $26.0$    &   $2.1$       
                   &   $17.1$    &   $2.3$       \\ 
$K \omega$ &
$11$     &   $2$   &   $14.8$    &   $4.7$        
                   &   $22.5$    &   $5.2$       \\ 
$K f_{0}(1370)$ &
 $3$     &   $2$   &   \multicolumn{2}{c}{N/A}   
                  &    \multicolumn{2}{c}{N/A}  \\ 
\end{tabular*}
\end{ruledtabular}
\end{table}

\subsubsection{Mass and width of the $K_{1}(1270)$}

As shown in Sec.~\ref{amplitude:section_results},
the data favor 
a smaller mass and a larger width
for the $K_{1}(1270)$
than the Particle Data Group (PDG) values.
This is mainly due 
to the excess of $K^{*}(892)$ and $\rho$ 
at low $M^{2}(K\pi\pi)$,
as can be ascertained by 
comparing the first row of plots in 
Figs.~\ref{amplitude:fig_signal_pdg_slices_jkpp} 
and \ref{amplitude:fig_signal_float_slices_jkpp}.
The measured mass and width 
agree remarkably well with Ref.~\onlinecite{astier:1969}
and are also consistent with Ref.~\onlinecite{asner:2000}.

\subsubsection{Limitations of the method}

There 
are large uncertainties in the masses and widths of many 
of the states included in the fits,
as can be seen in Table~\ref{amplitude:table_masses_and_widths}.
Although this is taken into account in calculating the systematic error,
it nonetheless limits the accuracy of the model.

In $B^{+} \rightarrow \psi^{\prime} K^{+} \pi^{+} \pi^{-}$,
the small sample size and the kinematic cutoff 
limit the conclusions that can be drawn 
about the signal components.
In $B^{+} \rightarrow J/\psi K^{+} \pi^{+} \pi^{-}$,
the sample size is larger, but
a further limitation is imposed by 
the increase in computation time 
as more parameters are added to the fit.
Each additional decay channel that is included in the signal function
contributes a modulus and possibly a phase 
to be varied in the fit.
Since the normalization integral 
of the signal function in Eq.~\ref{amplitude:eq_pdf} 
depends on the values of the parameters $\vec a$, 
the integration must be performed for each set of parameters 
attempted by the fitter.
While the step size used in the numerical integration can be increased
to speed up the process, it must be small enough to allow the 
PDF to resolve the structures in the data.  
In particular, the $\rho$-$\omega$ interference pattern
can be fitted with a step size of $0.01~{\mathrm{GeV}}^{2}/c^{4}$, 
but not with a step size of $0.02~{\mathrm{GeV}}^{2}/c^{4}$.
As a consequence of the finite 
processor speed,
not every possible decay channel can be included in the fit.
The model is necessarily incomplete. 

The large nonresonant component seen in both 
$B^{+} \rightarrow J/\psi K^{+} \pi^{+} \pi^{-}$
and
$B^{+} \rightarrow \psi^{\prime} K^{+} \pi^{+} \pi^{-}$
may be an indication 
of contributions from additional wide kaon excitations.
It may also incorporate some misreconstructed resonant signal.
While the nonresonant component is assumed in this analysis to be
distributed according to phase space,
this assumption may be inaccurate.  There are currently no accepted 
models of nonresonant $B$-meson decays.

It is difficult,
in an analysis like the one presented here,
to determine the significance of a given
component of the signal.
An improvement in the likelihood 
upon the addition of a new resonance
to the signal function
indicates 
only that the model is incomplete,
not necessarily that
the data contain the particular resonance.
Furthermore, unless the model is accurate in every other way, 
floating the mass and width of a 
particle in the fit
may not yield a reliable result,
as the fitter may set these parameters
to compensate for the model's deficiencies.
This is especially important in the high-$M^{2}(K\pi\pi)$ region, 
where the statistics are limited 
and 
there are large uncertainties 
in the masses and widths of the 
resonances included in the signal function.
Thus, although the 
$K(1600)$ component of the signal function for
$B^{+} \rightarrow J/\psi K^{+} \pi^{+} \pi^{-}$ 
greatly improves the quality of the fit,
it is difficult to claim that it is a single particle,
let alone measure its mass and width.

%%%%% Conclusion

\section{Conclusions}
\label{section_conclusion}

Using data recorded by the Belle detector,
we have measured branching fractions for the decays
$B^{+} \rightarrow J/\psi   K^{+} \pi^{+} \pi^{-}$ and
$B^{+} \rightarrow \psi^{\prime} K^{+} \pi^{+} \pi^{-}$ 
with improved precision
(see Sec.~\ref{inclusive:section_results}).
We have also performed amplitude analyses
in three dimensions---$M^{2}(K\pi\pi)$, 
$M^{2}(K\pi)$, and $M^{2}(\pi\pi)$---to 
determine the resonant structure of the
$K^{+} \pi^{+} \pi^{-}$ final state in these decays
(see Sec.~\ref{amplitude:section_results}).

We have shown that 
the $K_{1}(1270)$, 
which is the dominant component of
the $K^{+} \pi^{+} \pi^{-}$ final state 
in
$B^{+} \rightarrow J/\psi   K^{+} \pi^{+} \pi^{-}$,
is also prominent in
$B^{+} \rightarrow \psi^{\prime} K^{+} \pi^{+} \pi^{-}$.
The large sample available for the former decay
reveals a small peak at
$M(K\pi\pi) \approx 1.4~{\mathrm{GeV}}/c^{2}$.
Our three-dimensional fits represent
a first attempt to determine the components of this peak.

Performing an unbinned fit in three dimensions
exploits practically all of the information available in the data.  
While it is relatively easy to obtain a good fit in one dimension,
requiring a fit that succeeds in three dimensions
greatly restricts the class of successful models.
With high statistics, it is possible to use interference effects 
and the spin-dependent angular distribution of the final state
to distinguish 
overlapping resonances.
In particular, we have shown that $\rho$-$\omega$ interference 
cannot be neglected in studying 
$K_{1}(1270)$ decays to $K \pi^{+} \pi^{-}$ final states.

The large size of the
$B^{+} \rightarrow J/\psi   K^{+} \pi^{+} \pi^{-}$ data sample
allows us to measure
the mass and width of the $K_{1}(1270)$
with improved precision 
(see Eqs.~\ref{amplitude:eq_k1270_mass} and
\ref{amplitude:eq_k1270_width}).
These values differ considerably from previously-published
values~\cite{pdg08}.
The analysis of these data
also provides information on
the relative strengths of $K_{1}(1270)$ decays
to $K \rho$, $K \omega$, $K^{*}(892) \pi$, and $K_{0}^{*}(1430) \pi$ 
final states (see Table~\ref{amplitude:table_k1270_bf}). 
While the results 
are consistent with previous measurements for the first three modes, 
they indicate a much smaller rate of decay to $K_{0}^{*}(1430) \pi$
than previously accepted.

Although more data 
are required to clarify the structure of the
high $M^{2}(K^{+}\pi^{+} \pi^{-})$ region
in both  
$B^{+} \rightarrow J/\psi K^{+} \pi^{+} \pi^{-}$ and
$B^{+} \rightarrow \psi^{\prime} K^{+} \pi^{+} \pi^{-}$,
we have shown that this region contains broad resonances
that decay to $K^{*}(892) \pi$ and $K \rho$ final states.

The analysis presented in this paper demonstrates that the decay modes
$B^{+} \rightarrow J/\psi   K^{+} \pi^{+} \pi^{-}$ and
$B^{+} \rightarrow \psi^{\prime} K^{+} \pi^{+} \pi^{-}$ 
can 
provide clean laboratories 
for the spectroscopy of excited kaon states.
Many of these states still require confirmation or
more precise mass and width measurements.
As more data become available at future super-$B$ factories, analyses
similar to the one presented here can 
further elucidate the higher regions of the kaon spectrum.

%%%%% Acknowledgements

\begin{acknowledgments}
We thank the KEKB group for the excellent operation of the
accelerator, the KEK cryogenics group for the efficient
operation of the solenoid, and the KEK computer group and
the National Institute of Informatics for valuable computing
and SINET3 network support.  We acknowledge support from
the Ministry of Education, Culture, Sports, Science, and
Technology (MEXT) of Japan, the Japan Society for the 
Promotion of Science (JSPS), and the Tau-Lepton Physics 
Research Center of Nagoya University; 
the Australian Research Council and the Australian 
Department of Industry, Innovation, Science and Research;
the National Natural Science Foundation of China under
Contract No.~10575109, 10775142, 10875115 and 10825524; 
the Ministry of Education, Youth and Sports of the Czech 
Republic under Contract No.~LA10033 and MSM0021620859;
the Department of Science and Technology of India; 
the BK21 and WCU program of the Ministry Education Science and
Technology, National Research Foundation of Korea,
and NSDC of the Korea Institute of Science and Technology Information;
the Polish Ministry of Science and Higher Education;
the Ministry of Education and Science of the Russian
Federation and the Russian Federal Agency for Atomic Energy;
the Slovenian Research Agency;  the Swiss
National Science Foundation; the National Science Council
and the Ministry of Education of Taiwan; and the U.S.\
Department of Energy.
This work is supported by a Grant-in-Aid from MEXT for 
Science Research in a Priority Area (``New Development of 
Flavor Physics''), and from JSPS for Creative Scientific 
Research (``Evolution of Tau-lepton Physics'').
\end{acknowledgments}

%%%%% Appendix

\appendix

\section*{Appendix: local maxima}
\label{appendix_local_maxima}

This appendix summarizes
the results of the 
local-maximum test,
in which each of the three signal-region fits was repeated
$100$ times 
with randomly selected starting values for the parameters.
In each case,
the best likelihood obtained
coincided with the 
solution presented in 
Tables~\ref{amplitude:table_fit_pdg_jkpp}-\ref{amplitude:table_fit_float_jkpp}.
In the following, these solutions are 
referred to as the ``global maxima.''

For the $B^{+} \rightarrow J/\psi K^{+} \pi^{+} \pi^{-}$ fit
with the $K_{1}(1270)$ mass and width fixed to their
values in Table~\ref{amplitude:table_masses_and_widths},
the local maximum closest to the global maximum presented in 
Table~\ref{amplitude:table_fit_pdg_jkpp}
had a likelihood of $-10608.6$,
which is
$8.2\sigma$ away from the global maximum.

For the $B^{+} \rightarrow \psi^{\prime} K^{+} \pi^{+} \pi^{-}$ fit,
two local maxima were found:
one with a likelihood of $635.4$
and the other with a likelihood of $635.9$;
these are
$2.4\sigma$ and $2.2\sigma$ away from the global maximum,
respectively.
The former had an unphysically large decay fraction for
$K_{1}(1270) \rightarrow K \omega$ and was discarded.
For the latter, all the parameters were within statistical error 
of the values presented in 
Table~\ref{amplitude:table_fit_pdg_pkpp},
with the exception of the 
$K_{1}(1270) \rightarrow K^{*}(892) \pi$ amplitude and decay fraction,
which were higher by $1.4$ times the statistical error.

For the $B^{+} \rightarrow J/\psi K^{+} \pi^{+} \pi^{-}$ fit
with the $K_{1}(1270)$ mass and width allowed to float,
two local maxima were found,
both with a likelihood of
$-10528.8$,
which is 
$2.7\sigma$ away from the global maximum.
The fitted parameters and decay fractions
for these local maxima are presented in
Tables~\ref{appendix:table_fit_float_jkpp_lm1}
and 
\ref{appendix:table_fit_float_jkpp_lm2}, respectively.
The fitted mass and width are
\begin{eqnarray}
M_{K_{1}(1270)}      
 & = & (1241.9 \pm 3.2)~{\mathrm{MeV}}/c^{2}, 
 \nonumber\\
\Gamma_{K_{1}(1270)} 
 & = & (128.3 \pm 5.8)~{\mathrm{MeV}}/c^{2}
 \nonumber
\end{eqnarray}
for the former, and
\begin{eqnarray}
M_{K_{1}(1270)}      
 & = & (1244.3 \pm 3.3)~{\mathrm{MeV}}/c^{2}, 
 \nonumber\\
\Gamma_{K_{1}(1270)} 
 & = & (129.0 \pm 5.7)~{\mathrm{MeV}}/c^{2} 
 \nonumber
\end{eqnarray}
for the latter.

\begin{table*}[htbp]
\caption{Fitted parameters 
   and decay fractions 
   corresponding to
   the first local maximum
   for the mode
   $B^{+} \rightarrow J/\psi K^{+} \pi^{+} \pi^{-}$,
   with the $K_{1}(1270)$ mass and width floated.
   The errors are statistical.}
\label{appendix:table_fit_float_jkpp_lm1}
\renewcommand{\arraystretch}{1.1}
\begin{ruledtabular}
\begin{tabular*}{17.8cm}
  {@{\hspace{0.7cm}}c@{\hspace{0.7cm}}l@{\hspace{1.6cm}}l@{$\,\pm\,$}l@{\hspace{1.6cm}}r@{$\,\pm\,$}l@{\hspace{1.6cm}}l@{$~\pm~$}l}
$J_{1}$
  & Submode
    & \multicolumn{2}{c}{Modulus}         \hspace{1.6cm} 
    & \multicolumn{2}{c}{Phase (radians)} \hspace{1.6cm}
    & \multicolumn{2}{c}{Decay Fraction}  
  \\ 
\colrule
  & Nonresonant $K^{+}\pi^{+}\pi^{-}$
    & \multicolumn{2}{c}{$1.0$ (fixed)}   \hspace{1.6cm} 
    & \multicolumn{2}{c}{$0$ (fixed)}     \hspace{1.6cm}
    & $0.139$   &  $0.015$ 
  \\ 
\colrule
  & $K_{1}(1270) \rightarrow K^{*}(892) \pi$ 
    &  $0.662$  &  $0.056$ 
    & \multicolumn{2}{c}{$0$ (fixed)}     \hspace{1.6cm}
    & $0.090$   &  $0.012$
  \\ 
  & $K_{1}(1270) \rightarrow K \rho$ 
    &  $2.22$   &  $0.13$
    & $-0.96$   &  $0.13$
    & $0.438$   &  $0.021$
  \\ 
$1^{+}$
  & $K_{1}(1270) \rightarrow K \omega$ 
    &  $0.301$  &  $0.043$
    &  $0.94$   &  $0.18$ 
    &  $0.0078$ &  $0.0022$
  \\ 
  & $K_{1}(1270) \rightarrow K_{0}^{*}(1430) \pi$ 
    &  $0.88$   &  $0.18$  
    &  $2.39$   &  $0.27$ 
    & $0.0116$  &  $0.0047$
  \\ 
  & $K_{1}(1400) \rightarrow K^{*}(892) \pi$ 
    &  $0.258$  &  $0.083$
    &  $2.35$   &  $0.43$
    &  $0.017$  &  $0.011$
  \\ 
\colrule
$1^{-}$
  & $K^{*}(1410) \rightarrow K^{*}(892) \pi$ 
    &  $0.755$  &  $0.099$ 
    & \multicolumn{2}{c}{$0$ (fixed)}     \hspace{1.6cm}
    & $0.091$   &  $0.023$
  \\ 
\colrule
  & $K_{2}^{*}(1430) \rightarrow K^{*}(892) \pi$ 
    &  $0.384$  &  $0.081$
    & \multicolumn{2}{c}{$0$ (fixed)}     \hspace{1.6cm}
    &  $0.027$  &  $0.012$
  \\ 
  & $K_{2}^{*}(1430) \rightarrow K \rho$ 
    & \multicolumn{2}{c}{$0.214$ (fixed)} \hspace{1.6cm}
    &  $2.63$   &  $0.60$ 
    & \multicolumn{2}{c}{$0.0071$ (fixed)}
  \\ 
$2^{+}$
  & $K_{2}^{*}(1430) \rightarrow K \omega$ 
    & \multicolumn{2}{c}{$0.023$ (fixed)} \hspace{1.6cm}
    & $-2.6$    &  $1.1$ 
    & \multicolumn{2}{c}{$0.00011$ (fixed)}
  \\
  & $K_{2}^{*}(1980) \rightarrow K^{*}(892) \pi$ 
    &  $0.659$  &  $0.059$
    &  $0.42$   &  $0.48$ 
    & $0.0487$  &  $0.0064$
  \\ 
  & $K_{2}^{*}(1980) \rightarrow K \rho$ 
    &  $0.733$  &  $0.057$
    &  $2.92$   &  $0.40$
    &  $0.0689$ &  $0.0071$
  \\ 
\colrule
  & $K(1600) \rightarrow K^{*}(892) \pi$ 
    &  $0.175$  &  $0.066$
    & \multicolumn{2}{c}{$0$ (fixed)}     \hspace{1.6cm}
    & $0.031$   &  $0.021$
  \\ 
  & $K(1600) \rightarrow K \rho$ 
    &  $0.169$  &  $0.020$
    &  $0.61$   &  $0.50$
    &  $0.0297$ &  $0.0070$
  \\ 
  & $K_{2}(1770) \rightarrow K^{*}(892) \pi$ 
    &  $0.138$  &  $0.066$
    &  $2.9$    &  $1.3$
    &  $0.019$  &  $0.017$
  \\ 
\raisebox{1.5ex}[0pt]{$2^{-}$}
  & $K_{2}(1770) \rightarrow K_{2}^{*}(1430) \pi$ 
    &  $0.305$  &  $0.055$
    &  $2.55$   &  $0.68$
    &  $0.0104$ &  $0.0032$
  \\ 
  & $K_{2}(1770) \rightarrow K f_{2}(1270)$ 
    &  $0.515$  &  $0.090$
    &  $2.81$   &  $0.79$
    & $0.0153$  &  $0.0045$
  \\ 
  & $K_{2}(1770) \rightarrow K f_{0}(980)$ 
    &  $0.121$  &  $0.031$
    &  $2.54$   &  $0.69$
    &  $0.0036$ &  $0.0019$
  \\ 
\end{tabular*}
\end{ruledtabular}
\end{table*}

\begin{table*}[htbp]
\caption{Fitted parameters 
   and decay fractions 
   corresponding to
   the second local maximum
   for the mode
   $B^{+} \rightarrow J/\psi K^{+} \pi^{+} \pi^{-}$,
   with the $K_{1}(1270)$ mass and width floated.
   The errors are statistical.}
\label{appendix:table_fit_float_jkpp_lm2}
\renewcommand{\arraystretch}{1.1}
\begin{ruledtabular}
\begin{tabular*}{17.8cm}
  {@{\hspace{0.7cm}}c@{\hspace{0.7cm}}l@{\hspace{1.6cm}}l@{$\,\pm\,$}l@{\hspace{1.6cm}}r@{$\,\pm\,$}l@{\hspace{1.6cm}}l@{$~\pm~$}l}
$J_{1}$
  & Submode
    & \multicolumn{2}{c}{Modulus}         \hspace{1.6cm} 
    & \multicolumn{2}{c}{Phase (radians)} \hspace{1.6cm}
    & \multicolumn{2}{c}{Decay Fraction}  
  \\ 
\colrule
  & Nonresonant $K^{+}\pi^{+}\pi^{-}$
    & \multicolumn{2}{c}{$1.0$ (fixed)}   \hspace{1.6cm} 
    & \multicolumn{2}{c}{$0$ (fixed)}     \hspace{1.6cm}
    & $0.136$   &  $0.013$ 
  \\ 
\colrule
  & $K_{1}(1270) \rightarrow K^{*}(892) \pi$ 
    &  $0.701$  &  $0.060$ 
    & \multicolumn{2}{c}{$0$ (fixed)}     \hspace{1.6cm}
    & $0.100$   &  $0.013$
  \\ 
  & $K_{1}(1270) \rightarrow K \rho$ 
    &  $2.24$   &  $0.13$
    & $-1.00$   &  $0.10$
    & $0.447$   &  $0.018$
  \\ 
$1^{+}$
  & $K_{1}(1270) \rightarrow K \omega$ 
    &  $0.293$  &  $0.043$
    &  $0.91$   &  $0.16$ 
    &  $0.0075$ &  $0.0020$
  \\ 
  & $K_{1}(1270) \rightarrow K_{0}^{*}(1430) \pi$ 
    &  $0.91$   &  $0.18$  
    &  $2.40$   &  $0.25$ 
    &  $0.0123$ &  $0.0045$
  \\ 
  & $K_{1}(1400) \rightarrow K^{*}(892) \pi$ 
    &  $0.162$  &  $0.076$
    &  $2.53$   &  $0.45$
    &  $0.0066$ &  $0.0060$
  \\ 
\colrule
$1^{-}$
  & $K^{*}(1410) \rightarrow K^{*}(892) \pi$ 
    &  $0.739$  &  $0.093$ 
    & \multicolumn{2}{c}{$0$ (fixed)}     \hspace{1.6cm}
    & $0.086$   &  $0.020$
  \\ 
\colrule
  & $K_{2}^{*}(1430) \rightarrow K^{*}(892) \pi$ 
    &  $0.687$  &  $0.082$
    & \multicolumn{2}{c}{$0$ (fixed)}     \hspace{1.6cm}
    &  $0.085$  &  $0.018$
  \\ 
  & $K_{2}^{*}(1430) \rightarrow K \rho$ 
    & \multicolumn{2}{c}{$0.384$ (fixed)} \hspace{1.6cm}
    &  $3.01$   &  $0.27$ 
    & \multicolumn{2}{c}{$0.022$ (fixed)}
  \\ 
$2^{+}$
  & $K_{2}^{*}(1430) \rightarrow K \omega$ 
    & \multicolumn{2}{c}{$0.041$ (fixed)} \hspace{1.6cm}
    & $-1.4$   &  $1.2$ 
    & \multicolumn{2}{c}{$0.00034$ (fixed)}
  \\
  & $K_{2}^{*}(1980) \rightarrow K^{*}(892) \pi$ 
    &  $0.814$  &  $0.063$
    & $-1.52$   &  $0.19$ 
    & $0.0731$  &  $0.0077$
  \\ 
  & $K_{2}^{*}(1980) \rightarrow K \rho$ 
    &  $0.791$  &  $0.059$
    &  $0.81$   &  $0.25$
    &  $0.0788$ &  $0.0070$
  \\ 
\colrule
  & $K(1600) \rightarrow K^{*}(892) \pi$ 
    &  $0.253$  &  $0.043$
    & \multicolumn{2}{c}{$0$ (fixed)}     \hspace{1.6cm}
    & $0.063$   &  $0.019$
  \\ 
  & $K(1600) \rightarrow K \rho$ 
    &  $0.152$  &  $0.021$
    &  $0.86$   &  $0.29$
    &  $0.0236$ &  $0.0062$
  \\ 
  & $K_{2}(1770) \rightarrow K^{*}(892) \pi$ 
    &  $0.221$  &  $0.050$
    & $-2.56$   &  $0.26$
    &  $0.048$  &  $0.020$
  \\ 
\raisebox{1.5ex}[0pt]{$2^{-}$}
  & $K_{2}(1770) \rightarrow K_{2}^{*}(1430) \pi$ 
    &  $0.326$  &  $0.048$
    &  $2.98$   &  $0.40$
    &  $0.0117$ &  $0.0030$
  \\ 
  & $K_{2}(1770) \rightarrow K f_{2}(1270)$ 
    &  $0.541$  &  $0.073$
    & $-2.89$   &  $0.34$
    & $0.0166$  &  $0.0037$
  \\ 
  & $K_{2}(1770) \rightarrow K f_{0}(980)$ 
    &  $0.121$  &  $0.032$
    &  $2.98$   &  $0.48$
    &  $0.0035$ &  $0.0018$
  \\ 
\end{tabular*}
\end{ruledtabular}
\end{table*}

%%%%% Bibliography

\end{document}